\documentclass{article}
\usepackage[T1]{fontenc}

\usepackage[numbers]{natbib}
\usepackage{chapterbib}
\usepackage{etex}

\usepackage{graphicx}
\usepackage{subfig}
\usepackage[tableposition=top,margin=0.33cm]{caption}
\usepackage{color,colortbl}
\usepackage{multirow}
\usepackage{multicol}
\usepackage{float}
\usepackage{xr}
\usepackage{makecell}
\usepackage{tikz}
\usepackage{dingbat}
\usepackage{listings}

\usepackage{wrapfig}

\usepackage{pdflscape}

\usepackage{booktabs}
\usepackage{array}
\usepackage{multicol}

\newcolumntype{P}[1]{>{\centering\arraybackslash}p{#1}}
\newcolumntype{C}[1]{>{\centering\let\newline\\\arraybackslash\hspace{0pt}}m{#1}}
\newcolumntype{M}[1]{>{\centering\arraybackslash}m{#1}}
\newcolumntype{L}[1]{>{\arraybackslash}m{#1}}

\definecolor{Gray}{gray}{0.9}

\def\pagewidth{170mm}
\def\pageheight{257mm}

\usepackage{geometry}
\title{Quantifying the Impact of Energy System Model Resolution on Siting, Cost, Reliability, and Emissions}
\author{Anna F. Jacobson$^1$\\Denise L. Mauzerall$^{2,3}$\\Jesse D. Jenkins$^{4,5}$}

\graphicspath{{Figures/}} 

\begin{document}

%
%
\newcommand\coc{carbon penalty~}
\newcommand\Coc{Carbon penalty~}
\newcommand\Refc{Business-as-usual~}
\newcommand\refc{business-as-usual~}
\newcommand\refca{BAU~}
\newcommand\coca{CP~}
\newcommand\opone{capacity expansion}
\newcommand\optwo{operations}
\newcommand\Opone{Capacity expansion}
\newcommand\Optwo{Operations}

\maketitle
\begin{center}
    {\small
    1. Lewis Sigler Institute for Integrative Genomics, Princeton University \\
    2. Center for Policy Research on Energy and the Environment, Princeton University \\
    3. Department of Civil and Environemntal Engineering, Princeton University \\
    4. Department of Mechanical and Aerospace Engineering, Princeton University \\
    5. Andlinger Center for Energy and the Environment, Princeton University}
\end{center}

\section*{Abstract}
Energy systems models, critical for power sector decision support, incur non-linear memory and runtime penalties when scaling up under typical formulations. Even hardware improvements cannot make large models tractable, requiring omission of detail which affects siting, cost, and emission outputs to an unknown degree. Recent algorithmic innovations have enabled large scale, high resolution modeling. Newly tractable, granular systems can be compared with coarse ones for better understanding of inaccuracies from low resolution. Here we use a state of the art model to quantify the impact of resolution on results salient to policymakers and planners, affording confidence in decision quality. We find more realistic siting in recommendations from high resolution energy systems models, improving emissions, reliability, and price outcomes. Errors are generally stronger from low spatial resolution. When models have low resolution in multiple dimensions, errors are introduced by the coarser of temporal or spatial resolution. We see no diminishing returns in accuracy for several key metrics when increasing resolution. We recommend using computationally efficient techniques to maximize granularity and allocating resolution without leaving any aspect (spatial, temporal, operational) of systems unduly coarse.

\subsection*{Key Words}
Energy systems models, resolution, structural uncertainty, error, capacity expansion, benchmarks

\begin{table}[H]
    \centering \small
    \setlength\tabcolsep{0pt}
    \begin{tabular}{L{0.1\linewidth}L{0.4\linewidth}L{0.1\linewidth}L{0.4\linewidth}}
        
        \toprule
        Acronym & Definition & Acronym & Definition \\
        \midrule

        BAU & Business-as-Usual &
        CONUS & Continental United States \\
        CP & Carbon Penalty &
        ESM & Energy Systems Model \\
        HR & High Resolution &
        HRB & High Resolution Baseline \\
        LCOE & Levelized Cost of Electricity &
        LP & Linear Programming \\
        MILP & Mixed-Integer Linear Programming &
        MSE & Mean Squared Error \\
        NG & Natural Gas &
        NSE & Nonserved Energy \\
        SCO & Site Capacity Overlap &
        SI & Supplemental Information \\
        UC & Unit Commitment &
        VRE & Variable Reneawable Energy \\

        \bottomrule

    \end{tabular}

    \caption{Relevant acronyms}
       
\end{table}

\section{Introduction}
\label{introduction}

To address climate change, we must decarbonize the electricity sector while meeting increasing demands due to electrification in transportation, building heating and cooling, industrial processes, and fuels\cite{nature_decisive}. Energy systems models (ESMs) explore capacity siting, reliability, costs, curtailment, and more. Accurate modeling is critical in supporting consistent, affordable, clean electricity\cite{brown2018response,pfenninger2014energy}.

Most ESMs are formulated as linear programming problems (LP) or mixed-integer linear programming problems (MILP)\cite{cho2022,ringkjob2018review}. Because runtime and memory usage of LP scales quadratically with model size (MILP even worse,) many high resolution ESMs are intractable\cite{frew2016temporal, jacobson2023benders}, even on modern hardware. To ease burden, researchers often decrease model size by looking at operations in fewer representative hours, aggregating geographic regions, or ignoring physical aspects of systems. Structural uncertainties, errors from inaccuracies in how models represent the world\cite{bojke2009characterizing}, inherently result from omitted information.

Decreasing temporal resolution makes it difficult for models to capture the full covariance of weather patterns and their impacts on wind, solar power, and demand. Ignoring physical characteristics of systems, like ramping limits on thermal powerplants, omits key operational constraints and allows models to operate with more flexibility than is realistic. Decreasing spatial resolution both makes it difficult to fully capture weather patterns and blinds the planning model to potentially binding transmission constraints that impact operations. In other words, structural uncertainties arise from two primary sources: operational and spatial coarseness may result in unexpected (omitted) operational constraints when systems are implemented, while temporal and spatial coarseness may fail to capture the full temporal and spatial covariance structure of weather-dependent timeseries data.

Energy sector investments and their climate benefits are in the hundreds of billions of dollars\cite{bistline2023emissions}, making errors in recommendations to decisionmakers costly. Low spatial\cite{balta2015spatially,brinkerink2024role,frysztacki2021strong,frysztacki2023inverse} and temporal\cite{frew2016temporal,jacobson2023benders} granularity are shown to impact system cost, reliability, investments, and more. Biases may be more likely in systems with high penetrations of variable renewable energy (VRE)\cite{aryanpur2021review} or undergoing transmission expansion\cite{frysztacki2023inverse}. Mid-transition systems contain a non-optimized mix of older thermal units and early stage VRE\cite{grubert2022designing}; due to resources' diverse modeling demands, these systems may be at high risk for error.

Ramifications of ESM inaccuracies can be both consequential and unpredictable. A model that overestimates wind availability, for example, may recommend suboptimally low onshore wind buildout by underestimating capacity required to meet demand. Alternatively, overestimating production could lead to overvaluation of turbines and excessive buildout. Biases impacting wind may lead to difficulties optimizing storage capacity, thermal retirement, nonserved energy (NSE), and more. Because output is an emergent property of inputs and problem structure, analysis of input variations cannot be extrapolated to strength or direction of bias on outputs\cite{frew2016temporal, jacobson2023benders} including capacity, cost, reliability, and emissions. Planners may have difficulty justifying decisions should models be placed under scrutiny, or should different models' uncertainties lead to conflicting recommendations.

Modern algorithmic innovations have improved ESM efficiency\cite{jacobson2023benders,lara2018deterministic,li2022mixed,lohmann2017tailored}, enabling high-resolution modeling. Our recent ESM used Benders Decomposition for linear scaling of runtime and memory with temporal resolution\cite{jacobson2023benders,pecci2024regularized} unlocking high resolution benchmark (HRB) systems to compare with coarse models. In absence of other methods of quantifying output error\cite{frew2016temporal}, HRBs are a prime means of exploring model configurations' impacts on policy-relevant outcomes.

Here we use our state of the art model\cite{jacobson2023benders,pecci2024regularized} to create HRBs to explore the impact of granularity on costs, emissions, reliability, and capacity. We explore the synergistic impacts of multiple forms of resolution: spatial, temporal, and operational. To our knowledge, this study is the first to probe three dimensions of resolution; leverage decomposition for an HRB; examine accuracy of resource siting within regions; and use a two-phase methodology, described below, to examine operations in high resolution. Our results guide where modelers should allocate computational resources for maximal accuracy. We demonstrate the utility of recent advances for ESMs. We see no diminishing returns for many metrics in improvement of accuracy with resolution, and find models are inhibited by their coarser dimension between spatial or temporal. We recommend modelers leverage state of the art algorithms to maximize granularity and subsequently allocate computational resources without neglecting any aspect (spatial, temporal, operational) of resolution.

\section{Methodology}

\begin{table}[ht]
    \centering
    \begin{tabular}{C{0.05\linewidth}C{0.1\linewidth}C{0.075\linewidth}L{0.55\linewidth}C{0.15\linewidth}}

        \toprule
        Step & & Section & Description & Output \\
        \midrule

        1 & \includegraphics[width=\linewidth]{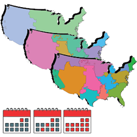} &
        \ref{systems_tested} &
        Use PowerGenome\cite{schivley2021powergenome} to create 24 cases of GenX input with varying levels of spatial, temporal, and operational resolution for a \refc and \coc case, Table~\ref{cases_tested}. & 
        GenX inputs, varying resolution \\

        \midrule

        2 & \includegraphics[width=\linewidth]{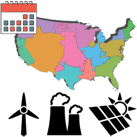} &
        \ref{genx_inv} &
        Run the GenX\cite{genX_github} capacity expansion model to optimize each case. Discard operations and return optimal investments for generators, storage, and transmission. &
        GenX outputs, varying resolution \\
        
        \midrule

        3 & \includegraphics[width=\linewidth]{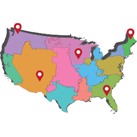} &
        \ref{sites_downscaling} &
        Translate investments of lower resolution cases into individual investment sites and transmission lines within the United States. &
        Individual resource sites \\

        \midrule

        4 & \includegraphics[width=\linewidth]{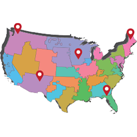} &
        \ref{sites_upscaling} &
        Turn investment sites into high resolution (HR) GenX inputs, generating 48 directly comparable investment portfolios with the same numbers of resources and identical load curves and VRE profiles. &
        GenX inputs, HR\\

        \midrule

        5 & \includegraphics[width=\linewidth]{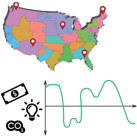} &
        \ref{genx_op} &
        Optimize operations on the HR investment portfolios derived above. &
        GenX outputs, HR \\

        \midrule

        6 & \includegraphics[width=\linewidth]{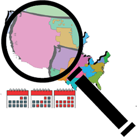} &
        \ref{results} &
        Analyze the impact and wider implications of resolution; how do results differ case by case? &
        Results and discussion \\

        \bottomrule

    \end{tabular}
    {\captionof{figure}{Graphical methodology. High resolution (HR) cases are run at 26-zone, 52-week resolution with unit commitment (UC) constraints active.\label{methods_steps}}}
    
\end{table}

We use GenX\cite{genX_github}, a detailed open-source ESM, to examine the impact of spatial, temporal, and operational resolution on outputs. We use PowerGenome\cite{schivley2021powergenome}, a case generation software, to create 24 continental United States (CONUS) power systems of varying resolution (Table~\ref{resolution_table}) for the year 2045 under two pathways:

\begin{enumerate}
    \item \textit{\Refc (\refca\unskip):} Current policies without additional interventions. Incentives introduced via the \textit{Inflation Reduction Act} of 2022 are included.
    \item \textit{\Coc (\coca\unskip):} Carbon emissions are abated exogenously (e.g., via direct air capture or land use management) for net-zero emissions. The cost of abatement is assumed to be \$200 per ton\cite{erans2022direct,friedmann2020levelized} and is implemented as a fee for exogenous abatement.
\end{enumerate}

\subsection{A Two-Phase Methodology}
We run two ESMs in serial per case: we first optimize a \opone~phase for investments, and subsequently an \optwo~phase for operations. \Opone~model operational outputs are not directly comparable between cases, as emissions, profits, and generation occur in different sets of regions and timesteps with no direct mapping between them. Instead, we translate investments from each \opone~simulation to 26-zone portfolios. We then optimize operations with hourly resolution for a full year in an \optwo~phase. Emissions, profits, and reliability outputs of these equally granular, high resolution operational outputs can be compared between cases to explore the operational impacts of differing investments from various levels of resolution. Running two optimizations thus allows us to examine the impact of resolution on both where resources are built and on how they are run.

Our methods are summarized in Figure~\ref{methods_steps}.

%
%
\begin{table}[ht]
    \centering
    \begin{tabular}{C{0.14\linewidth}L{0.25\linewidth}L{0.45\linewidth}}
        \toprule
        Dimension & General Description & Resolution Implementation \\
        \midrule
        Spatial & Spatial characteristics of systems & Geographic zones, intraregionally homogeneous \\
        Temporal & Time-dependent aspects of the systems (e.g., load, weather) & Days or weeks (timesteps,) reweighted to represent the entire operational year \\
        Operational & Physical rules governing systems & Inclusion or omission of unit commitment (UC) constraints which constrain thermal power plants \\
        \bottomrule

    \end{tabular}
    \caption{Forms of resolution tested in this study.\label{resolution_table}}
    
\end{table}

\subsection{Case Creation}
\label{systems_tested}
GenX implements generators (e.g., utility scale solar, wind turbines, coal power plants,) as ``resource clusters.'' Each represents a number of real world sites (for VRE) or generating units (for thermal plants.) While sites and units are situated at co-ordinates in space, resource clusters themselves have no explicit location. As the number of regions in the system increases, the number of sites in (and implicit spatial extent covered by) each cluster shrinks, leading to a better approximation of location. Our methods for using the physical characteristics of the US energy system to curate regions (Fig~\ref{geography}) and the sites within clusters are included in the supplemental information (SI), section~\ref{supplemental}.

We sample temporal resolutions by subsetting a 52-week year into days or weeks with hourly resolution. Our weather year from PowerGenome is calibrated to 2012 with VRE profiles from Vibrant Clean Energy and demand from NREL\cite{schivley2021powergenome}. We use k-means clustering as built into GenX\cite{genX_github} to group timesteps based on their demand and VRE profiles, and then select a representative timestep per group. We forcibly include ``extreme'' timesteps (one minimum each for solar and wind, one maximum for load) in all cases except three spatial variations of the 15-day case with representative timesteps only. We test three spatial variations of cases that omit unit commitment (UC) constraints. The full suite of cases sampled is in Table~\ref{cases_tested}.

Because the 26-zone, 52-week case with UC is the largest run, it serves as our HRB. Computational advances may soon unlock higher resolution modeling; once feasible, users may wish to include more weather years or higher spatial resolution; 26-zones is still relatively coarse for a real-world transmission network with tens of thousands of nodes. Our HRB should not be considered ``full'' resolution outside the context of this report.

%
%
\begin{table}[h]
    \centering
    \begin{tabular}{l|cc|cccccc}
        \toprule
        Time Domain & UC & ET & 3-Zone & 7-Zone & 12-Zone & 16-Zone & 22-Zone & 26-Zone \\

        \midrule

        52-Week & + & NA & \checkmark & \checkmark & \checkmark & \checkmark & \checkmark & \checkmark\textsuperscript{*} \\

        52-Week & - & NA & \checkmark & & & \checkmark & & \checkmark \\

        30-Week & + & + & \checkmark & & & \checkmark & & \checkmark \\

        10-week & + & + & \checkmark & & & \checkmark & & \checkmark \\

        30-Day & + & + & \checkmark & & & \checkmark & & \checkmark \\

        15-Day & + & + & \checkmark & & & \checkmark & & \checkmark \\

        15-Day & + & - & \checkmark & & & \checkmark & & \checkmark \\
        
        \bottomrule

    \end{tabular}
    \caption{Systems tested. Spatial resolutions are rightmost columns, with an indicator for inclusion of unit commitment (UC) and whether extreme timesteps (ET) were used when downscaling temporal resolution. All cases were run with hourly resolution per-timestep. 24 cases (\checkmark) spanning resolutions are simulated for both the the \refc and \coc cases. \textsuperscript{*} This case serves as our high resolution benchmark (HRB).\label{cases_tested}}
\end{table}

\begin{figure}[h]
    \centering
    \subfloat[3-zone system]{\includegraphics[width=0.3\textwidth]{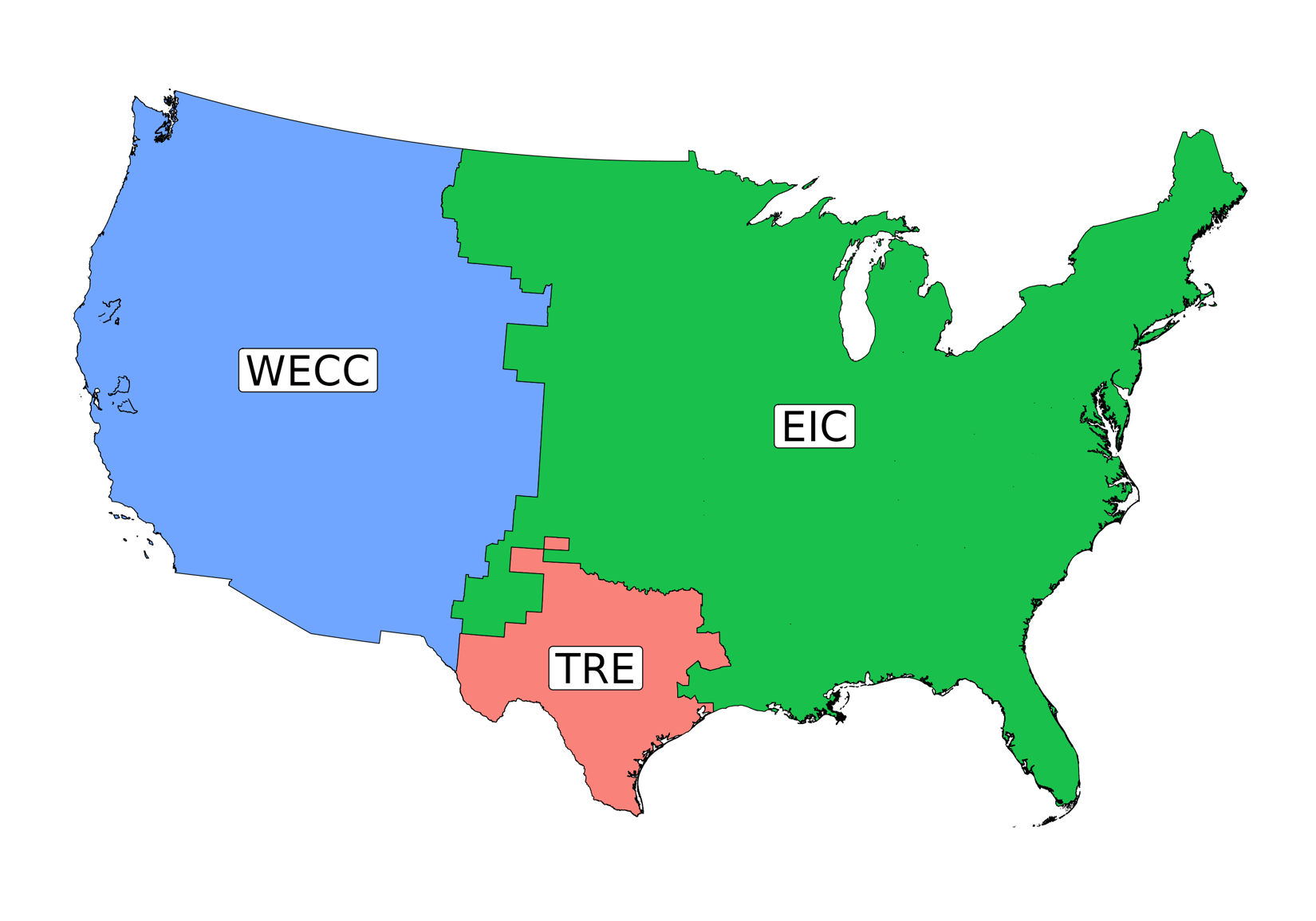}}
    \subfloat[7-zone system]{\includegraphics[width=0.3\textwidth]{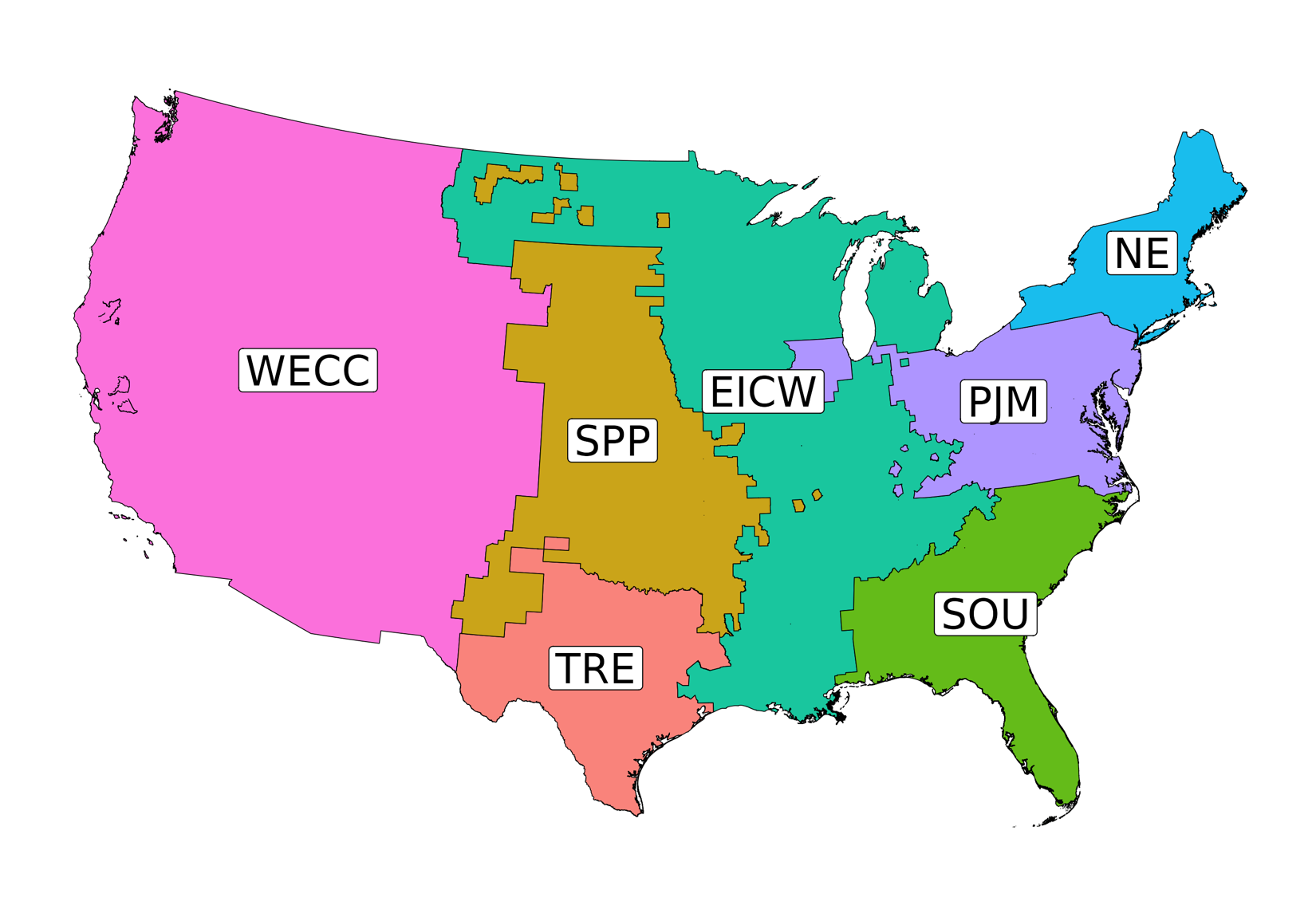}\label{geography_7}}
    \subfloat[12-zone system]{\includegraphics[width=0.3\textwidth]{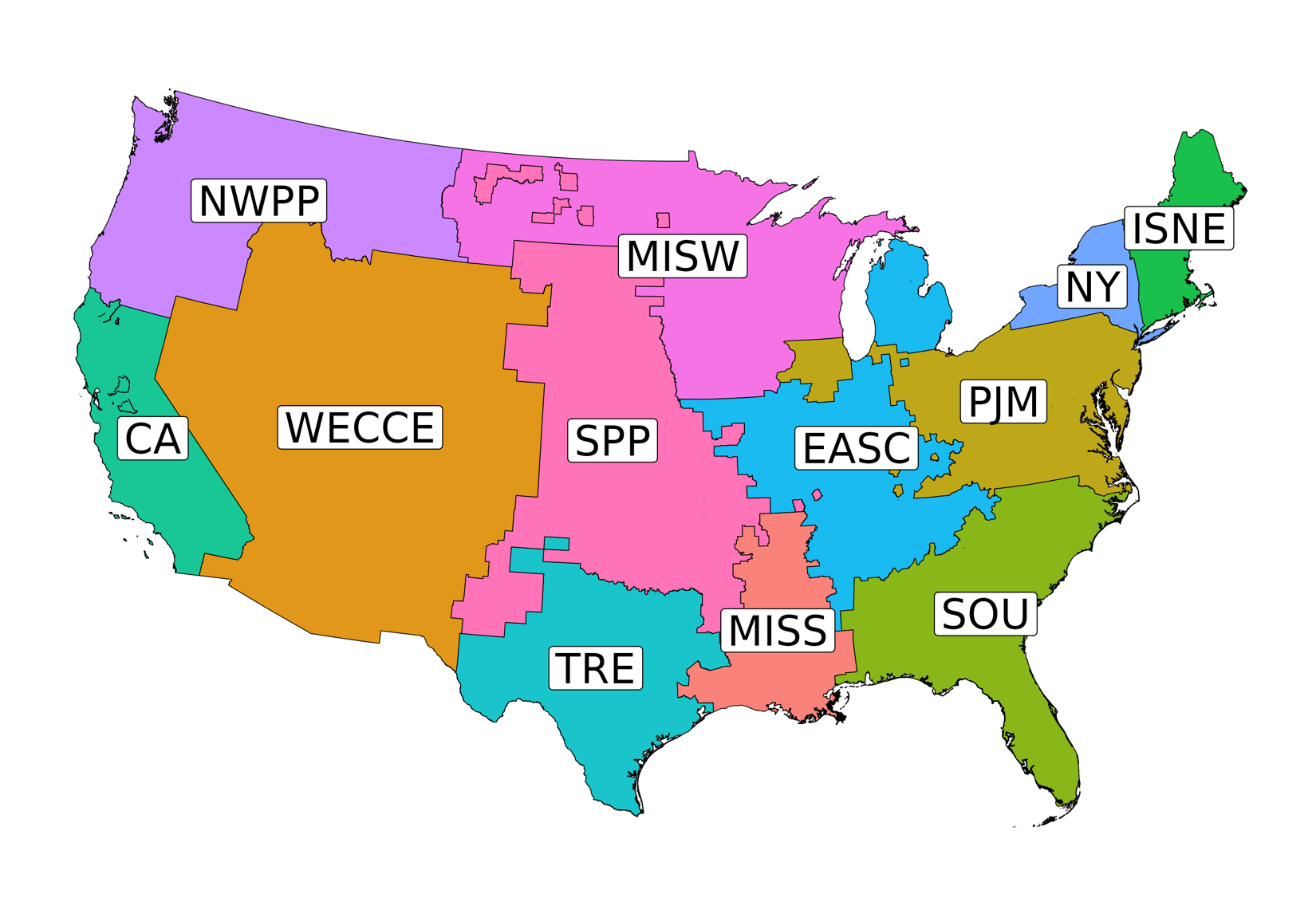}}\\
    \subfloat[16-zone system]{\includegraphics[width=0.3\textwidth]{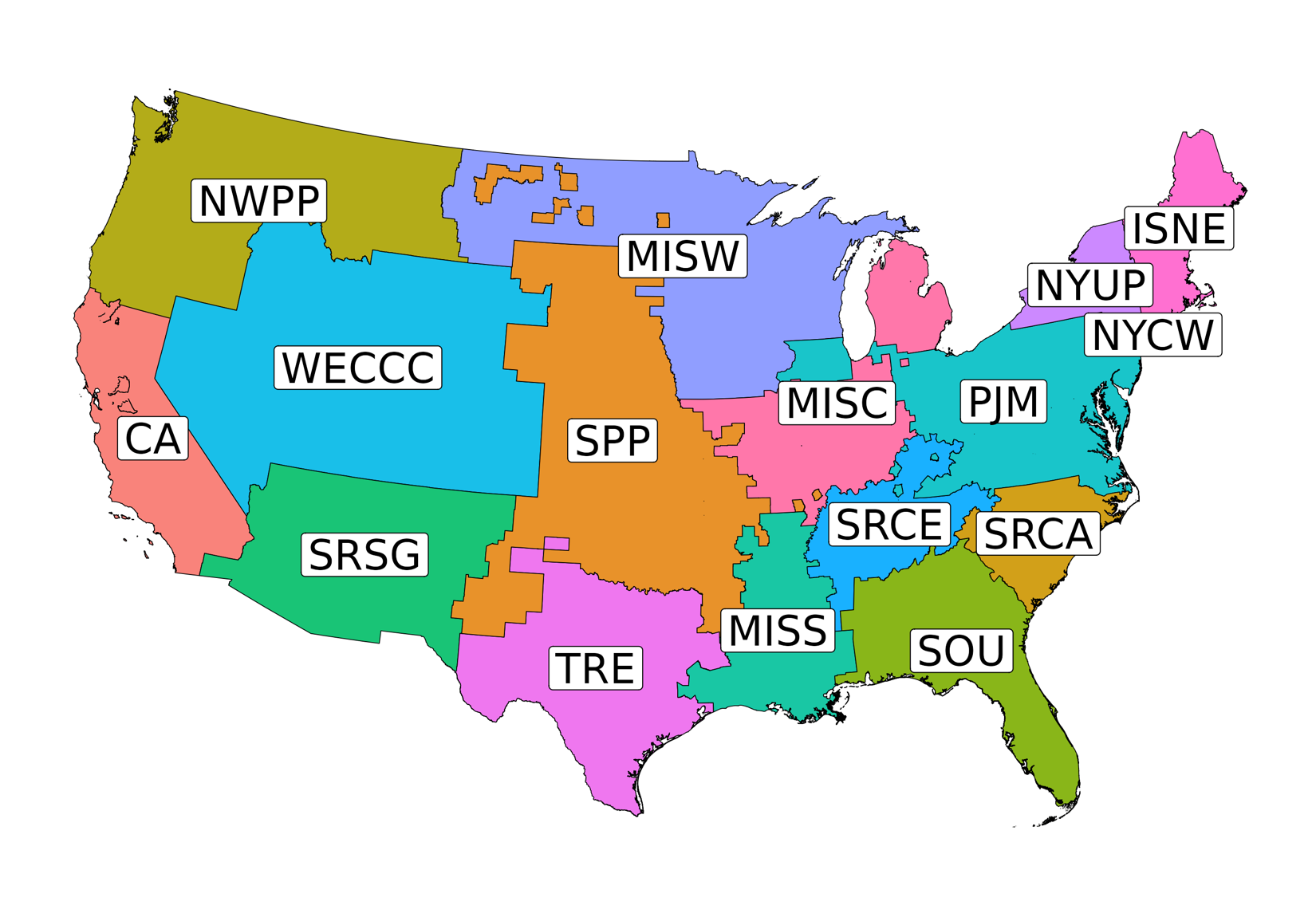}}
    \subfloat[22-zone system]{\includegraphics[width=0.3\textwidth]{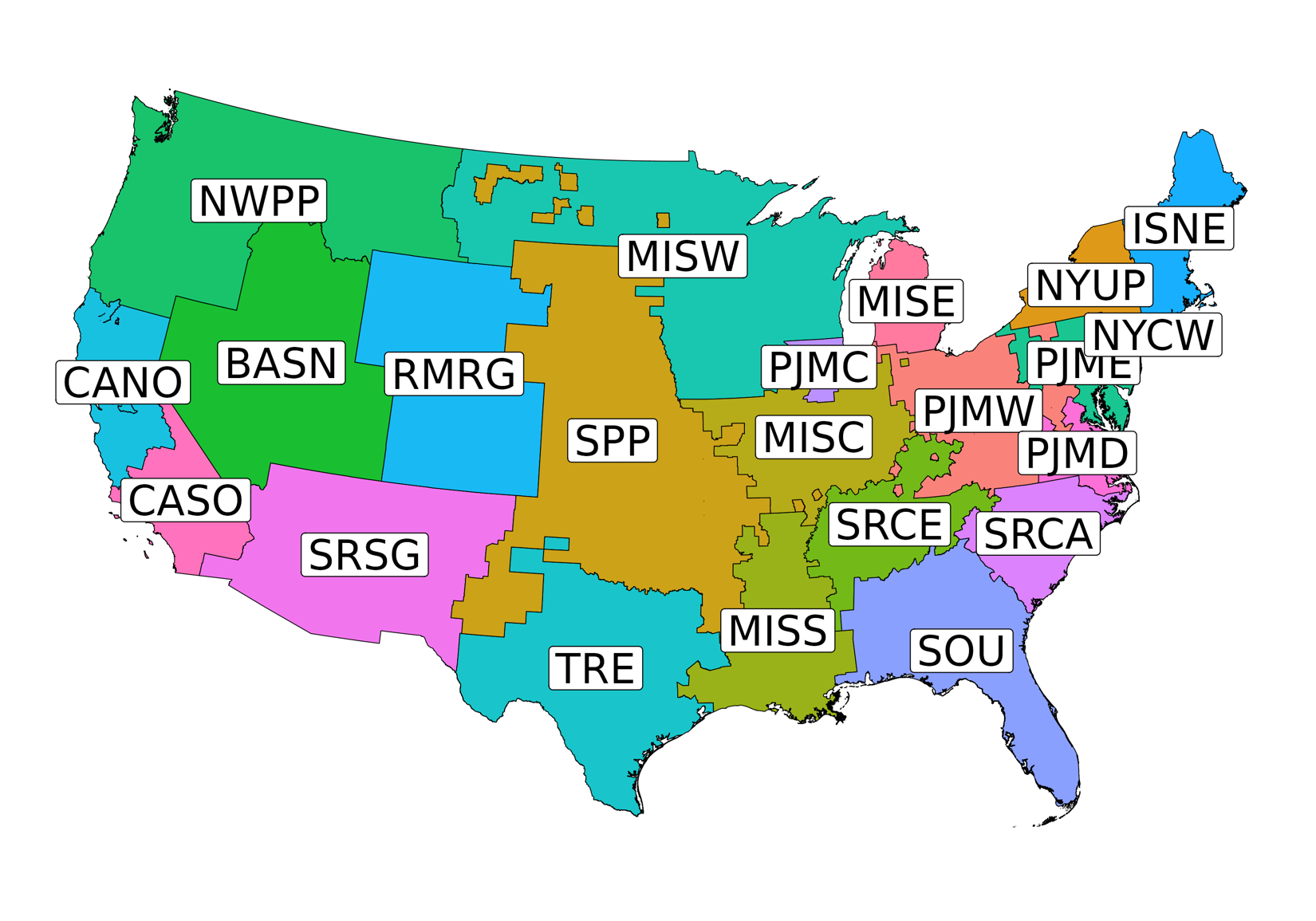}\label{geography_22}}
    \subfloat[26-zone system]{\includegraphics[width=0.3\textwidth]{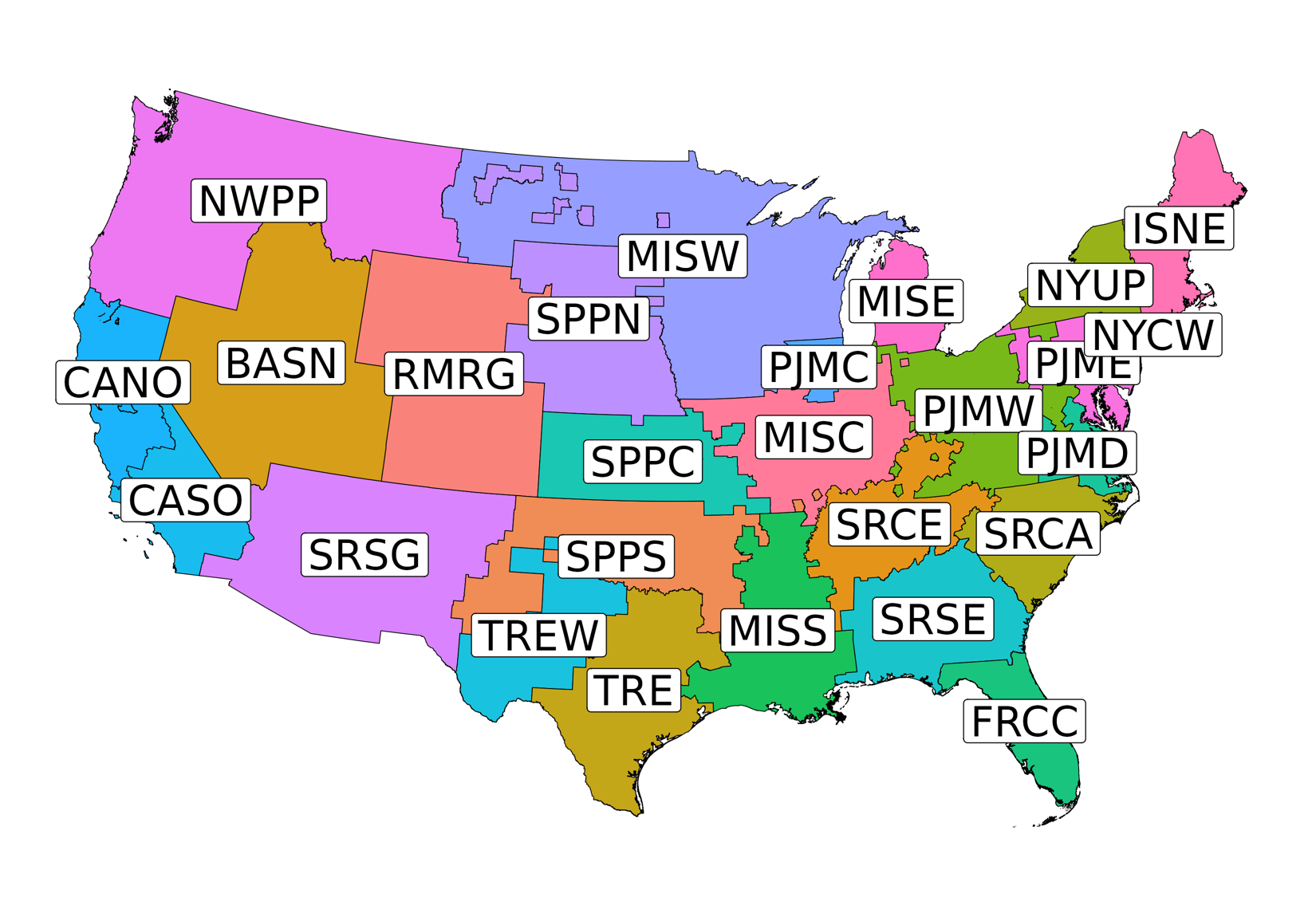}\label{geography_26}}
    \caption{Region aggregations used. Name abbreviations are listed in Table~\ref{region_glossary} in the SI.\label{geography}}
\end{figure}

\subsection{Investment Phase Optimization}
\label{genx_inv}
Simulations are run on the Della computer cluster at Princeton University with 200GB of memory across one node and 53 cores. The version of GenX used for optimizations in this phase and in section~\ref{genx_op} is implemented with a Benders Decomposition to separate investments and parallelize operations according to~\cite{jacobson2023benders,pecci2024regularized}; standard implementations of GenX are unable run large cases within 48 hours and 200GB of memory. We maintain capacity reserve margins per-region in the \opone~phase, to encourage systems to invest for maximum reliability.

\subsection{Site Selection}
\label{sites_downscaling}

Within GenX, a solar cluster may stand for hundreds of individual investment sites; a coal powerplant may be comprised of 10 discrete generating units; a transmission line may represent many real-world connections. To translate outputs from section~\ref{genx_inv} to 26 zones, we start by deriving information on individual sites from \opone~phase clusters using the following methods:

\textit{Variable Renewable Energy:} Select sites belonging to VRE resource clusters according to ascending levelized cost of electricity (LCOE). Sites providing cheaper power receive investment first. For offshore wind, fill fixed turbines' capacity first and allocate remaining sites amongst floating clusters. We allow partial investment in the final VRE site if its full capacity would, after a cumulative sum, exceed the amount of investment of the cluster.

\textit{Thermal Investments:} Allocate thermal investments proportionally to subregions' demand, where a region's ``subregions'' are any that, when going to a more spatially diaggregated case, lie geographically within it. (E.g., FRCC and SRSE in fig~\ref{geography_26} for SOU in fig~\ref{geography_22}.) If, for example, FRCC has twice the demand of SRSE, 66\% of thermal investment per resource in SOU is allocated to FRCC in this step, and SRSE receives the remainder.

\textit{Battery:} Battery storage largely offsets the variability of VRE. We assign storage to subregions proportionally to VRE capacity. Storage is reallocated identically to thermal investments, but weighting by VRE capacity instead of demand.

\textit{Thermal Retirements:} Because fuel used by each thermal sub unit within a resource cluster is identical, heat rates are inversely proportional to revenue; more efficient units are mobilized first. We retire the highest heat-rate (lowest-revenue / most inefficient) units per thermal plant according to a cumulative sum for retired capacity.

\begin{figure}[H]
    \centering
    \subfloat[3-zone case\label{trans_agg}]{\includegraphics[width=0.4\textwidth]{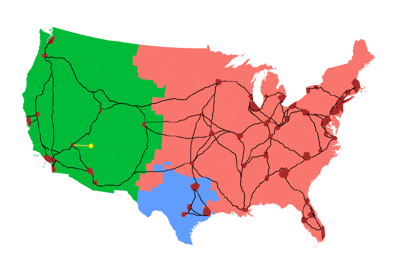}}
    \subfloat[26-zone case\label{trans_dagg}]{\includegraphics[width=0.4\textwidth]{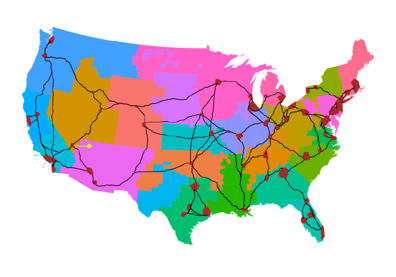}}
    \caption{Transmission topology under two spatial resolutions. When increasing spatial resolution, some intraregional capacity becomes interregional. Black lines are backbone transmission capacity. Yellow circle represents an example renewable resource investment site connecting via a spurline to an urban area (red polygon.) Real systems include tens of thousands of such sites for wind and hundreds of thousands for solar.\label{trans_aggs}}
\end{figure}

\textit{Transmission Investments:} Interregional transmission is included in GenX as pairwise connections between regions. Intraregional transmission, subdivided into ``spurlines'' and ``backbone lines,'' is considered when creating costs for VRE and interregional transmission but hidden from GenX itself. Spurlines, incorporated as investment costs for VRE clusters, connect individual VRE investment sites to urban areas with line capacity equal to that of the site. Backbone lines, a network of larger intraregional lines, ensure power can flow within zones to sites of demand; their capacity is assumed proportional to the interregional lines connecting to their region (Fig~\ref{trans_aggs}.)

We say a line is ``redistricted to interregional'' if it lies fully within one region in lower spatial resolution but would be in multiple were the boundaries of the 26-zone system enforced. The spurline connecting to the VRE site in Fig~\ref{trans_aggs}, for example, is redistricted interregionally when comparing Figs~\ref{trans_agg} and~\ref{trans_dagg}. For a line which is redistricted to interregional, its ``relevant interregional line'' is that which connects the 26-zone case regions at its endpoints.

When rescaling transmission, an interregional line's capacity is the sum of:
\begin{itemize}
    \item \textit{Spurline transmission:} For spurlines which are redistricted to interregional, add invested capacity to the relevant interregional line.
    \item \textit{Interregional transmission:} If several interregional lines in the 26-zone case correspond to one interregional line in the coarse case, interregional investment is allocated to 26-zone lines proportionally to the total population of urban areas at their ends.
    \item \textit{Backbone transmission:} For backbone lines that are redistricted to interregional, add invested capacity to the relevant interregional line.
\end{itemize}

\subsection{Creation of the High Resolution Operational Model}
\label{sites_upscaling}

Once sites, units, and lines have been selected as in section~\ref{sites_downscaling}, we translate to 26 zones. For VRE, we determine the cluster invested sites belong to in the 26-zone case, and add capacity accordingly. For thermal retirements, we find the 26-zone thermal clusters that units belong to, and subtract their capacity accordingly. Investments for transmission, thermal, and storage were reweighted and according to the metrics listed in section~\ref{sites_downscaling}. To derive new capacities, we incorporate these reweightings and allocate capacity accordingly.

\subsection{Operational Phase Optimization}
\label{genx_op}

We use the portfolios derived in section~\ref{sites_upscaling} in the decomposed version\cite{jacobson2023benders,pecci2024regularized} of GenX while preventing further retirements and investments, turning GenX into a production cost model. The \optwo~phase allows analysis of emissions, profit, cost, and reliability in section~\ref{results}. We requested identical computational resources for this phase as in section~\ref{genx_inv}. We omit capacity reserves in the \optwo~phase to allow systems to operate their resources as cost-effectively as possible.

\DeclareRobustCommand{\totalcapcaption}{Optimal installed capacity by region (\ref{total_r}) and technology (\ref{total_t}) for the 26-zone 52-week case for both the \refc (\refca\unskip) and \coc (\coca\unskip) cases. Carbon penalty increases total capacity 36.3\%, from 1580  to 2154 GW, increases solar capacity 67.7\%, and onshore wind capacity 124.2\%. Carbon penalty increases fraction of buildout on the east coast. Aggregations listed in \ref{total_r} are copied here from Fig~\ref{geography_7} and are listed in SI Table~\ref{region_glossary}.}

\DeclareRobustCommand{\totalcostcaption}{Total cost by spatial and temporal resolution. Cost is subdivided into fixed (e.g., investment,) variable (e.g., fuel) and non-served energy (NSE.) Total cost is \$19.5 B for the \coc (\coca\unskip) case and \$13.4 B for the \refc  (\refca\unskip) case. At highest resolution, fixed costs are 75\% and 56.8\% of total costs for the \coca and \refca cases, respectively due to a shift to low-emitting resources with heavy upfront investment and low variable costs. For the \coca case, lowest spatial resolution cases have 212.5\% of the cost of the highest-resolution model.}

\section{Results}
\label{results}

Results compare investments and operations for the \refca and \coca cases by spatial, temporal, and operational resolution. Recommendations from lower resolution systems are more costly, less reliable, and have higher CO$_2$ emissions. Effects are more significant locally than nationally. Systems with coarse granularity cannot predict optimal VRE placement or locations of transmission bottlenecks.

In~\ref{results:aggregate}, we look at the impacts of varying resolution on systemwide metrics (cost, installed capacity.) In~\ref{results:siting}, we analyze differences in recommended VRE and transmission siting locations. In~\ref{results:operations}, we look at how \optwo~results vary with \opone~model resolution. In~\ref{results:case}, we compare the two phases of optimization to determine what information was missed by low resolution simulations in absence of high resolution spatial and temporal constraints.

%
%

\subsection{Aggregate Results} \label{results:aggregate}
We see a shift in types of VRE built (e.g., solar to wind) and an overall decrease in VRE installation in lower resolution simulations. Investments recommended by low resolution simulations are more expensive to operate.

\begin{landscape}
    \begin{figure}[p]
        \begin{center}
        \subfloat[Total installed capacity by region\label{total_r}]{\includegraphics[width=0.25\linewidth]{7z_c1_co2/aggregations_7-Zone.png}\includegraphics[width=0.35\linewidth]{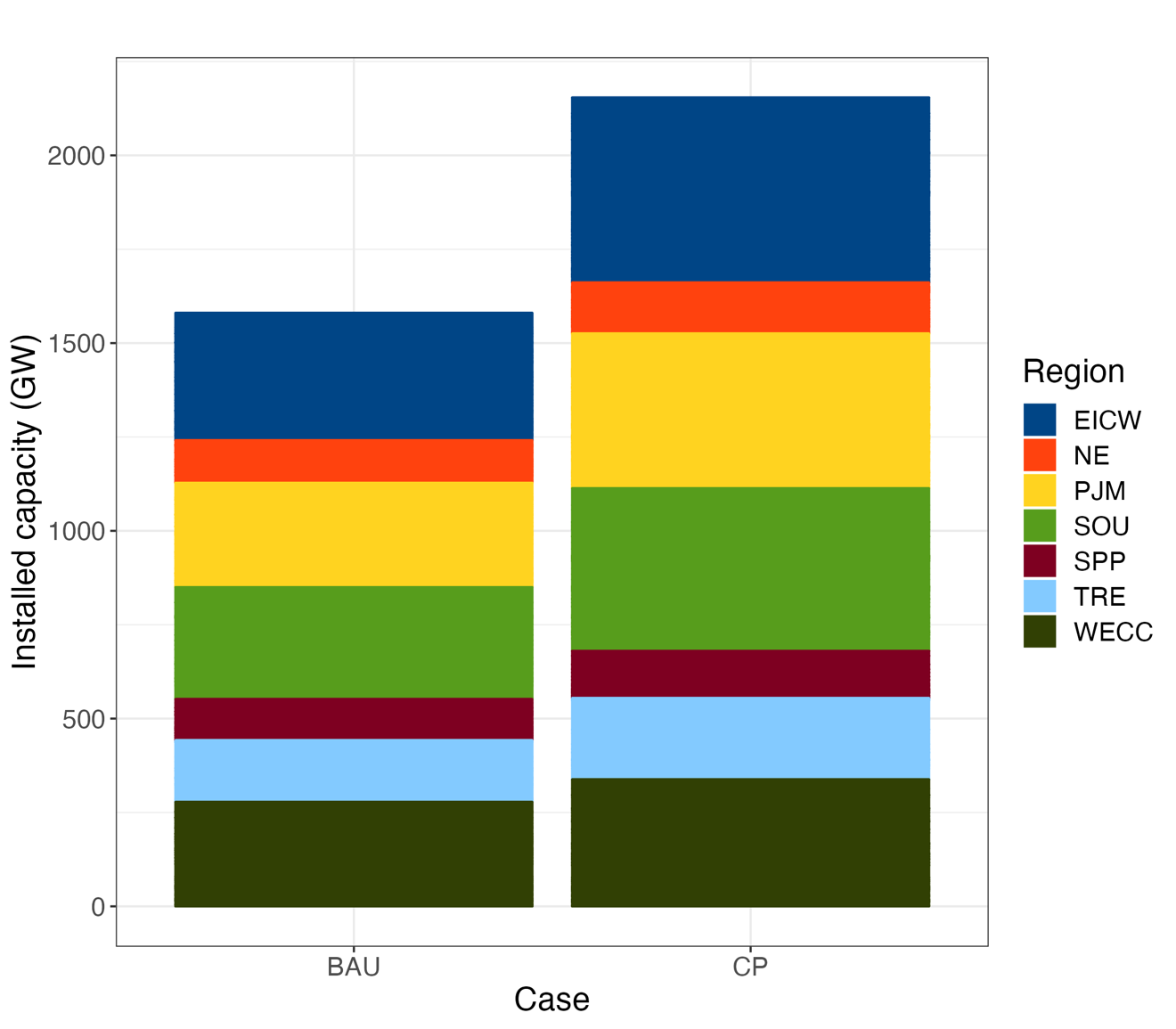}}
        \subfloat[Total installed capacity by technology\label{total_t}]{\includegraphics[width=0.35\linewidth]{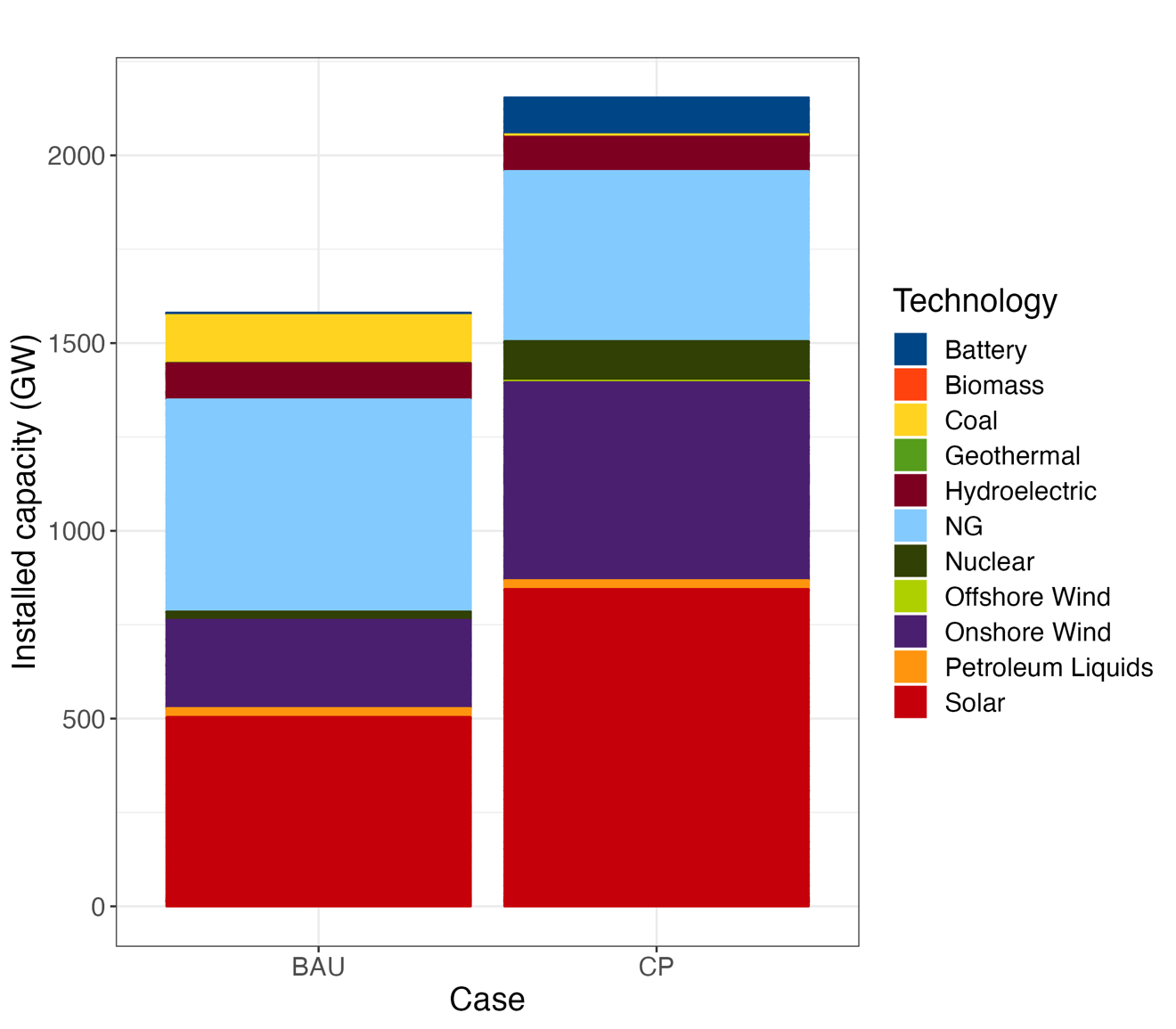}}\\
        \end{center}
        \caption{\totalcapcaption}
        \label{total}
    \end{figure}

    \begin{figure}[p]
        \begin{center}
        \subfloat[Zonal, \refc \label{diff_zone_ref}]{\includegraphics[width=0.5\linewidth]{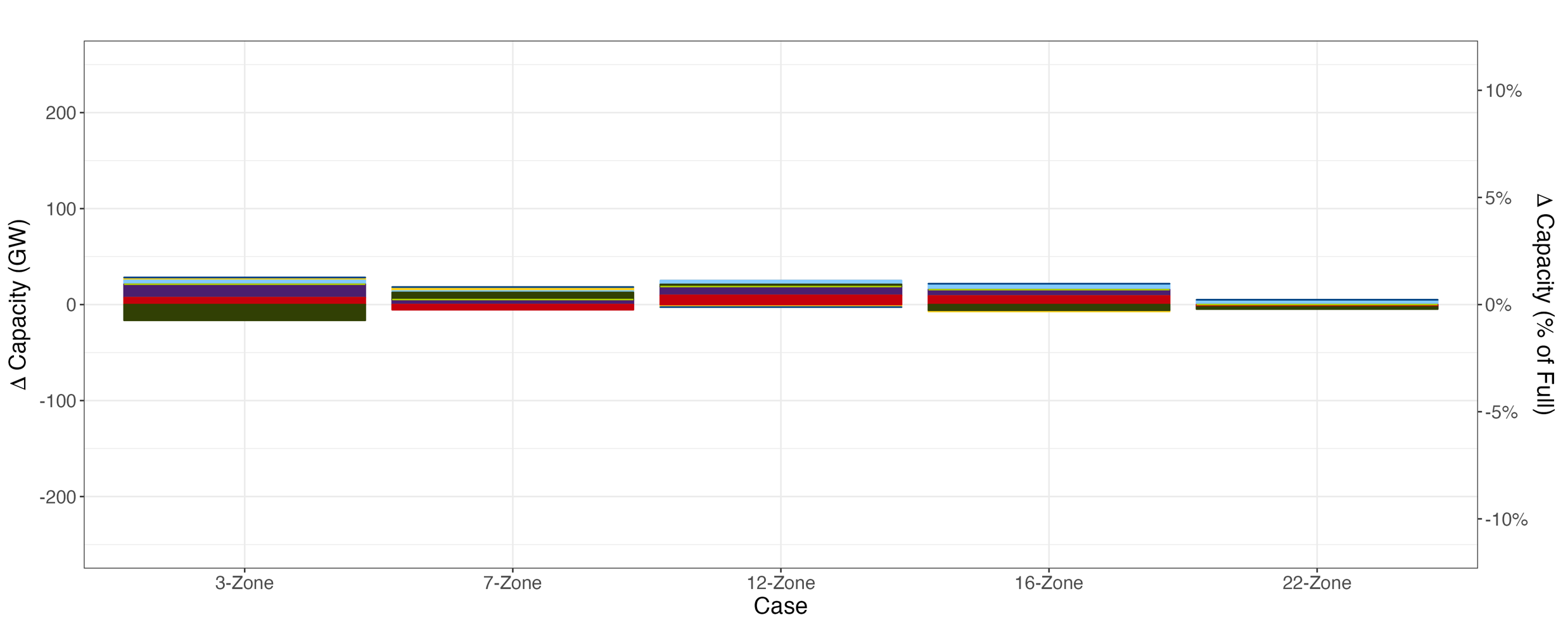}}
        \subfloat[Temporal / operational, \refc \label{diff_temp_ref}]{\includegraphics[width=0.5\linewidth]{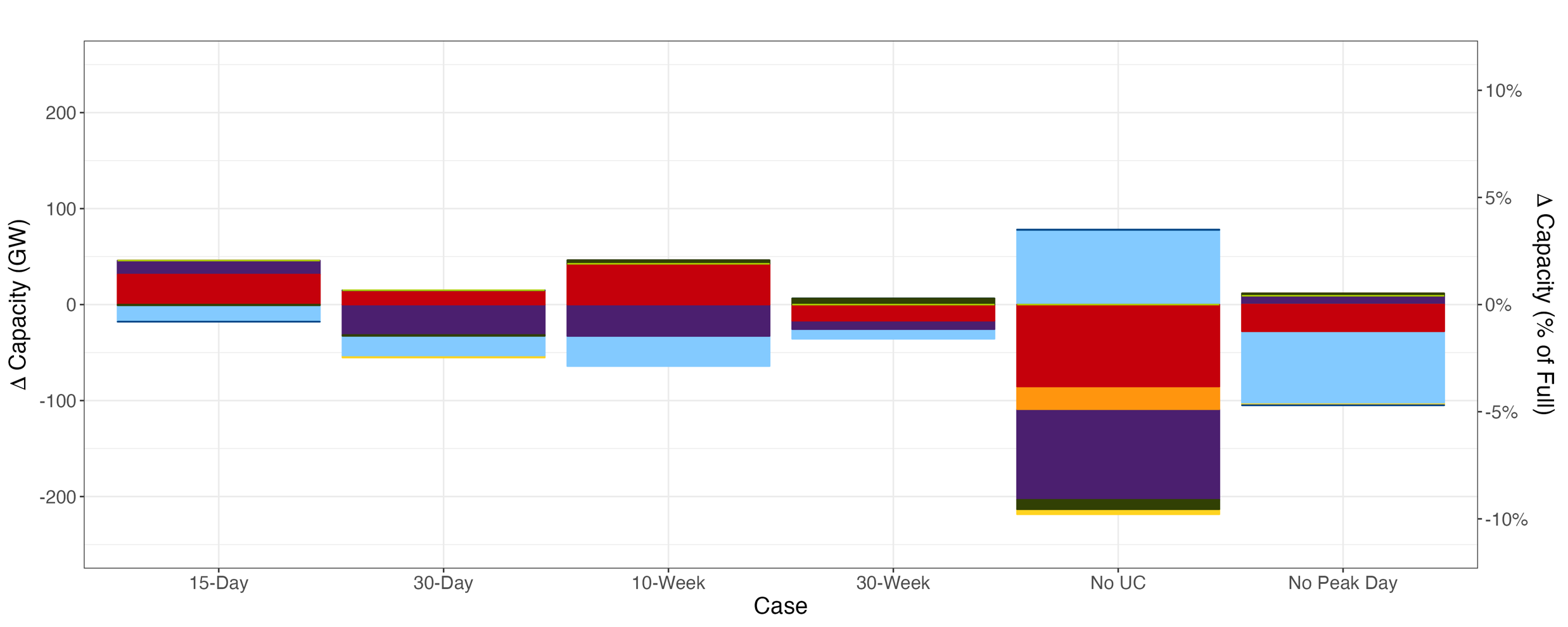}} \\
        \subfloat[Zonal, \coc \label{diff_zone_co2}]{\includegraphics[width=0.5\linewidth]{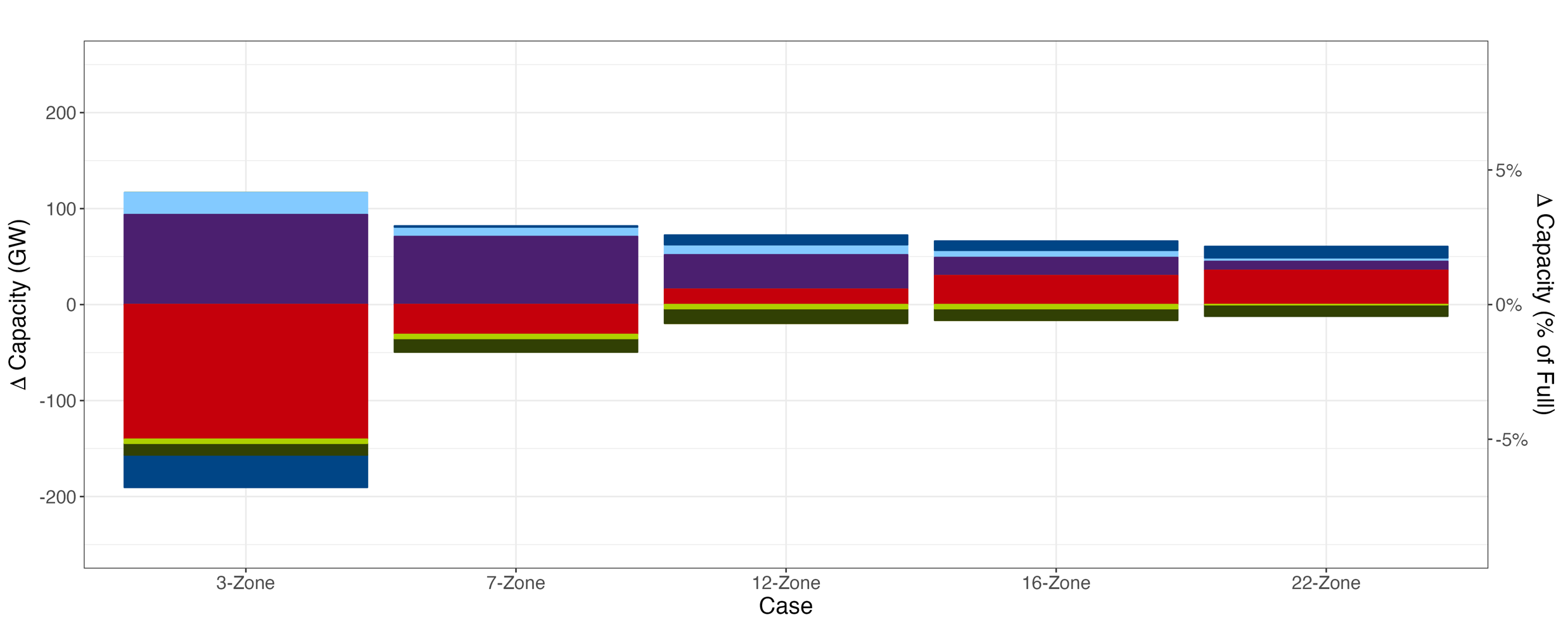}}
        \subfloat[Temporal / operational, \coc \label{diff_temp_co2}]{\includegraphics[width=0.5\linewidth]{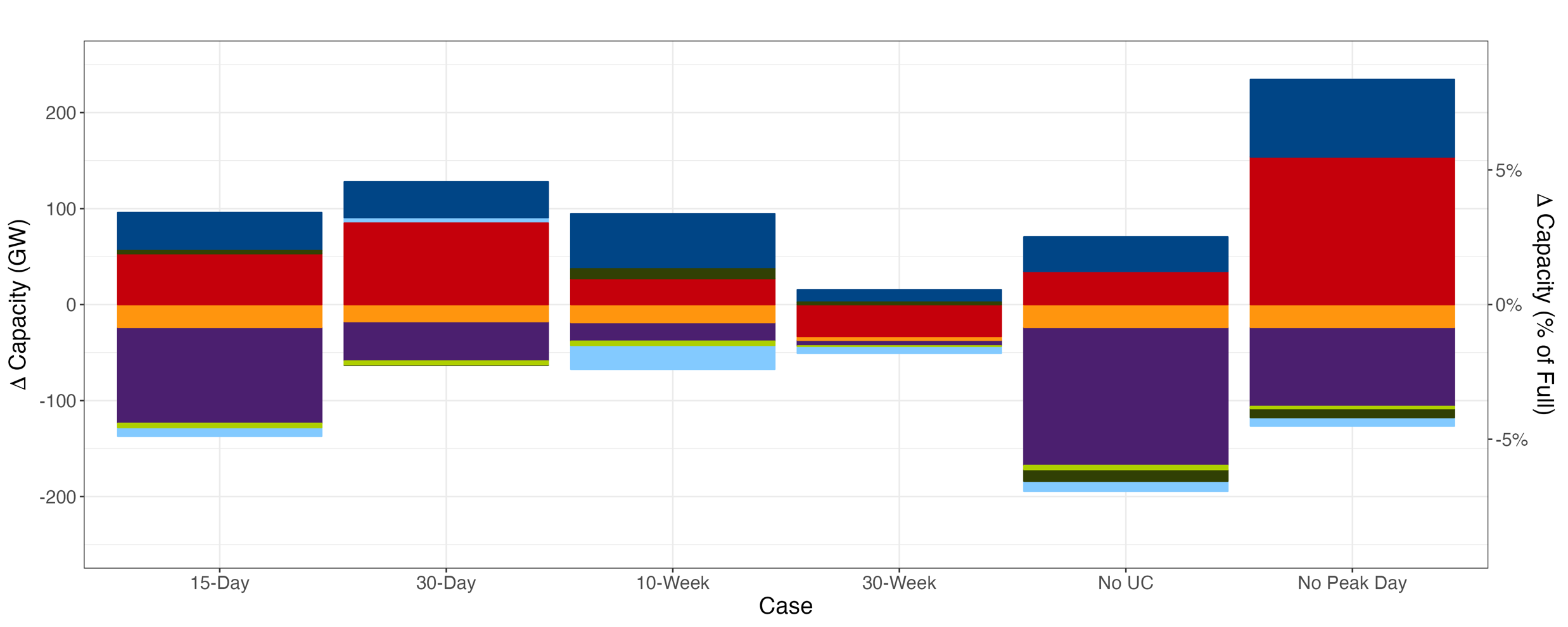}} \\  
        \subfloat{\includegraphics[width=0.4\linewidth]{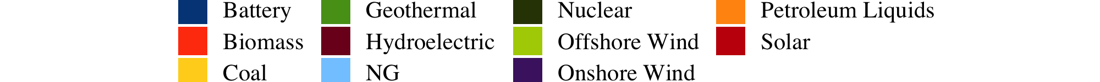}}
        \caption{Difference in installed capacity by technology relative to the high resolution baseline (HRB). Zonal resolutions are shown in \ref{diff_zone_co2}, \ref{diff_zone_ref}, spatial and temporal resolutions in \ref{diff_temp_co2}, \ref{diff_temp_ref}. Increasing spatial granularity generally increases solar buildout and decreases onshore wind. Trends are stronger for spatial resolution and the \coc cases. Omission of unit commitment (UC) decreases renewable energy buildout.\label{diffcap}}
        \end{center}
    \end{figure}  
\end{landscape}

\begin{figure}[h]
    \begin{center}
    \subfloat[Zonal, \refca\label{total_cost_zr}]{\includegraphics[width=0.25\linewidth]{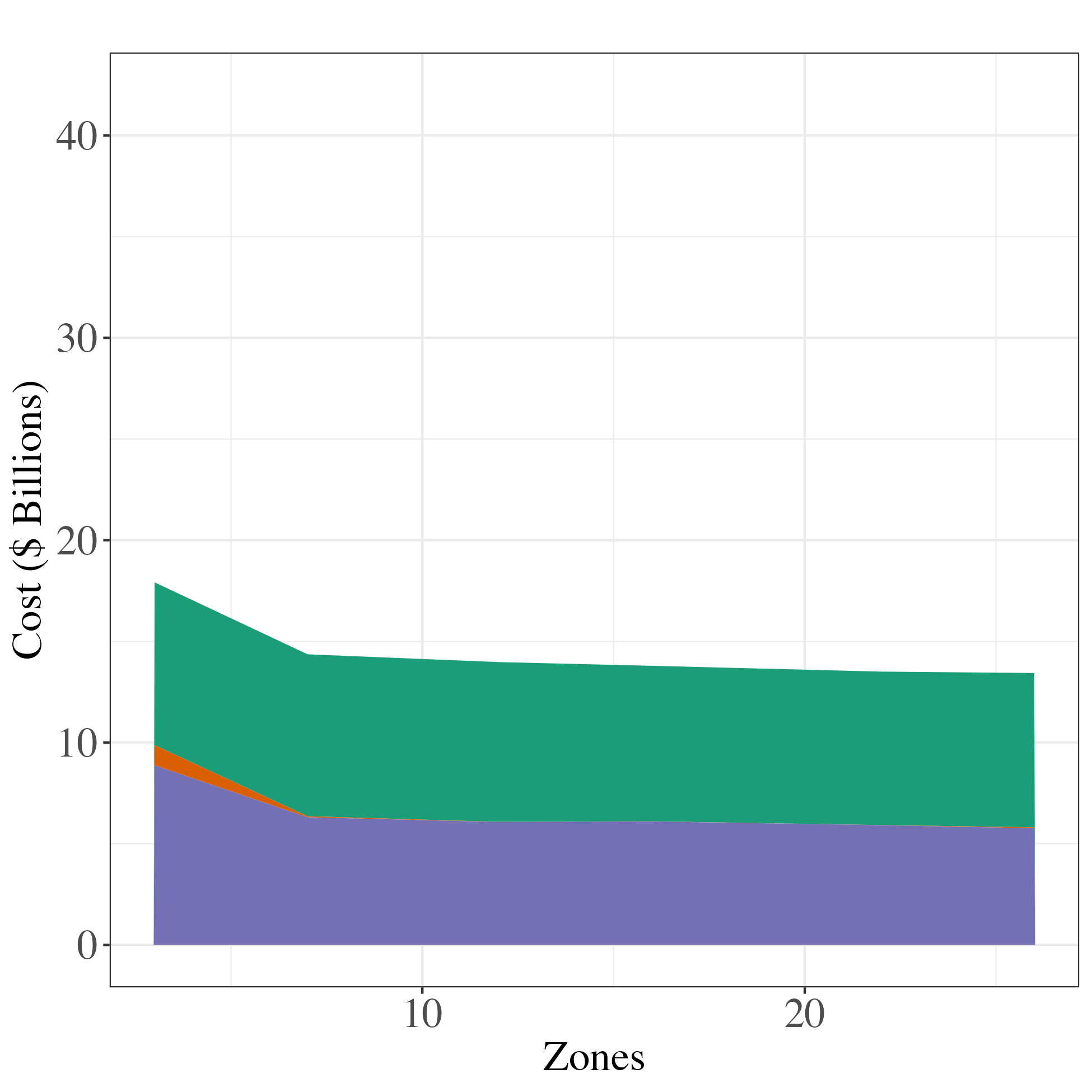}}
    \subfloat[Temporal, \refca\label{total_cost_tr}]{\includegraphics[width=0.25\linewidth]{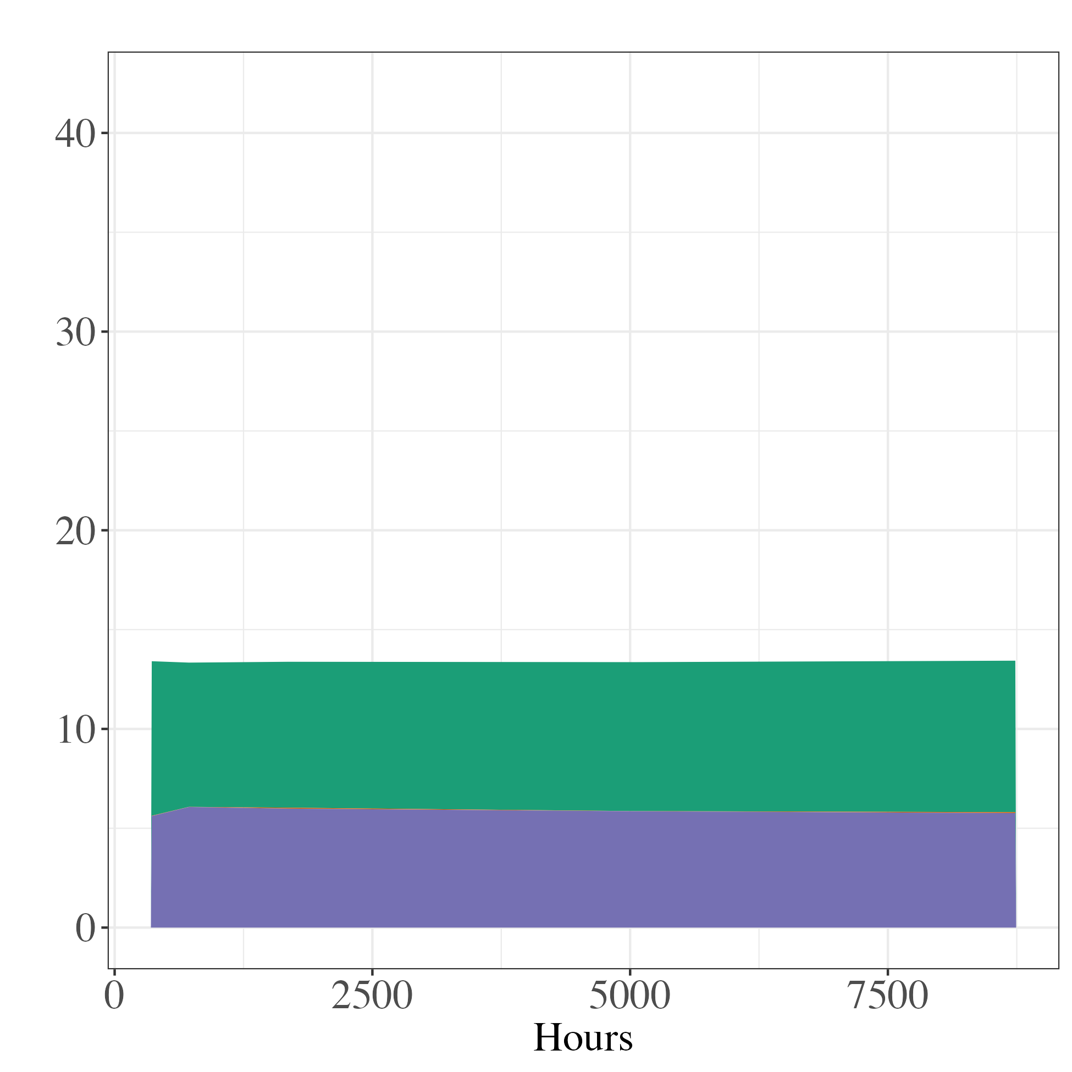}}
    \subfloat[Zonal, \coca\label{total_cost_zc}]{\includegraphics[width=0.25\linewidth]{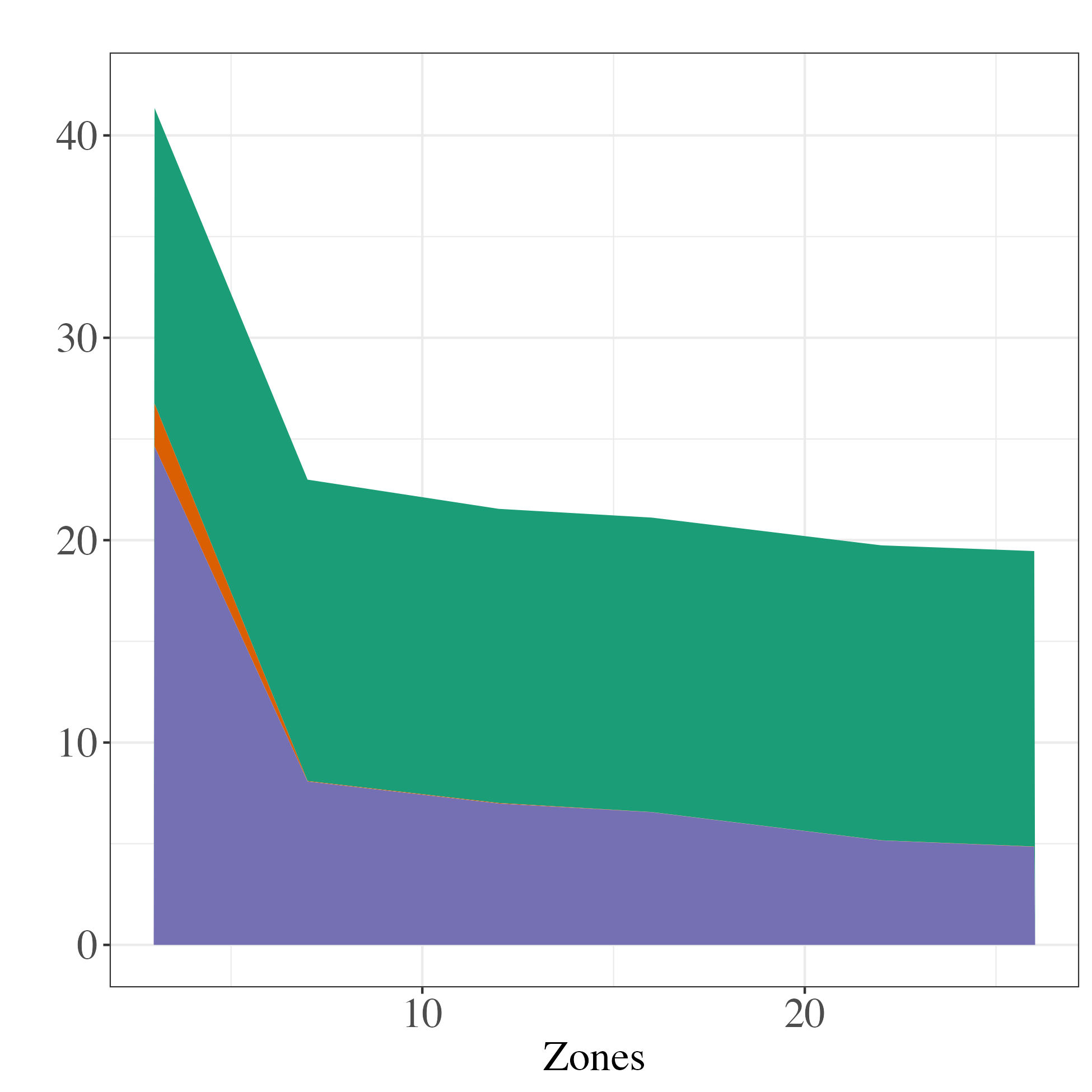}}
    \subfloat[Temporal, \coca\label{total_cost_tc}]{\includegraphics[width=0.25\linewidth]{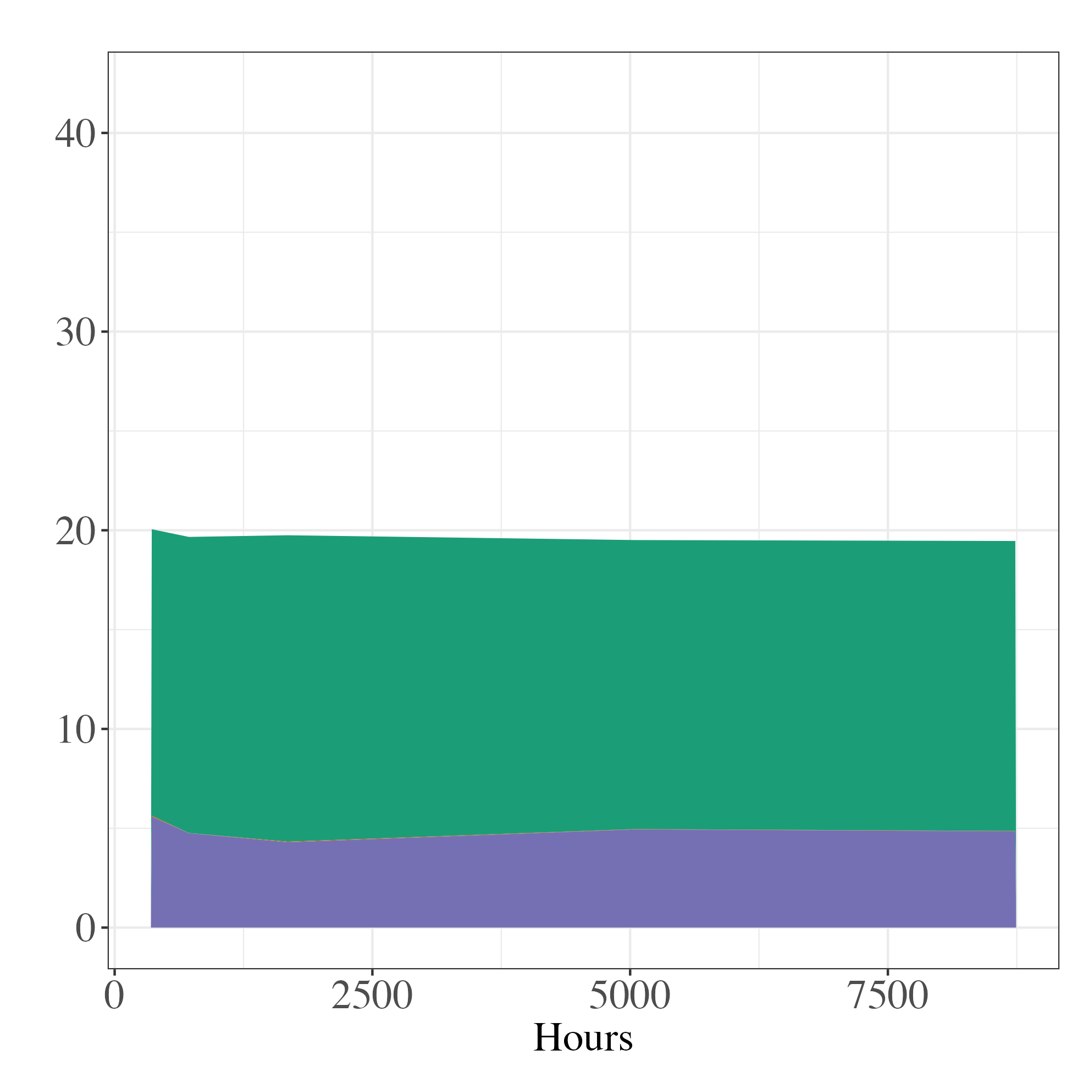}}\\
    \subfloat{\includegraphics[width=\linewidth]{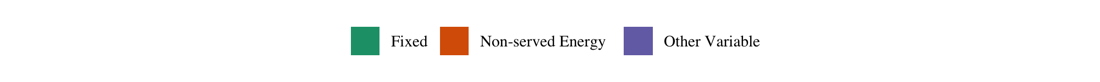}}
    \end{center}
    \caption{\totalcostcaption\label{total_cost}}
\end{figure}

Optimal capacity as recommended by the HRB for the \coca and \refca cases is shown in Fig~\ref{total}, both (Fig~\ref{total_r}) by region and (Fig~\ref{total_t}) by technology. Introduction of a carbon penalty in the \coca case increases solar and wind buildout, coal retirement, and the proportion of installed capacity on the east coast relative to \refca case.

Systems with lower spatial resolution recommend less VRE investment relative to the HRB by hundreds of GWs (\ref{diff_zone_ref},~Fig~\ref{diff_zone_co2}.) Large zones combine weather patterns across greater spaces, underestimating VRE variability. By omitting intraregional transmission constraints, large regions also fail to predict transmission bottlenecks over wider areas, leading to overestimation of resource deliverability. Both effects lead poorly spatially resolved models to overestimate VRE performance, thereby underestimating the amount of capacity needed to meet demand.

\begin{figure}[h]
    \begin{center}
    \subfloat[Mean capacity factor vs temporal resolution\label{cf_temp}]{\includegraphics[width=0.4\linewidth]{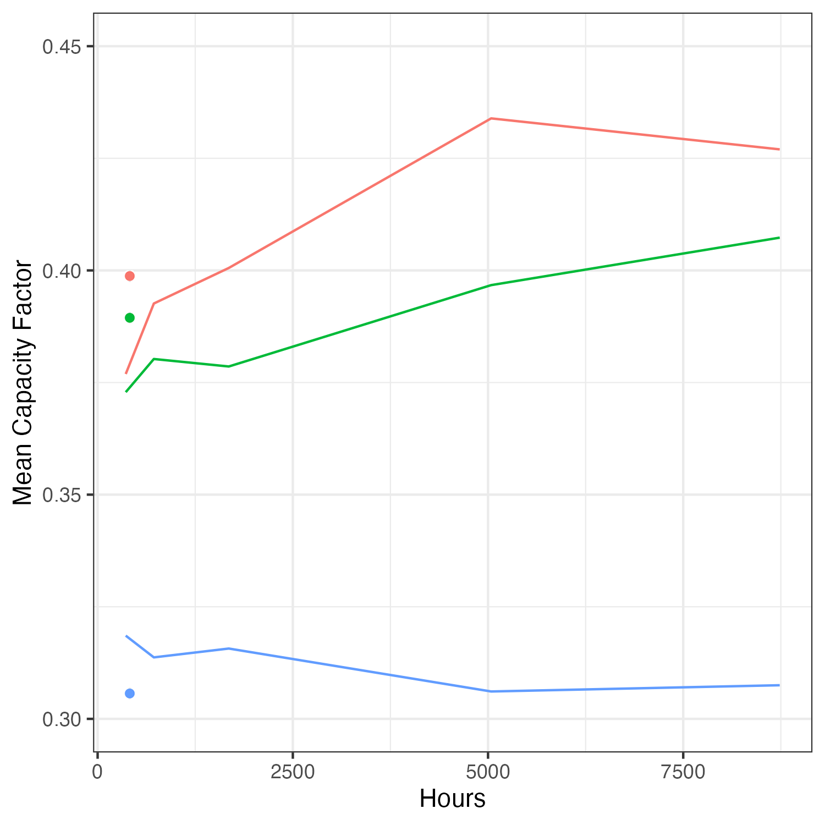}}
    \subfloat[Mean capacity factor vs spatial resolution\label{cf_zone}]{\includegraphics[width=0.4\linewidth]{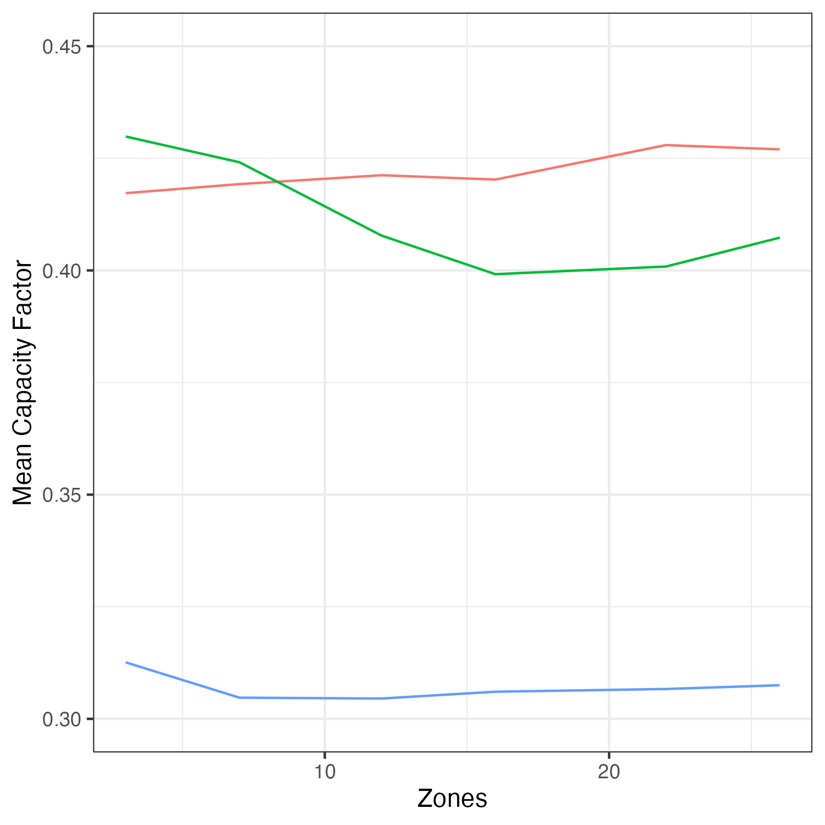}} \\
    \subfloat[]{\includegraphics[width=0.3\linewidth]{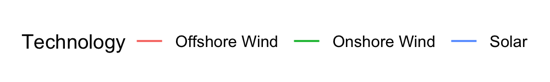}}
    \caption{Mean capacity factor (CF) per technology by resolution. CF shows the average amount of generation (per-MW of capacity) across the entire year that a given resource can expect to generate. CF of 0.5, for example, denotes that a resource is generating an average of 0.5 MW of generation per MW of capacity across the entire timeseries. In \ref{cf_temp}, the dots represent the 15-day case with no extreme timesteps included.\label{cf}}
    \end{center}
\end{figure}

Systems with lower temporal resolution recommend less onshore wind and more solar capacity, though trends are non-monotonic (Figs~\ref{diff_temp_co2},~\ref{diff_temp_ref}.) Using few timesteps misrepresents wind availability, solar availability, demand, and the temporal linkages between them. Our clustering method, when used on our data, has technology-dependent biases: SI Fig~\ref{cf_temp} shows that increasing temporal resolution inflates the overall availability of wind and decreases that of solar, explaining the increased installed wind capacity in \opone~models with lower temporal resolution. Excluding extreme timesteps (a minimal day for wind and solar, a maximum day for load) makes wind appear more available and solar less so (SI Fig~\ref{cf_temp}.) In the 26-zone, 15-day cases, excluding extreme timesteps encourages natural gas (NG) retirement in the \refca case due to underestimation of intermittency that must be overcome using thermal power (Fig~\ref{diff_temp_ref}.) In the 15-day \coca cases, excluding extreme timesteps increases onshore wind (due to increased availability) and battery buildout (Fig~\ref{diff_temp_co2}.)

Omission of UC constraints discourages retirement of NG in the 26-zone 52-week \refca case, as thermal plants can mobilize all capacity at once in the absence of ramping limits, performing better than is realistically feasible. In the \coca case, the ability to leverage thermal power to meet sudden spikes in demand leads the model to build less onshore wind capacity.

Investments suggested by cases with low spatial resolution in the \opone~phase are more costly to operate in the \optwo~phase (Figs~\ref{total_cost_zr},~\ref{total_cost_zc}.) Most added cost is variable (Figs~\ref{total_cost_zr},~\ref{total_cost_zc}.) An increase in fuel costs in particular indicates that \optwo~models, provided suboptimal investments, cannot meet demand cheaply when subjected to realistic operational constraints and must mobilize expensive resources to meet local demand. In section~\ref{results:case}, we confirm that \opone~models have unrealistically high expectations for VRE performance. Trends are weaker for systems with low temporal resolution (Figs~\ref{total_cost_tr},~\ref{total_cost_tc}.) While temporally coarse systems misrepresent weather patterns, they do not fail to see transmission constraints like spatially coarse systems do. This may explain the relative impact seen here.

\subsection{Siting Accuracy} \label{results:siting}
Accuracy of resource siting is strongly impacted by \opone~phase model resolution and is most impacted by the lowest level of resolution in the system. A model with coarse spatial representation, for example, cannot recover accuracy using high temporal resolution. We measure locational accuracy for VRE using the percentage of investment sites, weighted by capacity, that are selected by both the HRB and the lower resolution case. We call this metric site capacity overlap (SCO): high SCO implies strong agreement between a case and the HRB in terms of where investment should occur. To any local planner, SCO is a more relevant metric than systemwide in investments. While change in total capacity is constrained to under 10\% in all cases (Fig~\ref{diffcap}) SCO can be as low as 7\%.

\[SCO_{tech} = 100\% \cdot \frac{sites_{HRB,tech} \cap sites_{case,tech}}{sites_{HRB,tech} \cup sites_{case,tech}}\]

\begin{table}[H]
    \centering
    \setlength\tabcolsep{0pt}
    \begin{tabular}{cM{0.3\linewidth}M{0.3\linewidth}M{0.3\linewidth}}
        & \large 15-day & \large 10-week & \large 52-week \\
        
        \rotatebox[origin=c]{90}{\large 3-Zone} &
        \includegraphics[width=\linewidth]{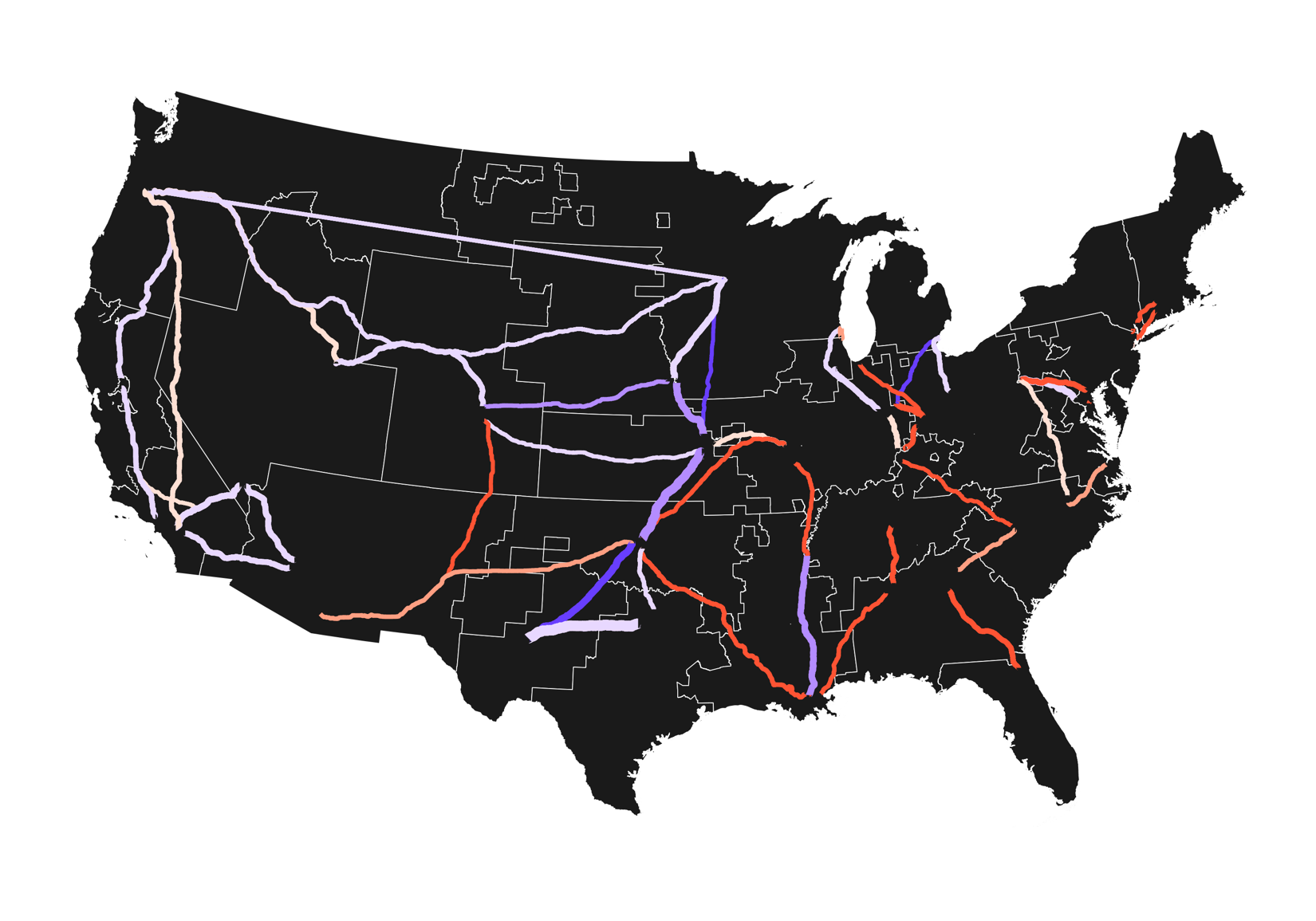} & 
        \includegraphics[width=\linewidth]{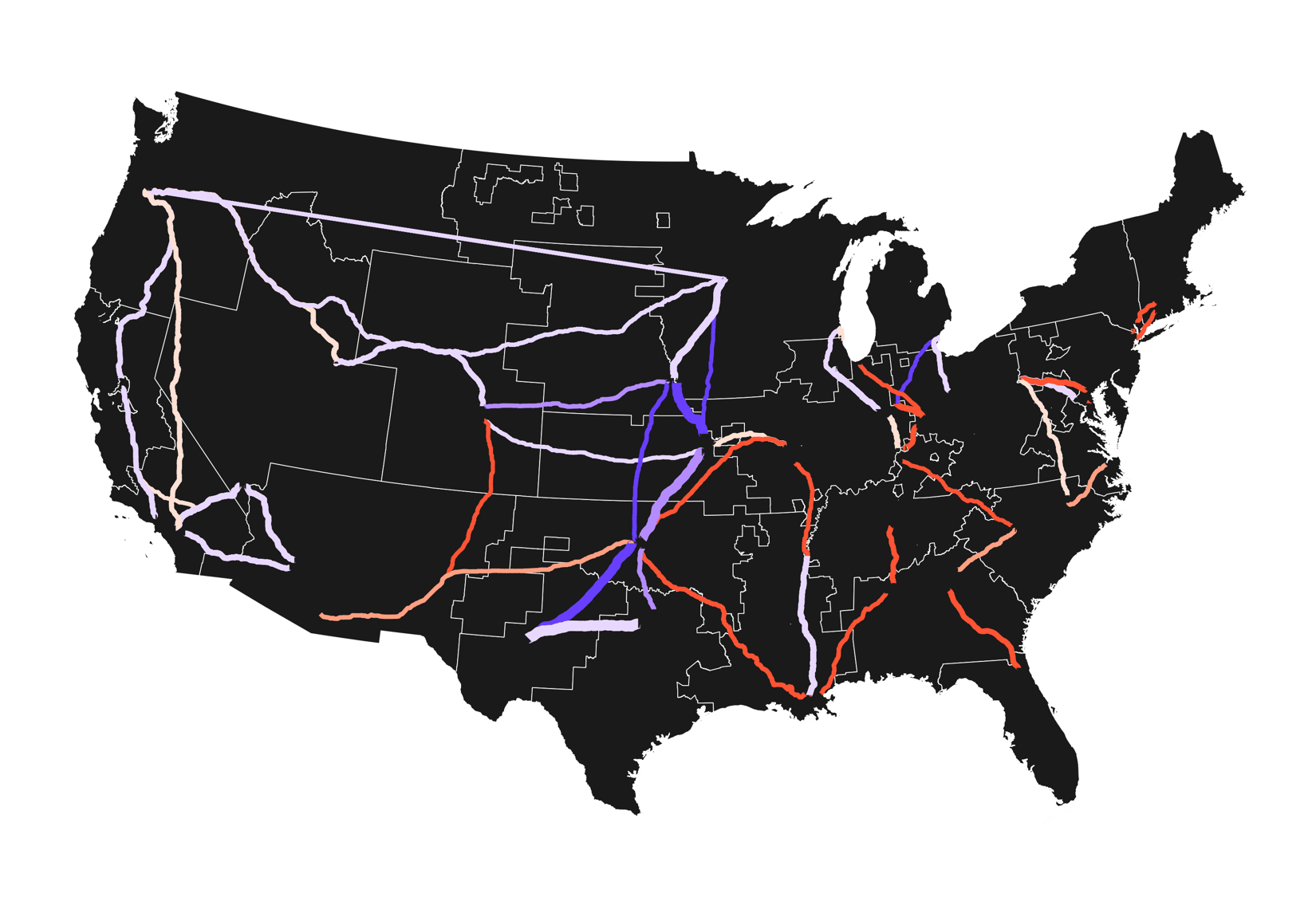} & 
        \includegraphics[width=\linewidth]{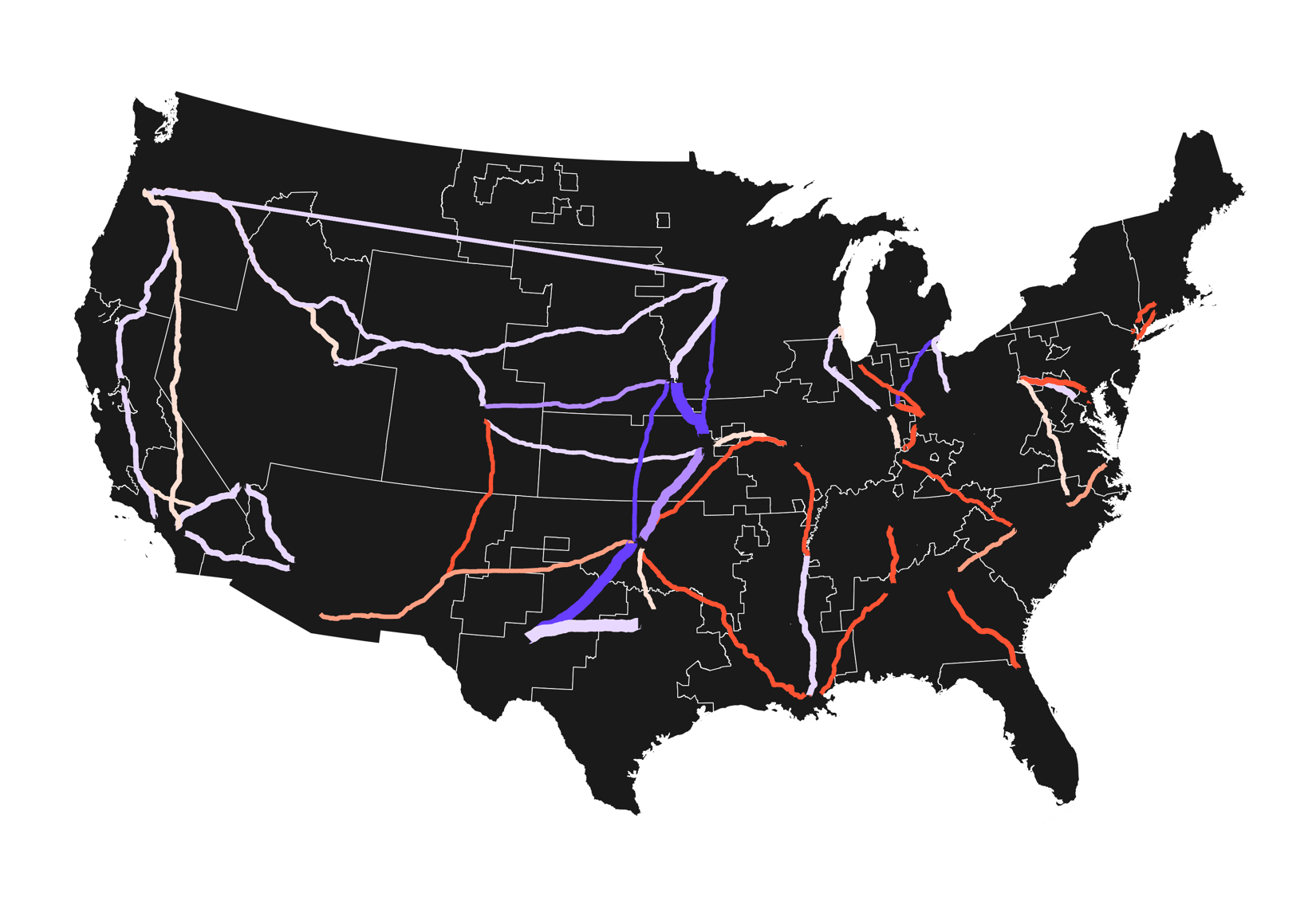} \\

        & {\footnotesize 162.9 GW, MSE: 0.3 GW} &
        {\footnotesize 177.6 GW, MSE: 0.4 GW} &
        {\footnotesize 190.2 GW, MSE: 0.5 GW} \\
    
        \rotatebox[origin=c]{90}{\large 16-Zone} &
        \includegraphics[width=\linewidth]{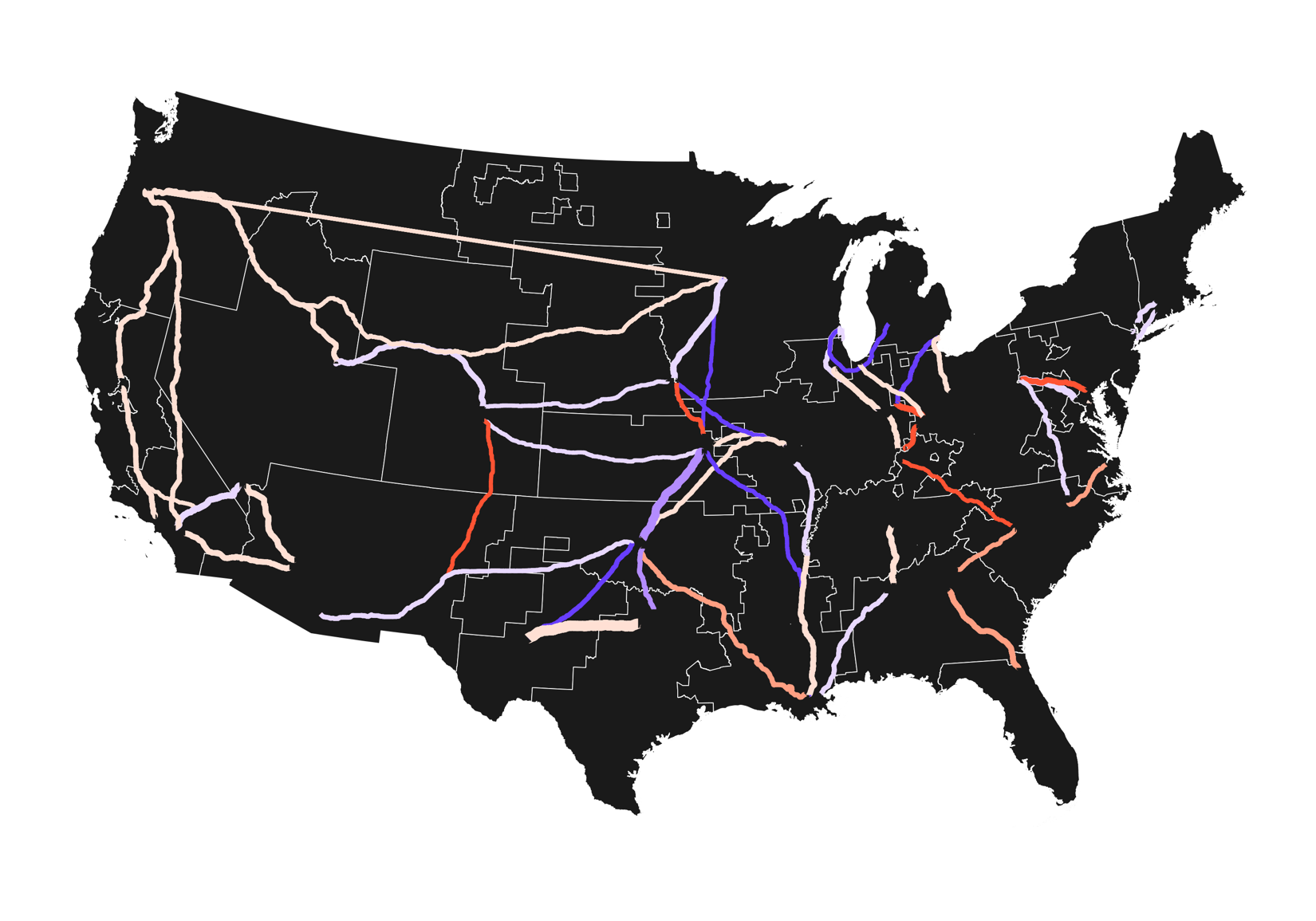} & 
        \includegraphics[width=\linewidth]{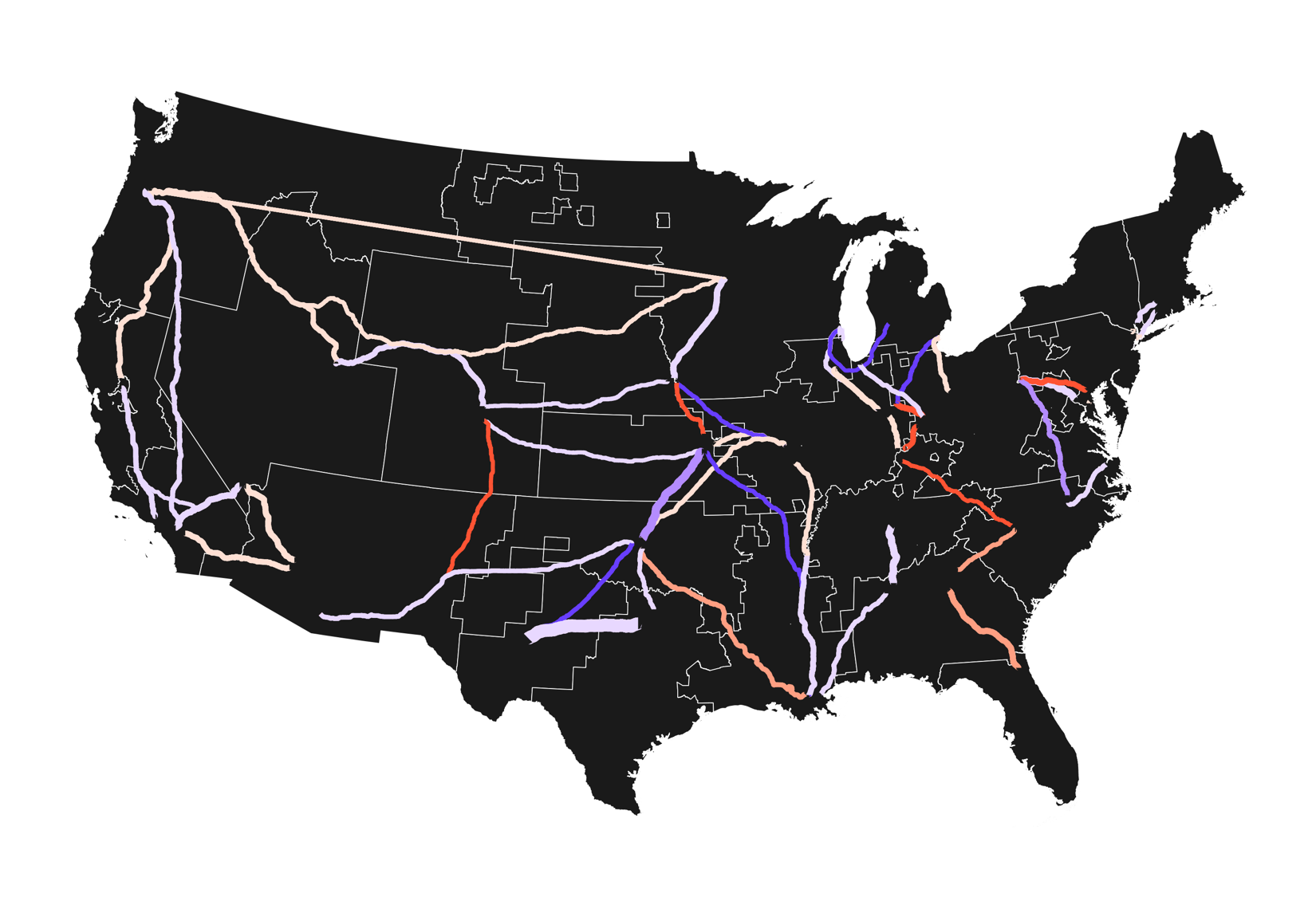} & 
        \includegraphics[width=\linewidth]{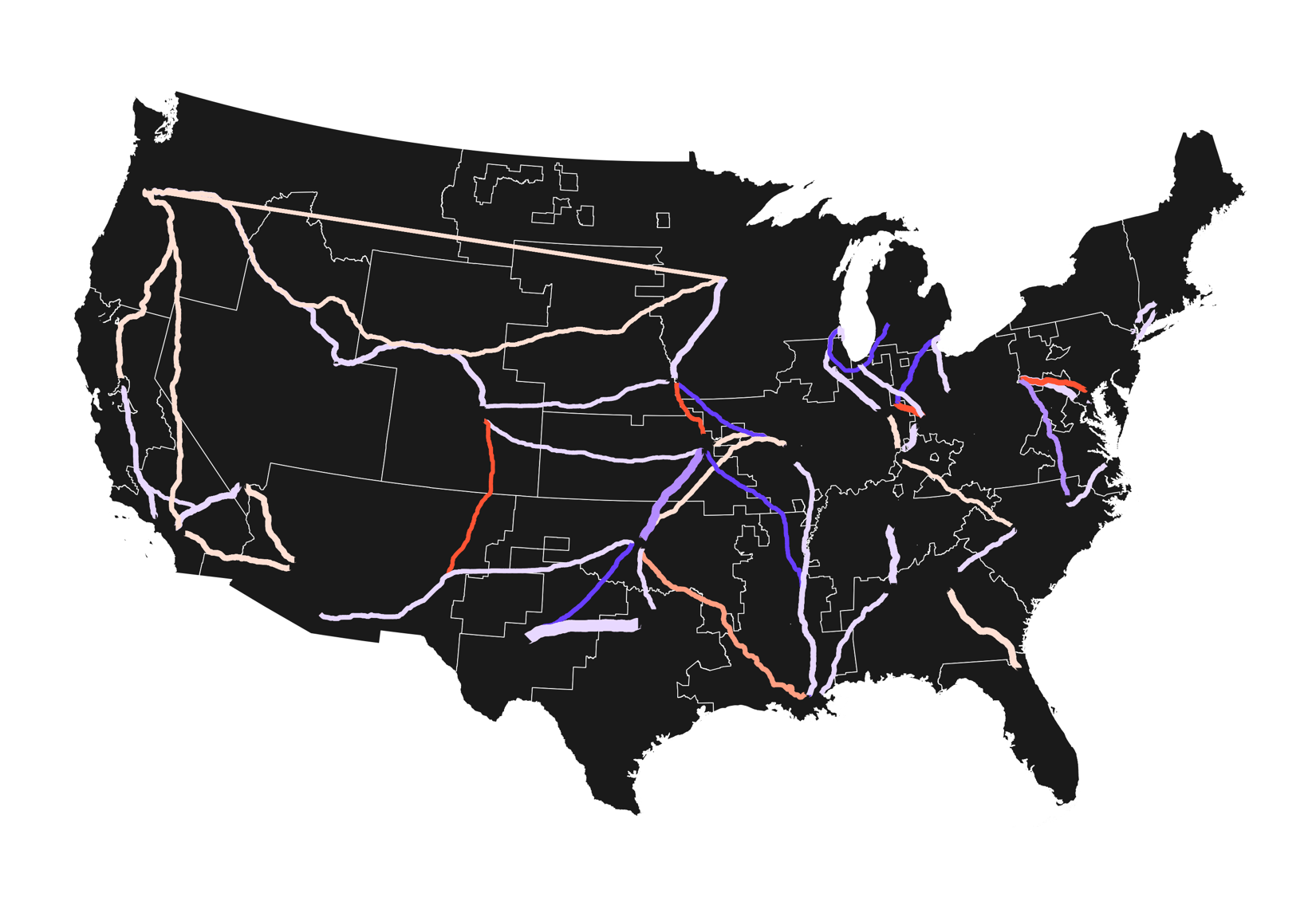} \\

        & {\footnotesize 168.6 GW, MSE: 0.2 GW} &
        {\footnotesize 178.1 GW, MSE: 0.2 GW} &
        {\footnotesize 187.1 GW, MSE: 0.2 GW} \\

    \end{tabular}
    \includegraphics[width=0.6\linewidth]{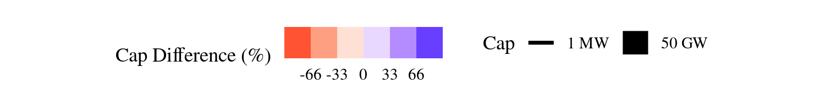}
    \captionof{figure}{Differential transmission capacity by resolution, \Coc case. 26-zone cases are omitted, as transmission is identical to the high resolution baseline (HRB) case. Color of line shows the difference~(\%) in capacity between the cases and the HRB: $cap_{GW} - cap_{GW,HRB}$. Blue indicates the low-resolution case overinvested in the given line, while red indicates underinvestment. Line thickness shows the capacity of the line in the given case. Metrics under given cases indicate the total capacity and the mean squared error (MSE) of capacity per line. $MSE_{cap} = \sqrt{\sum{(cap_{GW} - cap_{GW,HRB})^2}} \div \vert lines \vert$.}
    \label{map_trans_diff}
\end{table}

Total transmission capacity varies little with resolution, indicating that our methods for estimating transmission routing and costs at varying degrees of resolution (including intraregional backbone networks) are reasonably accurate (Fig~\ref{map_trans_diff}.) Still, transmission tends to be underbuilt near dense VRE investment (Figs~\ref{map_solar},~\ref{map_wind}.) Decreased transmission capacity in spatially coarse regions with heavy VRE investment is indicative that spurlines as incorporated are insufficient in capturing the dynamics of intraregional networks and the amount of transmission needed in a system with heavy investment in renewables.

\begin{landscape}
    \begin{table}[p]
        \centering
        \setlength\tabcolsep{0pt}
        \begin{tabular}{cM{0.195\linewidth}M{0.195\linewidth}M{0.195\linewidth}M{0.195\linewidth}M{0.195\linewidth}}
            & \large 15-day & \large 30-day & \large 10-week & \large 30-week & \large 52-week \\
            
            \rotatebox[origin=c]{90}{\large 3-Zone} &
            \includegraphics[width=\linewidth]{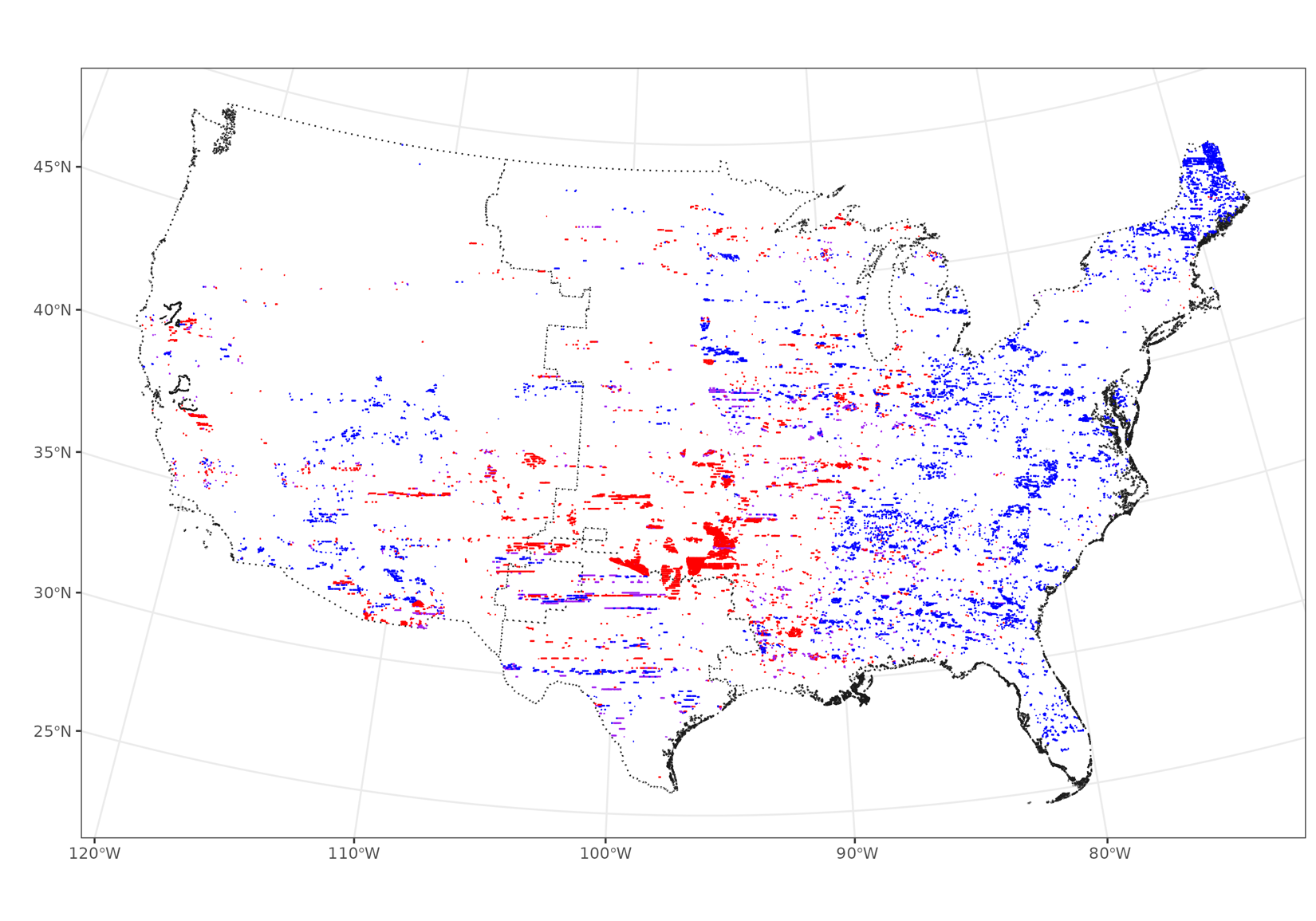} & 
            \includegraphics[width=\linewidth]{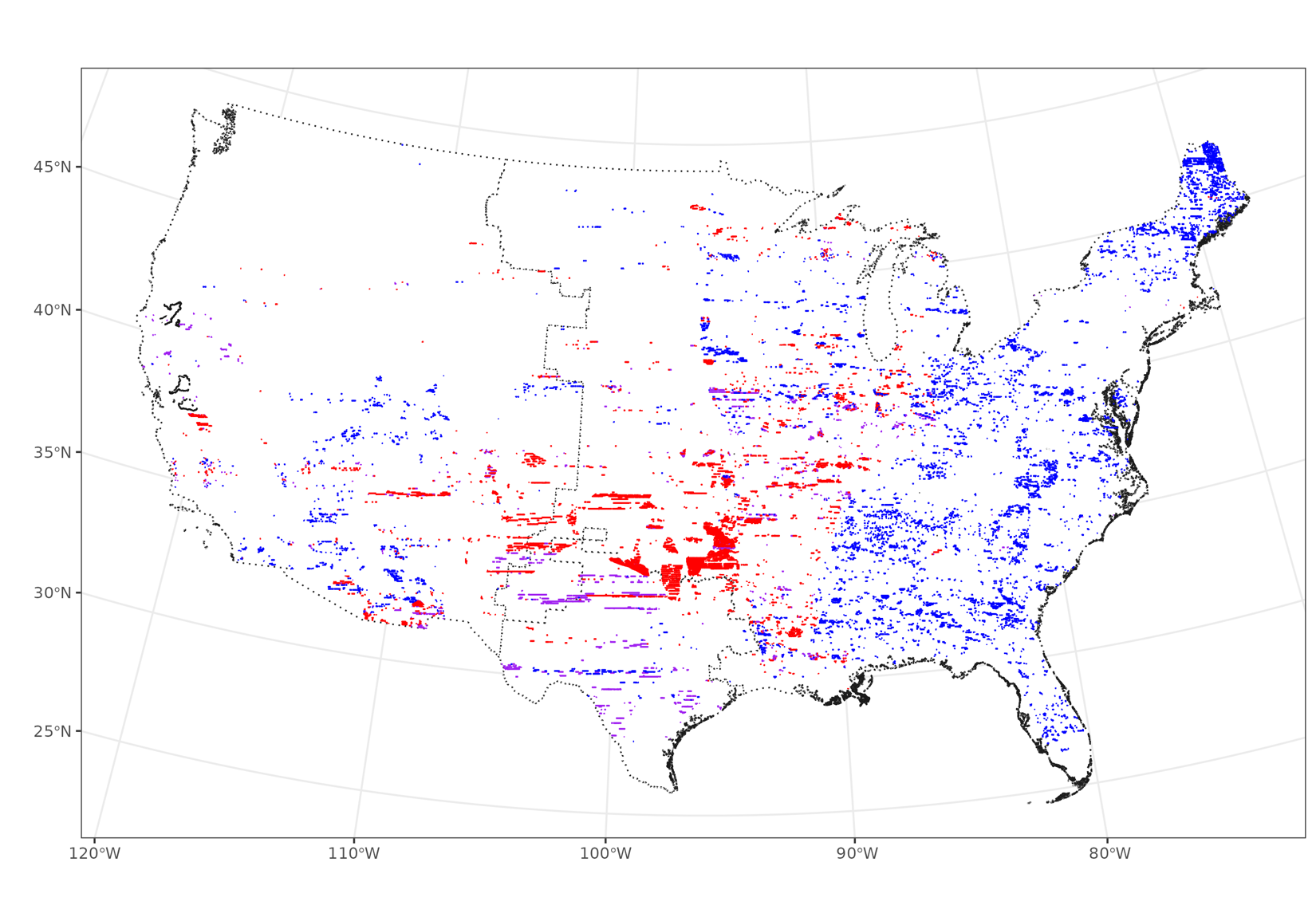} &
            \includegraphics[width=\linewidth]{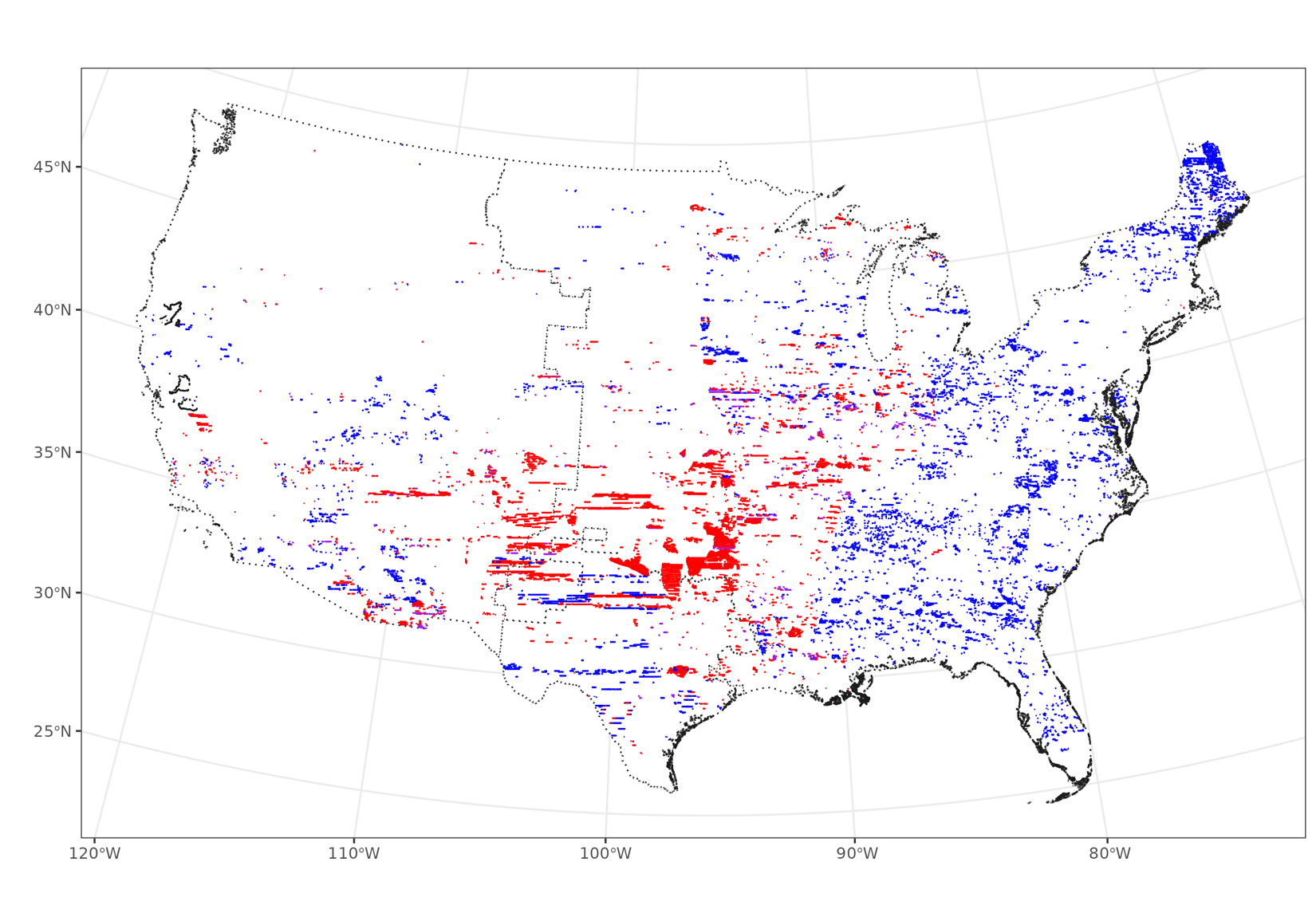} & 
            \includegraphics[width=\linewidth]{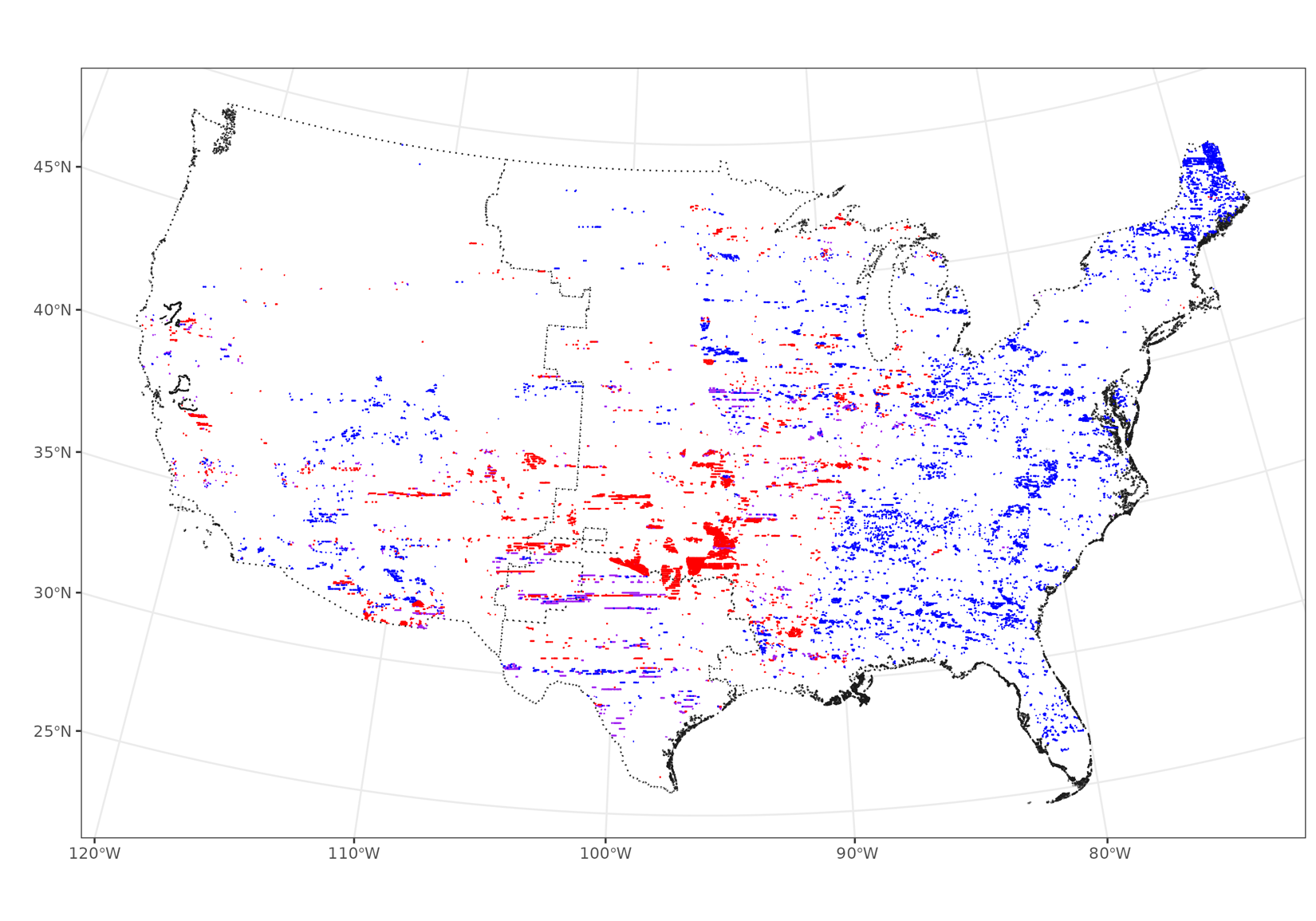} & 
            \includegraphics[width=\linewidth]{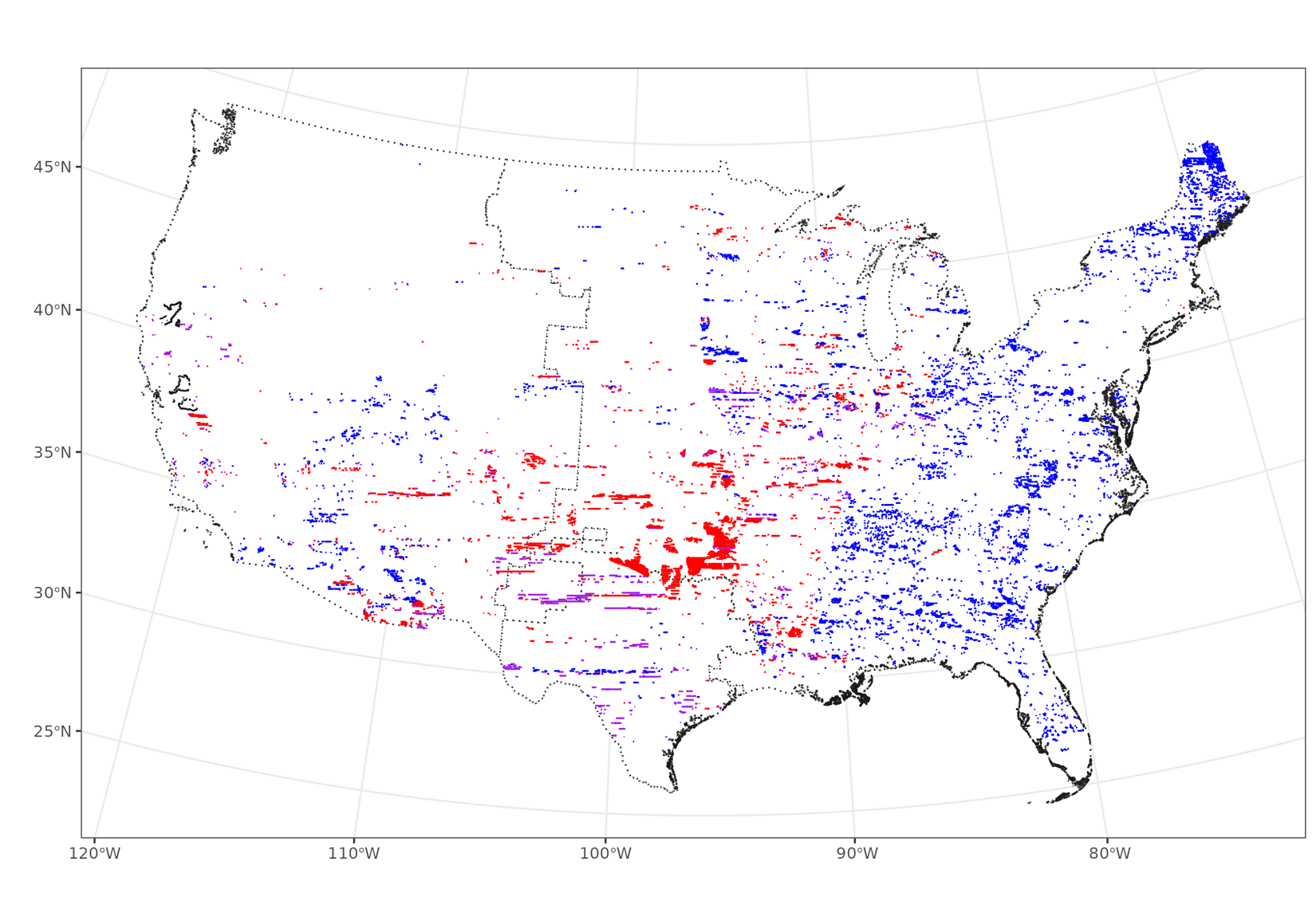} \\
    
            & {\footnotesize 13.6\% SCO} &
            {\footnotesize 13.9\% SCO} &
            {\footnotesize 7.1\% SCO} &
            {\footnotesize 13.2\% SCO} &
            {\footnotesize 14.5\% SCO} \\

            \rotatebox[origin=c]{90}{\large 16-Zone} &
            \includegraphics[width=\linewidth]{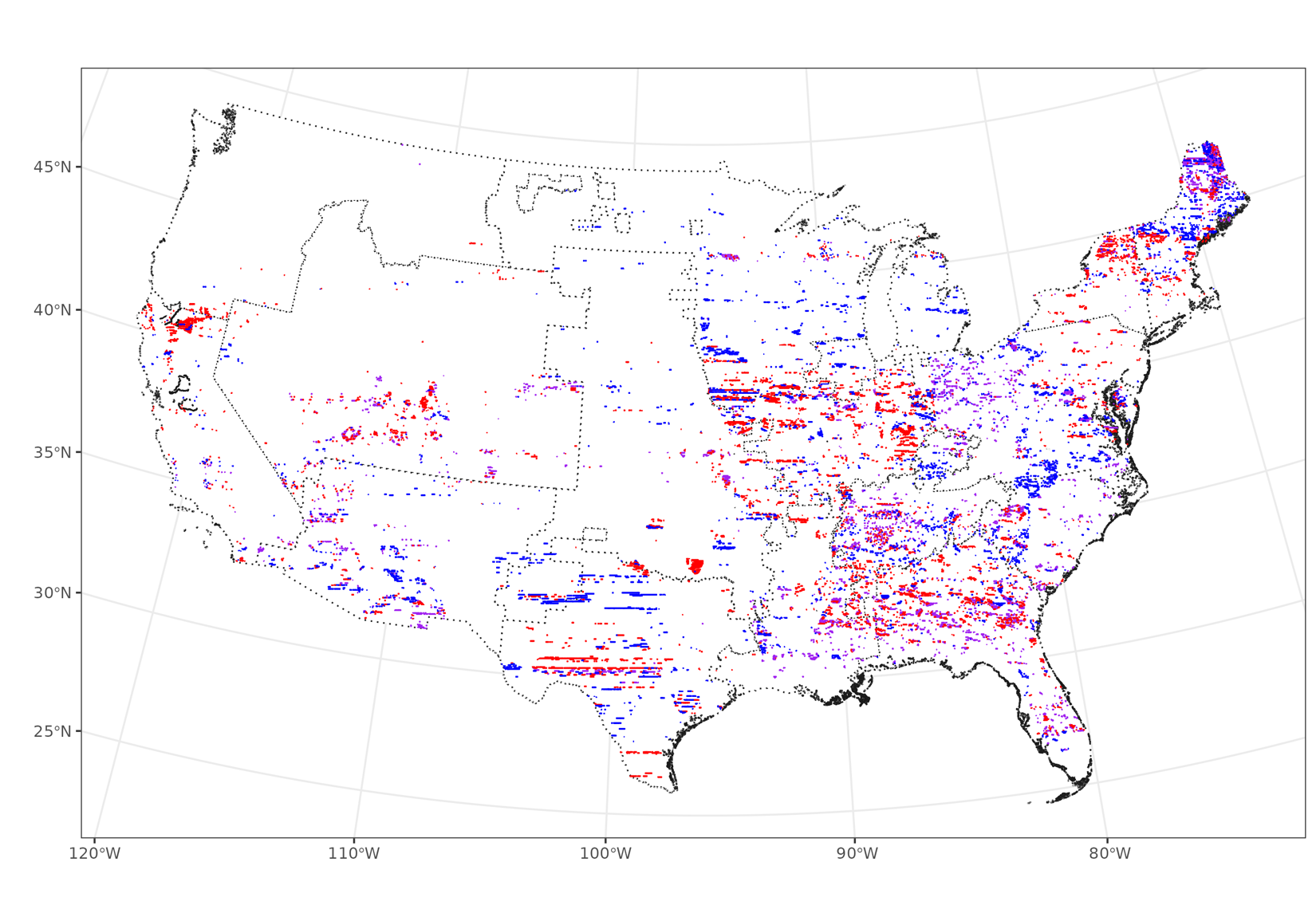} & 
            \includegraphics[width=\linewidth]{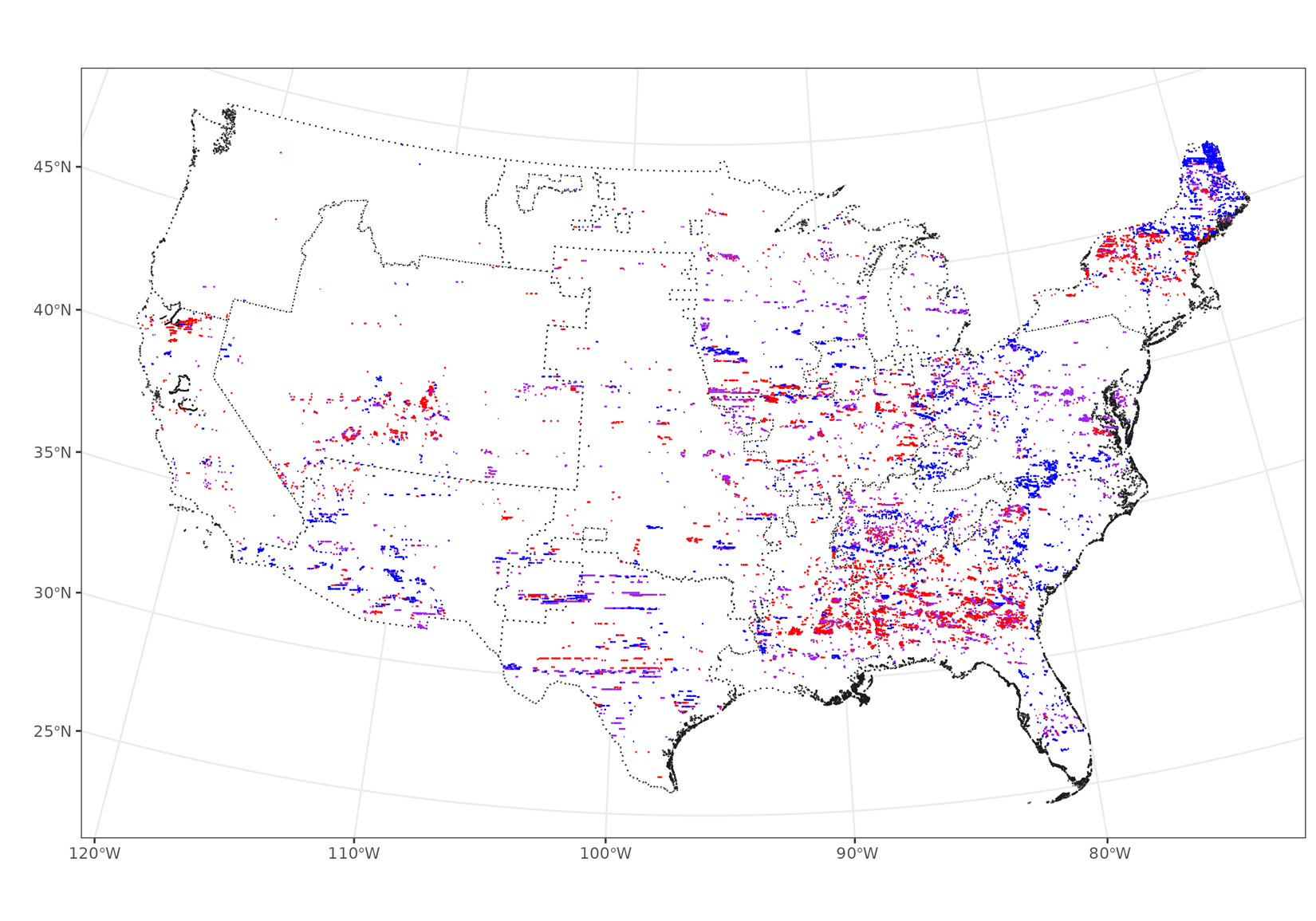} &
            \includegraphics[width=\linewidth]{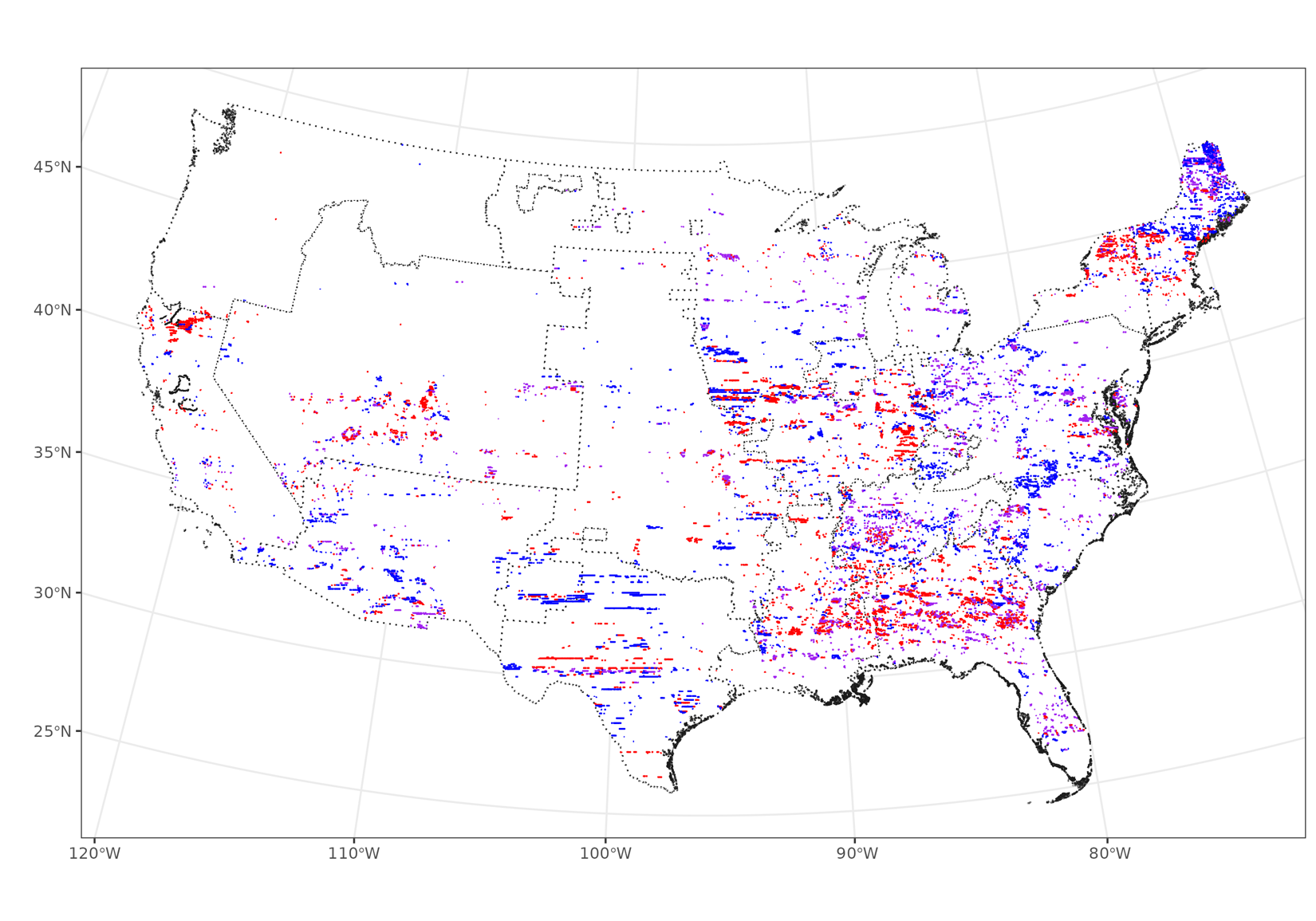} & 
            \includegraphics[width=\linewidth]{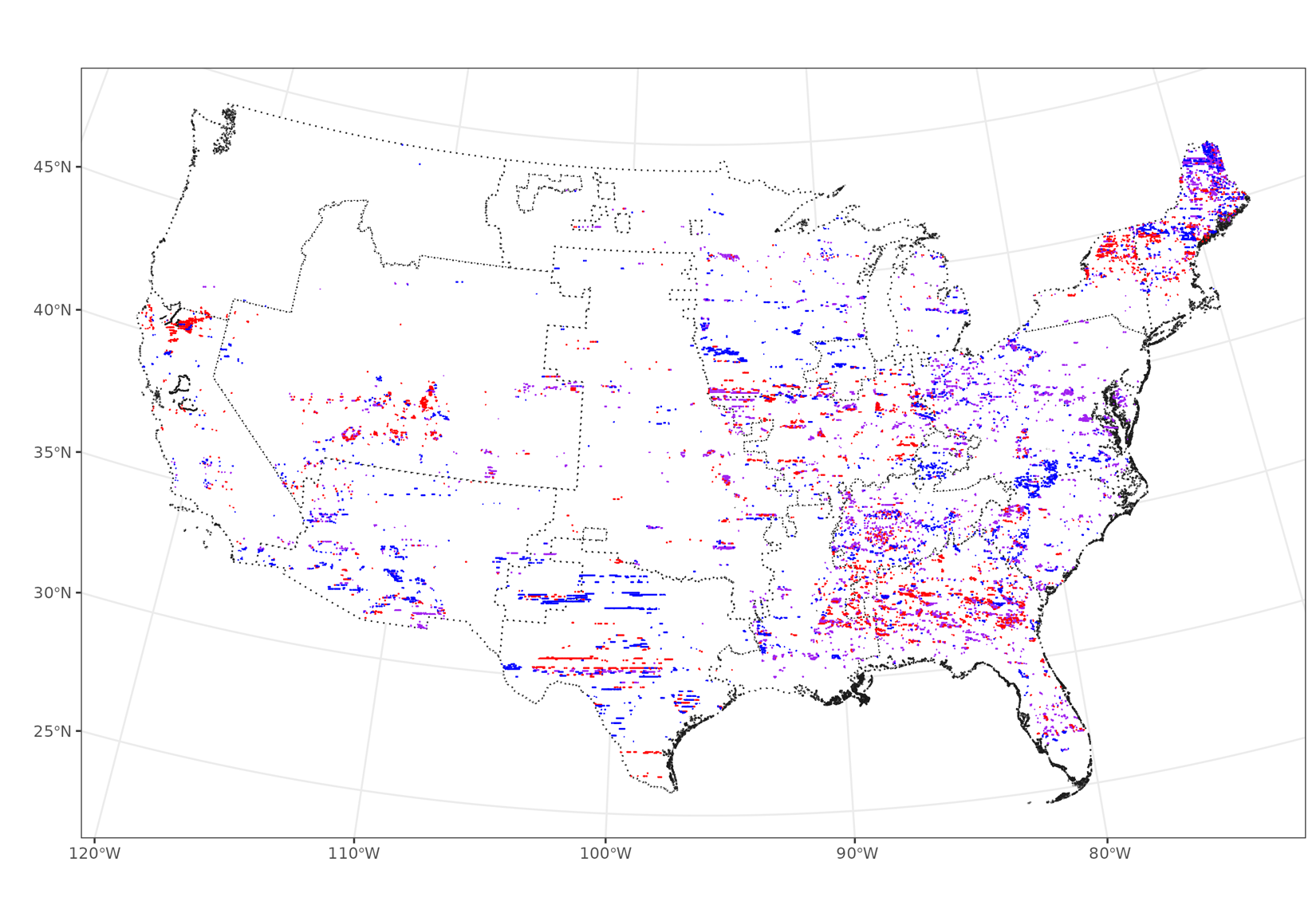} & 
            \includegraphics[width=\linewidth]{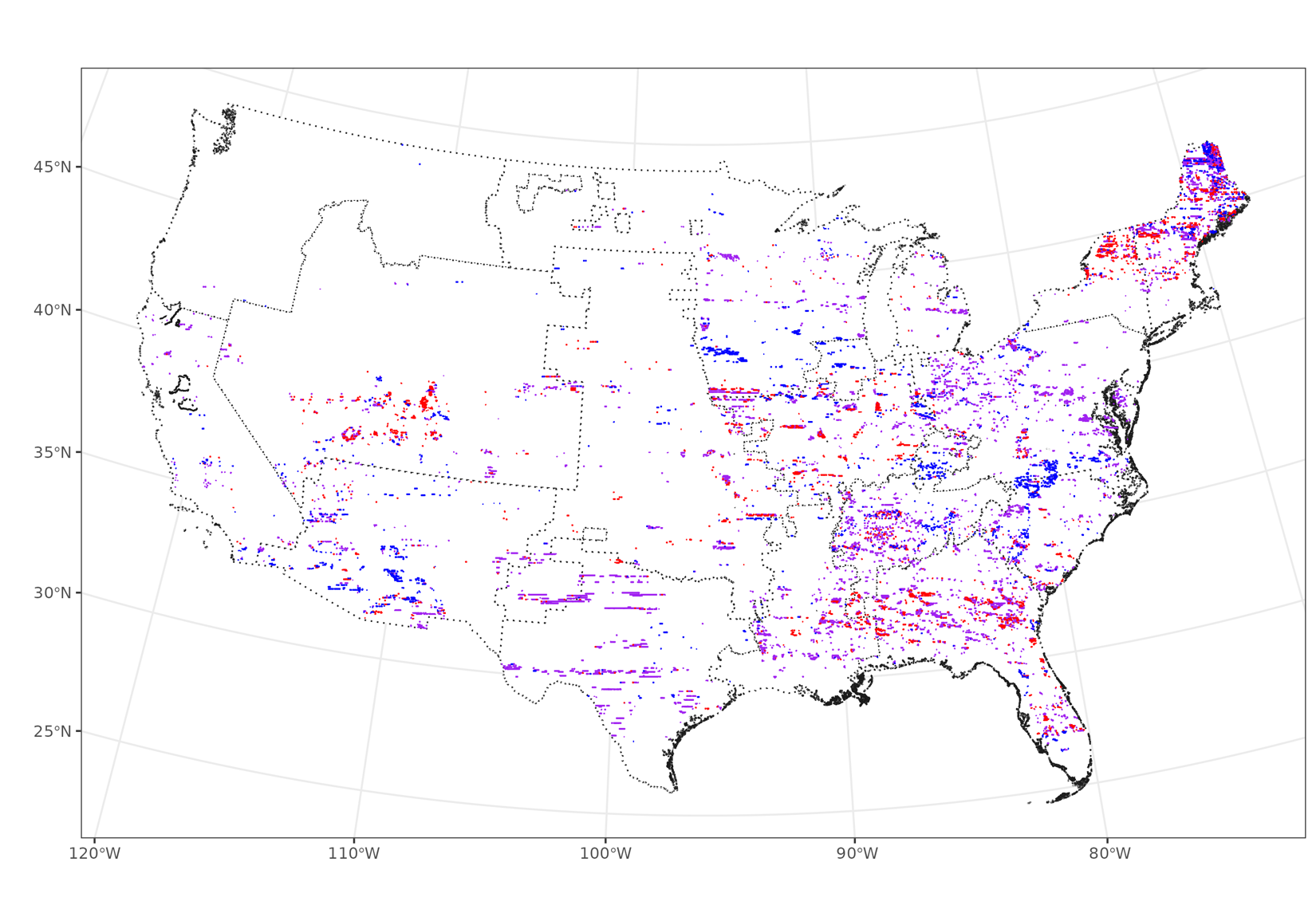} \\
    
            & {\footnotesize 25.8\% SCO} &
            {\footnotesize 35.9\% SCO} &
            {\footnotesize 28.2\% SCO} &
            {\footnotesize 39.3\% SCO} &
            {\footnotesize 59.1\% SCO} \\

            \rotatebox[origin=c]{90}{\large 26-Zone} &
            \includegraphics[width=\linewidth]{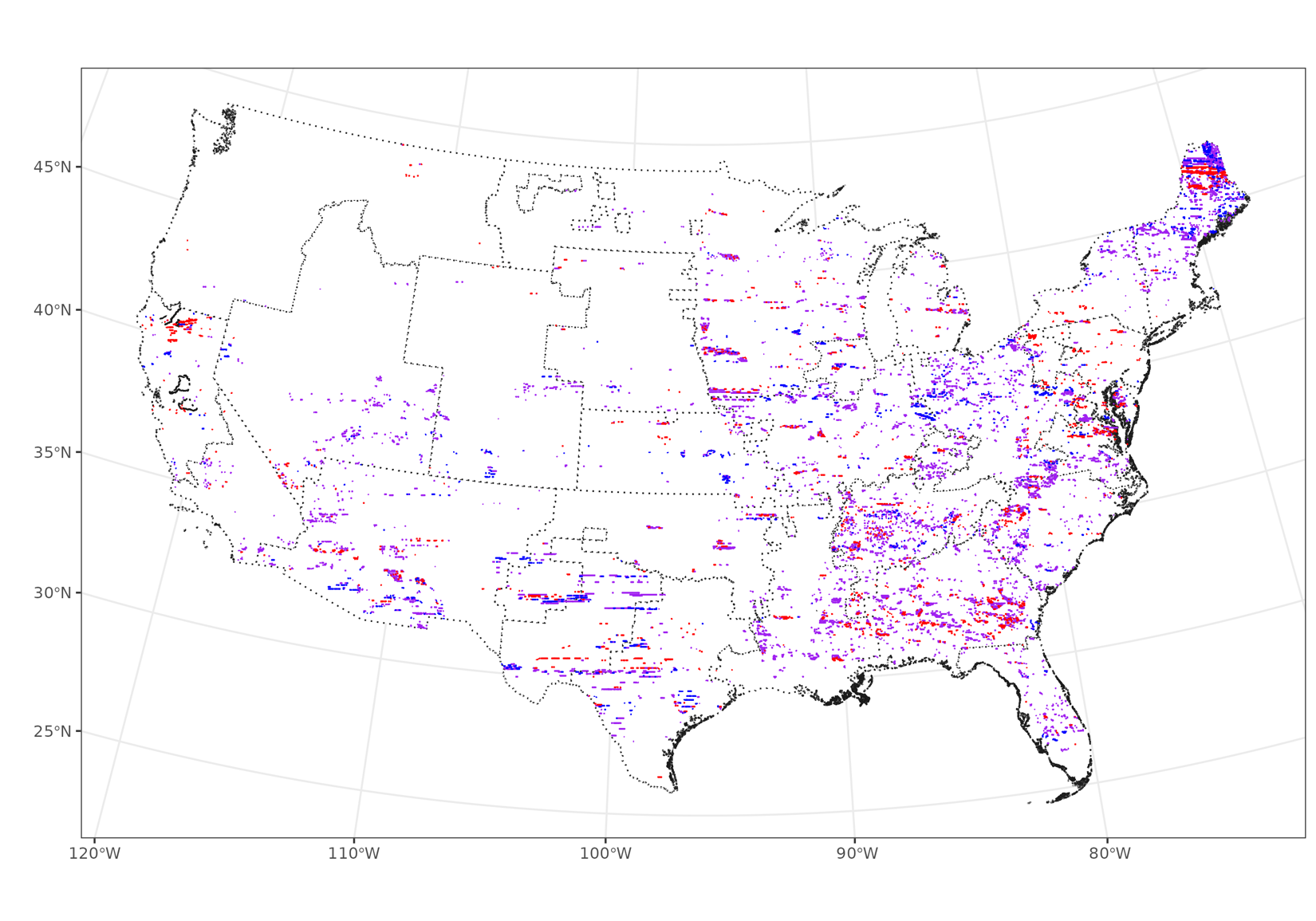} & 
            \includegraphics[width=\linewidth]{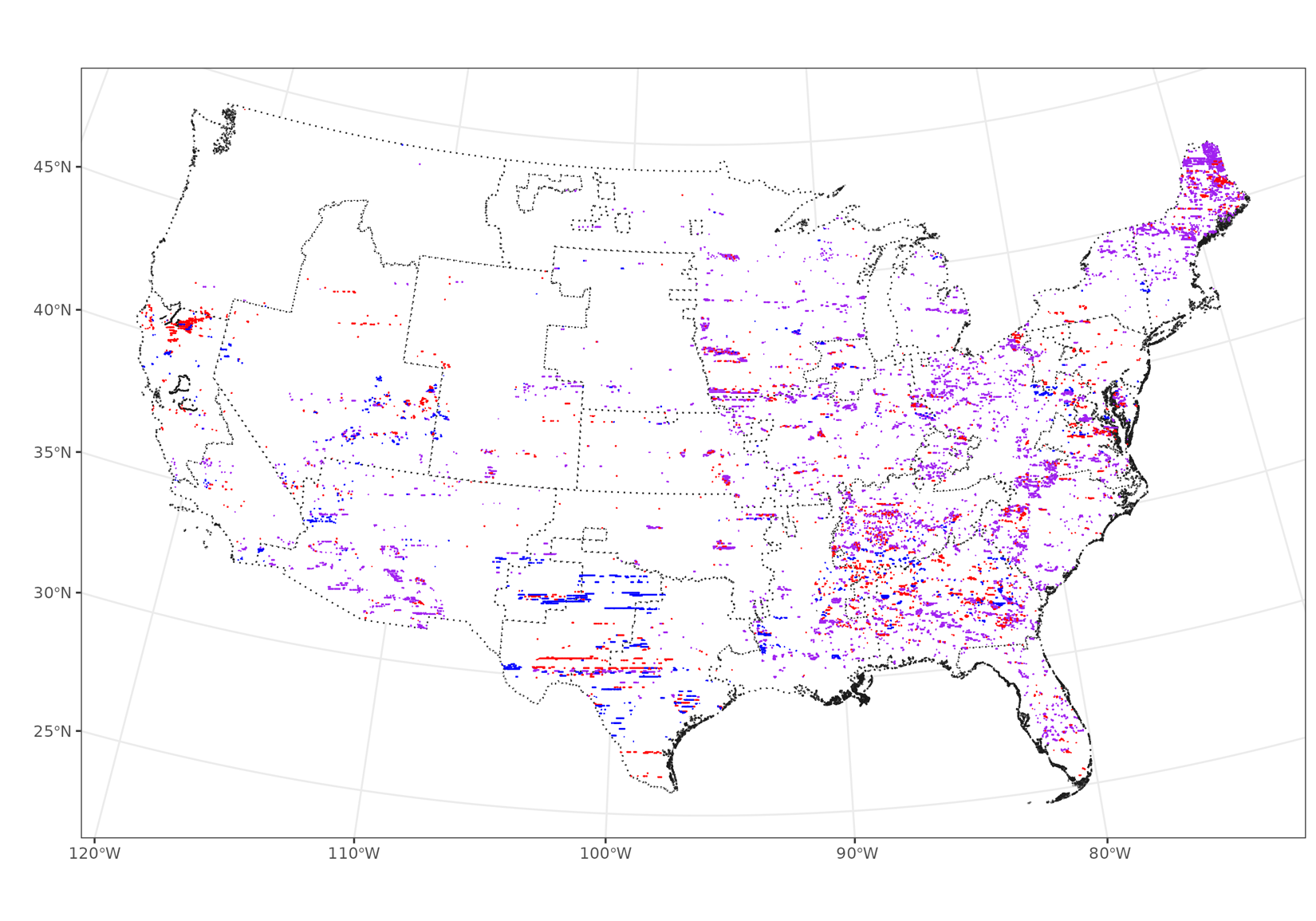} &
            \includegraphics[width=\linewidth]{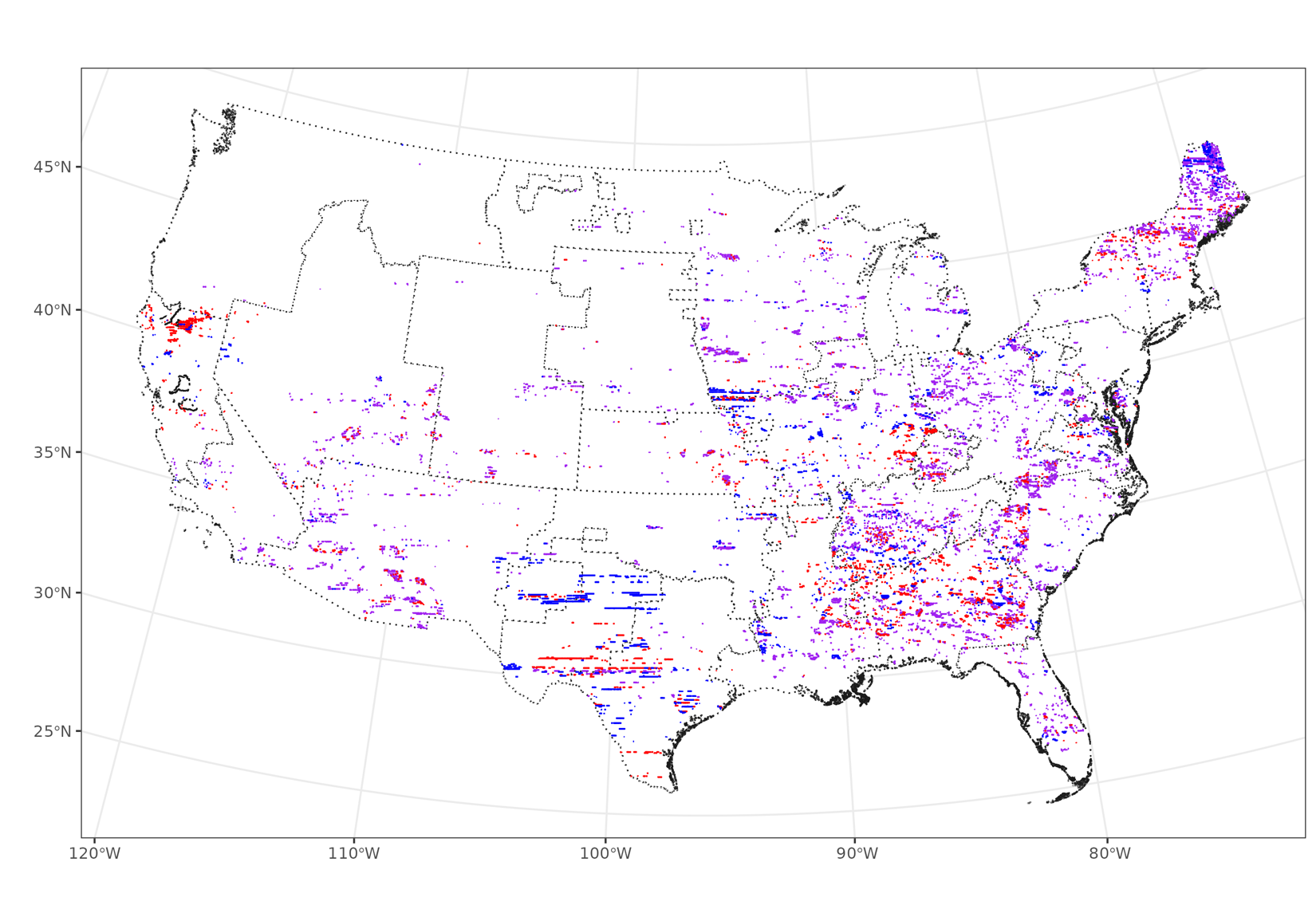} & 
            \includegraphics[width=\linewidth]{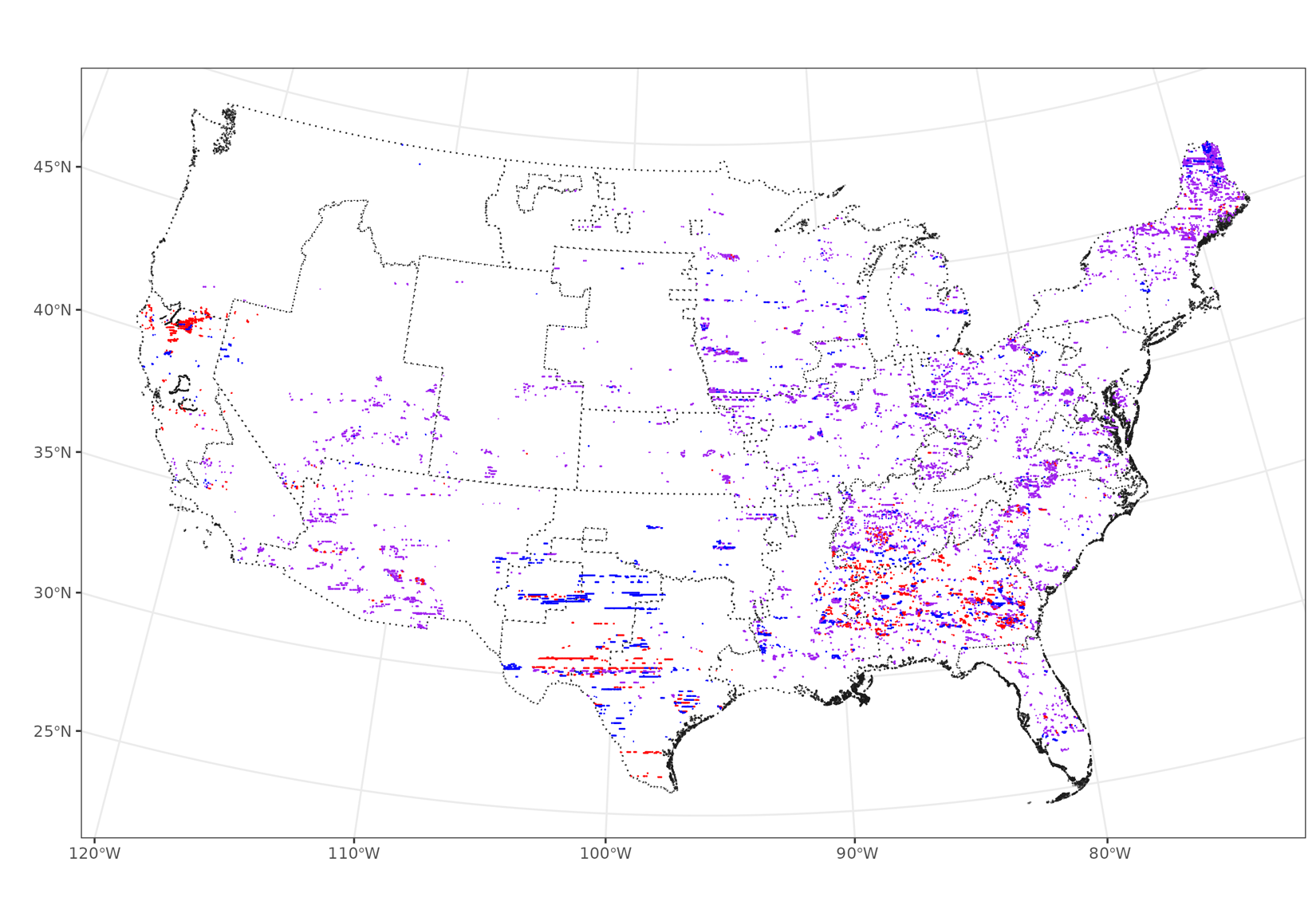} & 
            \includegraphics[width=\linewidth]{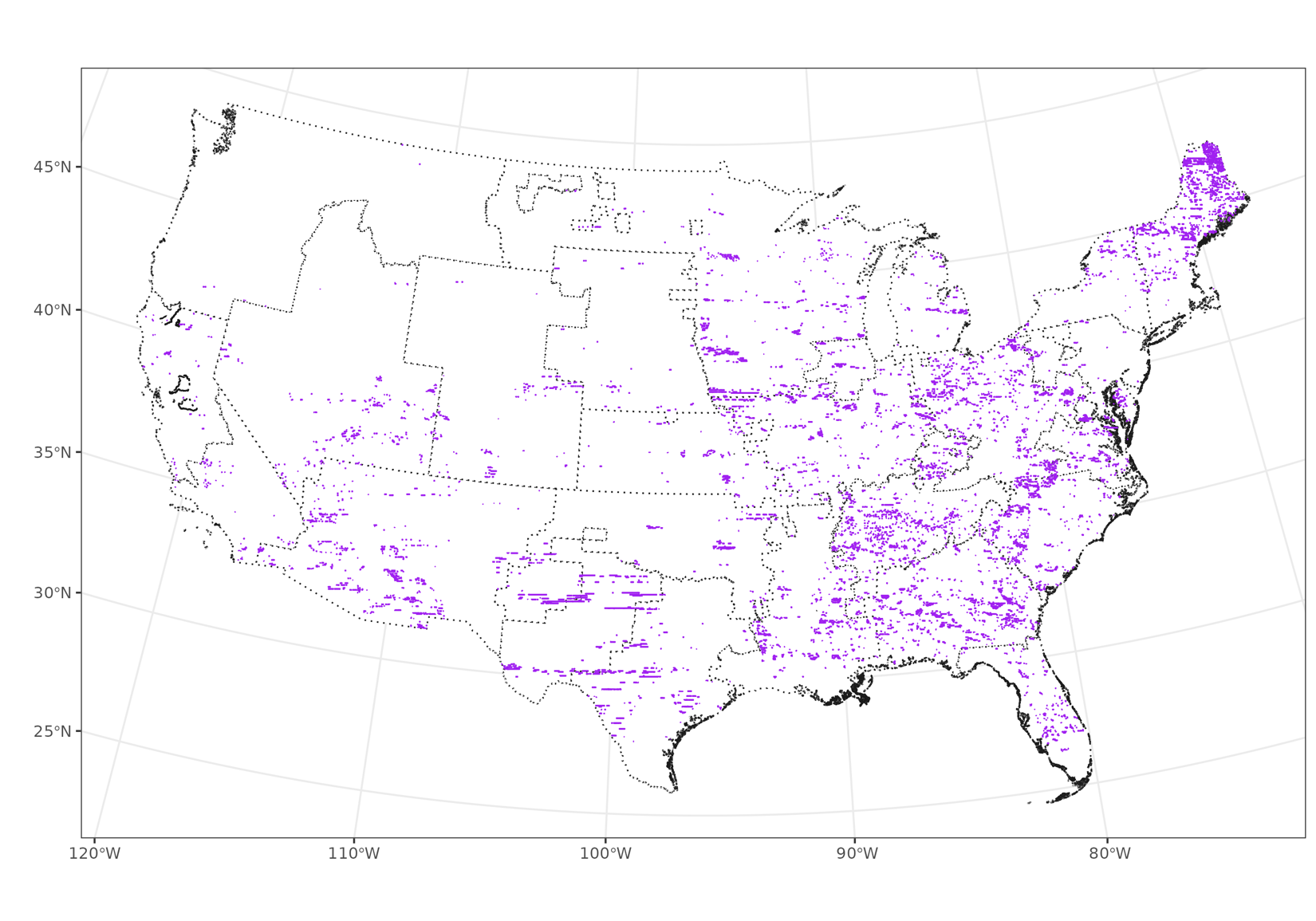} \\
        
            & {\footnotesize 67.1\% SCO} &
            {\footnotesize 61.8\% SCO} &
            {\footnotesize 57.1\% SCO} &
            {\footnotesize 64.2\% SCO} &
            {\footnotesize 100\% SCO} \\
    
        \end{tabular}
        \captionof{figure}{Difference plot for \textbf{solar capacity} by case. \Coc case. Sites selected by coarse cases are highlighted in red, sites selected by the highest resolution baseline (HRB) case are highlighted in blue. Overlapping sites are highlighted in purple. Let \textit{site capacity overlap} (SCO) be the percentage of capacity that is invested in by both the HRB and coarse-resolution case. Cases with increased spatial resolution and temporal resolution are better able to recapitulate siting results.}
        \label{map_solar}
    \end{table}

    \begin{table}[p]
        \centering
        \setlength\tabcolsep{0pt}
        \begin{tabular}{cM{0.195\linewidth}M{0.195\linewidth}M{0.195\linewidth}M{0.195\linewidth}M{0.195\linewidth}}
            & \large 15-day & \large 30-day & \large 10-week & \large 30-week & \large 52-week \\
            
            \rotatebox[origin=c]{90}{\large 3-Zone} &
            \includegraphics[width=\linewidth]{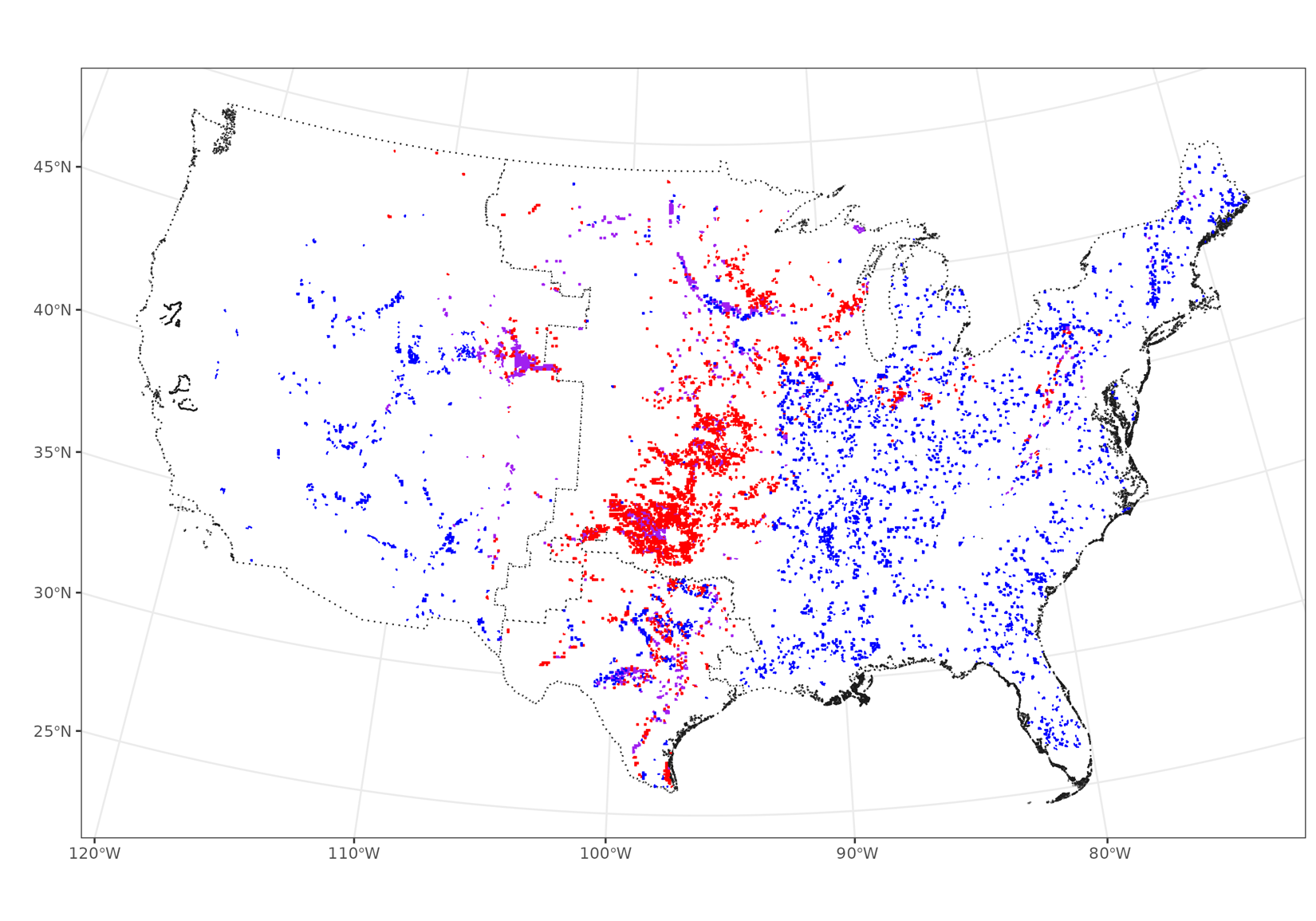} & 
            \includegraphics[width=\linewidth]{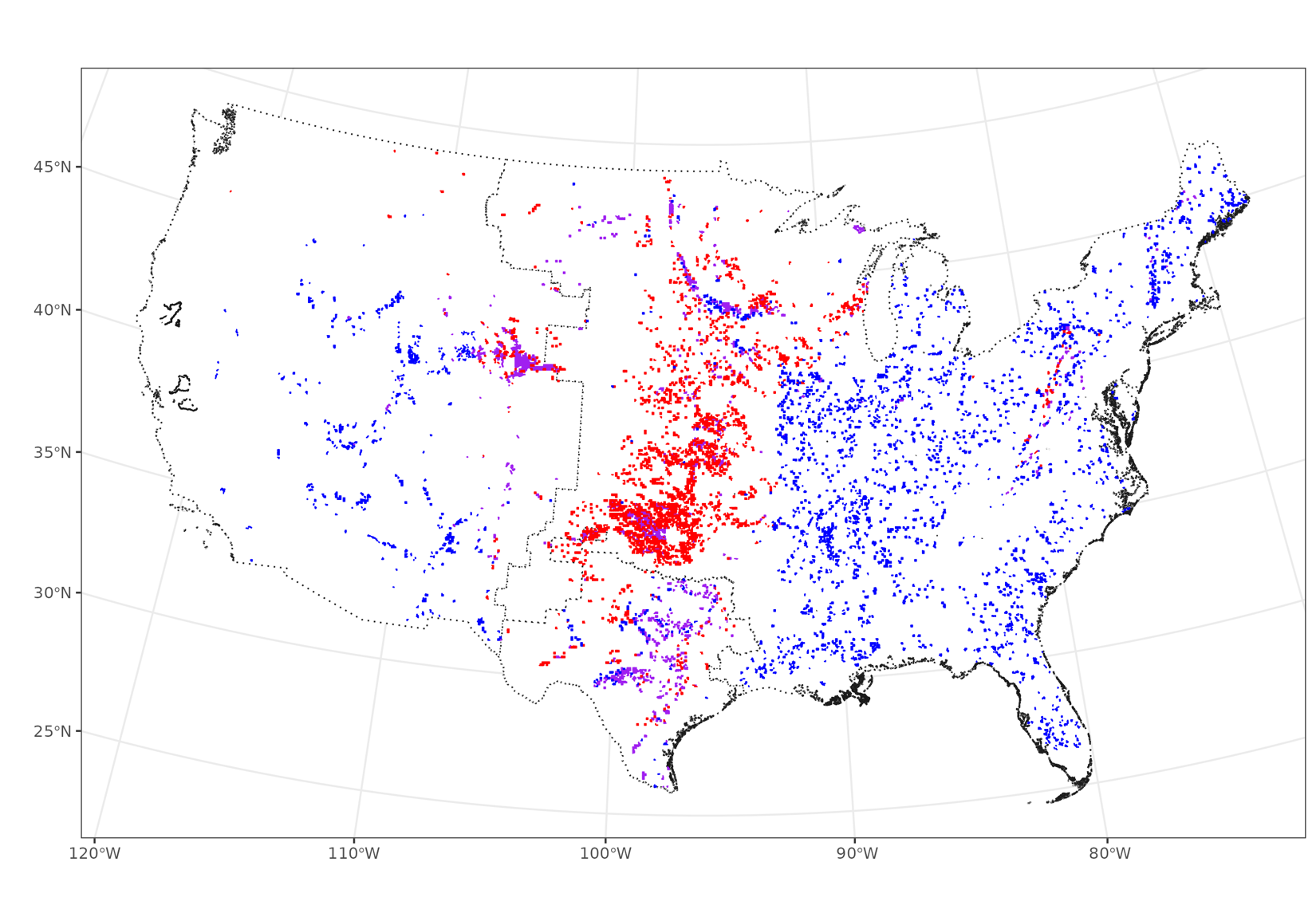} &
            \includegraphics[width=\linewidth]{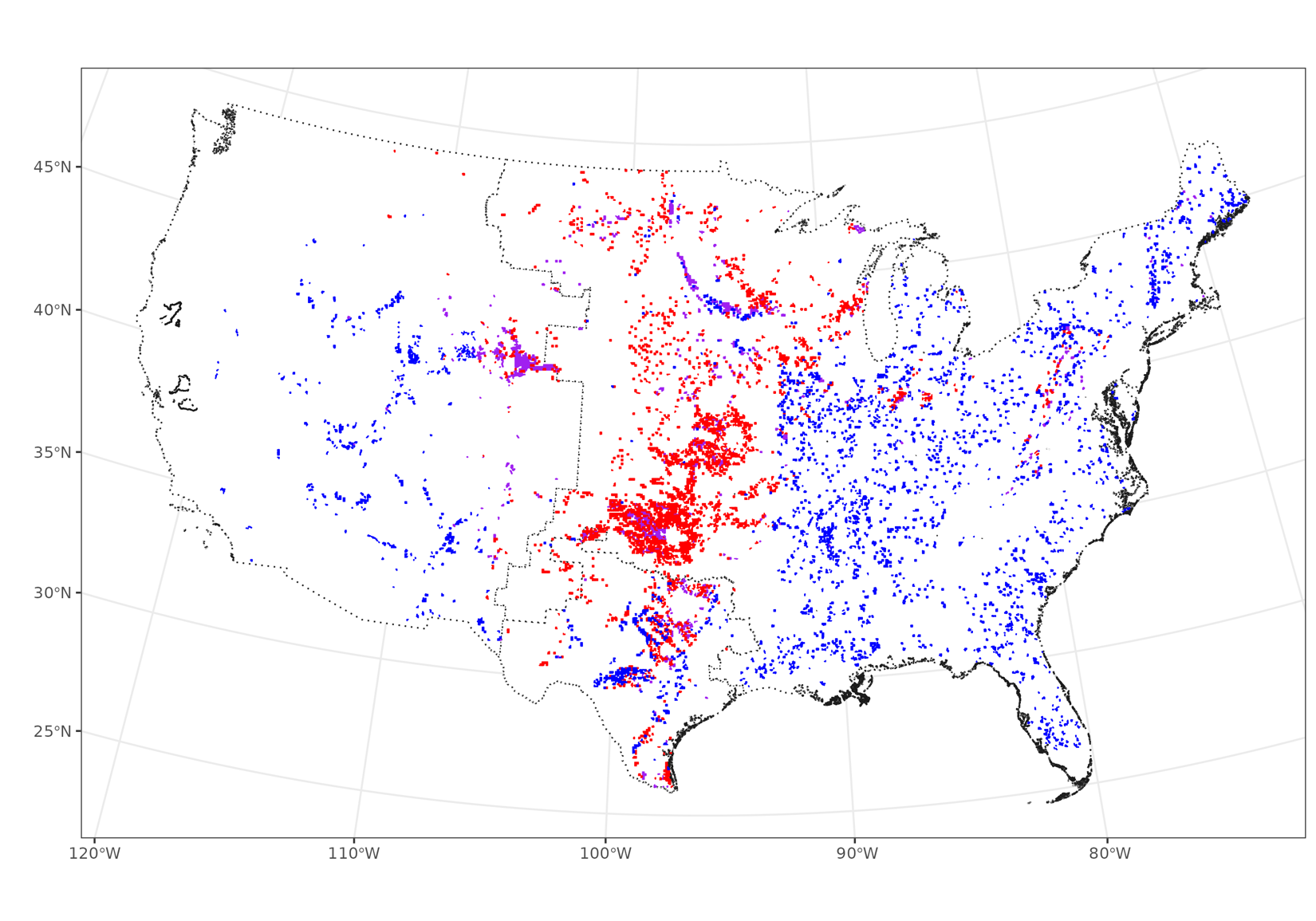} & 
            \includegraphics[width=\linewidth]{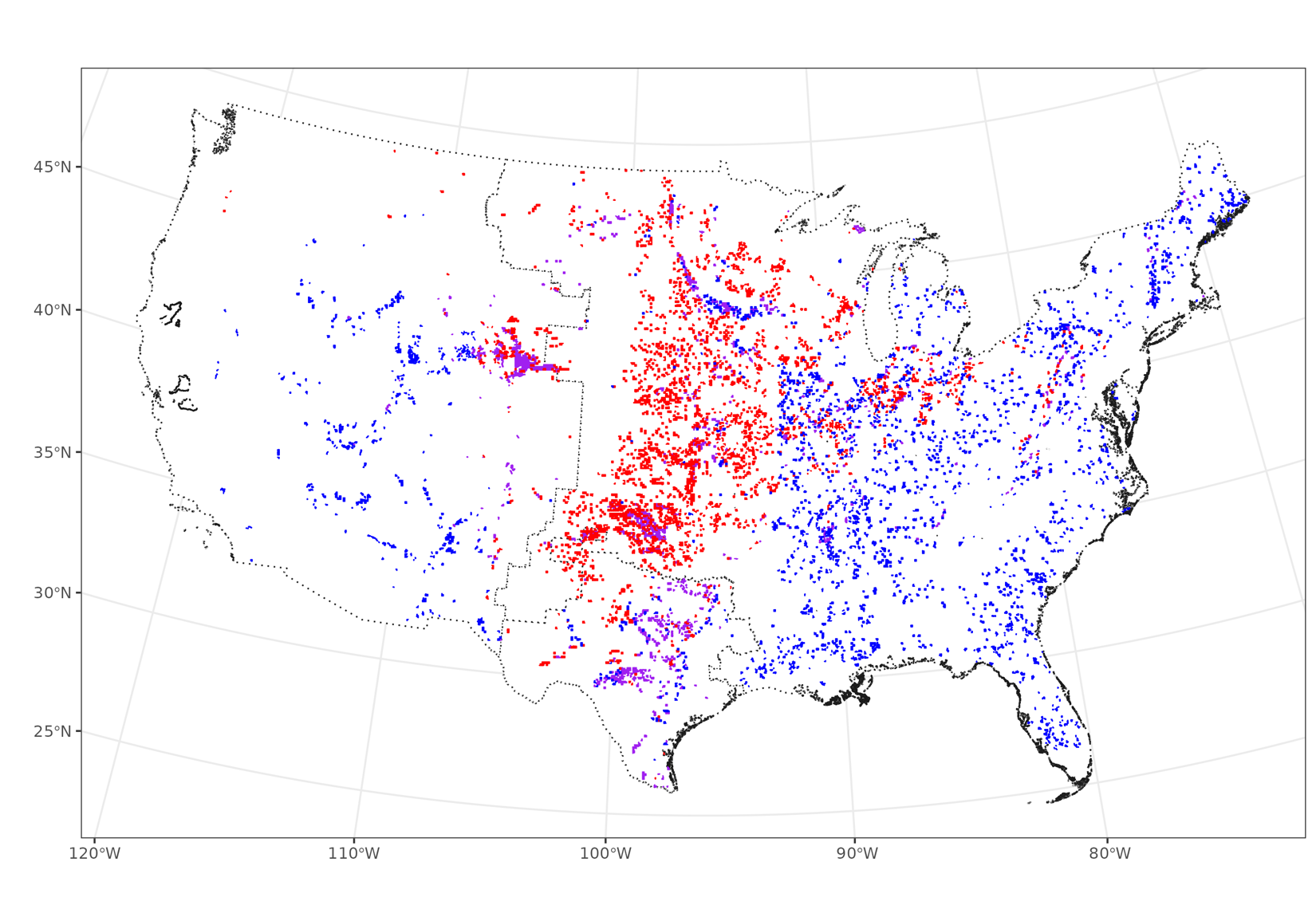} & 
            \includegraphics[width=\linewidth]{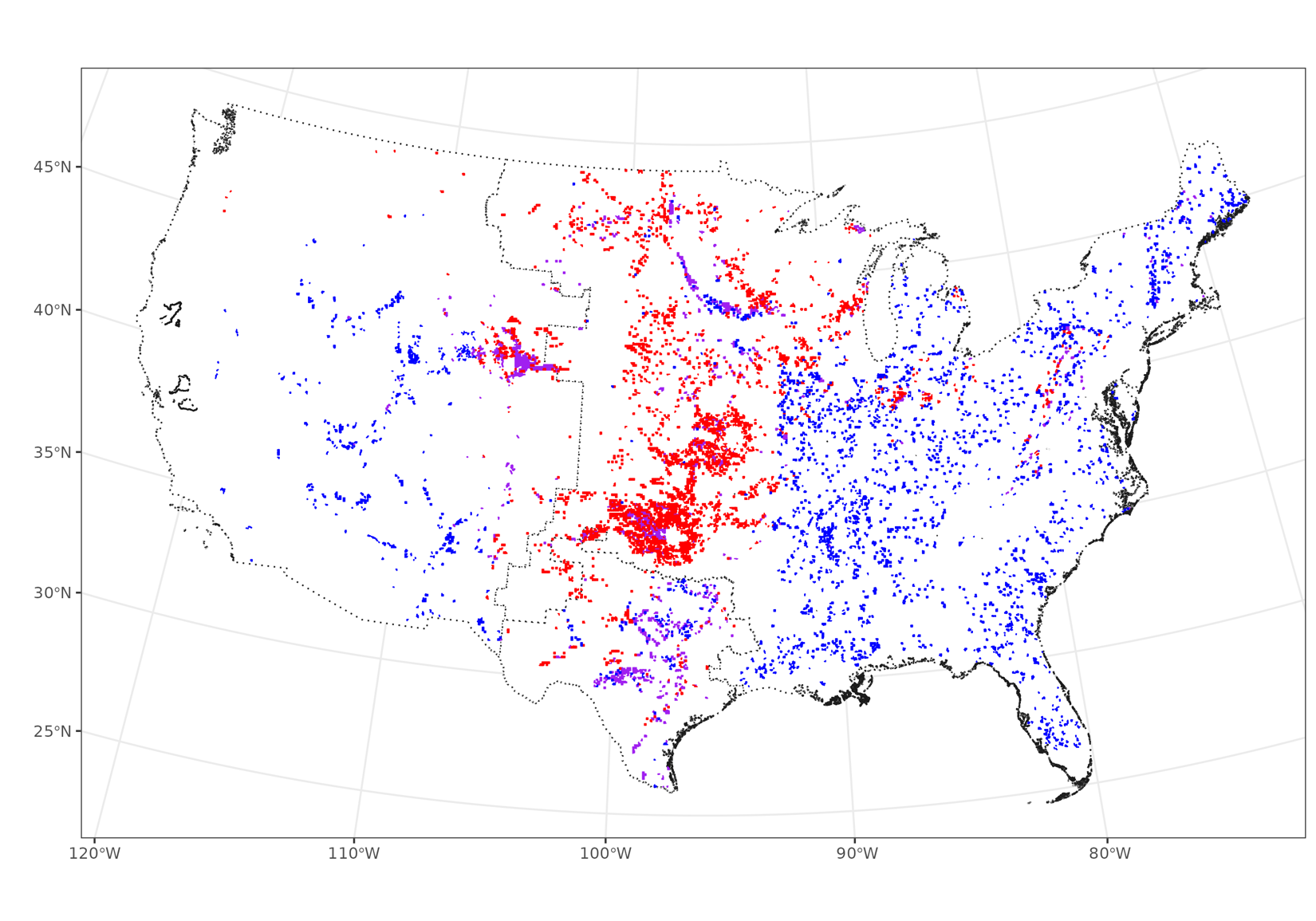} \\
    
            & {\footnotesize 17.1\% SCO} &
            {\footnotesize 19.1\% SCO} &
            {\footnotesize 14.6\% SCO} &
            {\footnotesize 18.5\% SCO} &
            {\footnotesize 17.8\% SCO} \\

            \rotatebox[origin=c]{90}{\large 16-Zone} &
            \includegraphics[width=\linewidth]{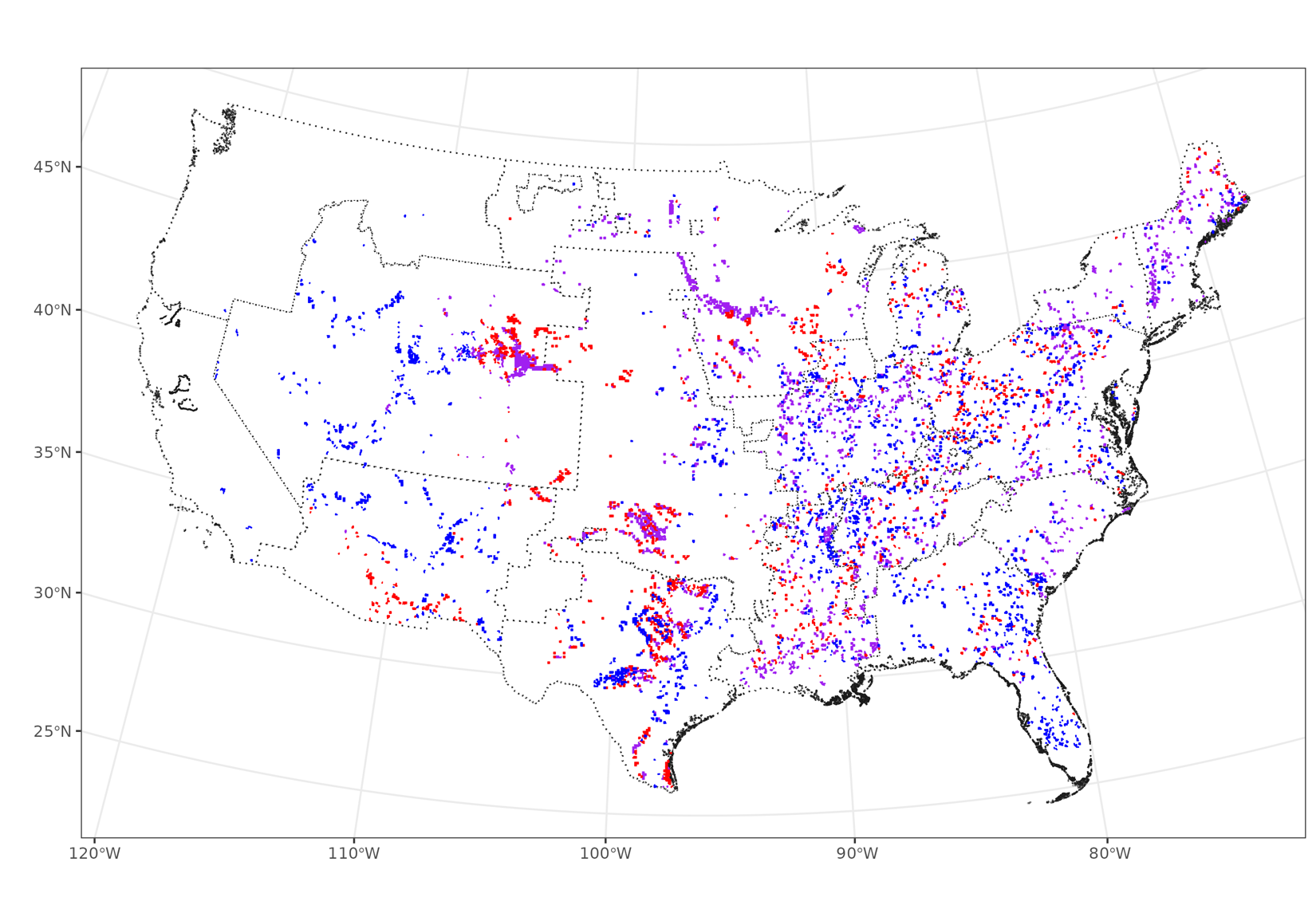} & 
            \includegraphics[width=\linewidth]{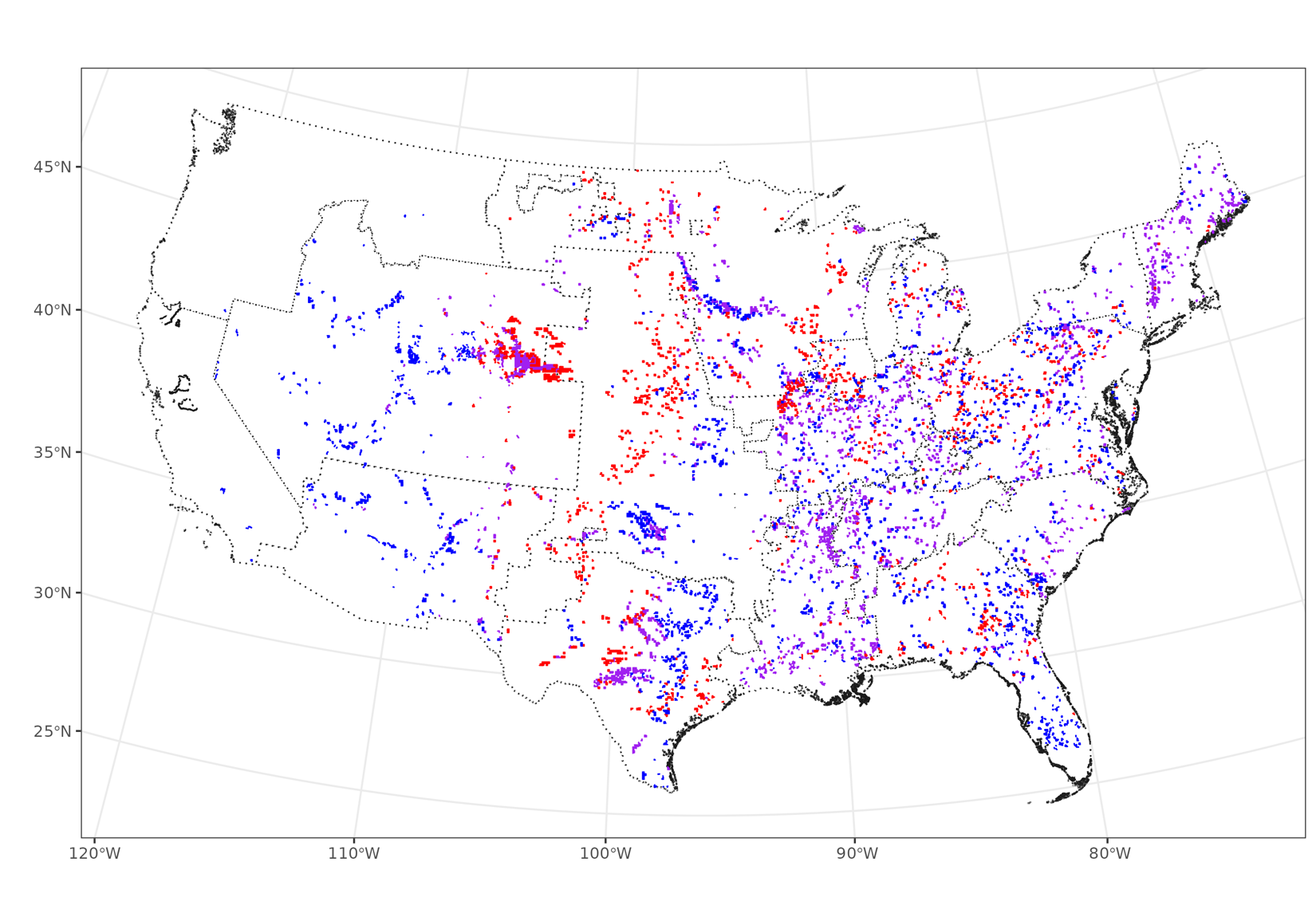} &
            \includegraphics[width=\linewidth]{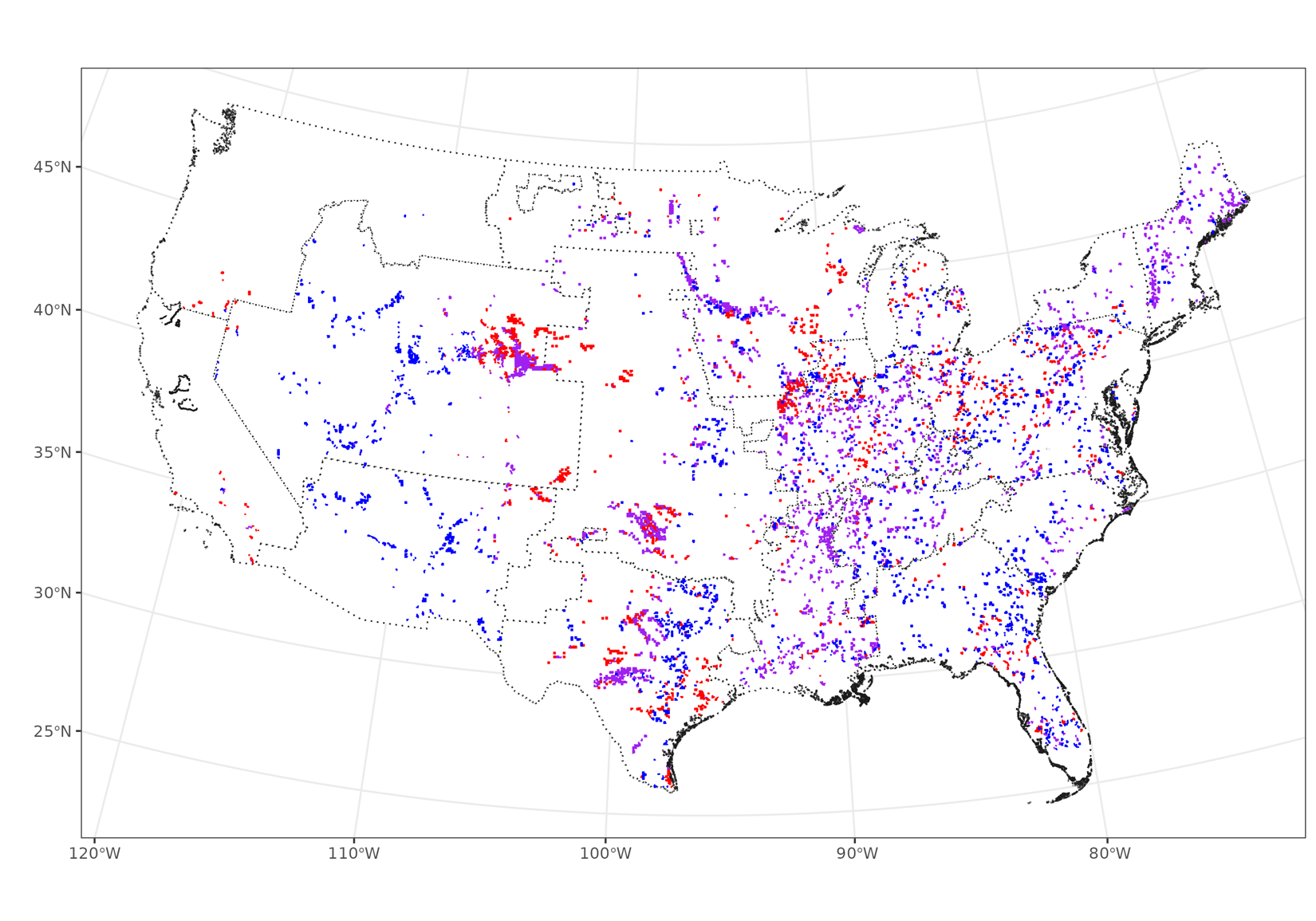} & 
            \includegraphics[width=\linewidth]{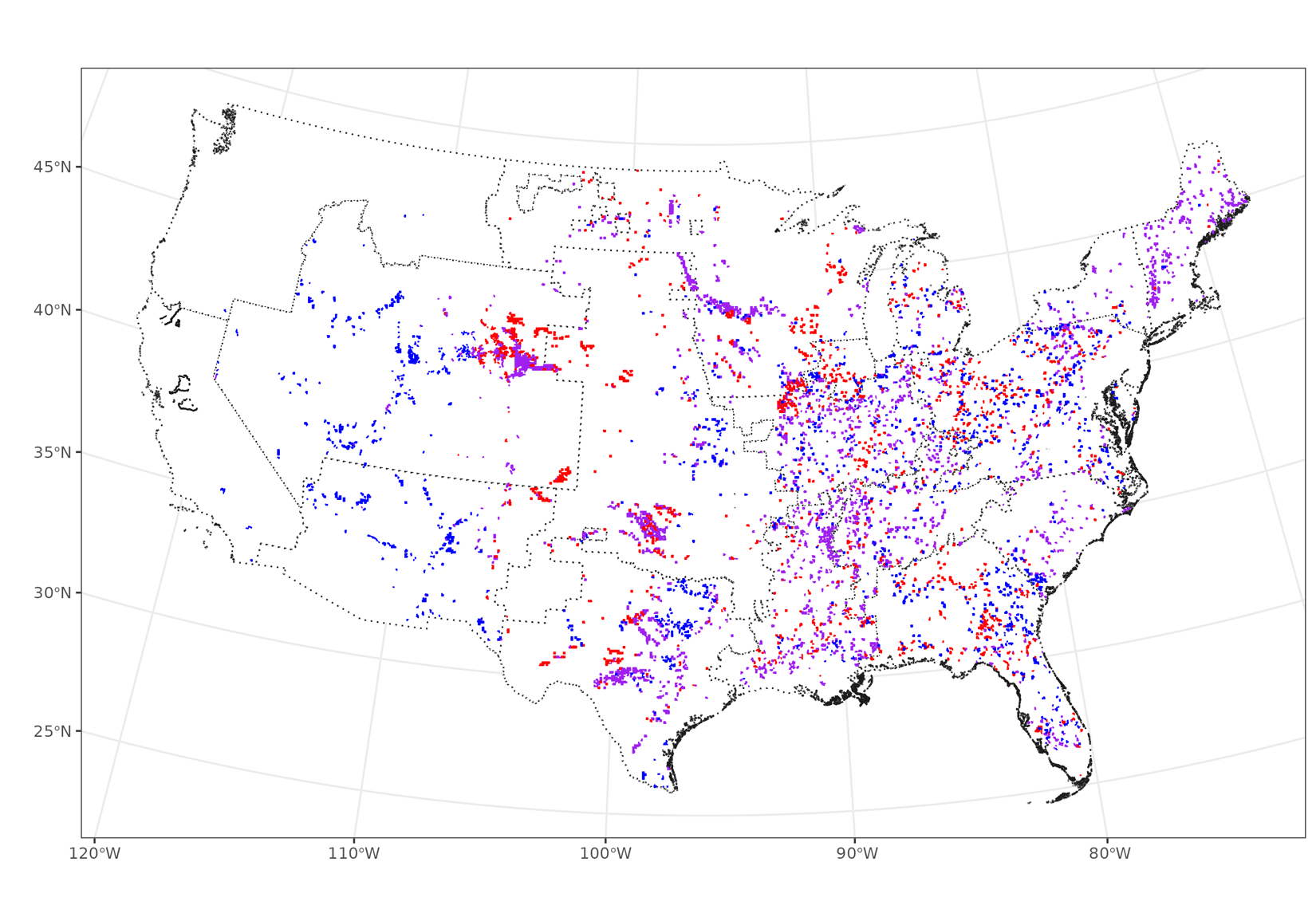} & 
            \includegraphics[width=\linewidth]{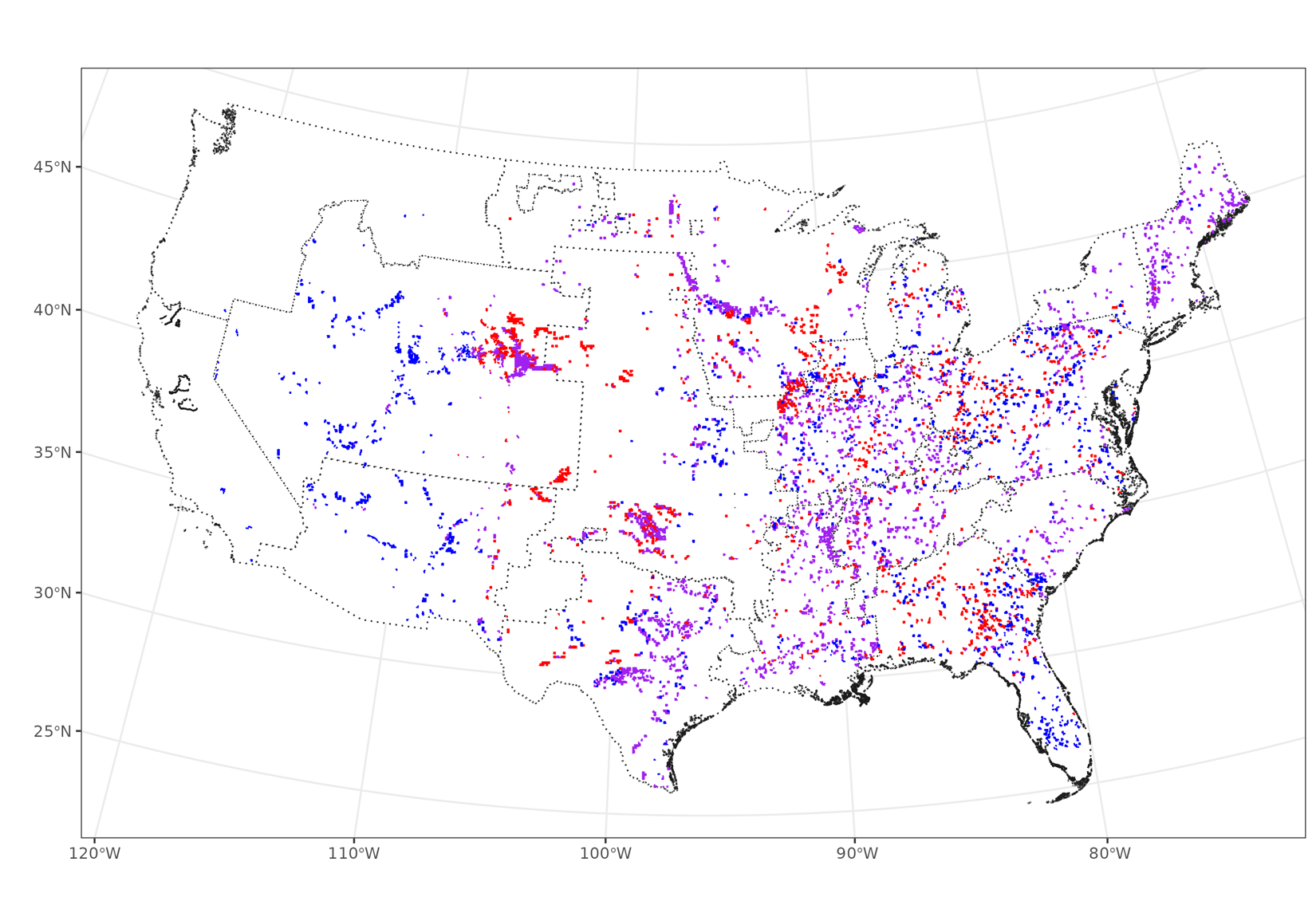} \\
    
            & {\footnotesize 37.1\% SCO} &
            {\footnotesize 37\% SCO} &
            {\footnotesize 42.9\% SCO} &
            {\footnotesize 46.2\% SCO} &
            {\footnotesize 48.2\% SCO} \\

            \rotatebox[origin=c]{90}{\large 26-Zone} &
            \includegraphics[width=\linewidth]{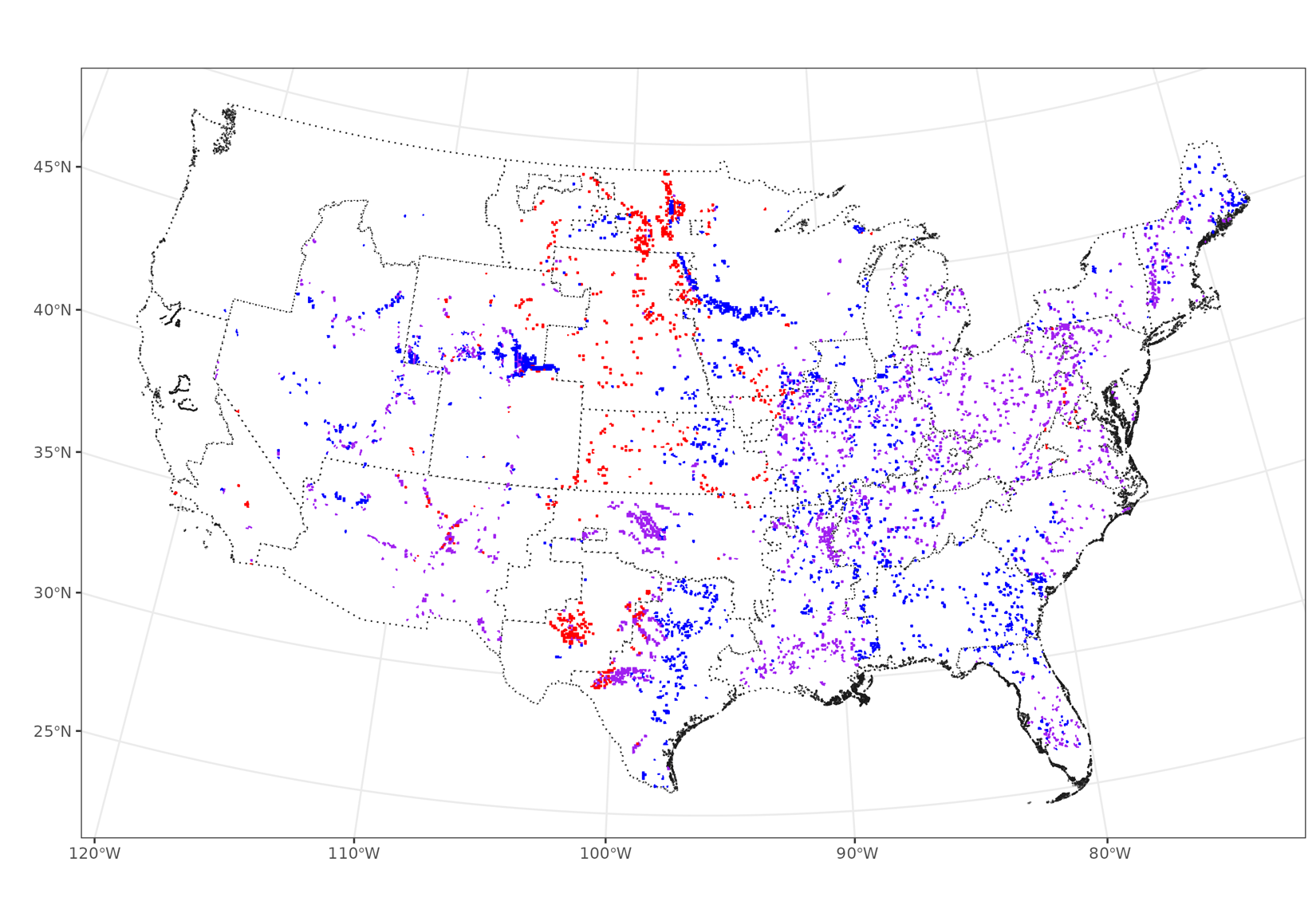} & 
            \includegraphics[width=\linewidth]{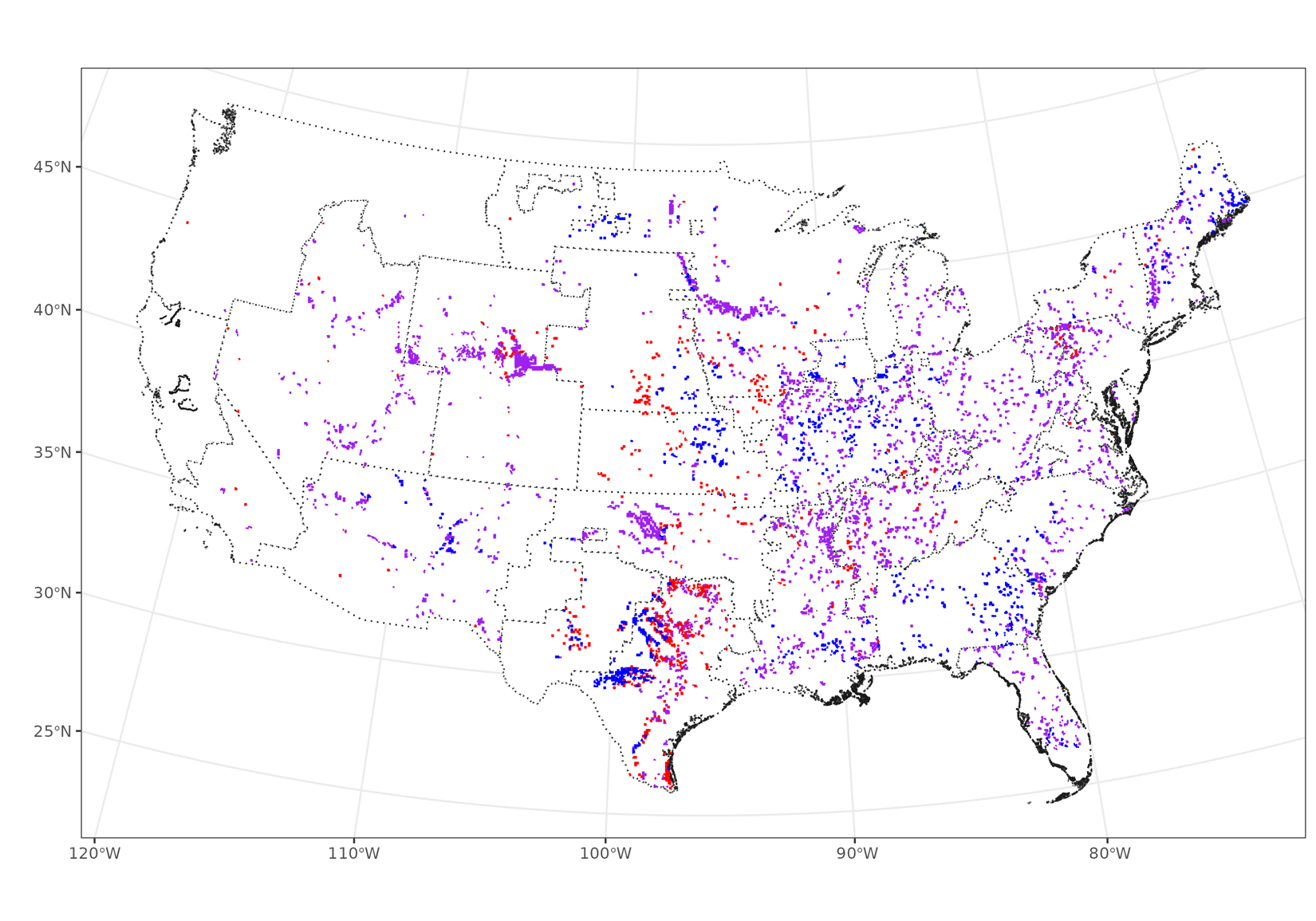} &
            \includegraphics[width=\linewidth]{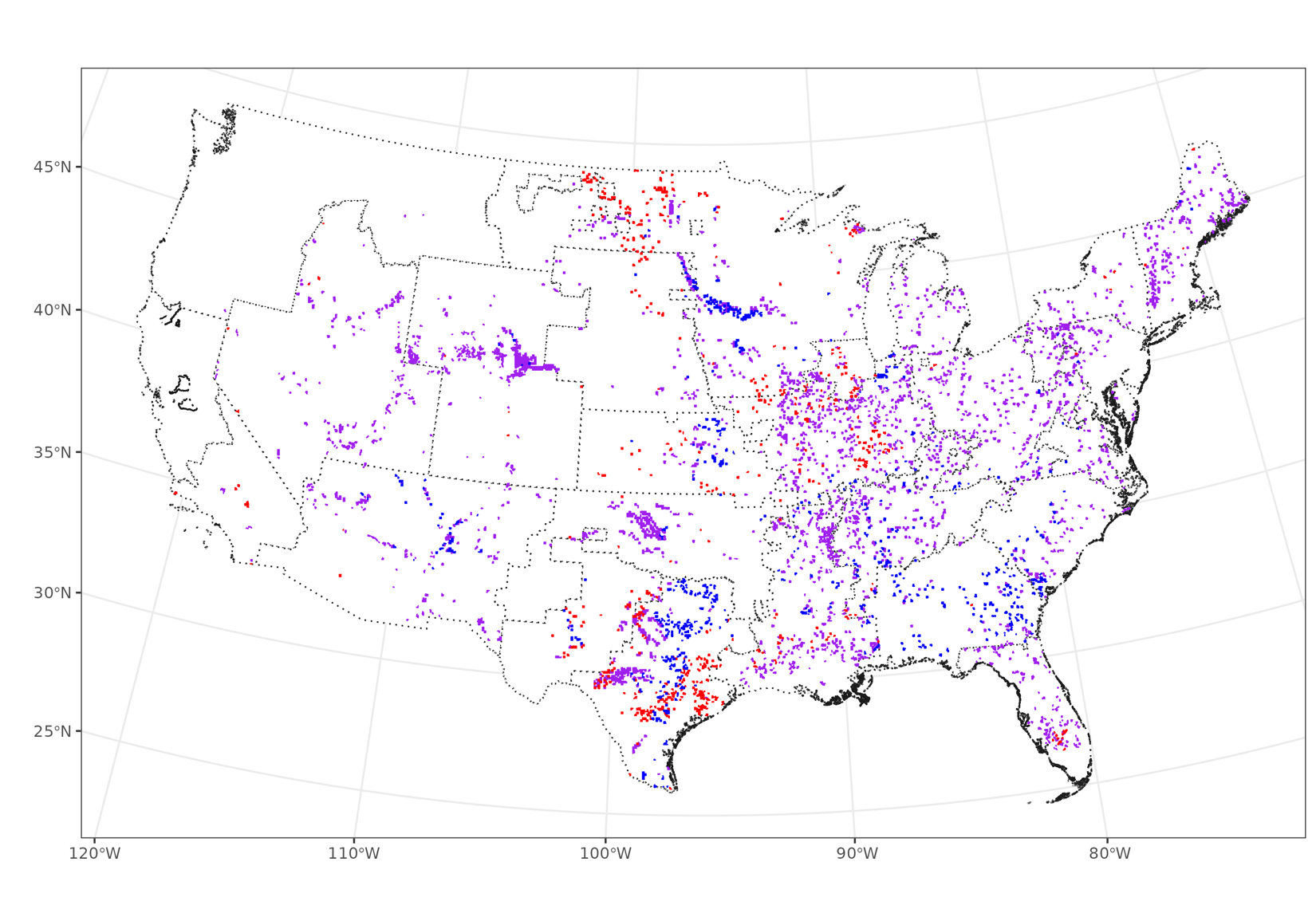} & 
            \includegraphics[width=\linewidth]{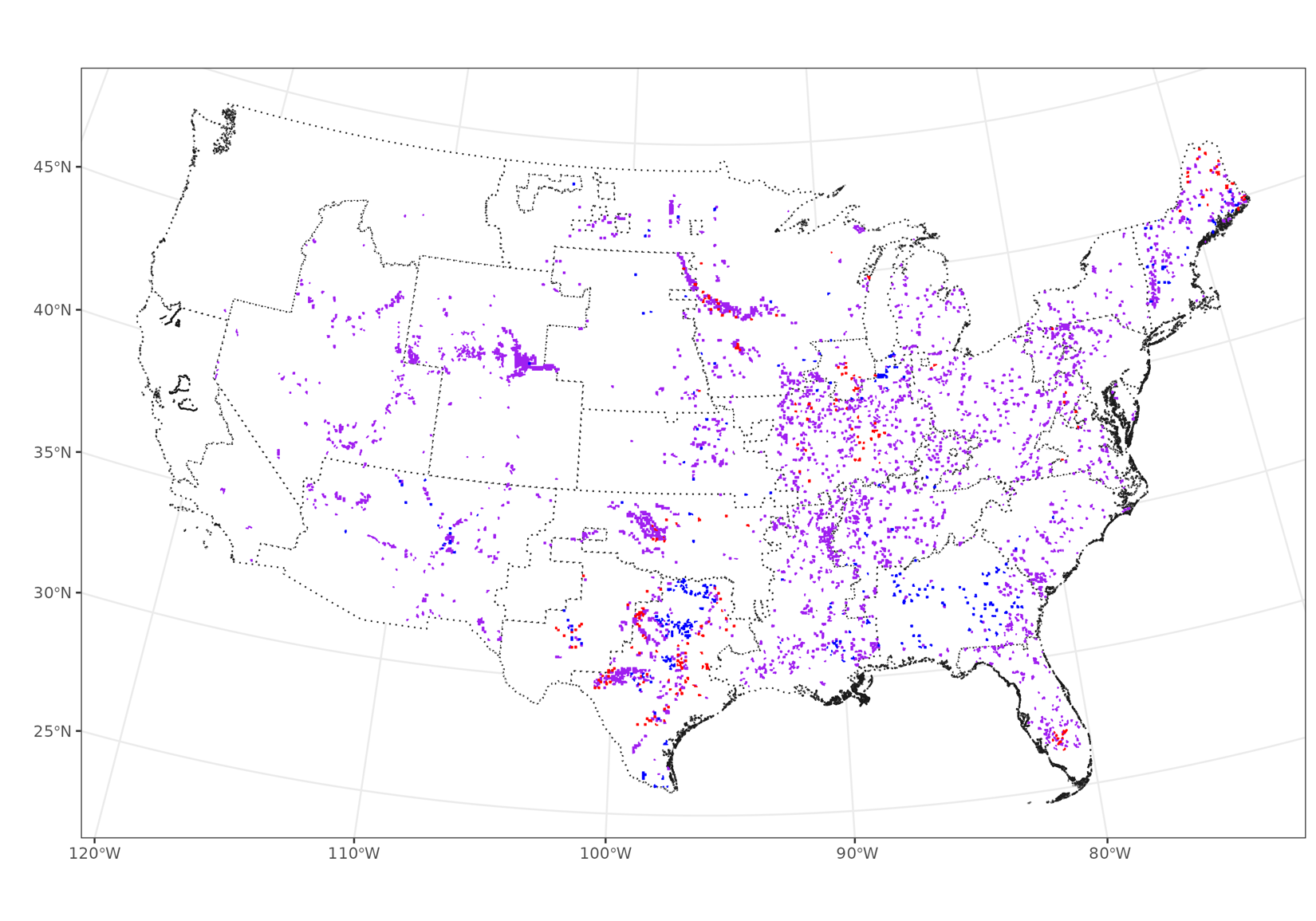} & 
            \includegraphics[width=\linewidth]{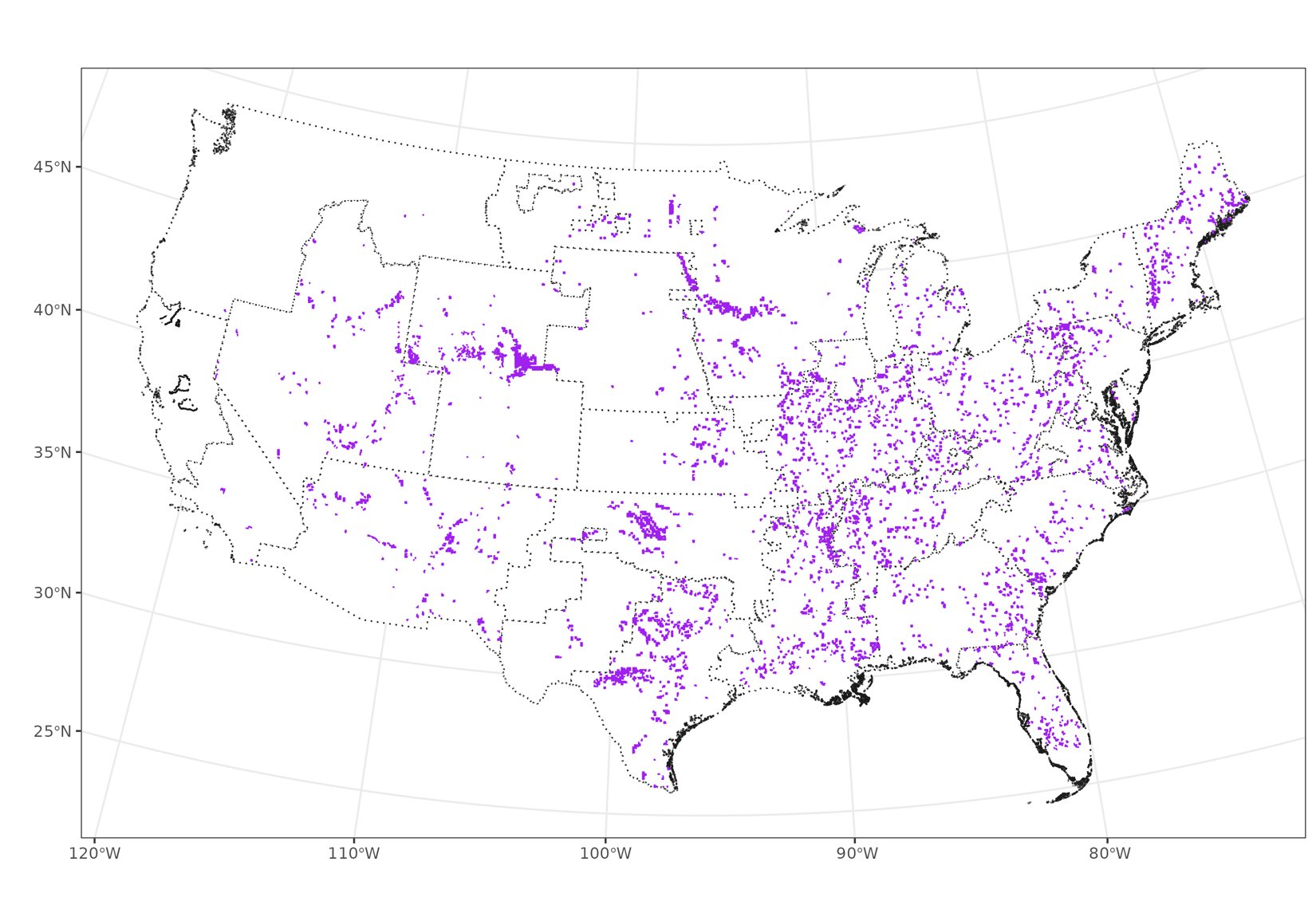} \\
        
            & {\footnotesize 43.8\% SCO} &
            {\footnotesize 63.7\% SCO} &
            {\footnotesize 70\% SCO} &
            {\footnotesize 85.7\% SCO} &
            {\footnotesize 100\% SCO} \\
    
        \end{tabular}
        \captionof{figure}{Difference plot for \textbf{onshore wind capacity} by case. \Coc case. Sites selected by coarse cases are highlighted in red, sites selected by the highest resolution baseline (HRB) case are highlighted in blue. Overlapping sites are highlighted in purple. Let \textit{site capacity overlap} (SCO) be the percentage of capacity that is invested in by both the HRB and coarse-resolution case. $SCO = 100\% \cdot \frac{sites_{HRB} \cap sites}{sites_{HRB} \cup sites}$ Cases with increased spatial resolution and temporal resolution are better able to recapitulate siting results.}
        \label{map_wind}
        
    \end{table}
    
\end{landscape}

\begin{figure}[h]
    \begin{center}
    \subfloat[Zonal, 52-week resolution\label{iso_win_z}]{\includegraphics[width=0.25\linewidth]{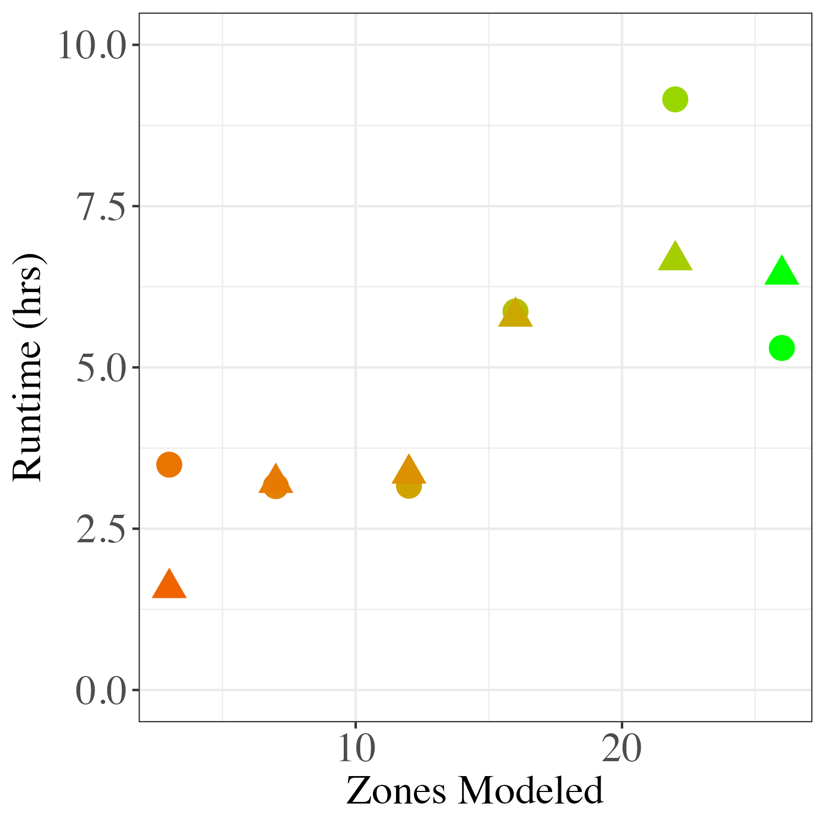}}
    \subfloat[Temporal, 26-zone resolution\label{iso_win_t}]{\includegraphics[width=0.25\linewidth]{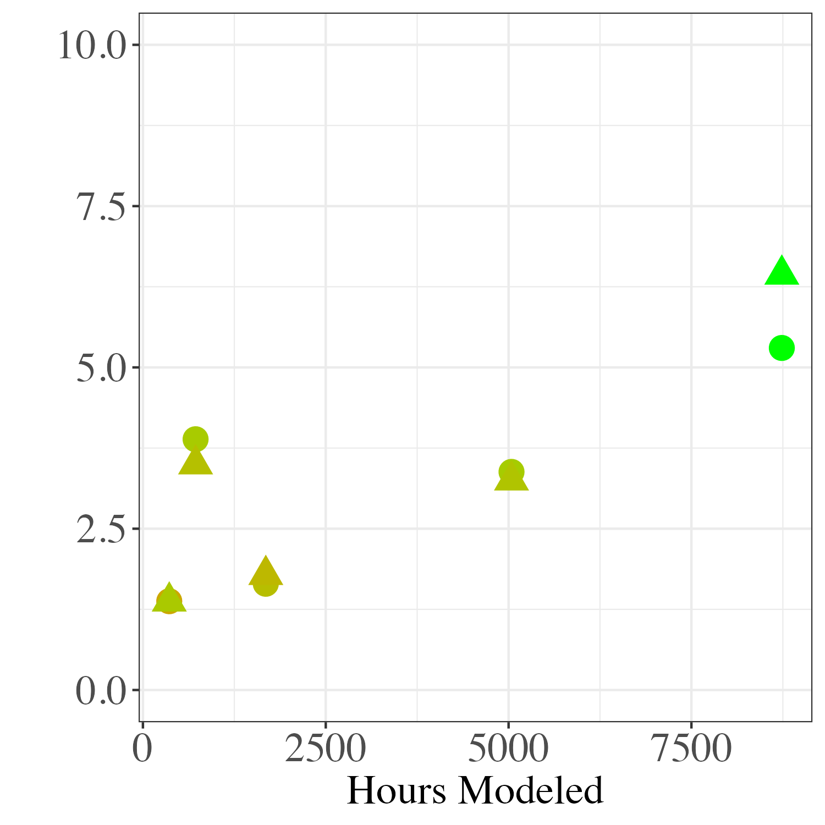}}
    \subfloat[Zonal, 15-day resolution\label{iso_win_zl}]{\includegraphics[width=0.25\linewidth]{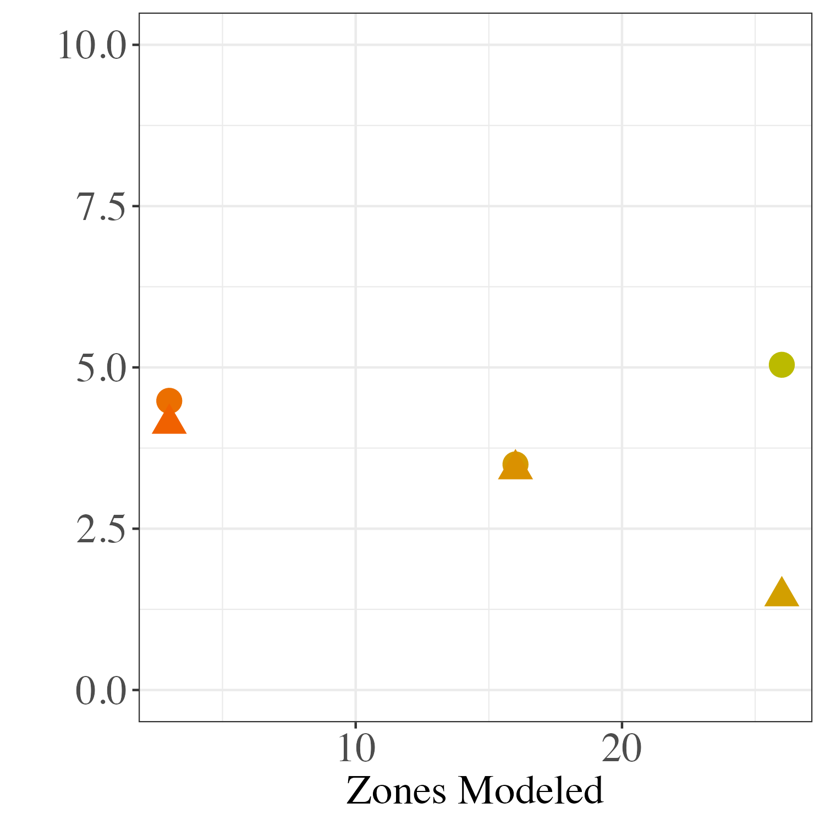}}
    \subfloat[Temporal, 3-zone resolution\label{iso_win_tl}]{\includegraphics[width=0.25\linewidth]{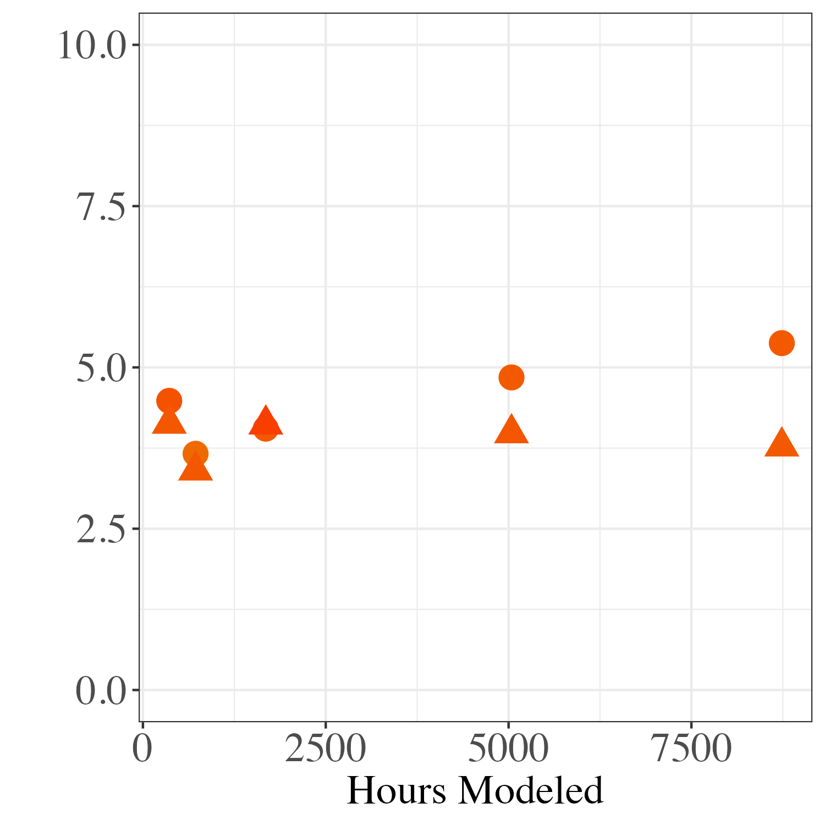}} \\
    \subfloat{\includegraphics[width=\linewidth]{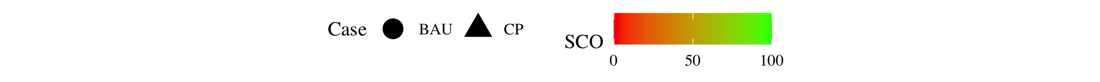}}
    \end{center}
    \caption{Resolution, accuracy, and runtime. X-axis is resolution, y-axis is runtime (hrs), and color is site capacity overlap of onshore wind, SCO\textsubscript{onshore}. An analagous figure using SCO\textsubscript{solar} is Fig~\ref{isoquant_sol} in the SI. Figures include both the \coc (\coca\unskip) and \refc (\refca\unskip) cases. Figs~\ref{iso_win_z} and~\ref{iso_win_t} have highest resolution for the dimension not on the x-axis. Figs~\ref{iso_win_zl} and~\ref{iso_win_tl} have lowest resolution for the excluded dimension. Increasing resolution improves accuracy, though this trend is masked in cases where the non-graphed dimension of resolution is low.}
    \label{isoquant}
\end{figure}

VRE buildout is densely concentrated in spatially coarse models. Figs~\ref{map_solar} and~\ref{map_wind} show the sites selected in the \coca case. Geographically disperse VRE is closer to multiple sources of demand and can take advantage of non-homogeneous national weather patterns. Spatially coarse models cannot tease out these benefits; 3-zone systems have 9x fewer degrees of freedom than the HRB when selecting sites. Poorly spatially resolved models cannot recapture local weather patterns and, as demand is modeled on the regional level, cannot parse the benefit of having VRE near multiple urban areas. Cheapest sites, often geographically close, receive investment per resource cluster; binning sites into fewer clusters leads to systemwide aggregation of selected sites. This explains the heavy onshore wind investment near Oklahoma (Fig~\ref{map_wind}) in the 3-zone case. Adding more VRE resource clusters per region would enable models to better parse weather patterns intraregionally. However, models would still be unable to predict transmission bottlenecks, and runtime impacts may occur. The ideal number of regions and clusters per region given system characteristics should be explored experimentally before any large scale study.

While VRE buildout shifts in temporally poorly resolved systems, it is not geographically aggregated as it is in spatially coarse systems. Locational inaccuracies in lower temporal resolutions are due to an inability to recapture weather patterns and correlations between VRE and demand. These systems do not have fewer degrees of freedom when selecting sites and do not ignore transmission infeasibilities. As a result, temporally coarse cases shift capacity around per cluster within regions but still result in relatively geographically dispersed capacity.

Omission of UC in the 26-zone case leads to 75.1\% SCO\textsubscript{solar}, 31.0\% SCO\textsubscript{onshore} for \refca and 72.8\% SCO\textsubscript{solar}, 60.5\% SCO\textsubscript{onshore} for \coca (SI Fig~\ref{map_uc}.) These inaccuracies are due partially to total shifts in installation (Fig~\ref{diff_temp_co2}.) In addition to total shifts in capacity, cases without UC decrease investment near areas with large thermal capacity; in the \refca case, VRE capacity decreases in the same regions of the mid-east with the highest coal generation in the \opone~phase. This is because these resources are operating better than is feasible and less VRE is needed near them to meet spikes in demand.

If one dimension of resolution is low, high granularity in other dimensions cannot overcome it; a model is able to increase its accuracy by increasing resolution in the spatial (temporal) dimension only if its resolution in the temporal (spatial) dimension is already high. Poorly spatially resolved \coca cases with full temporal resolution only reach 17.8\% SCO\textsubscript{onshore}, 14.5\% SCO\textsubscript{solar}. Poorly temporally resolved models at 26-zone resolution only reach 43.8\% SCO\textsubscript{onshore}, 67.1\% SCO\textsubscript{solar} (Fig~\ref{map_solar},~\ref{map_wind}.) Fig~\ref{isoquant} shows that SCO\textsubscript{onshore} improves with a given dimension of resolution only when the run is otherwise highly granular. Lowering (Fig~\ref{iso_win_zl}) spatial or (Fig~\ref{iso_win_tl}) temporal granularity decreases both SCO and the overall corrolation between resolution and SCO.

Offshore wind has high capital costs which may be prohibitive\cite{kota2015offshore}; a primary benefit of offshore is its consistency relative to onshore with high availability off the east coast. The HRB \coca case invested in 4 GW of offshore wind near New Jersey. The 30-week, 26-zone \coca case invested in 4 GW near New Jersey, 30 MW near Maine. No other cases invested in offshore wind. Lower temporal resolution, by underrepresenting onshore wind variability, sees less benefit to offshore resources. Spatially coarse cases overestimate the ease of transmitting VRE power to the northeast, decreasing need for local offshore wind. The trend that increasing spatial resolution leads to replacement of off- with onshore wind has been demonstrated previously\cite{frysztacki2021strong}. Because virtually all cases had SCO\textsubscript{offshore} of 0\% in the \coca case, (100\% for \refca\unskip,) a map for offshore wind is omitted here.

\subsection{Operational Impacts} \label{results:operations}
When investments from low resolution \opone~phase systems are tested in \optwo~models, they lead to higher costs, generator profits, NSE, and CO$_2$ emissions.

Investments from low resolution systems are frequently unable to meet demand. Investments from poorly spatially resolved systems incur higher NSE when operated (Fig~\ref{map_nse_co2},) indicating lower reliability. Investments from the 3-zone \coca cases have up to 6.3\% NSE with an average of 0.2\% of demand unmet across the entire timeseries. Much of the NSE in these cases occurs in the northeast, lining up with under-investment in VRE (Fig~\ref{map_solar},~\ref{map_wind}) in the zone ISONE and decreased transmission capacity (Fig~\ref{map_trans_diff}.)

Low resolution cases have higher generator profits and electricity prices. Fig~\ref{map_profit_co2} shows resource profits grouped by region for \coca cases. When demand exceeds available generation, power generators are able to inflate their prices while still finding markets for expensive electricity. This practice is called ``scarcity pricing.'' High NSE and profits are highly spatially correlated across Fig~\ref{map_nse_co2} and Fig~\ref{map_profit_co2}; SI Fig~\ref{price} demonstrates high electricity prices in low-resolution cases. These trends confirm scarcity pricing at low resolutions, indicating the cause of high profits.

Investments from spatially aggregated models lead to higher CO$_2$ emissions: going from 16- to 26-zones at full temporal resolution decreases CO$_2$ emissions in the \coca case by 30\% (Fig~\ref{map_emiss}.) According to our cost of exogenous carbon abatement, we expect to save \$12B in carbon mitigation from the carbon fee due to this improvement. When poorly sited resources from lower resolution cases cannot meet demand, models must either rely on high-emitting local resources or enforce blackouts. In our systems, NSE had costs of up to \$2,000~/~MWh, leading models to prioritize demand over preventing emissions, explaining the higher emissions seen here.

\begin{landscape}

    \begin{table}[p]
        \centering
        \setlength\tabcolsep{0pt}
        \begin{tabular}{cM{0.195\linewidth}M{0.195\linewidth}M{0.195\linewidth}M{0.195\linewidth}M{0.195\linewidth}}
            & \large 15-day & \large 30-day & \large 10-week & \large 30-week & \large 52-week \\
            
            \rotatebox[origin=c]{90}{\large 3-Zone} &
            \includegraphics[width=\linewidth]{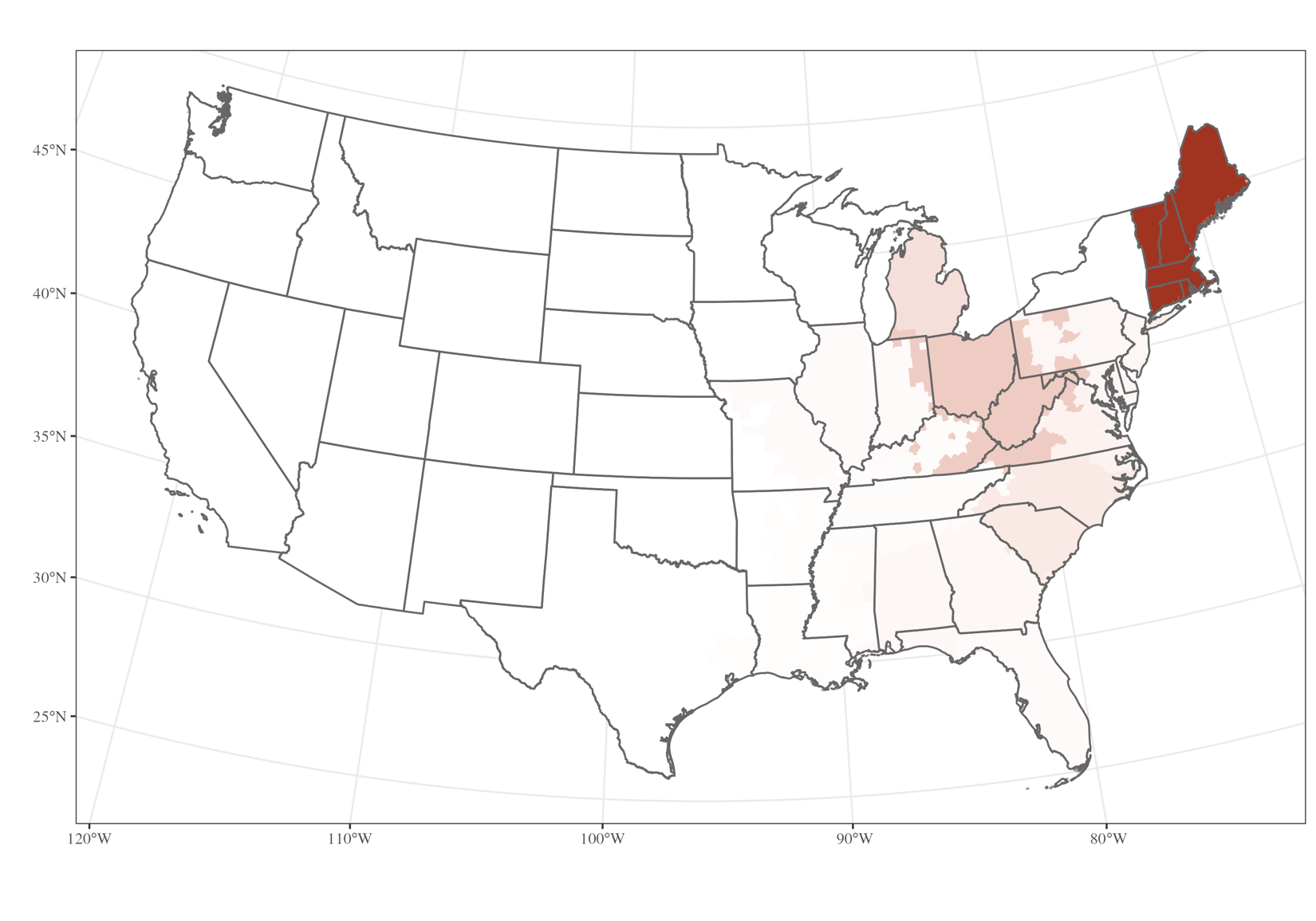} & 
            \includegraphics[width=\linewidth]{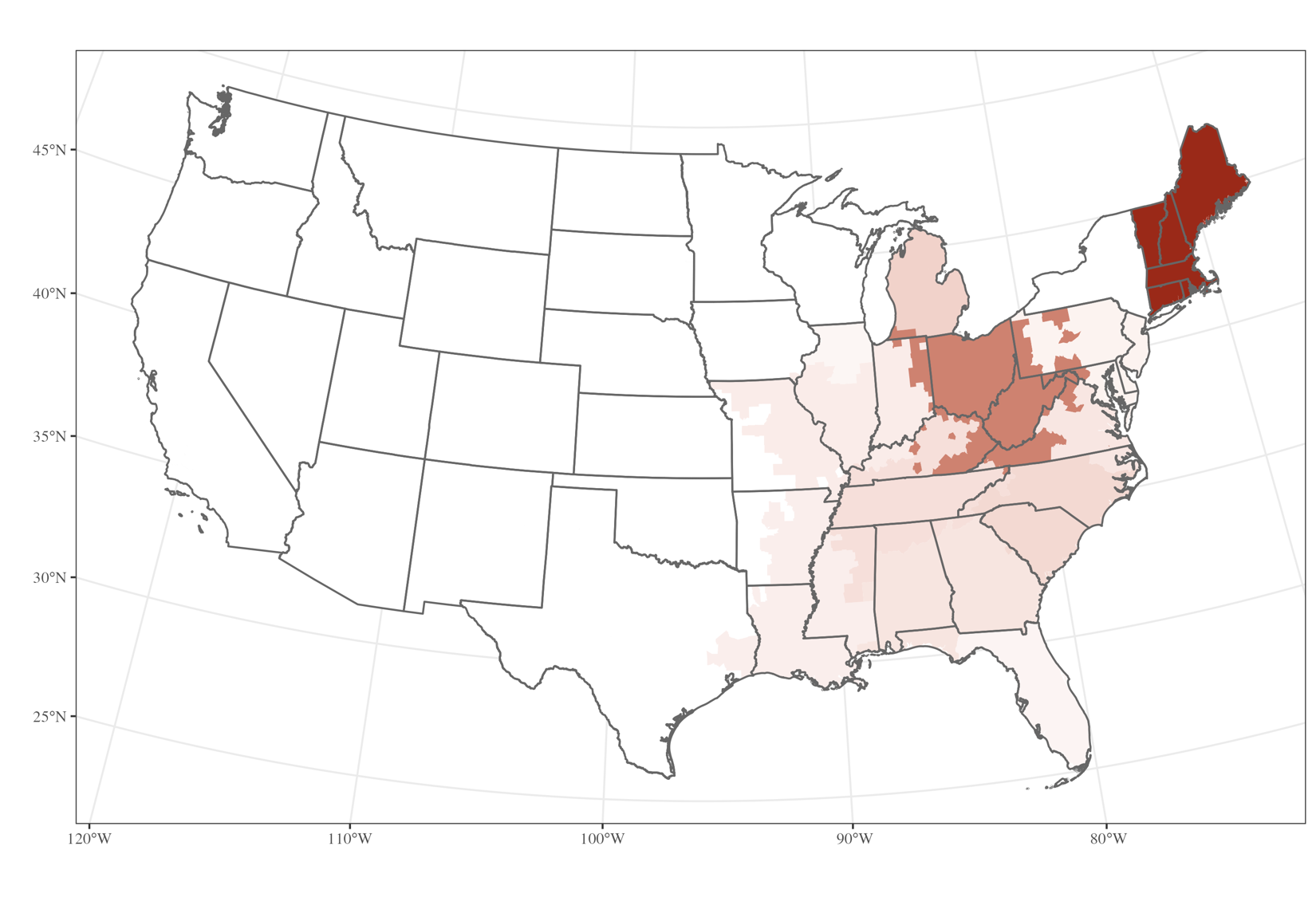} &
            \includegraphics[width=\linewidth]{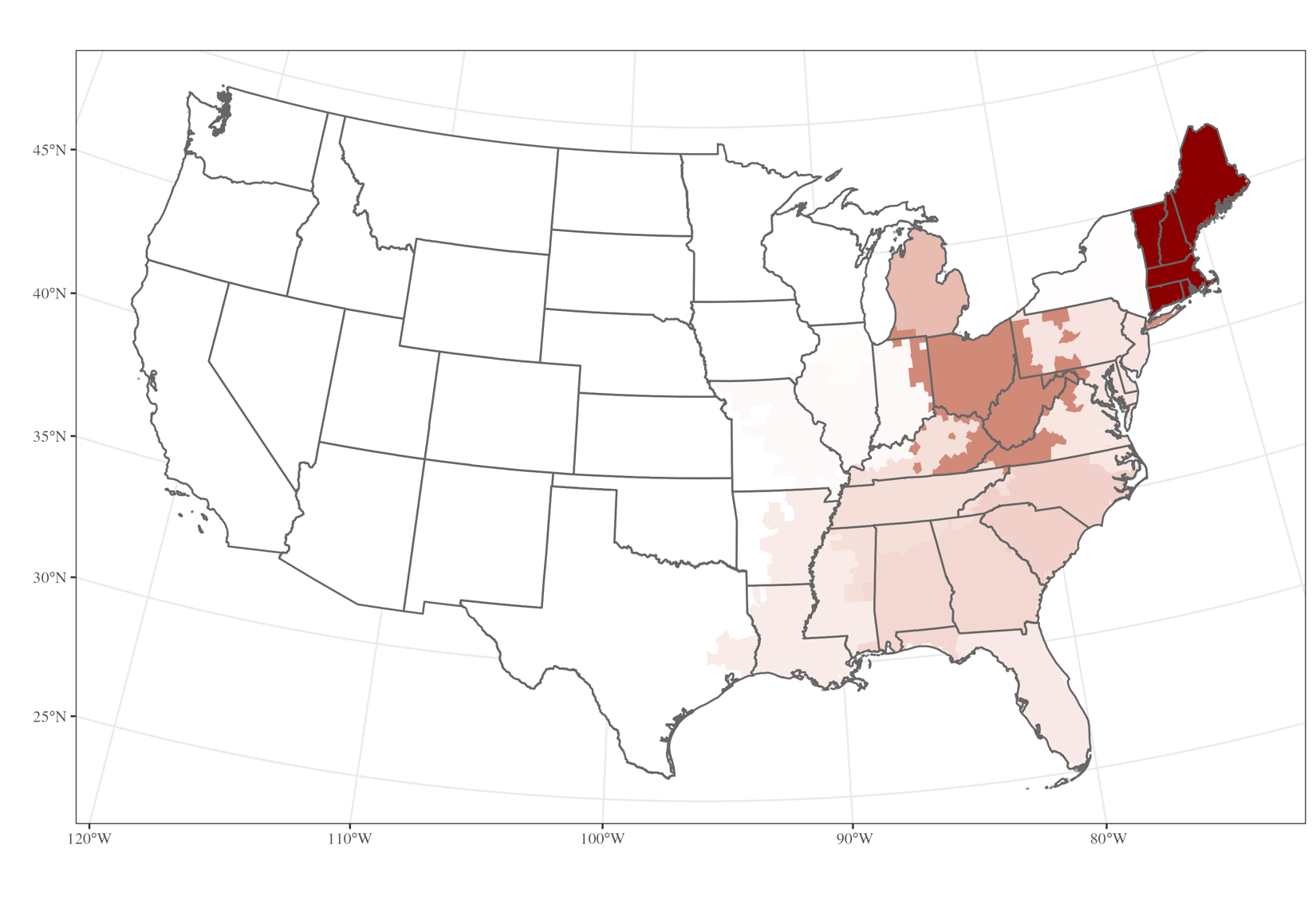} & 
            \includegraphics[width=\linewidth]{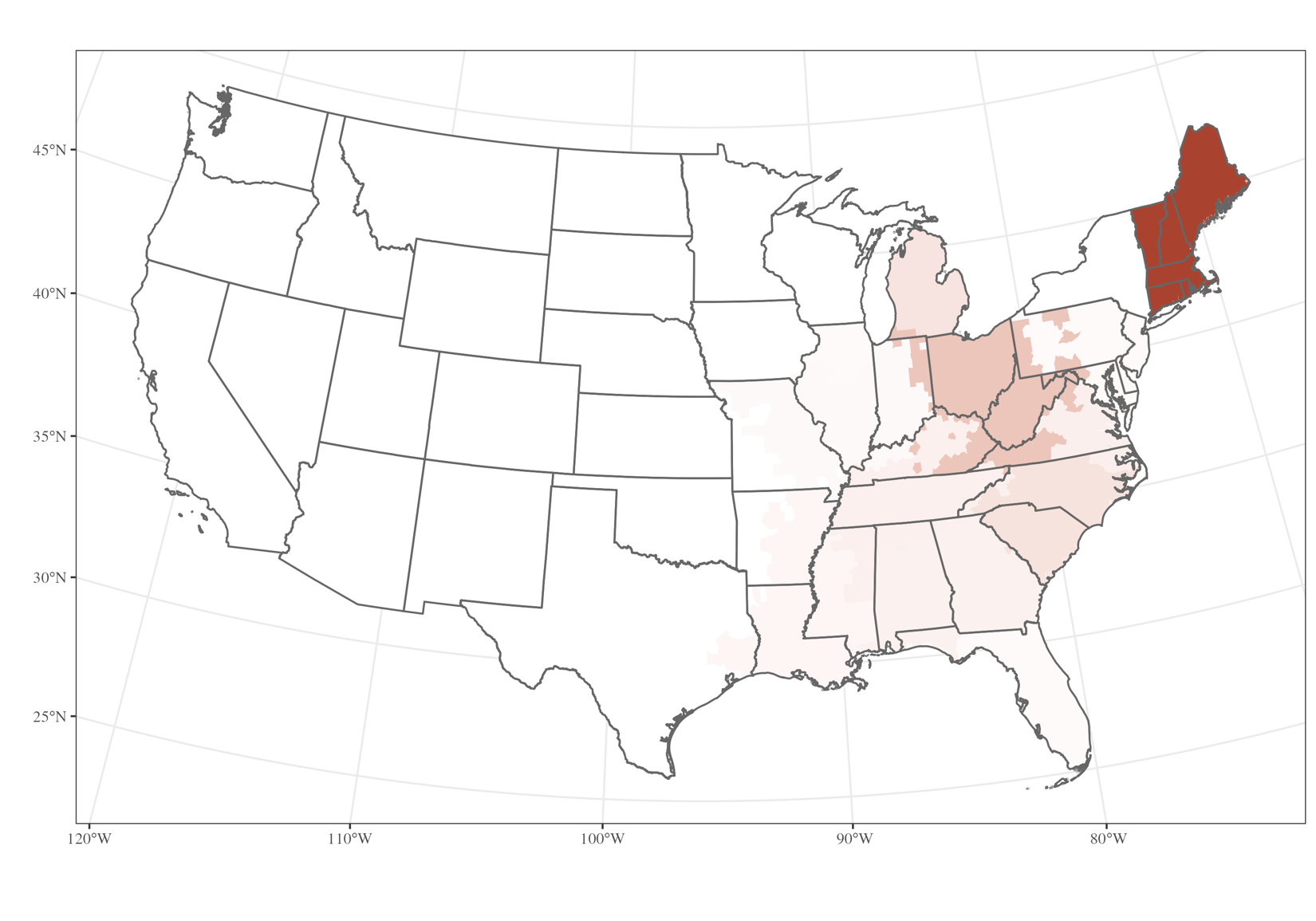} & 
            \includegraphics[width=\linewidth]{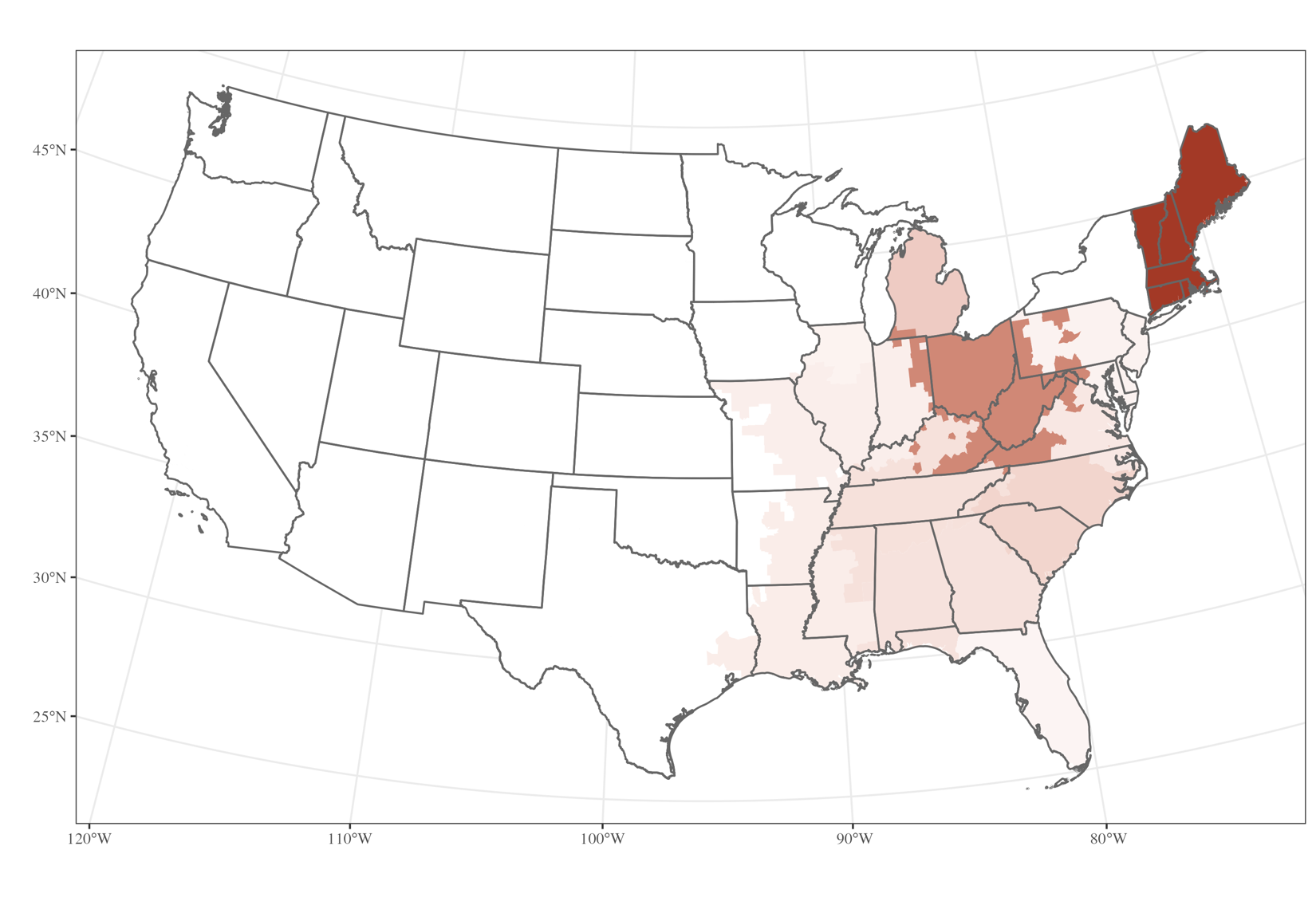} \\
    
            & {\footnotesize NSE: 9216.1 GW} &
            {\footnotesize NSE: 14133.9 GW} &
            {\footnotesize NSE: 18559.7 GW} &
            {\footnotesize NSE: 9356.9 GW} &
            {\footnotesize NSE: 14053.6 GW} \\

            \rotatebox[origin=c]{90}{\large 16-Zone} &
            \includegraphics[width=\linewidth]{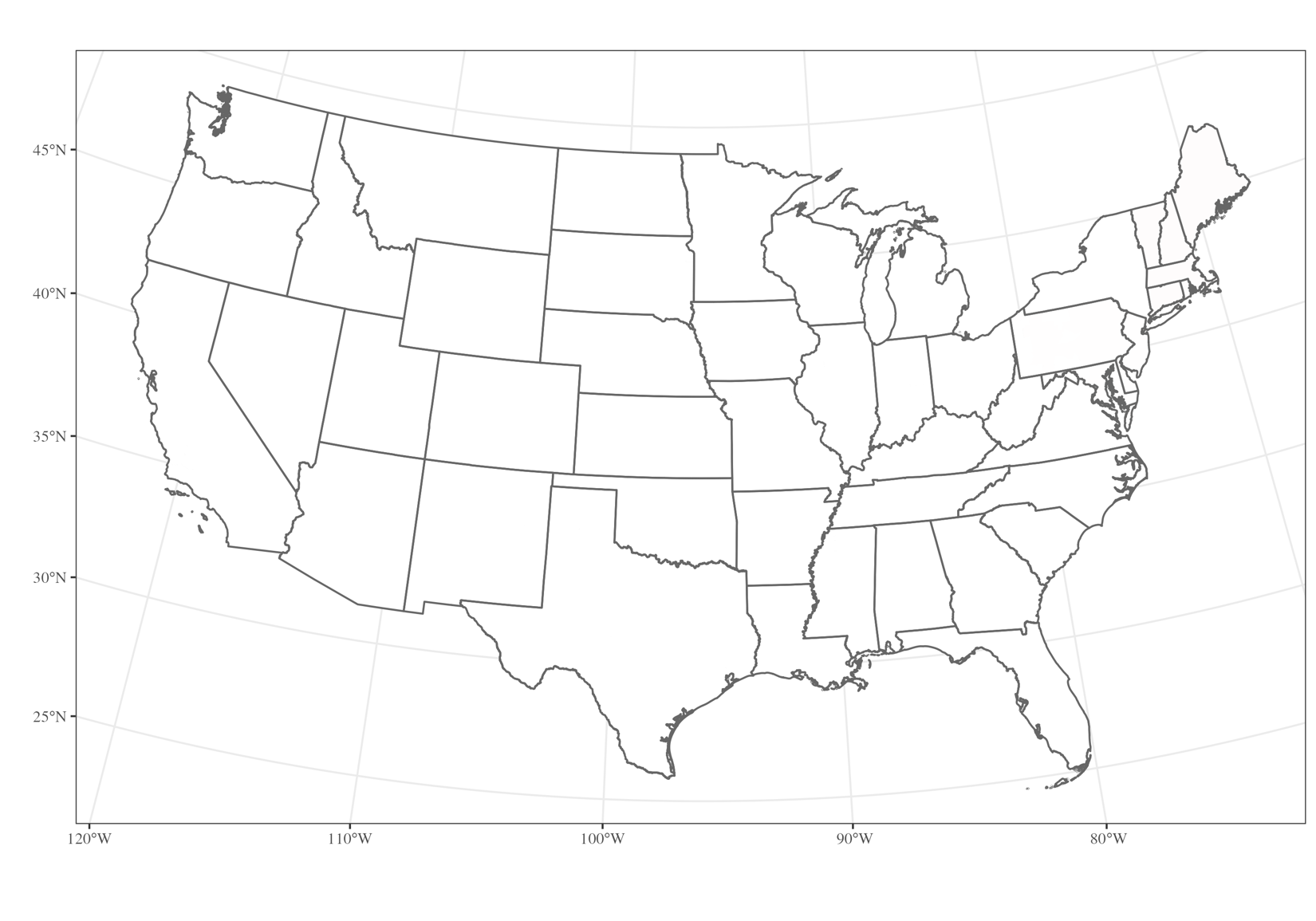} & 
            \includegraphics[width=\linewidth]{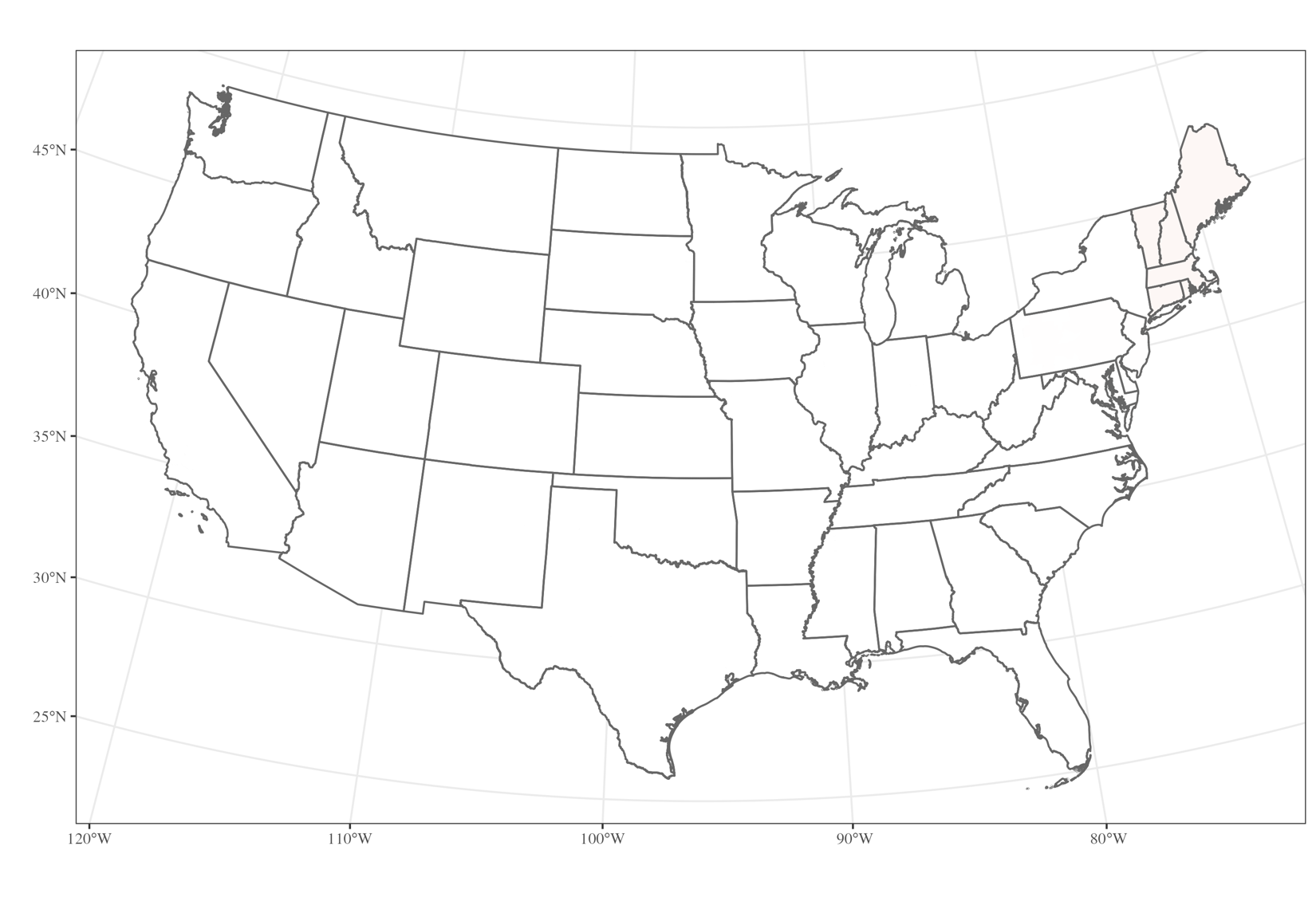} &
            \includegraphics[width=\linewidth]{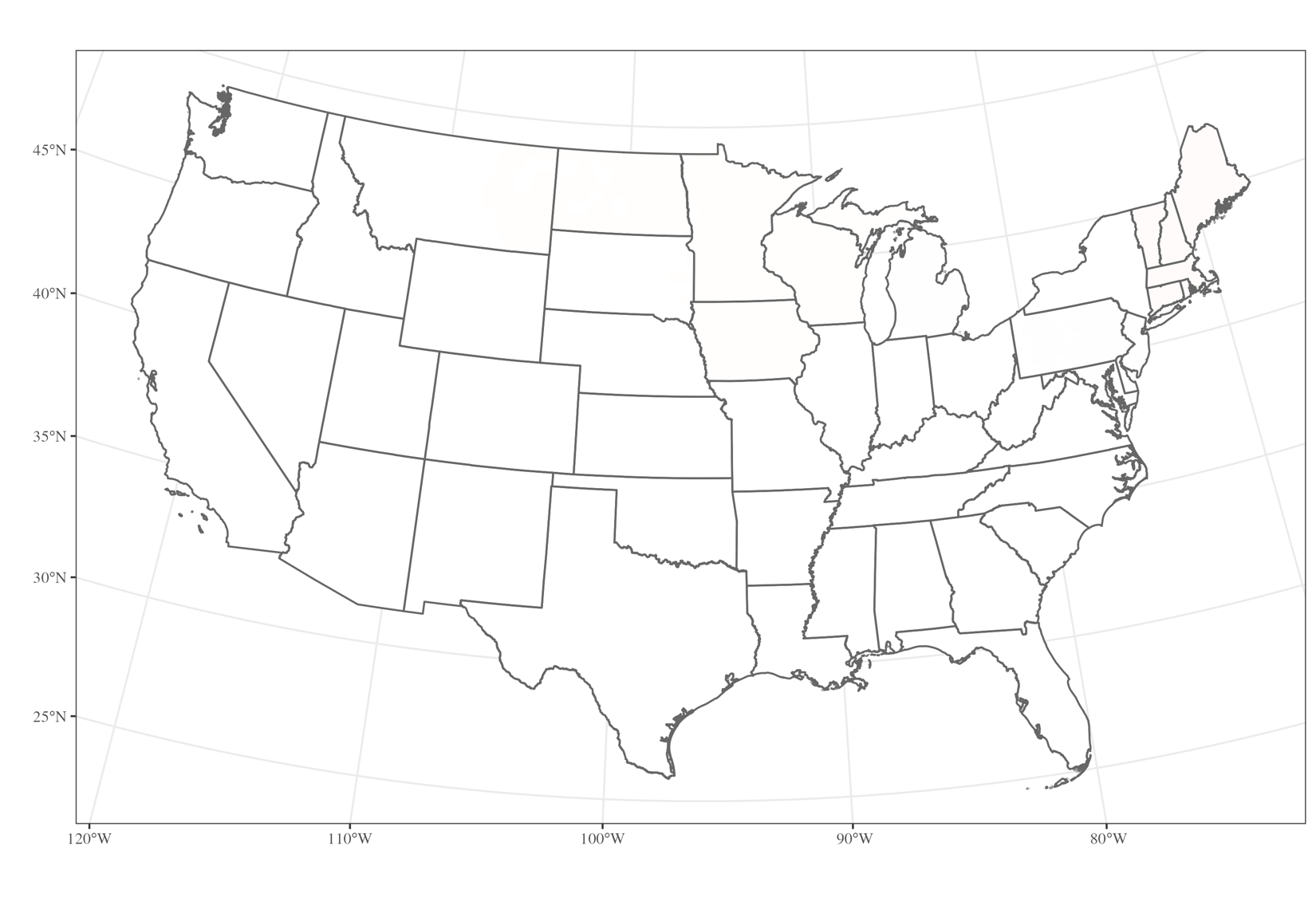} & 
            \includegraphics[width=\linewidth]{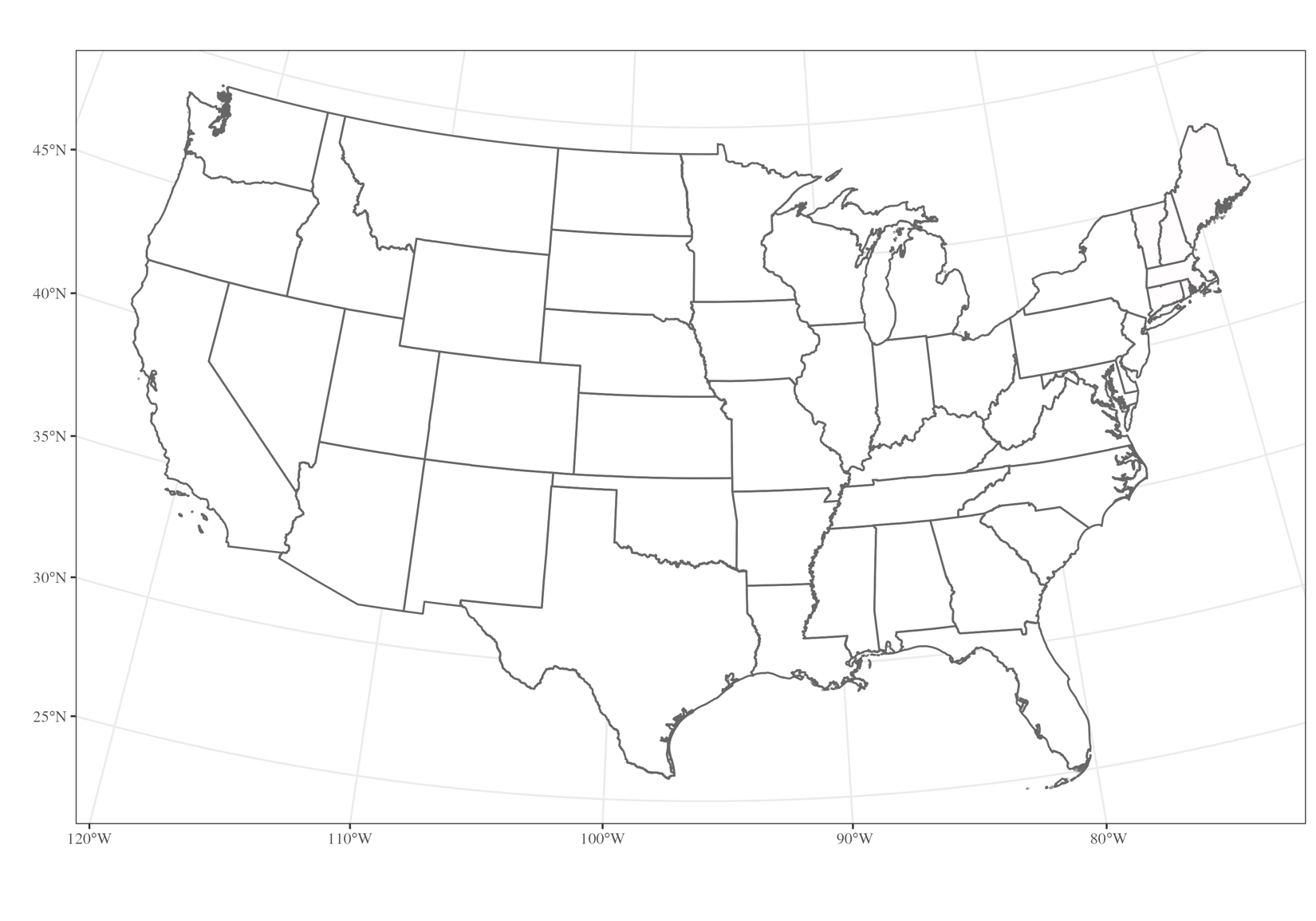} & 
            \includegraphics[width=\linewidth]{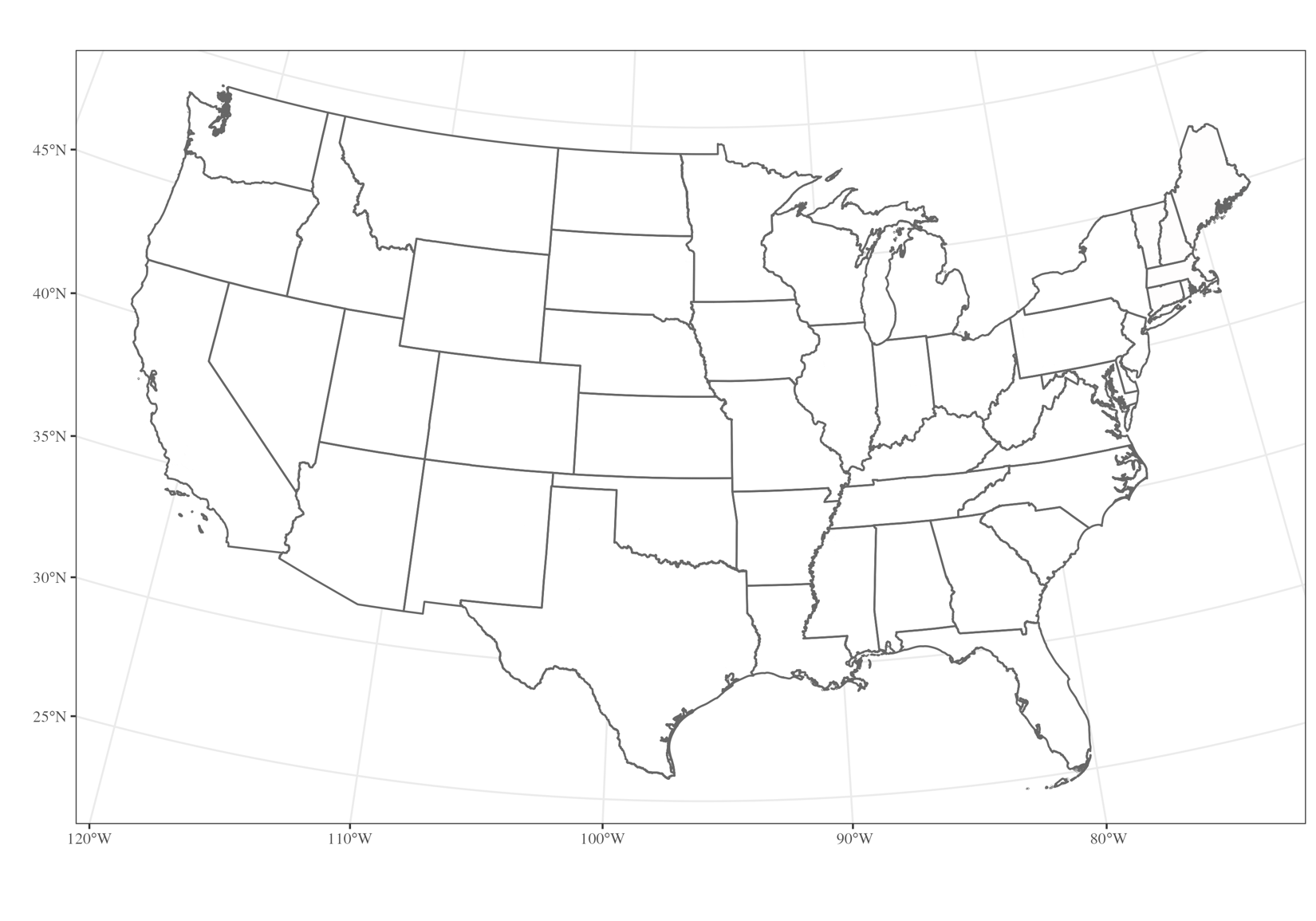} \\
    
            & {\footnotesize NSE: 93.3 GW} &
            {\footnotesize NSE: 238.5 GW} &
            {\footnotesize NSE: 238.5 GW} &
            {\footnotesize NSE: 48 GW} &
            {\footnotesize NSE: 53.5 GW} \\

            \rotatebox[origin=c]{90}{\large 26-Zone} &
            \includegraphics[width=\linewidth]{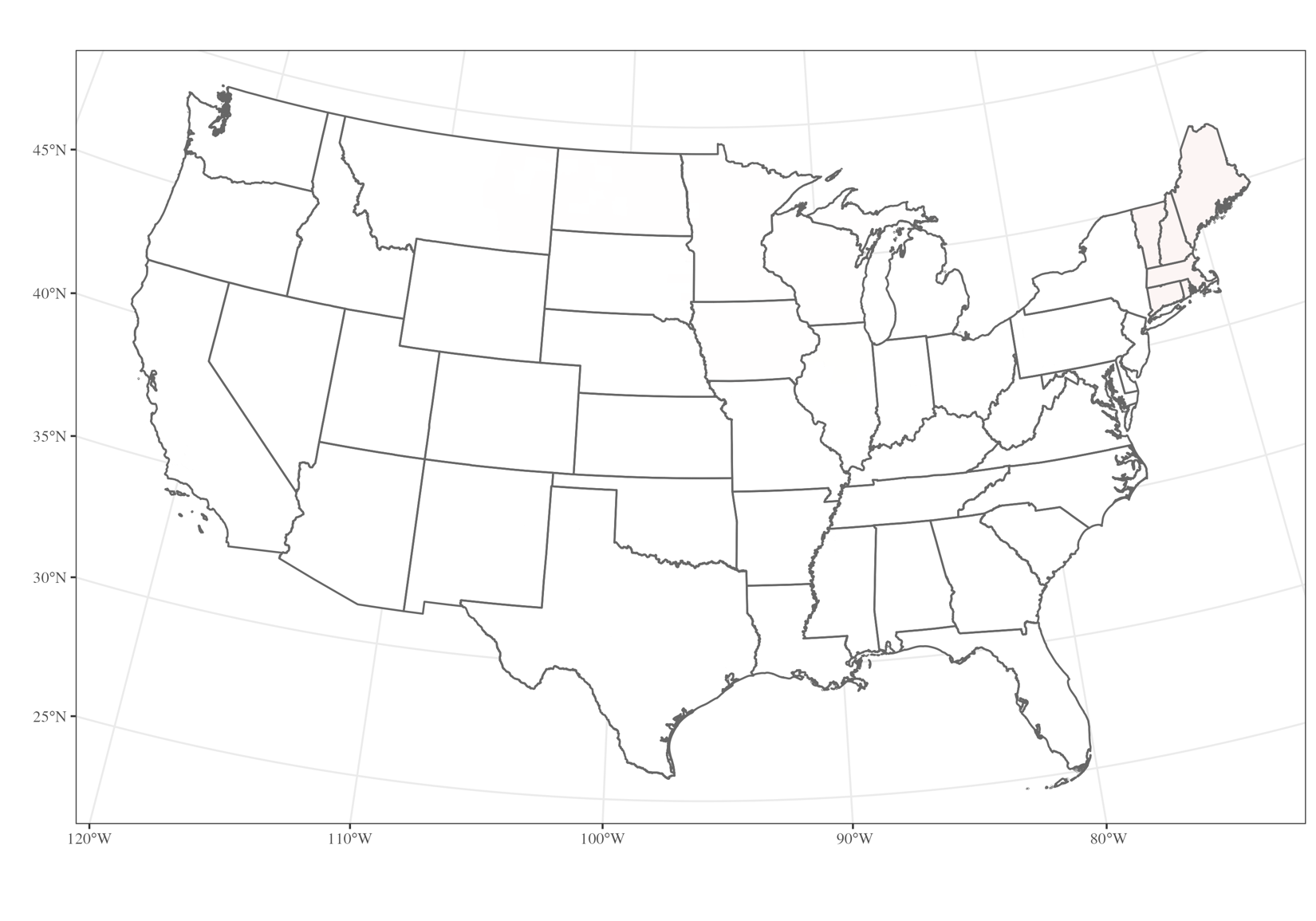} & 
            \includegraphics[width=\linewidth]{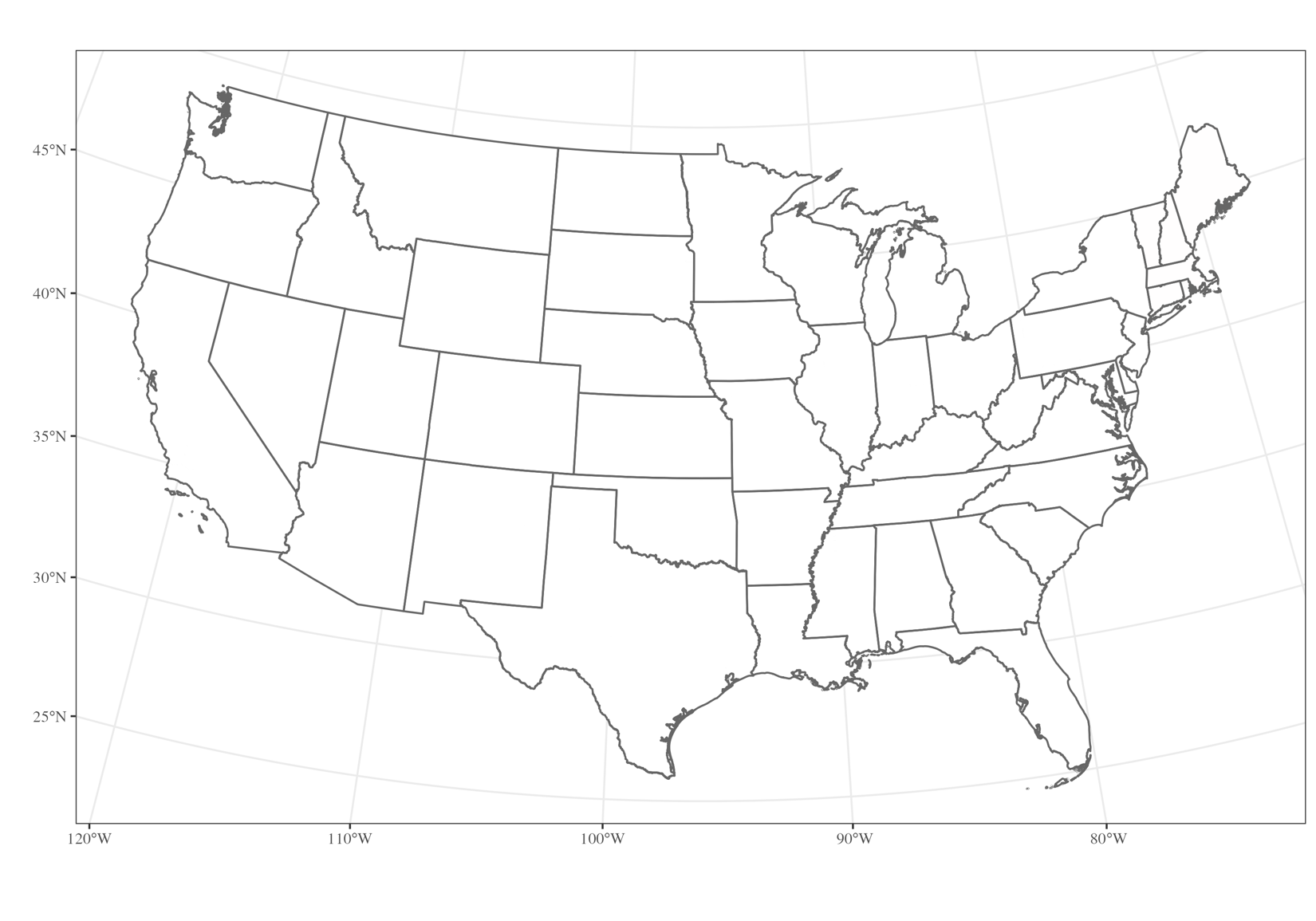} &
            \includegraphics[width=\linewidth]{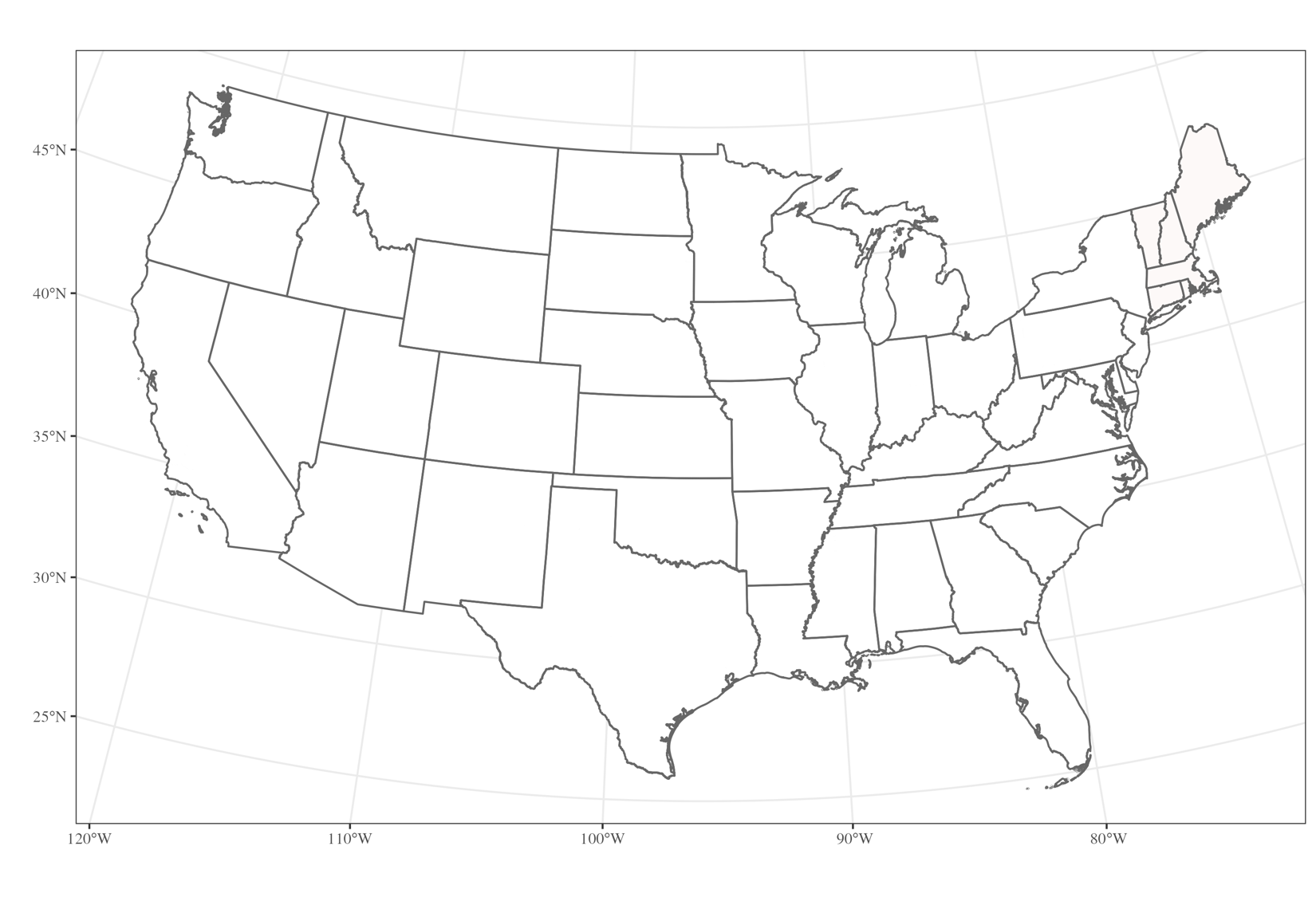} & 
            \includegraphics[width=\linewidth]{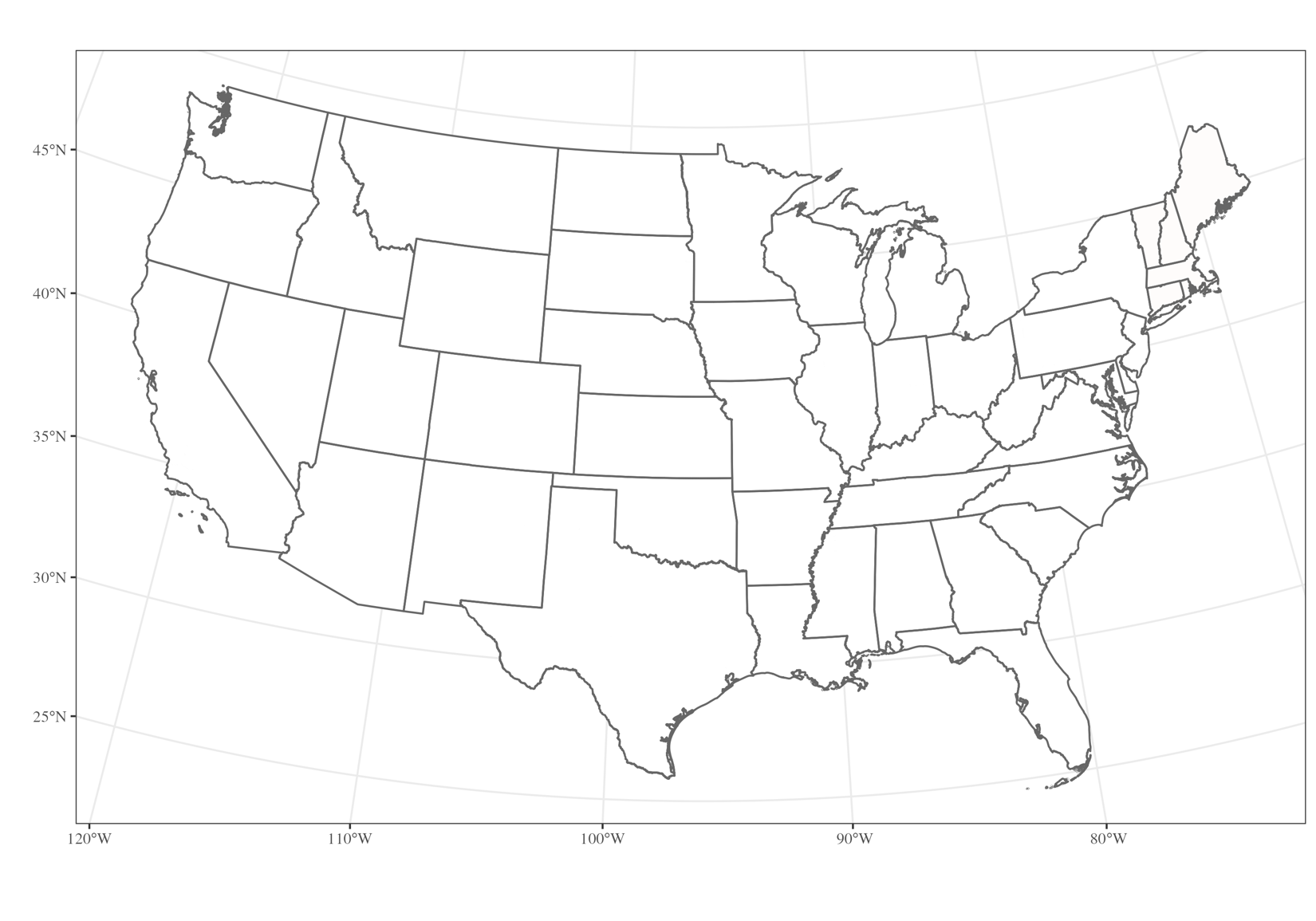} & 
            \includegraphics[width=\linewidth]{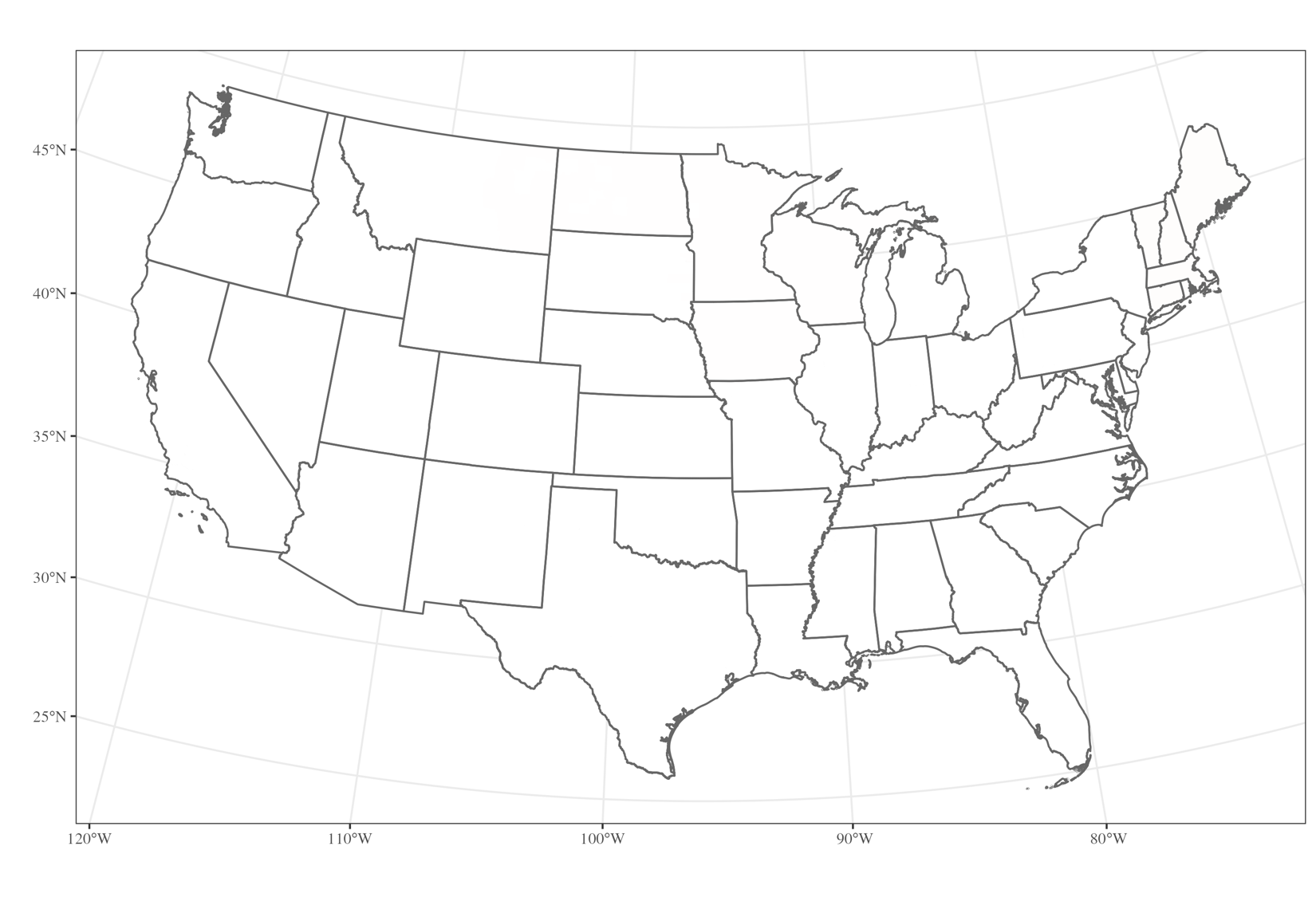} \\
        
            & {\footnotesize NSE: 300 GW} &
            {\footnotesize NSE: 33.9 GW} &
            {\footnotesize NSE: 194.8 GW} &
            {\footnotesize NSE: 88.4 GW} &
            {\footnotesize NSE: 144.9 GW} \\

            & & & & & \includegraphics[width=\linewidth]{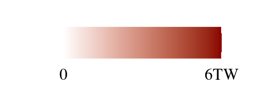} \\
    
        \end{tabular}
        \captionof{figure}{Systemwide nonserved energy (NSE) by spatial and temporal resolution. Excessive NSE in specific regions is likely due to suboptimal local renewable resource investment and underbuilt transmission capacity, causing low generation which may need to be transmitted long distances.\label{map_nse_co2}}
        
    \end{table}

    \begin{table}[p]
        \centering
        \setlength\tabcolsep{0pt}
        \begin{tabular}{cM{0.195\linewidth}M{0.195\linewidth}M{0.195\linewidth}M{0.195\linewidth}M{0.195\linewidth}}
            & \large 15-day & \large 30-day & \large 10-week & \large 30-week & \large 52-week \\
            
            \rotatebox[origin=c]{90}{\large 3-Zone} &
            \includegraphics[width=\linewidth]{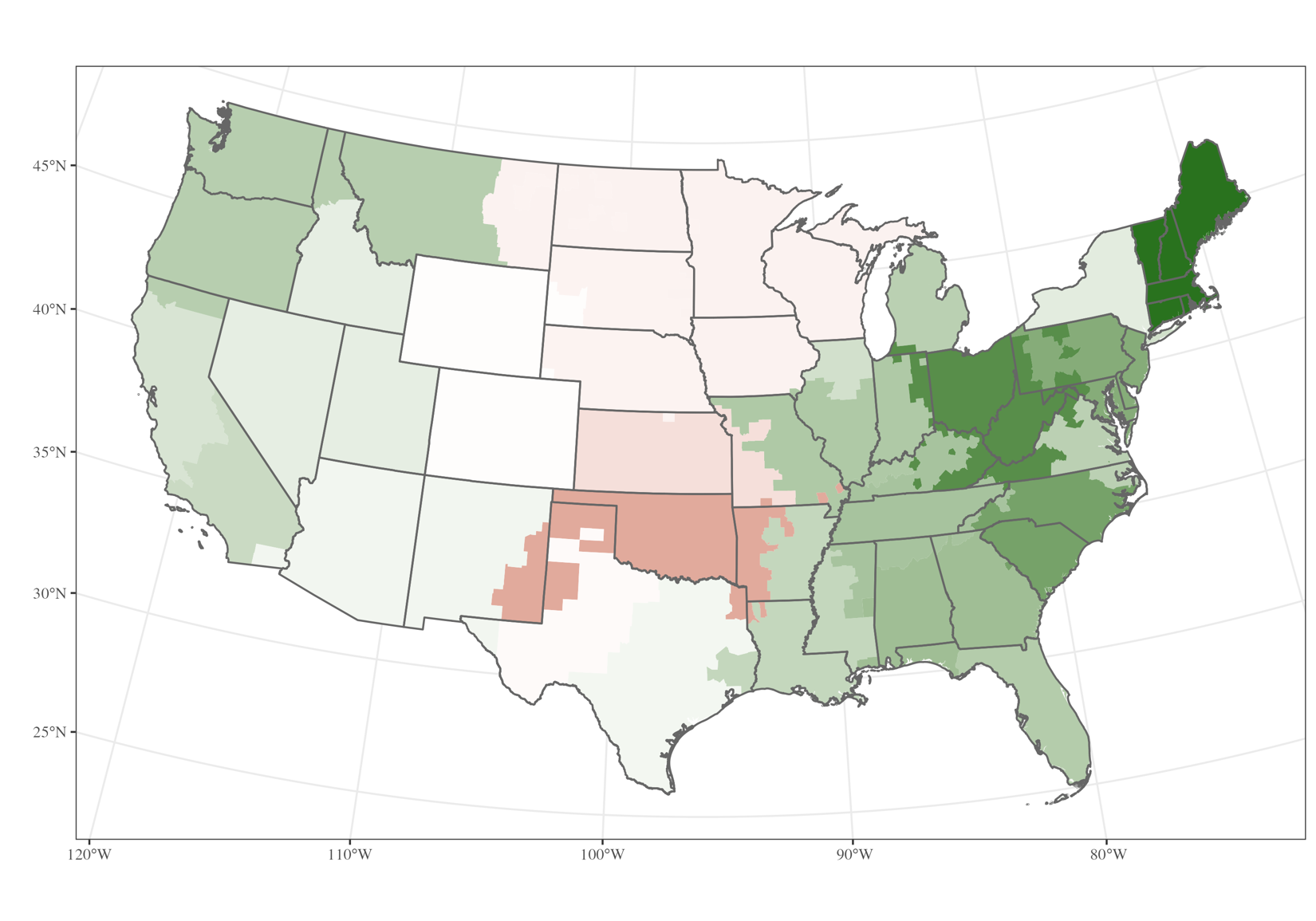} & 
            \includegraphics[width=\linewidth]{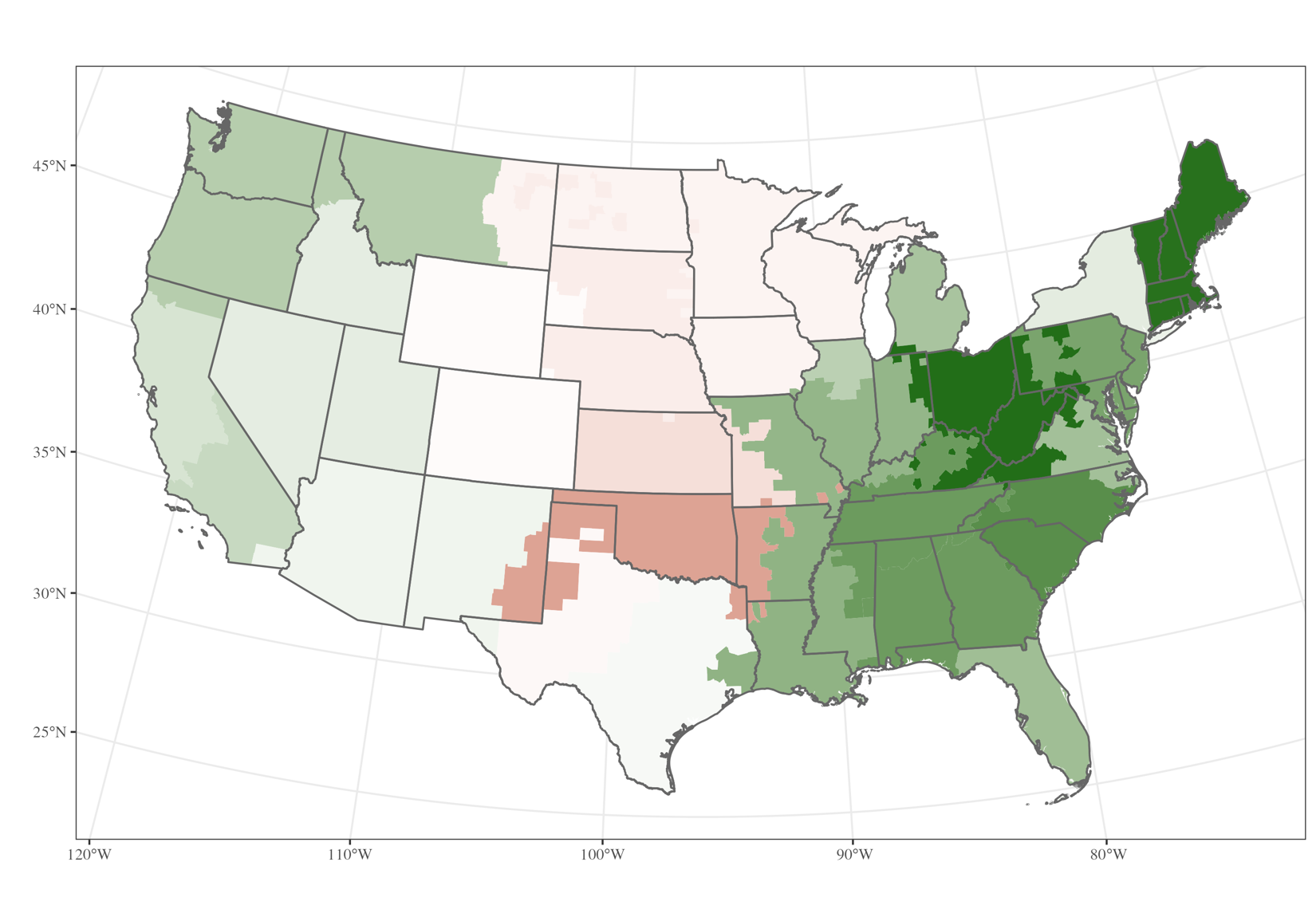} &
            \includegraphics[width=\linewidth]{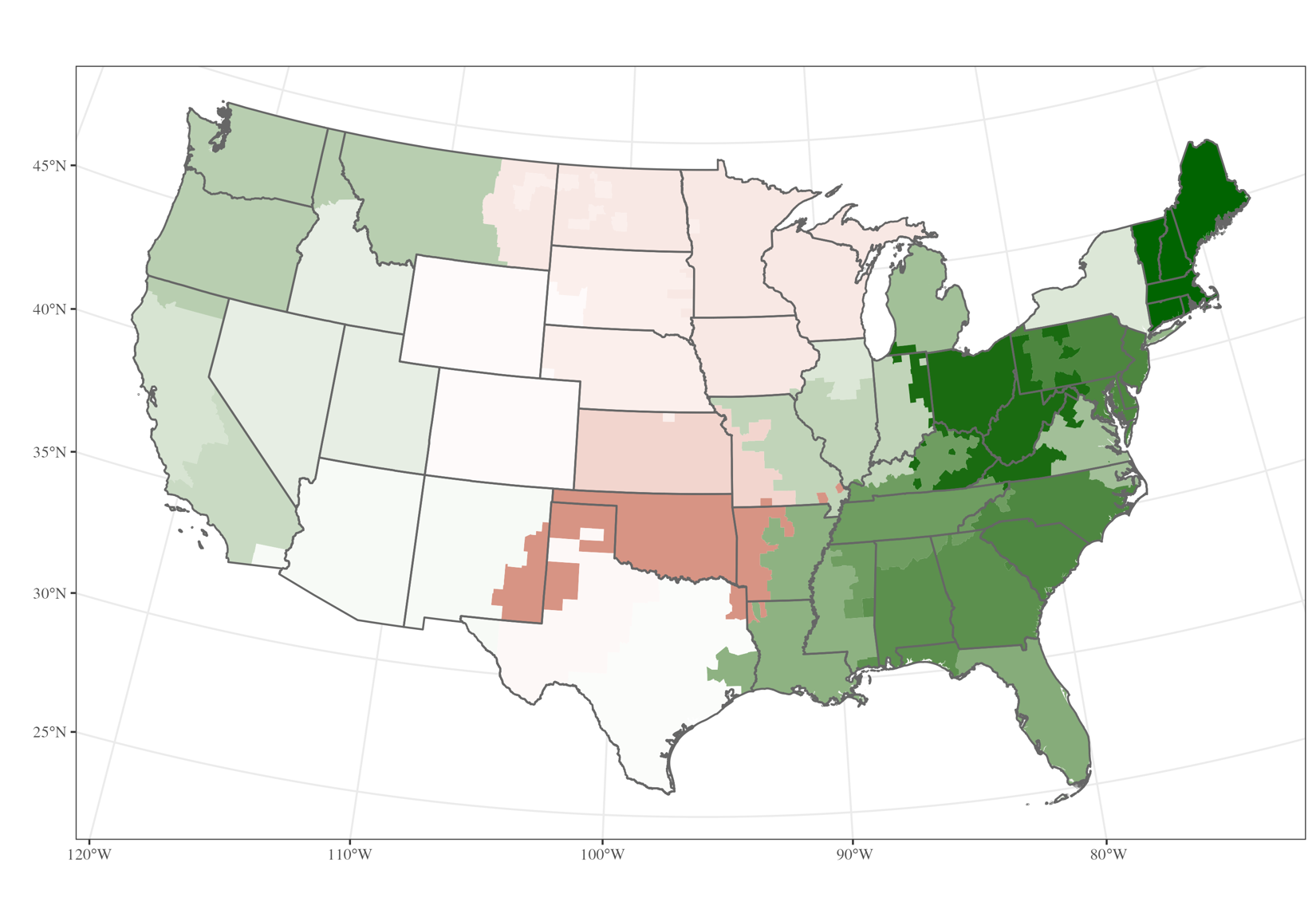} & 
            \includegraphics[width=\linewidth]{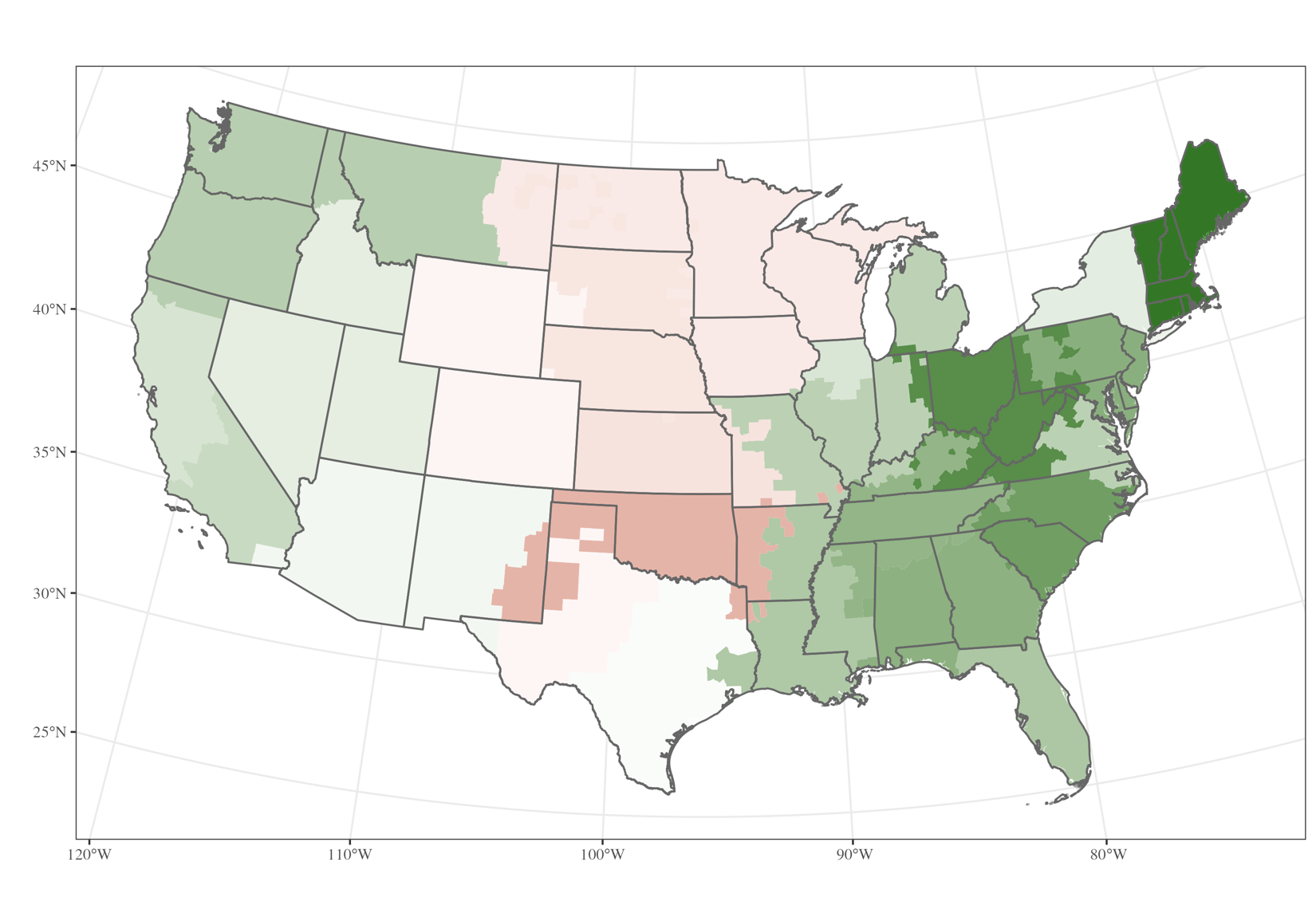} & 
            \includegraphics[width=\linewidth]{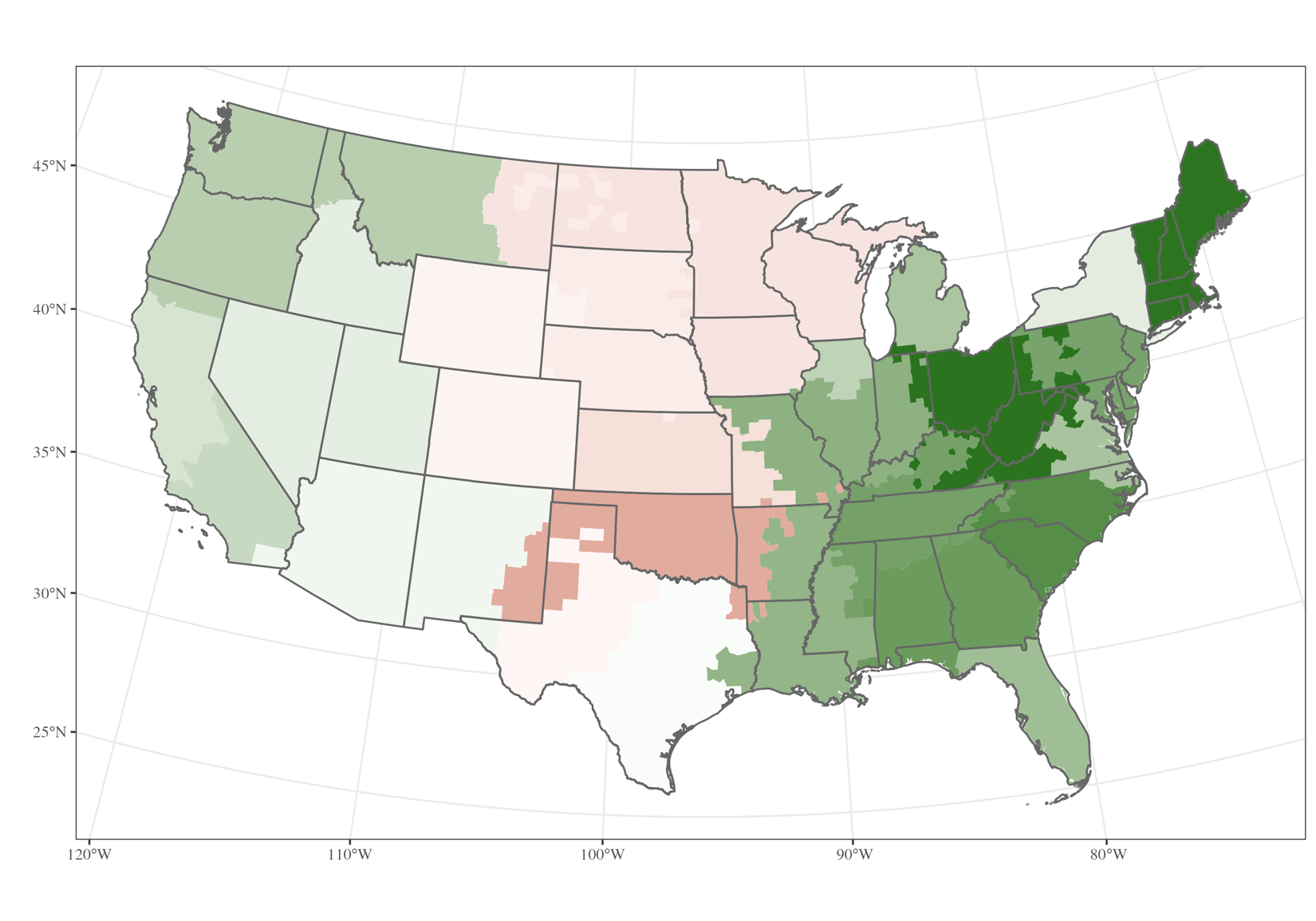} \\
    
            & {\footnotesize Profit: \$0.6 T, MSE: \$5.7 B} &
            {\footnotesize Profit: \$0.7 T, MSE: \$7.2 B} &
            {\footnotesize Profit: \$0.7 T, MSE: \$7.8 B} &
            {\footnotesize Profit: \$0.6 T, MSE: \$5.8 B} &
            {\footnotesize Profit: \$0.7 T, MSE: \$7.1 B} \\

            \rotatebox[origin=c]{90}{\large 16-Zone} &
            \includegraphics[width=\linewidth]{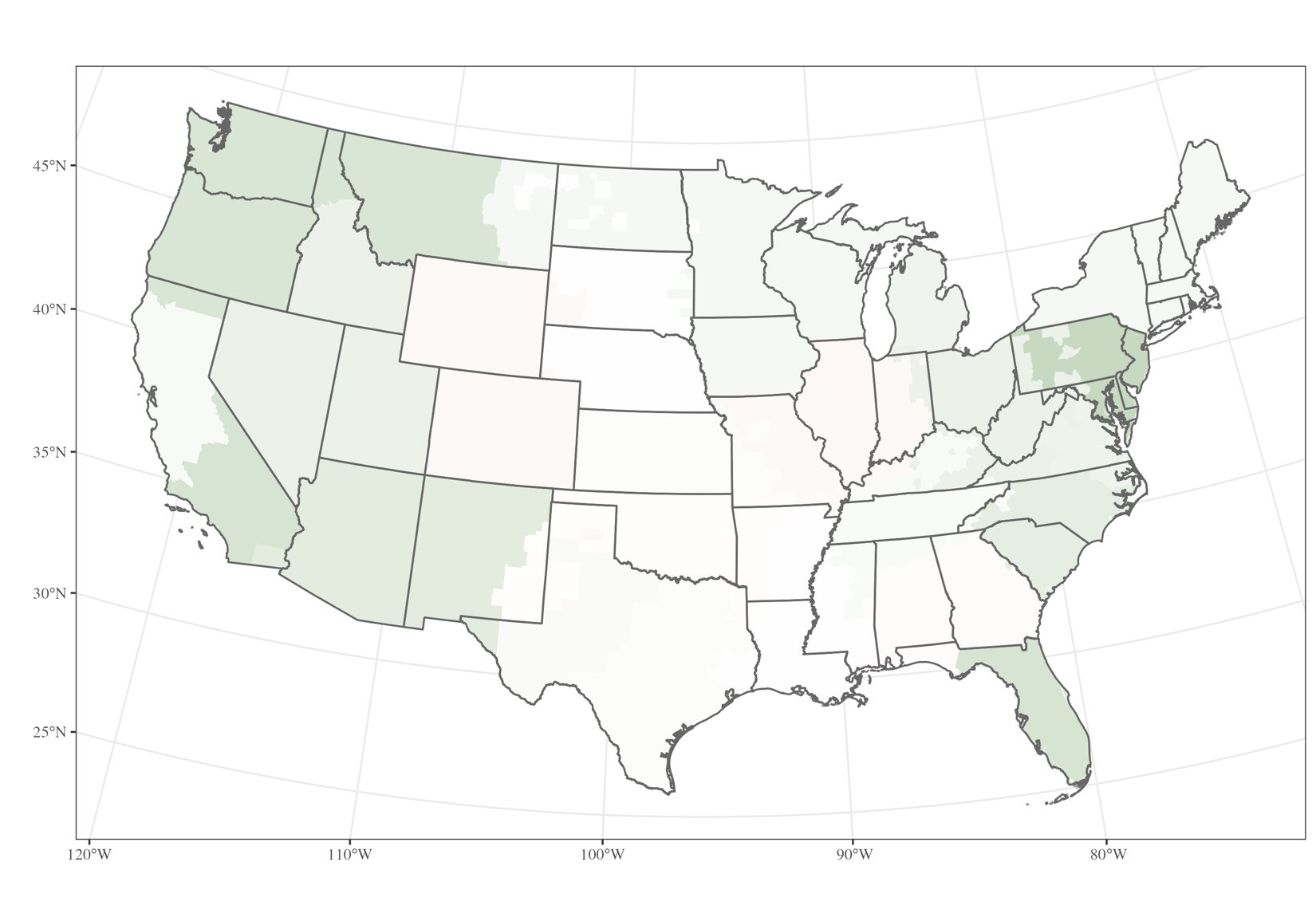} & 
            \includegraphics[width=\linewidth]{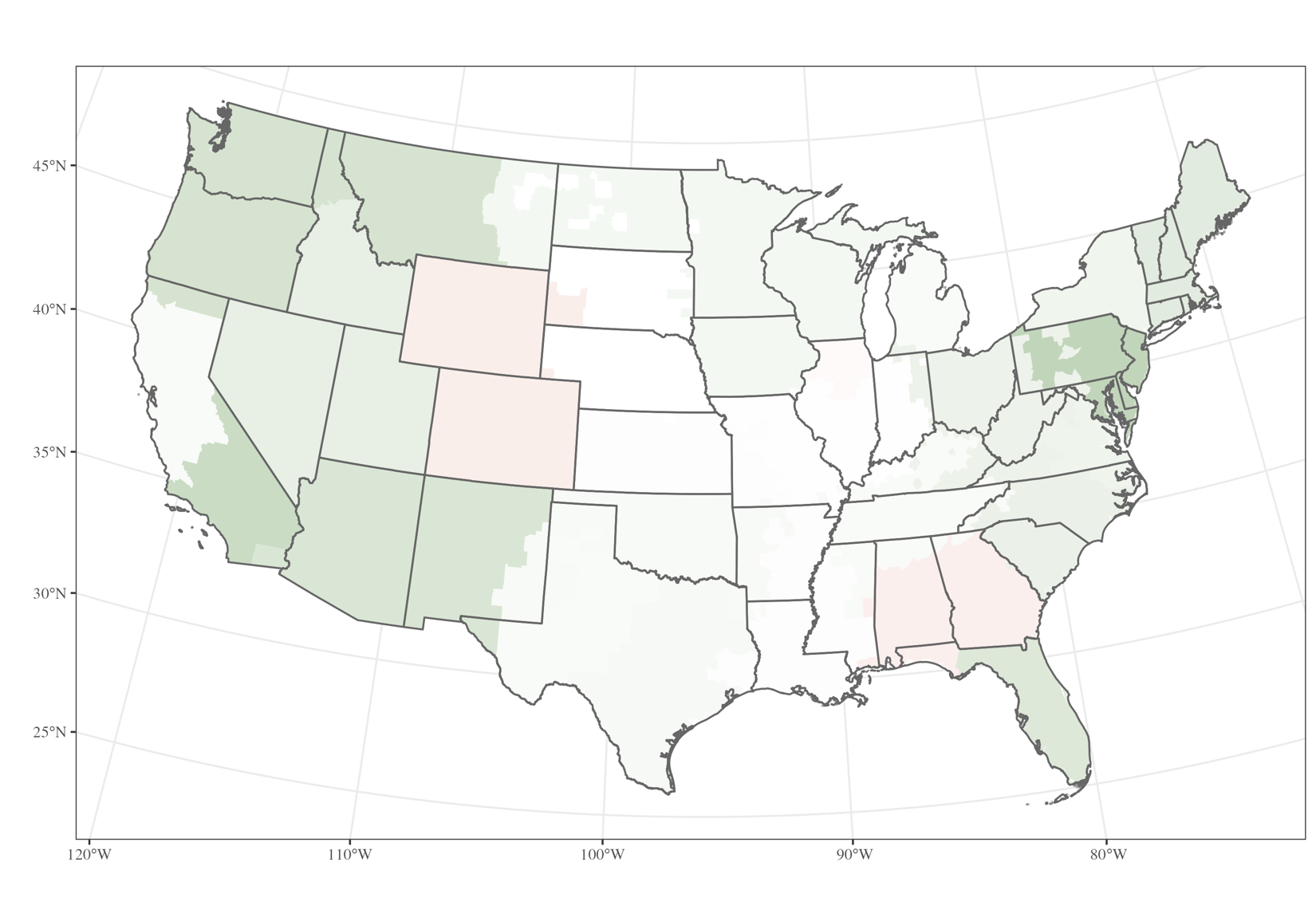} &
            \includegraphics[width=\linewidth]{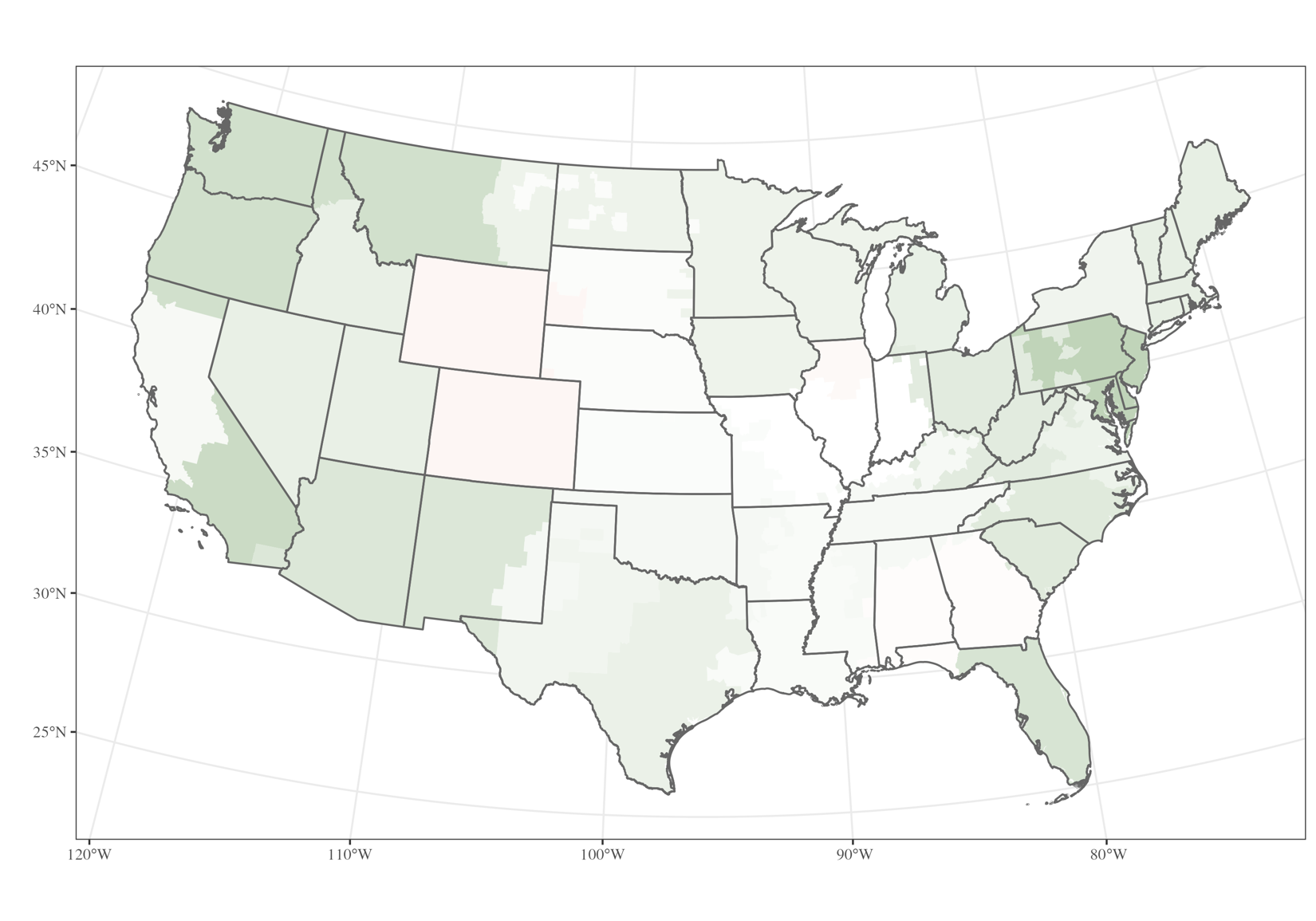} &
            \includegraphics[width=\linewidth]{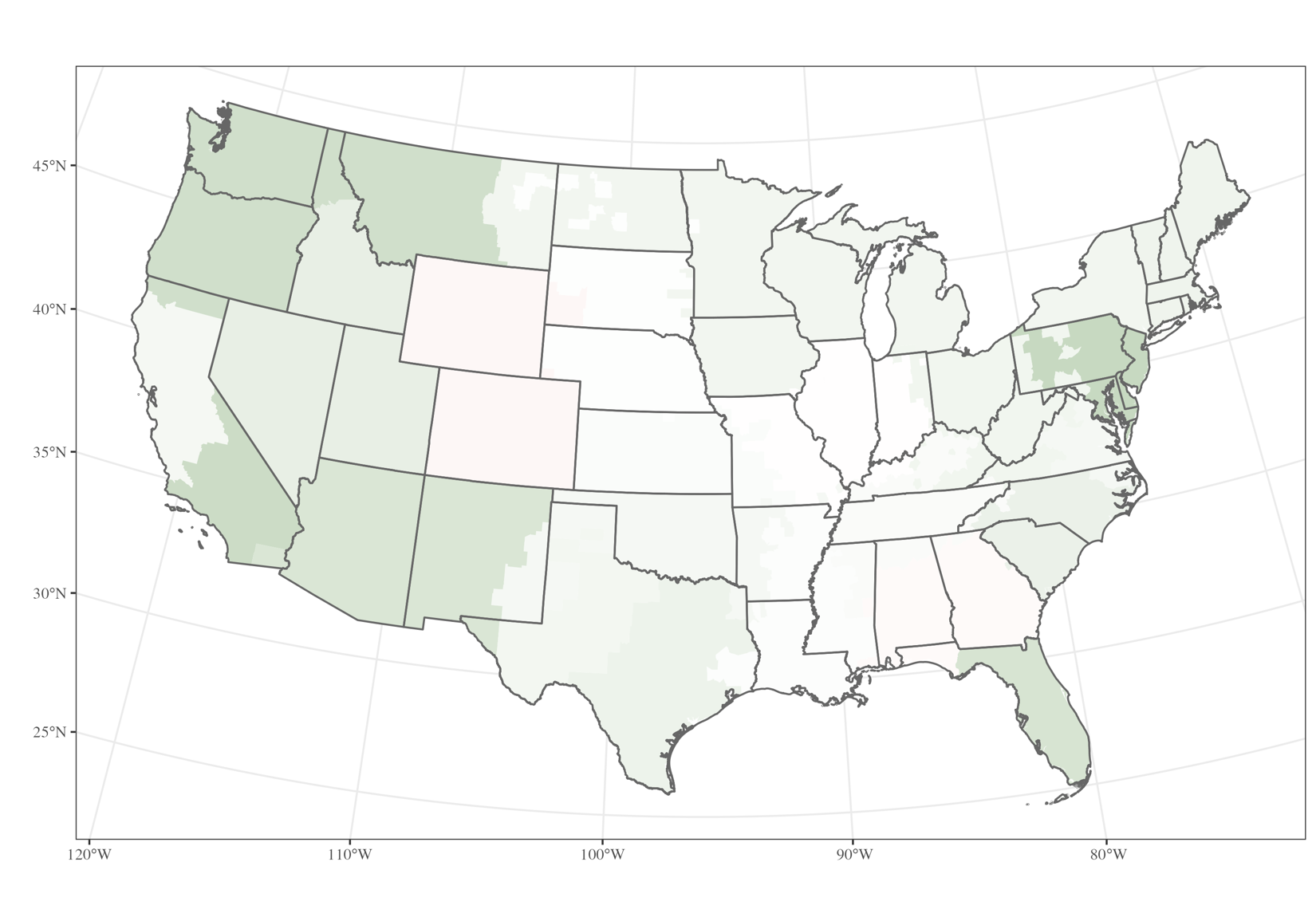} & 
            \includegraphics[width=\linewidth]{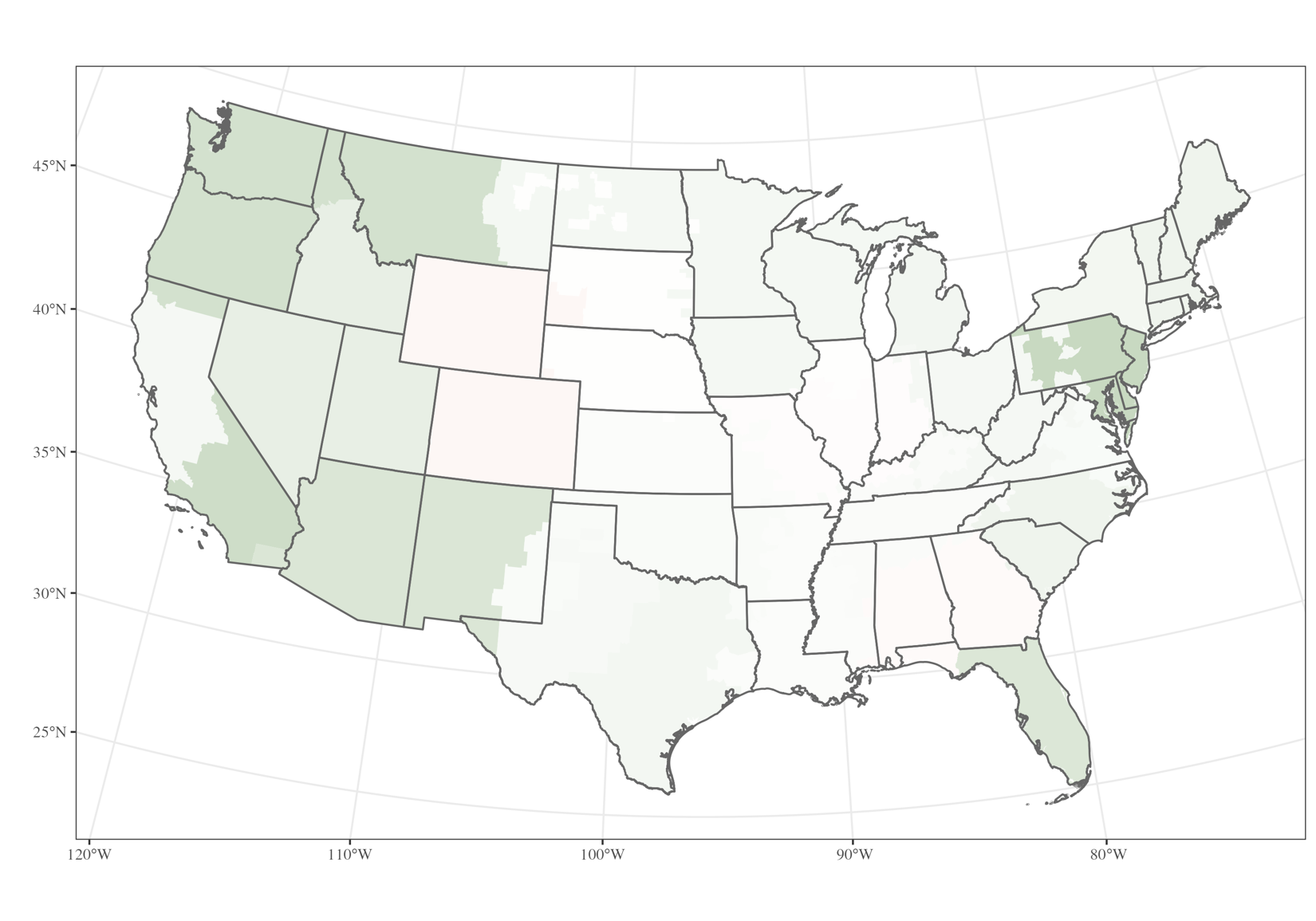} \\ 
    
            & {\footnotesize Profit: \$0.1 T, MSE: \$1 B} &
            {\footnotesize Profit: \$0.2 T, MSE: \$1.2 B} &
            {\footnotesize Profit: \$0.2 T, MSE: \$1.1 B} &
            {\footnotesize Profit: \$0.2 T, MSE: \$1.1 B} &
            {\footnotesize Profit: \$0.2 T, MSE: \$1 B} \\
    
            \rotatebox[origin=c]{90}{\large 26-Zone} &
            \includegraphics[width=\linewidth]{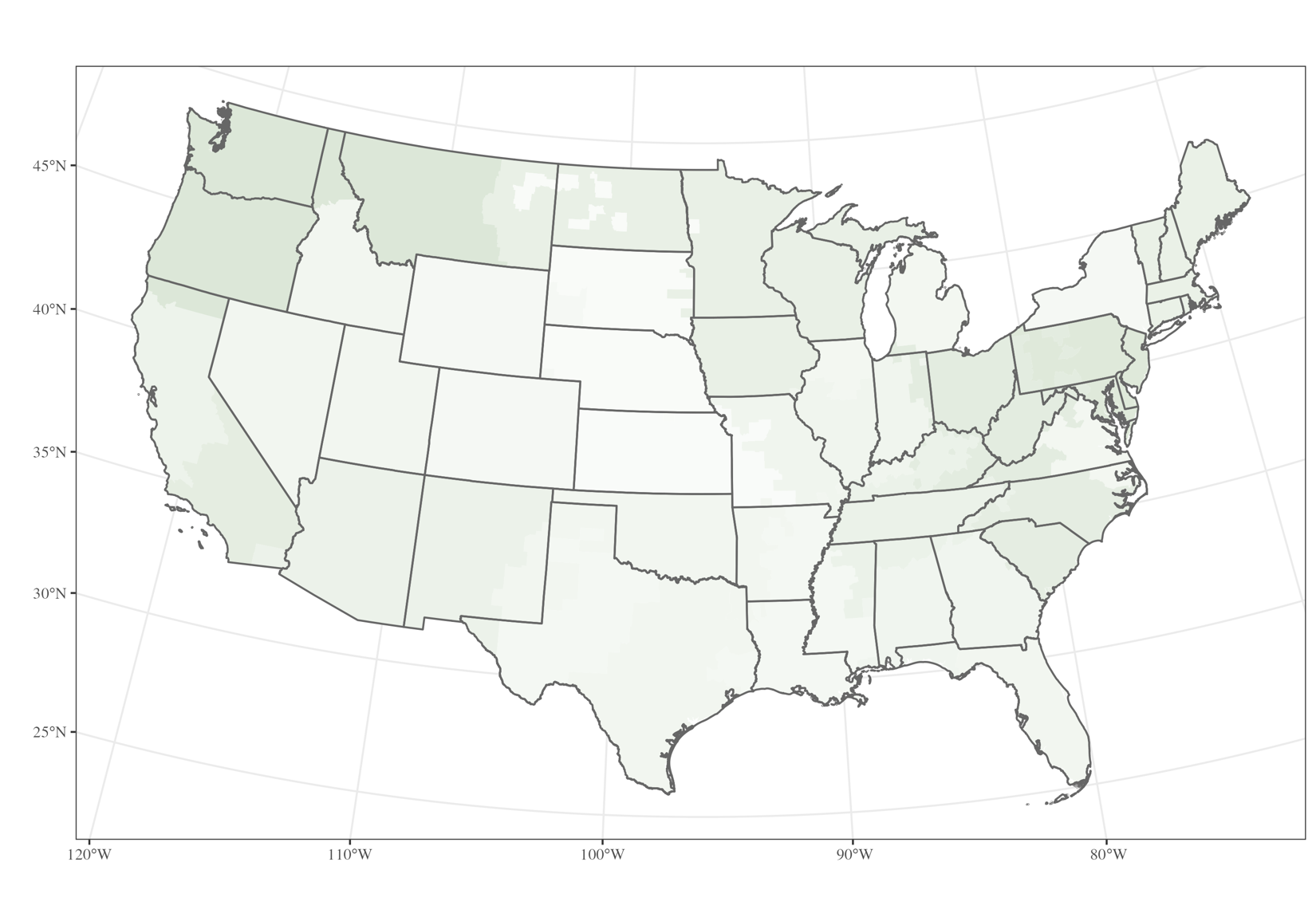} & 
            \includegraphics[width=\linewidth]{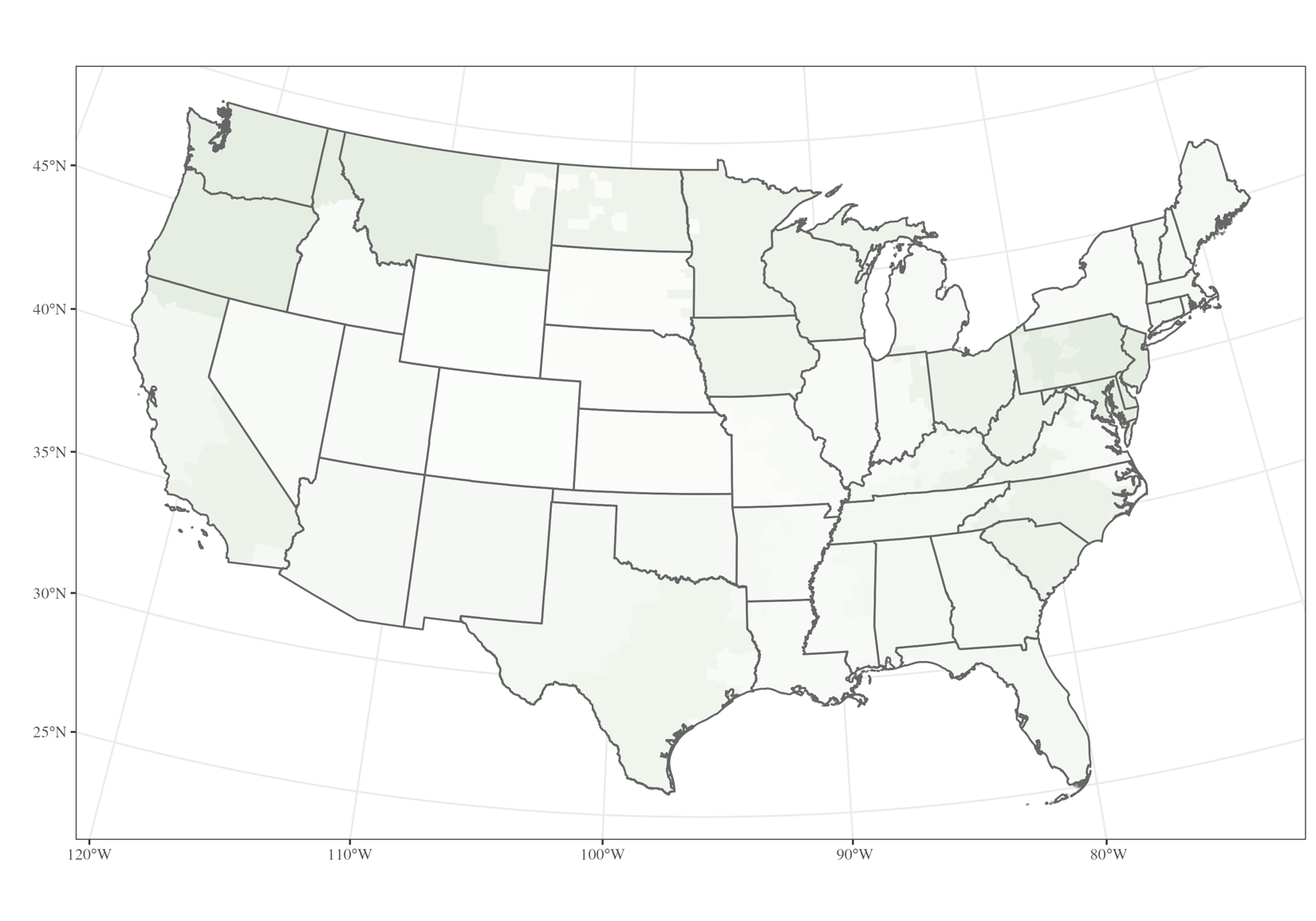} &
            \includegraphics[width=\linewidth]{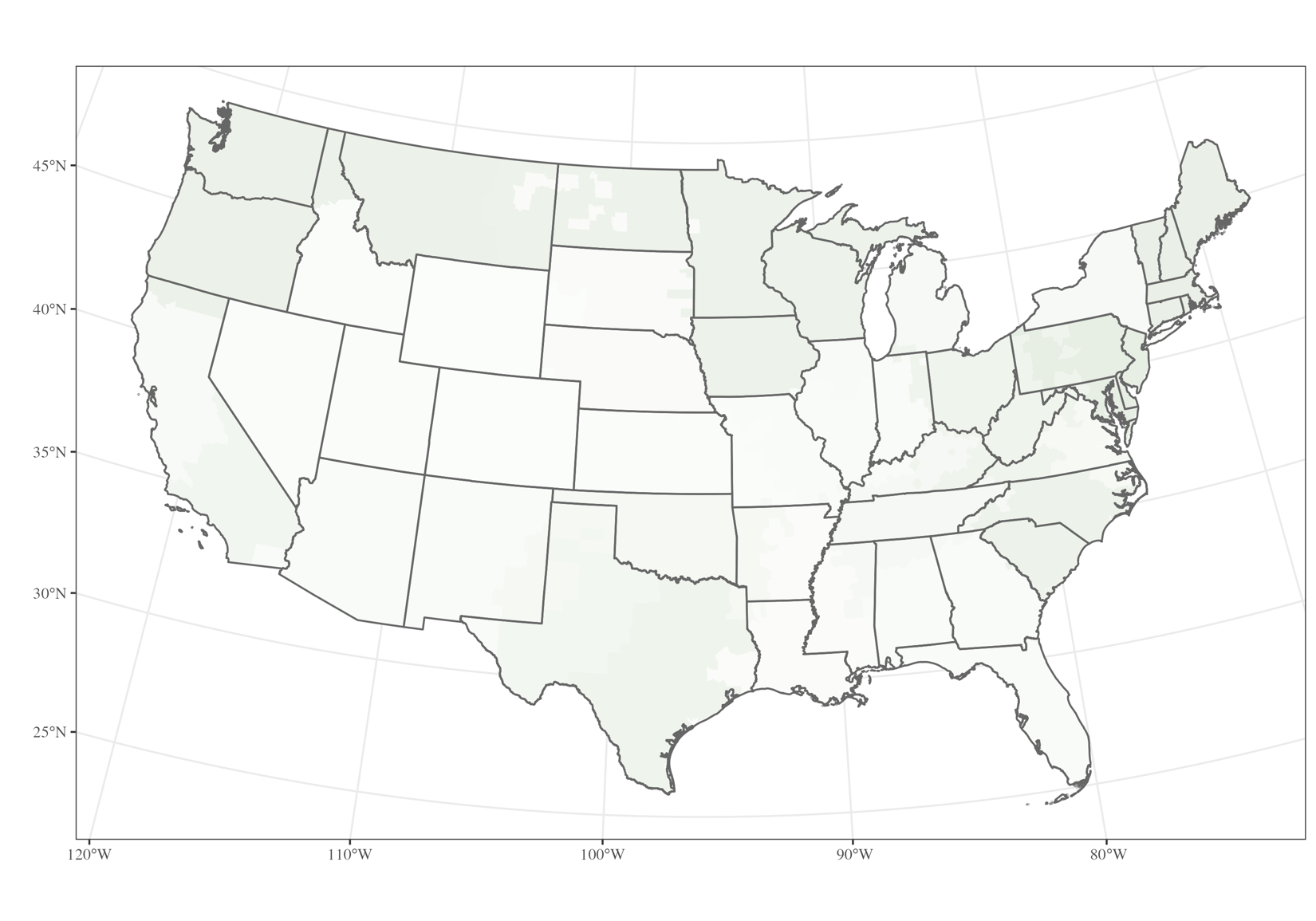} & 
            \includegraphics[width=\linewidth]{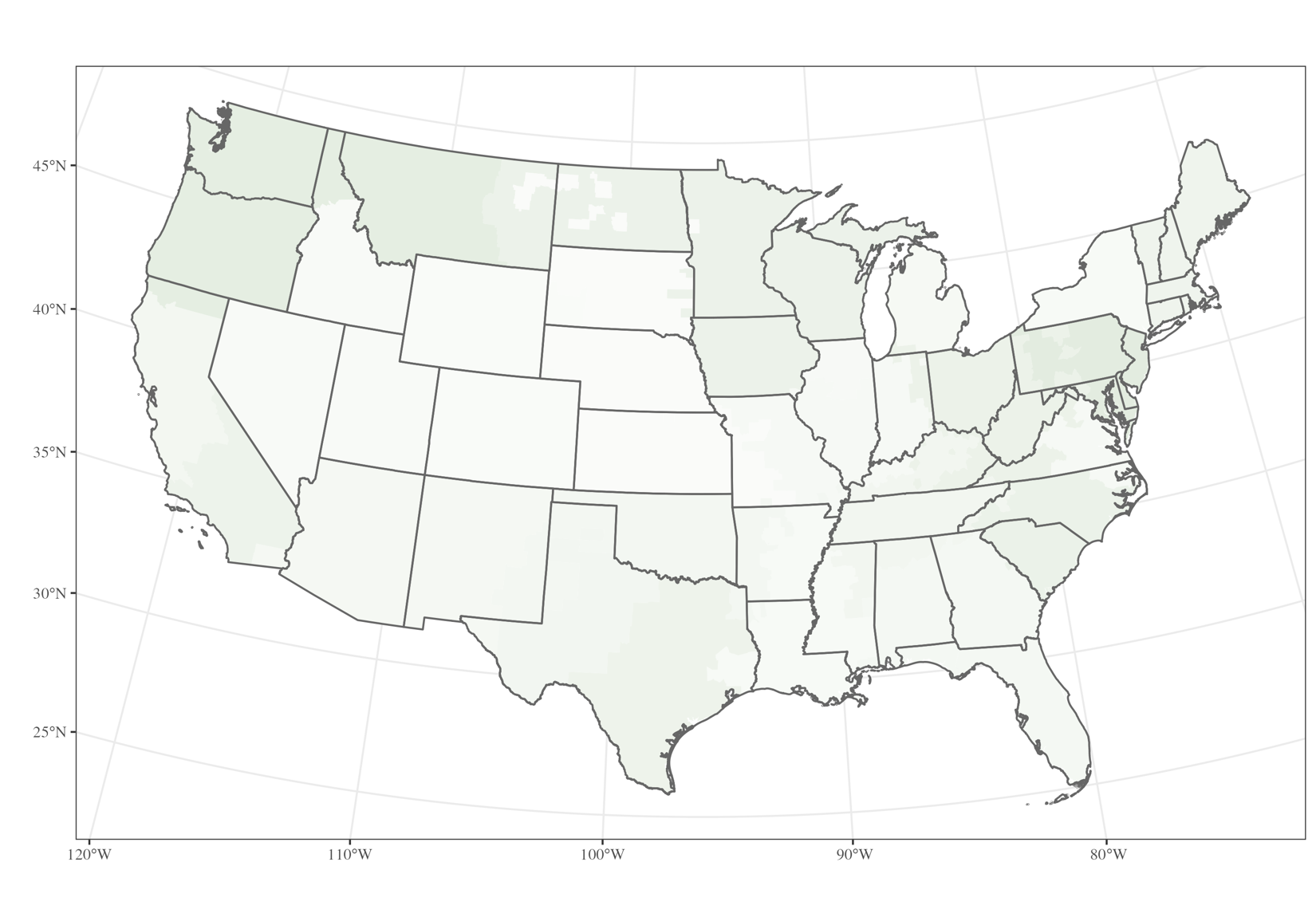} & 
            \includegraphics[width=\linewidth]{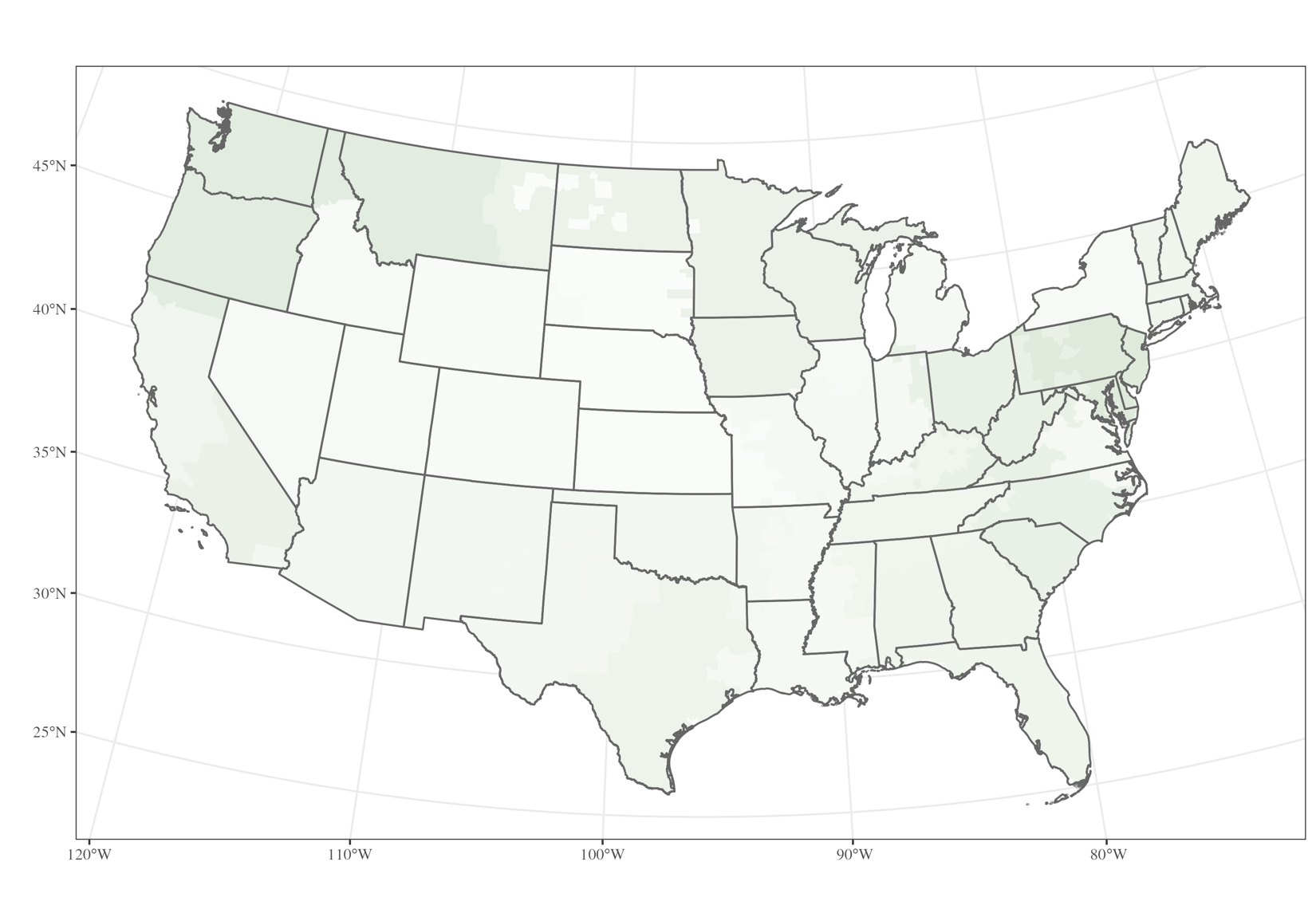} \\
        
            & {\footnotesize Profit: \$0.2 T, MSE: \$0.3 B} &
            {\footnotesize Profit: \$0.1 T, MSE: \$0.2 B} &
            {\footnotesize Profit: \$0.1 T, MSE: \$0.4 B} &
            {\footnotesize Profit: \$0.1 T, MSE: \$0.2 B} &
            {\footnotesize Profit: \$0.1 T, MSE: \$0 B} \\
    
            & & & & & \includegraphics[width=\linewidth]{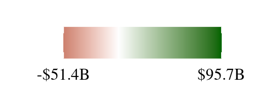} \\
        \end{tabular}
        \captionof{figure}{Total profit for all resources by region. \Coc case. Figure includes the total profit in USD per system, as well as the average per-region mean squared error (MSE) of profit when compared to the highest resolution case. $MSE_{profit} = \sqrt{\sum{(profit - profit_{HRB})^2}} \div 26$. Trends here are likely due to scarcity pricing.}
        \label{map_profit_co2}
    
    \end{table}

    \begin{table}[p]
        \centering
        \setlength\tabcolsep{0pt}
        \begin{tabular}{cM{0.195\linewidth}M{0.195\linewidth}M{0.195\linewidth}M{0.195\linewidth}M{0.195\linewidth}}
            & \large 15-day & \large 30-day & \large 10-week & \large 30-week & \large 52-week \\
            
            \rotatebox[origin=c]{90}{\large 3-Zone} &
            \includegraphics[width=\linewidth]{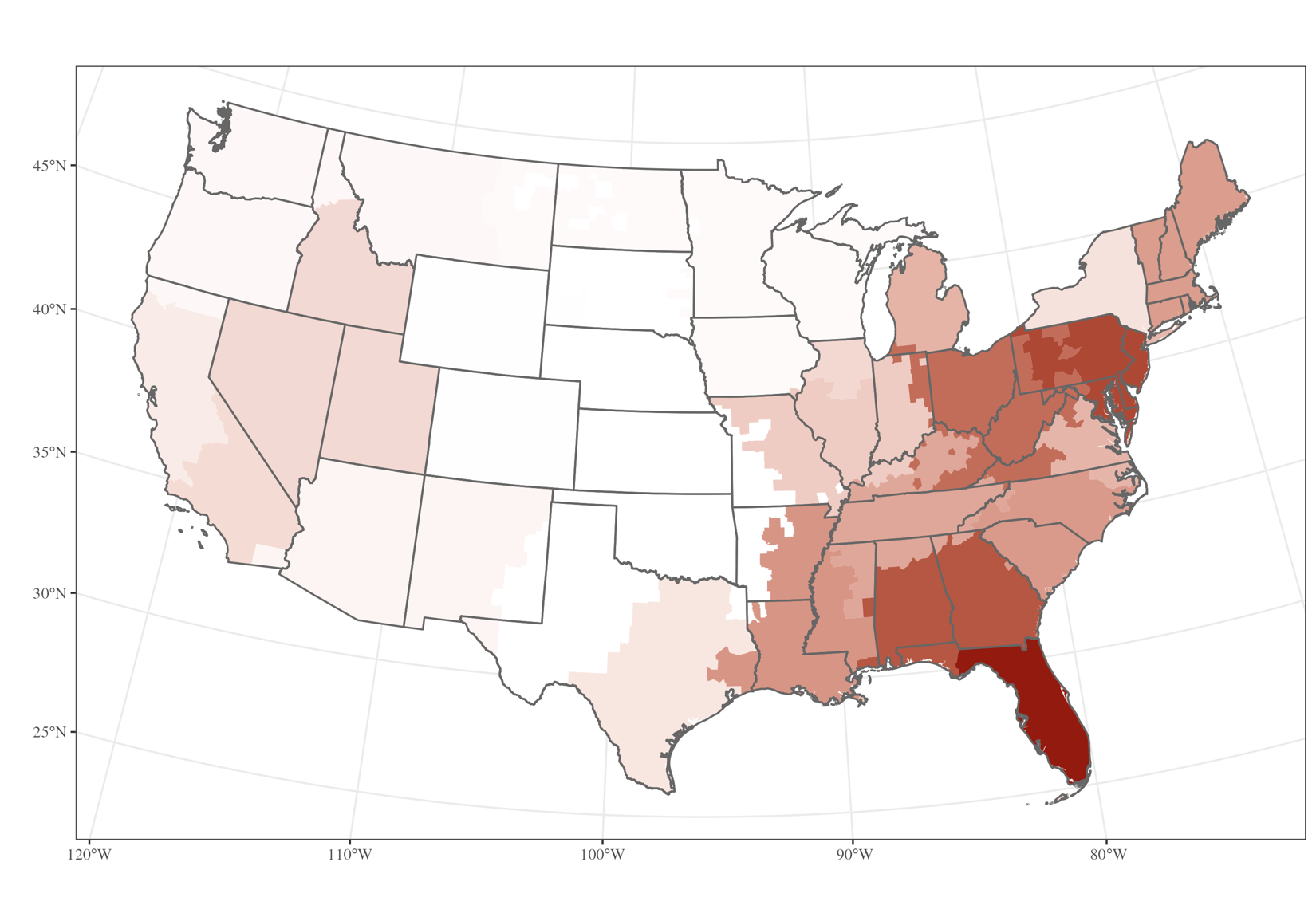} & 
            \includegraphics[width=\linewidth]{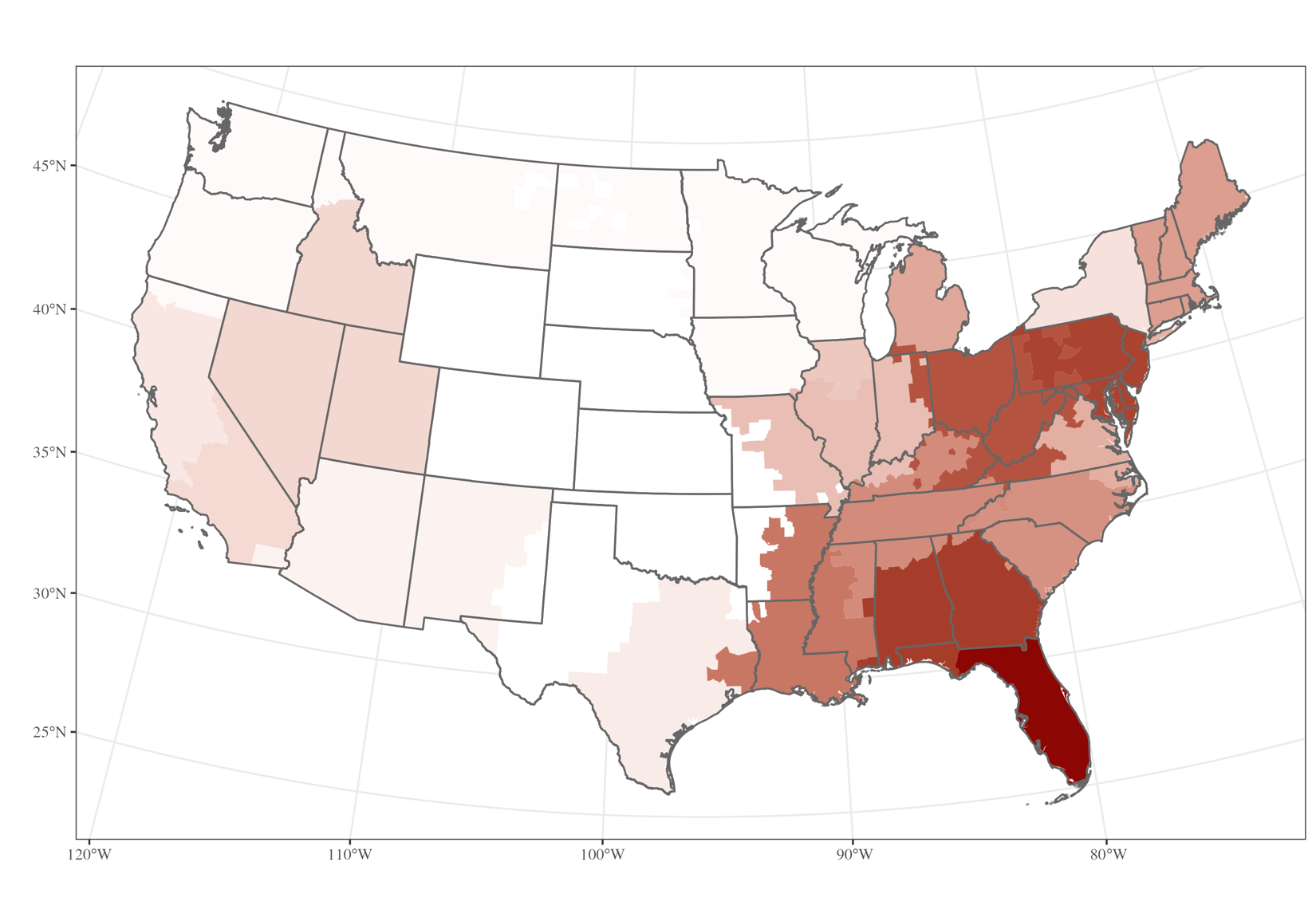} &
            \includegraphics[width=\linewidth]{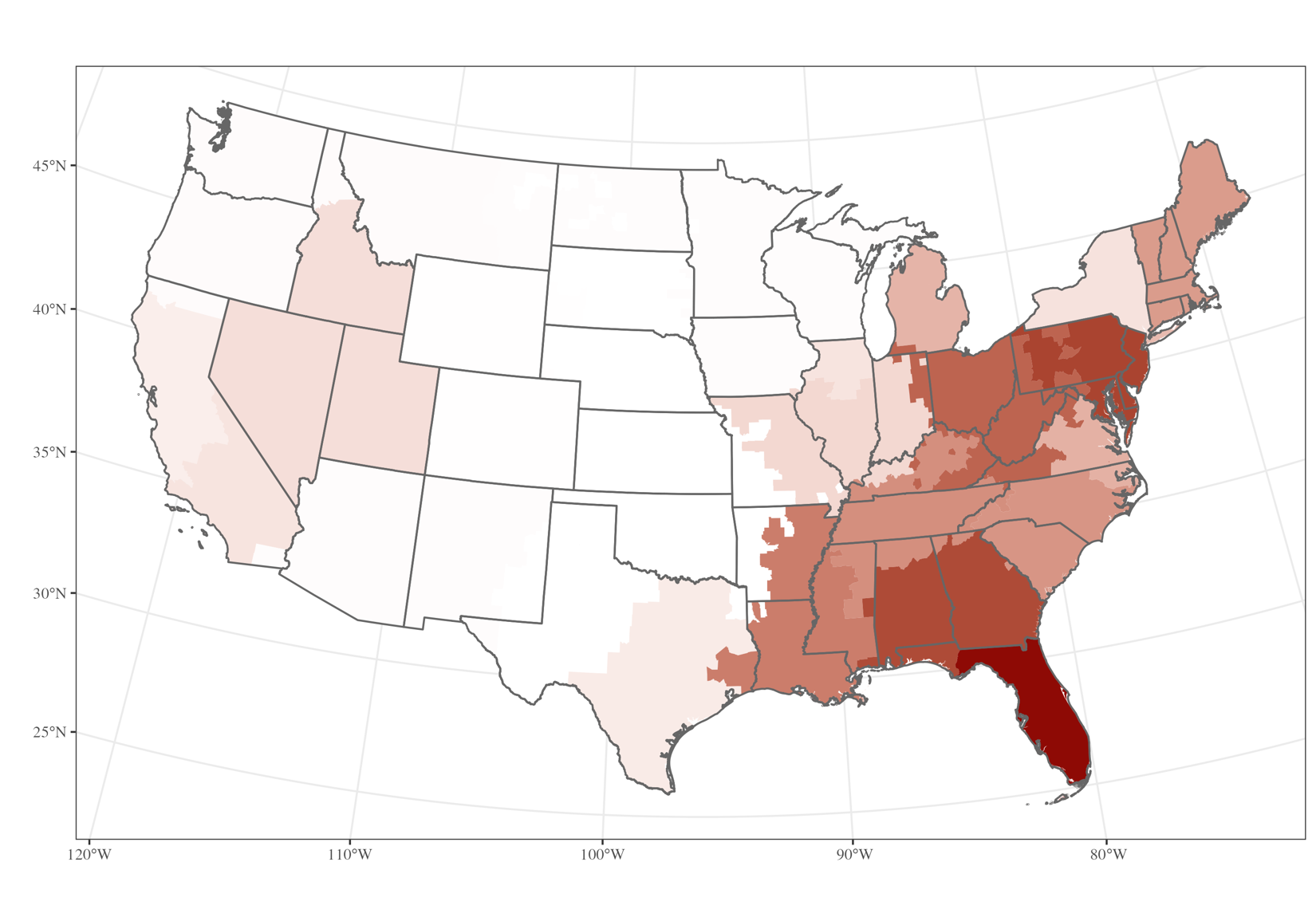} & 
            \includegraphics[width=\linewidth]{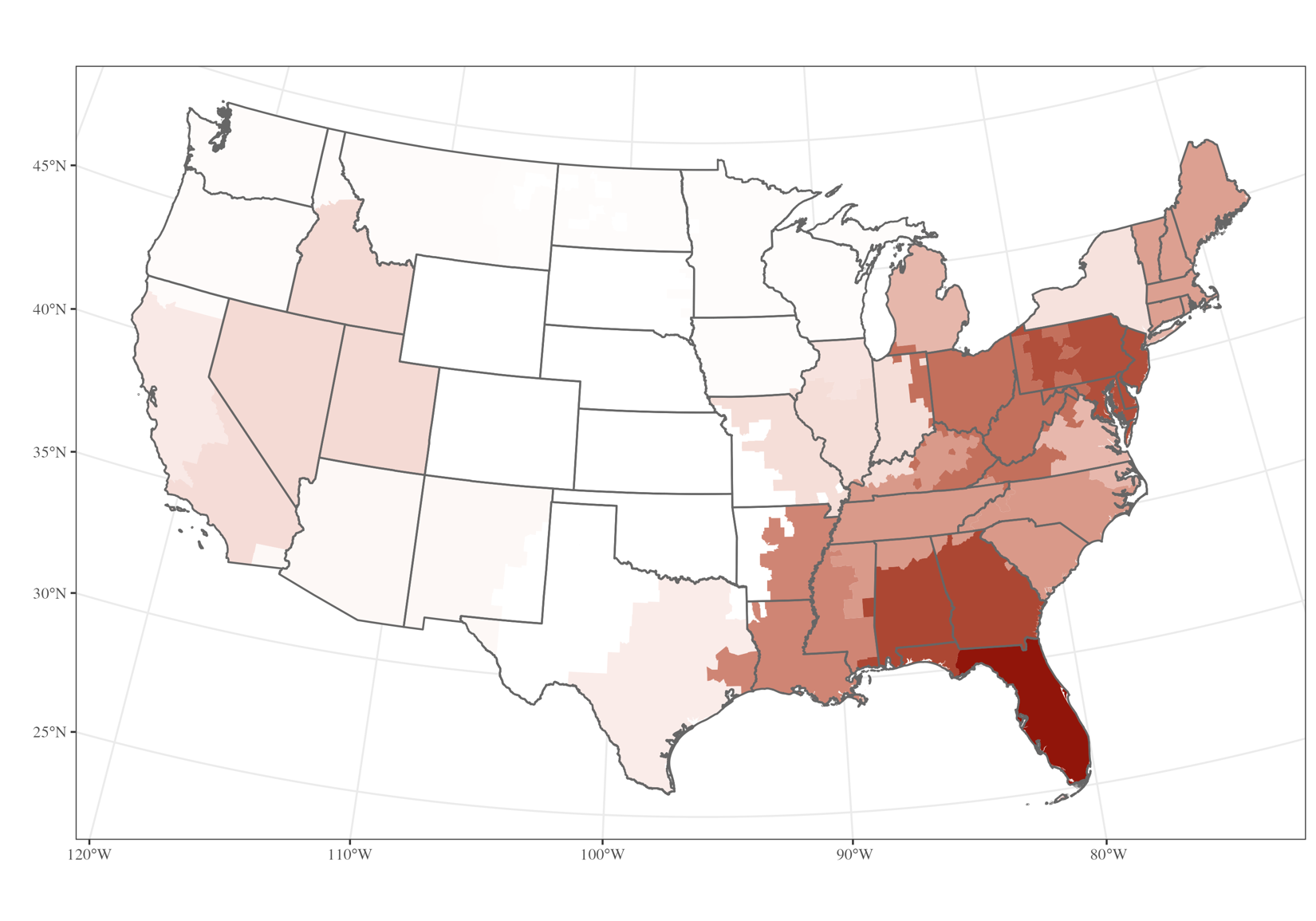} & 
            \includegraphics[width=\linewidth]{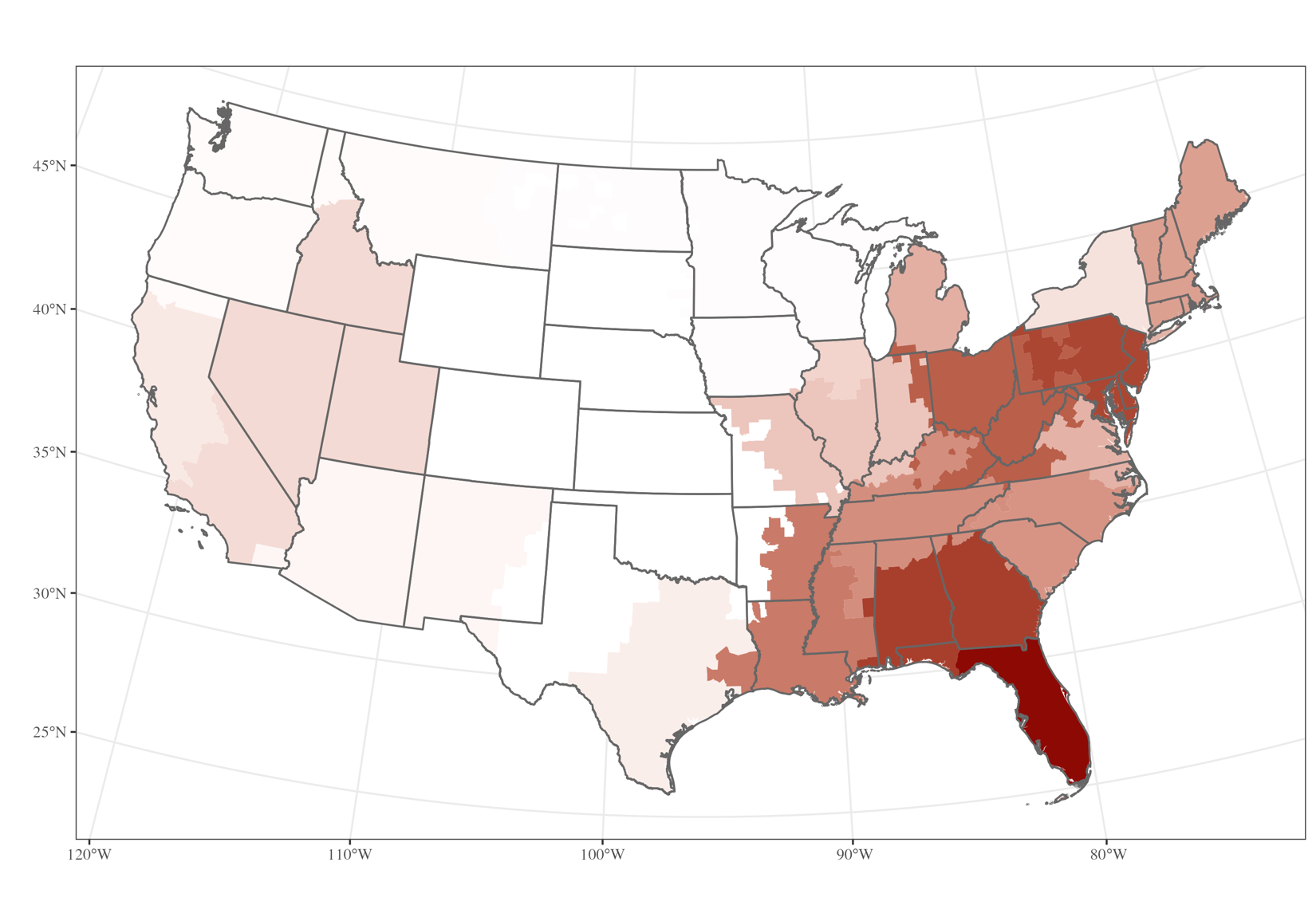} \\
    
            & {\footnotesize Emiss: 7.9e8 ,MSE: 7.3e6 tons CO$_2$} &
            {\footnotesize Emiss: 8.8e8 ,MSE: 8.2e6 tons CO$_2$} &
            {\footnotesize Emiss: 8.0e8 ,MSE: 7.7e6 tons CO$_2$} &
            {\footnotesize Emiss: 7.8e8 ,MSE: 7.4e6 tons CO$_2$} &
            {\footnotesize Emiss: 8.5e8 ,MSE: 8.0e6 tons CO$_2$} \\

            \rotatebox[origin=c]{90}{\large 16-Zone} &
            \includegraphics[width=\linewidth]{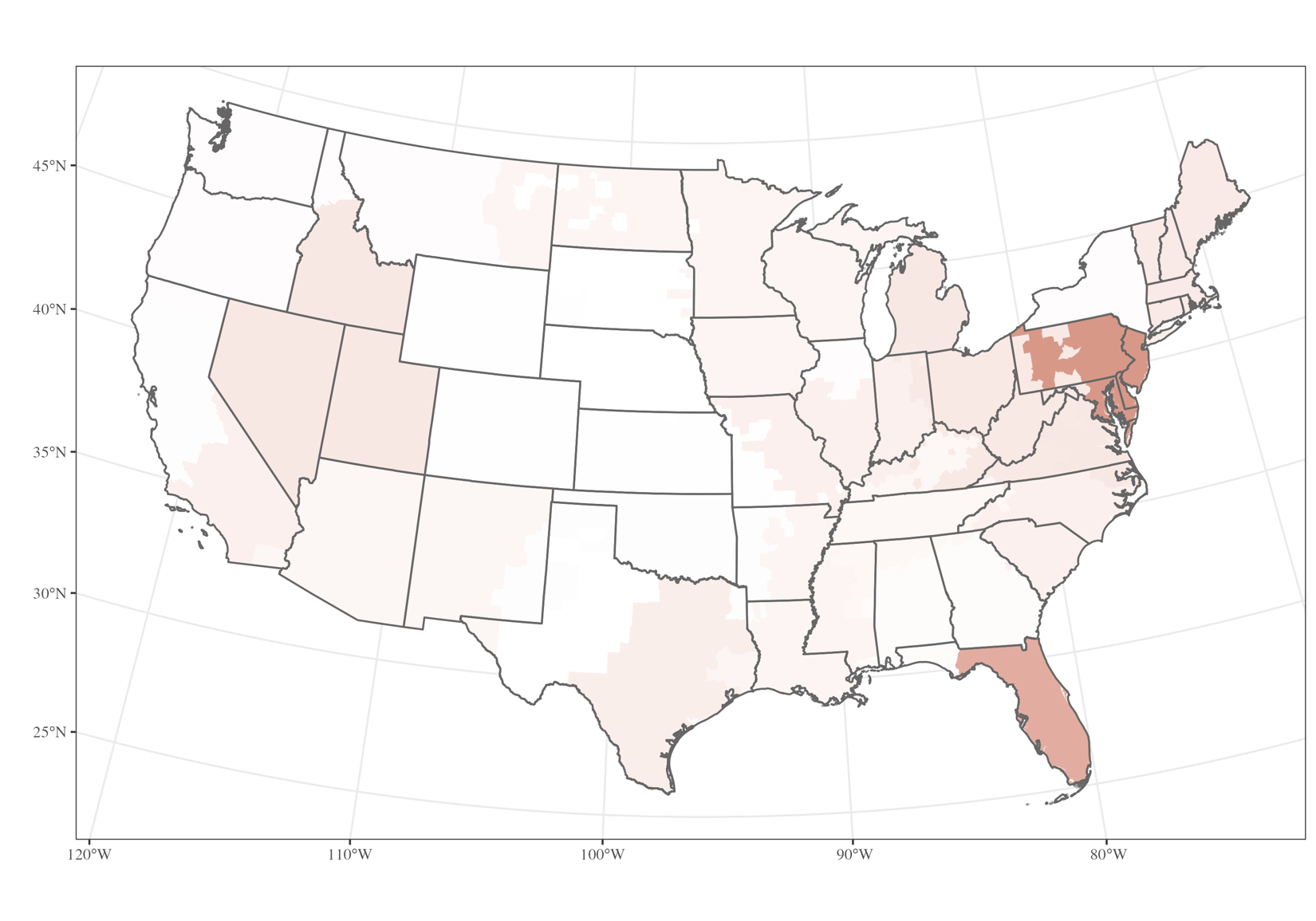} & 
            \includegraphics[width=\linewidth]{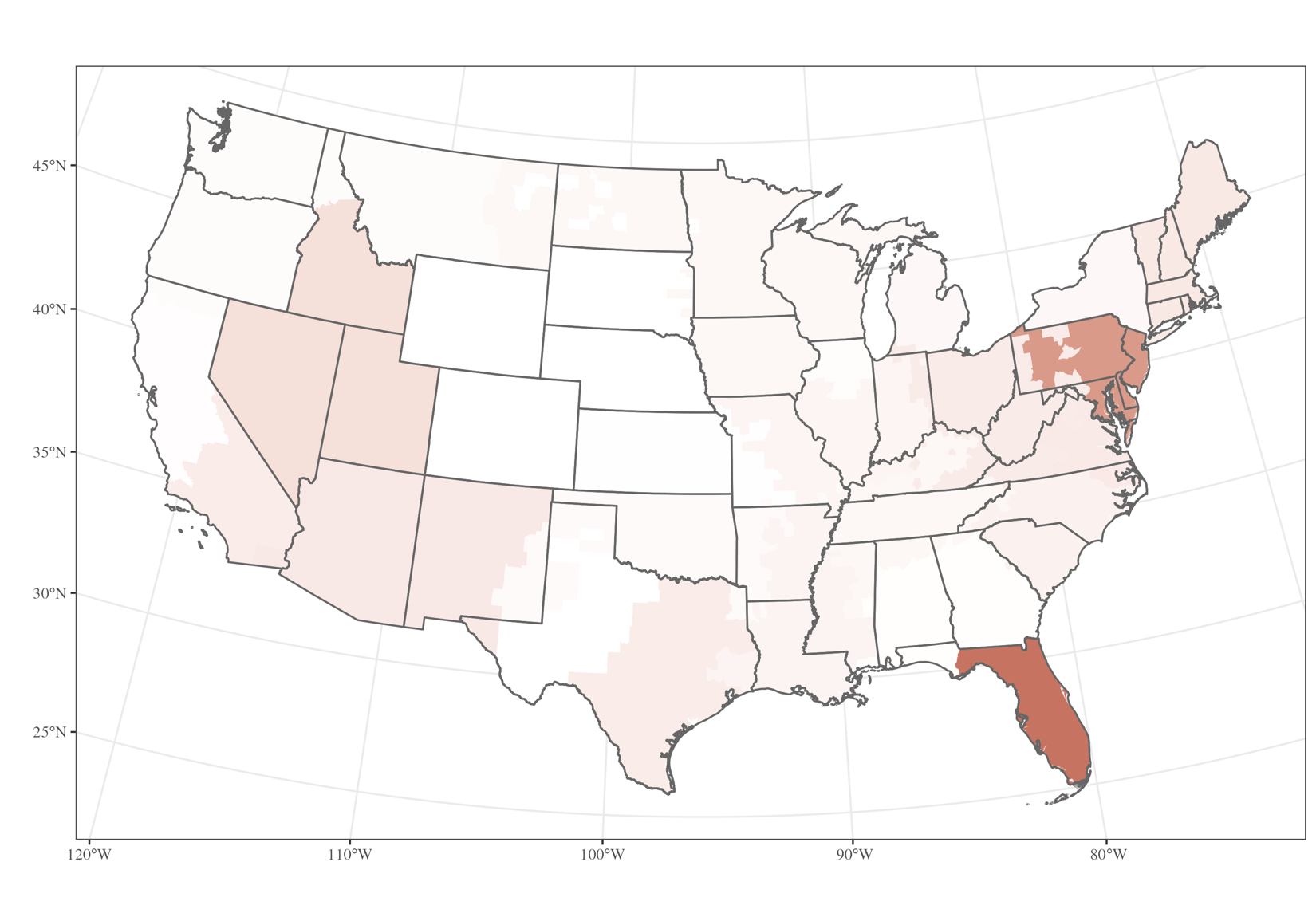} &
            \includegraphics[width=\linewidth]{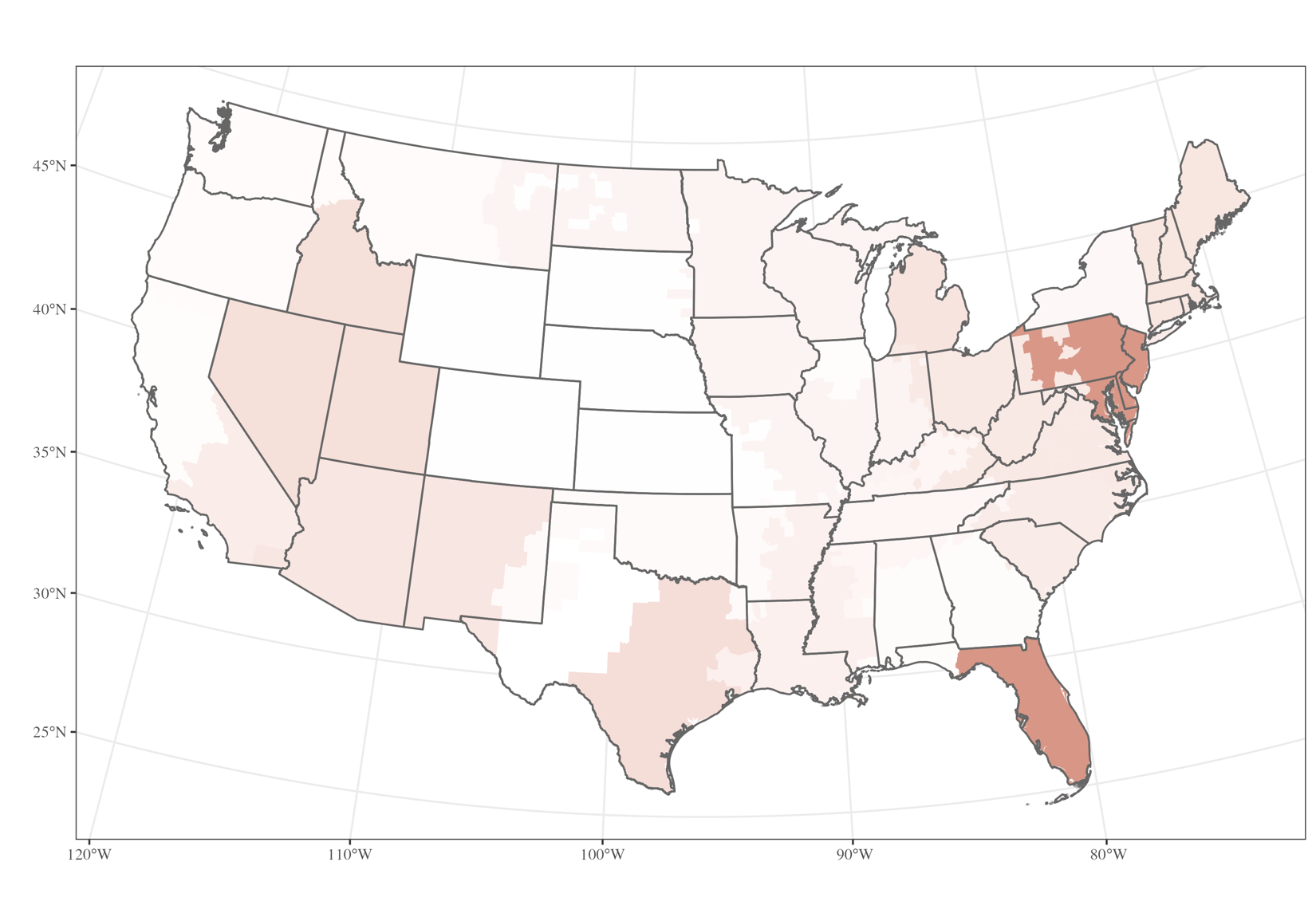} &
            \includegraphics[width=\linewidth]{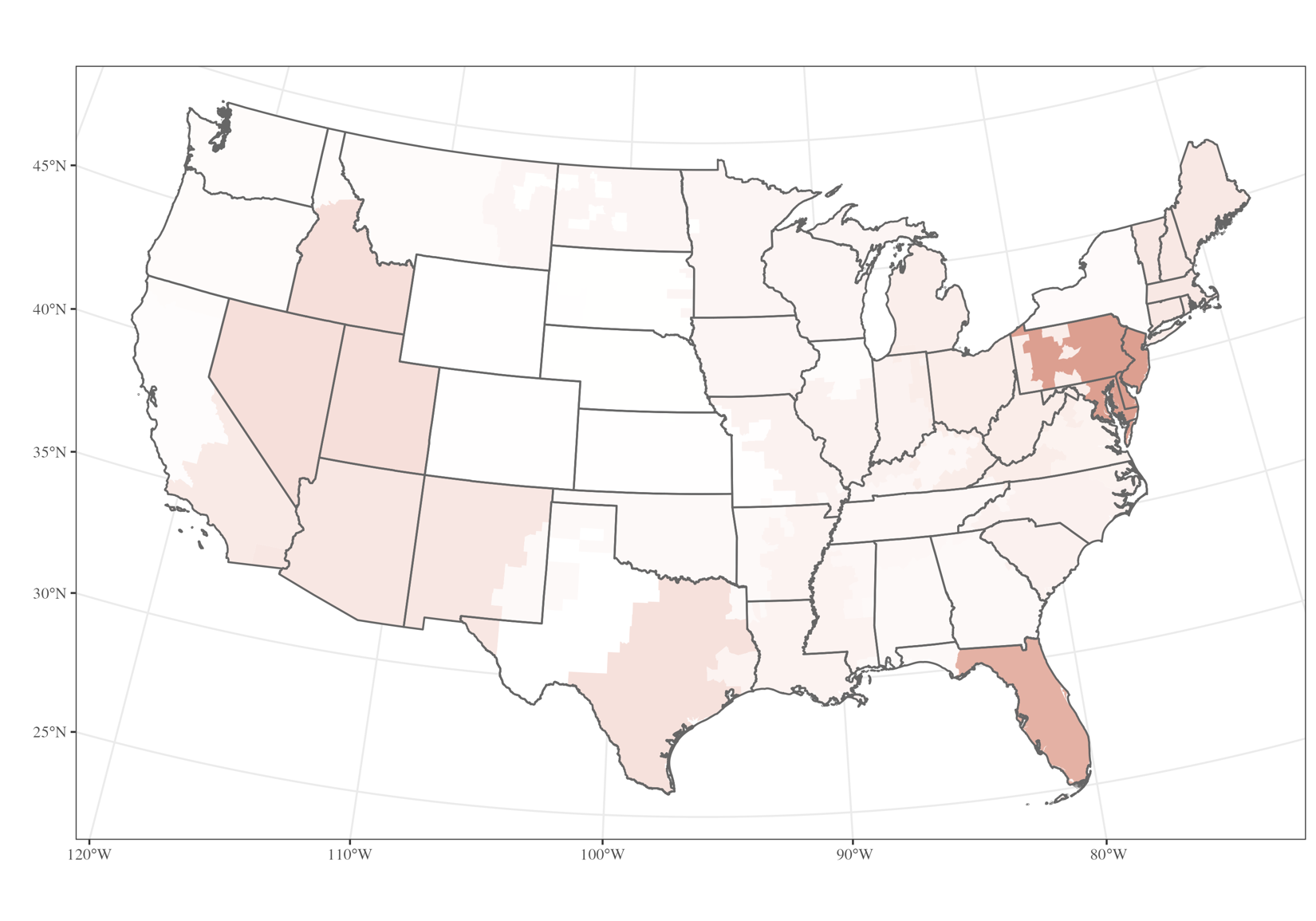} & 
            \includegraphics[width=\linewidth]{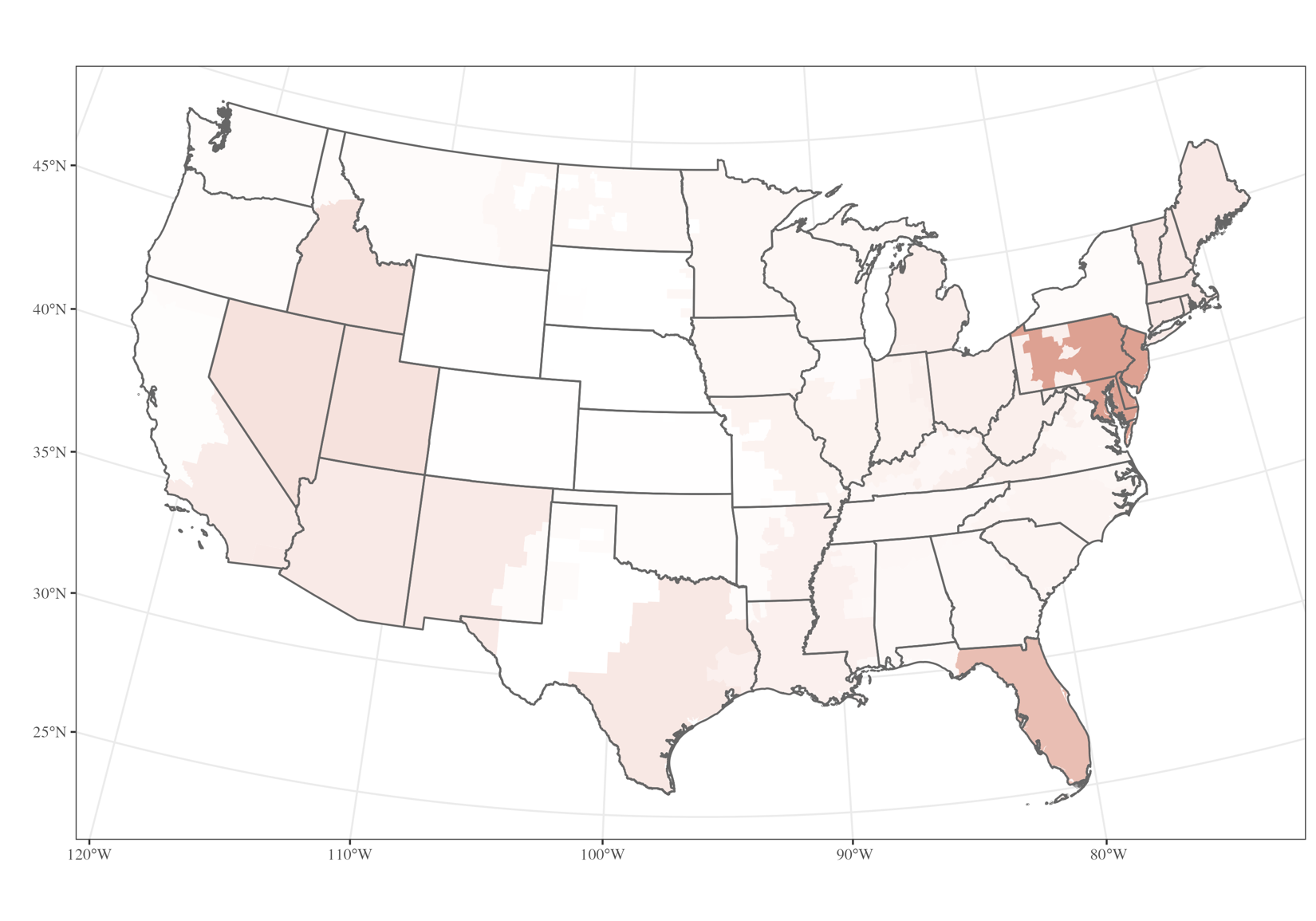} \\
    
            & {\footnotesize Emiss: 2.1e8 ,MSE: 1.9e6 tons CO$_2$} &
            {\footnotesize Emiss: 2.4e8 ,MSE: 2.8e6 tons CO$_2$} &
            {\footnotesize Emiss: 2.4e8 ,MSE: 2.4e6 tons CO$_2$} &
            {\footnotesize Emiss: 2.2e8 ,MSE: 1.8e6 tons CO$_2$} &
            {\footnotesize Emiss: 2.0e8 ,MSE: 1.6e6 tons CO$_2$} \\
    
            \rotatebox[origin=c]{90}{\large 26-Zone} &
            \includegraphics[width=\linewidth]{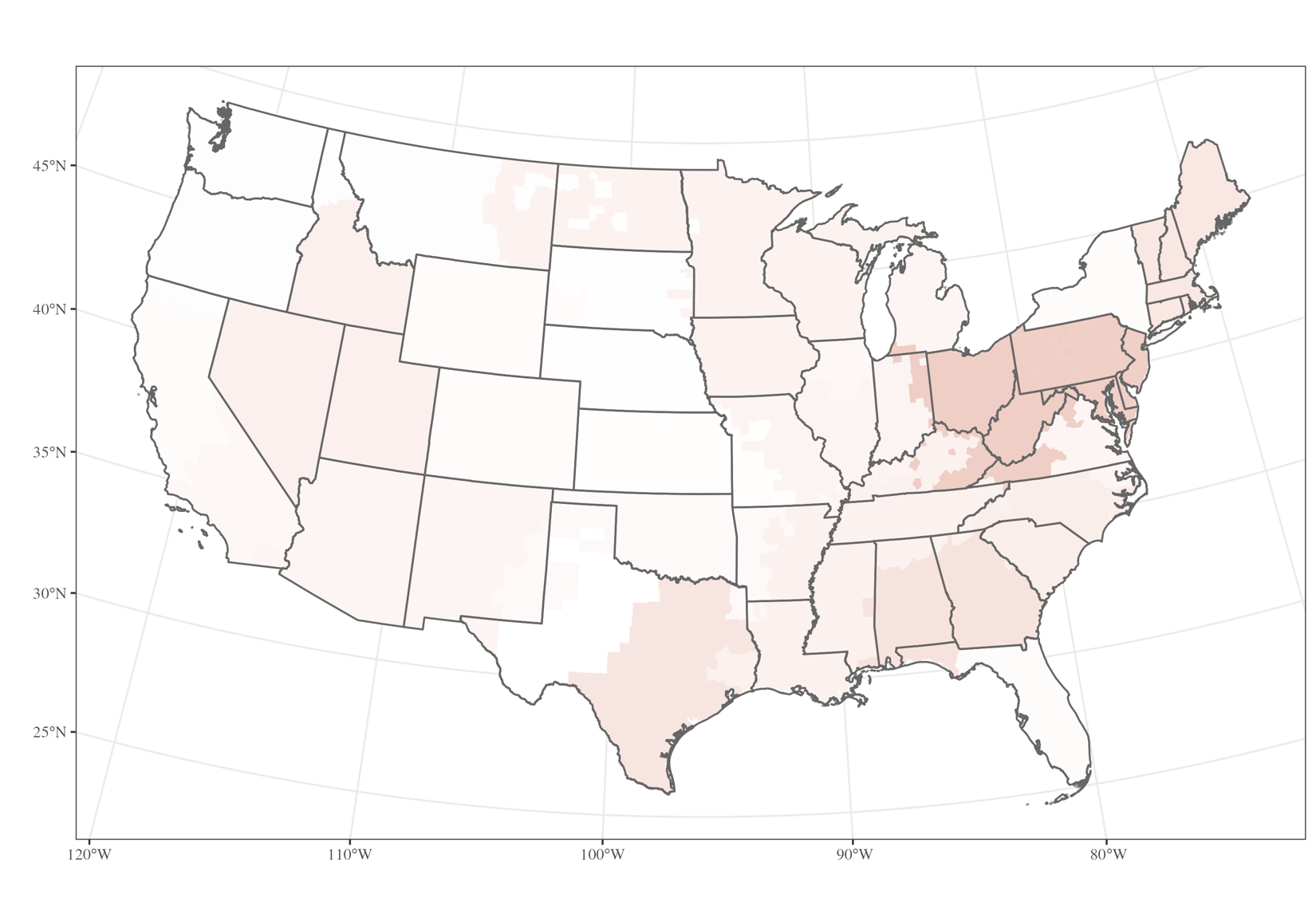} & 
            \includegraphics[width=\linewidth]{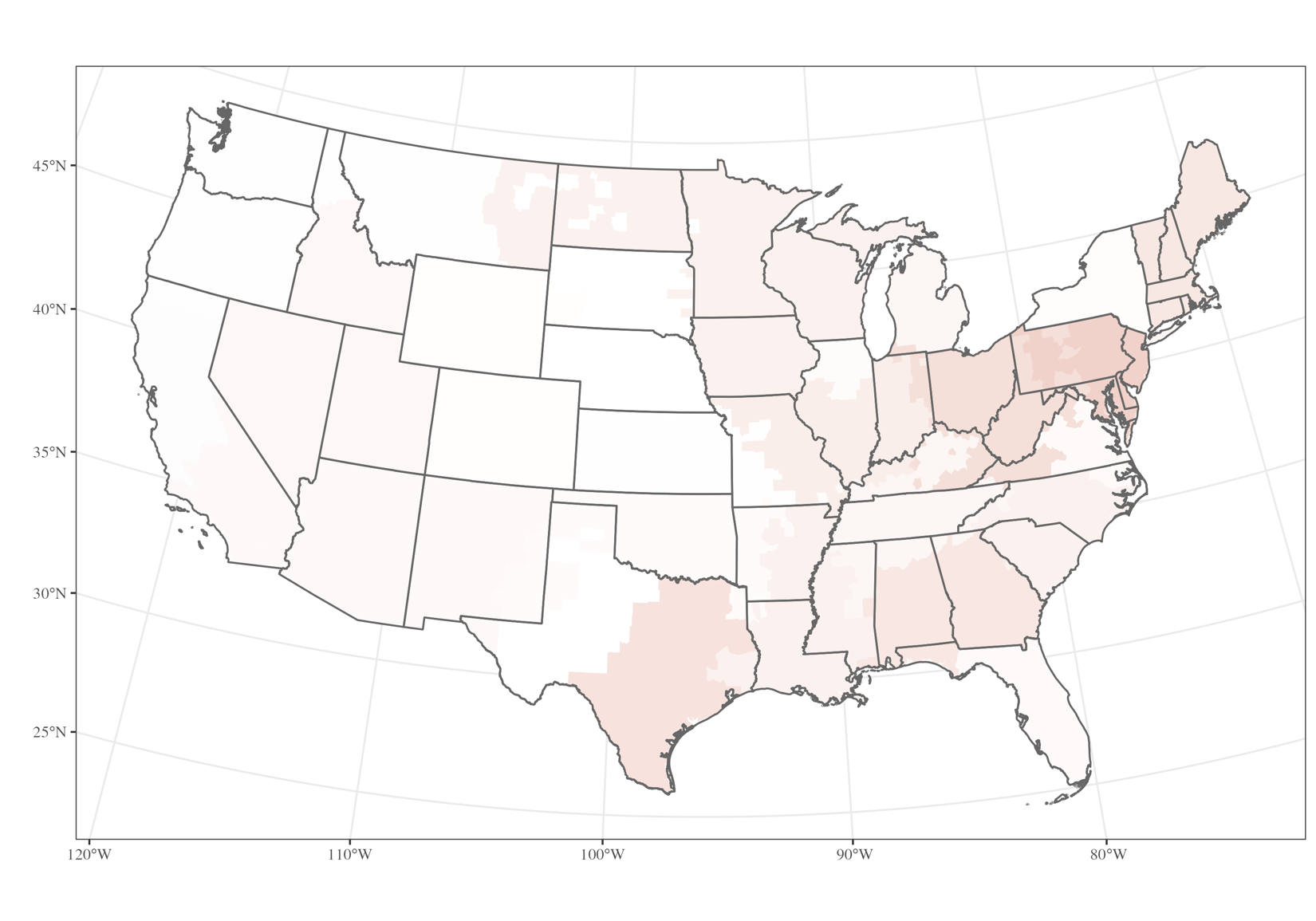} &
            \includegraphics[width=\linewidth]{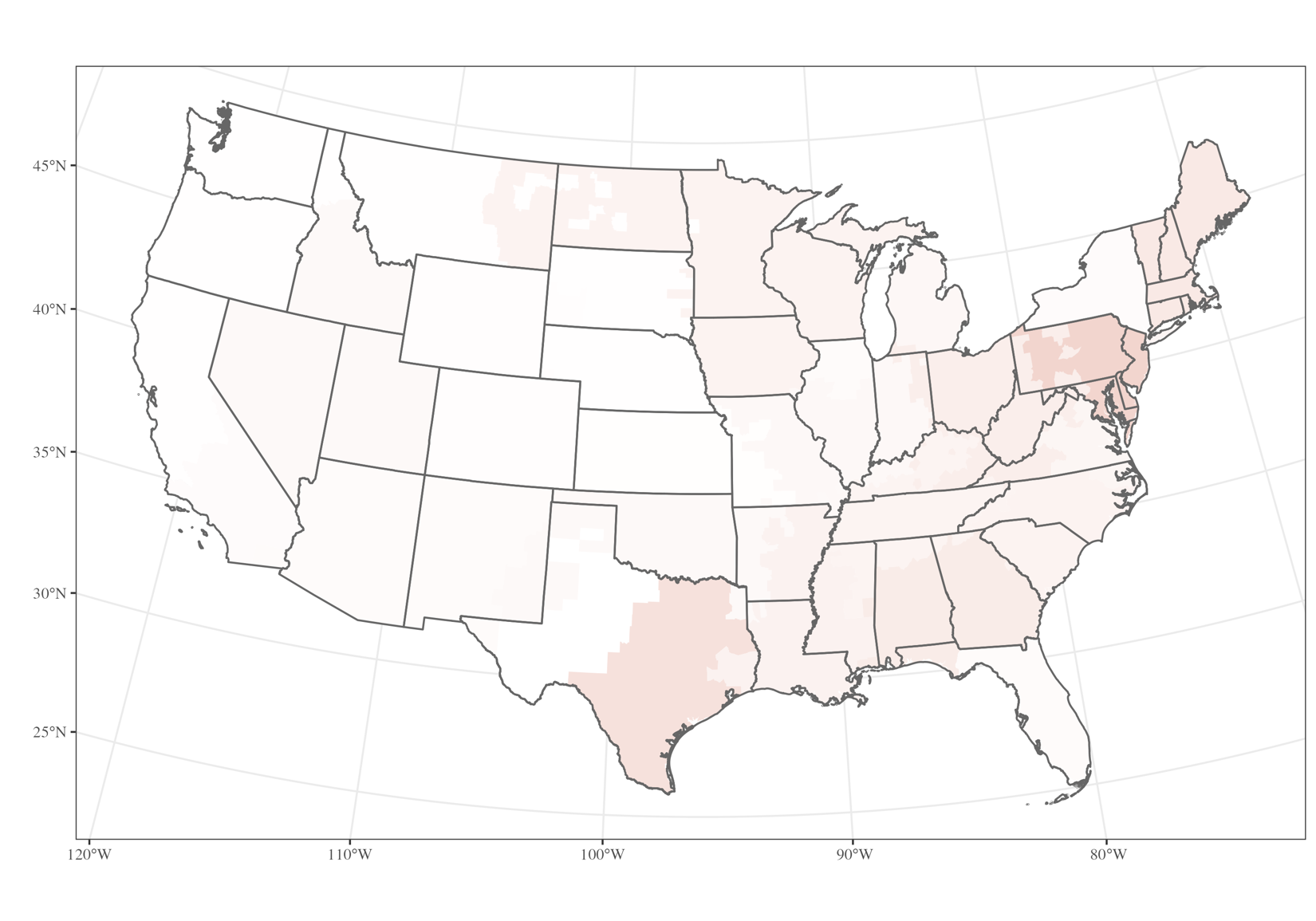} & 
            \includegraphics[width=\linewidth]{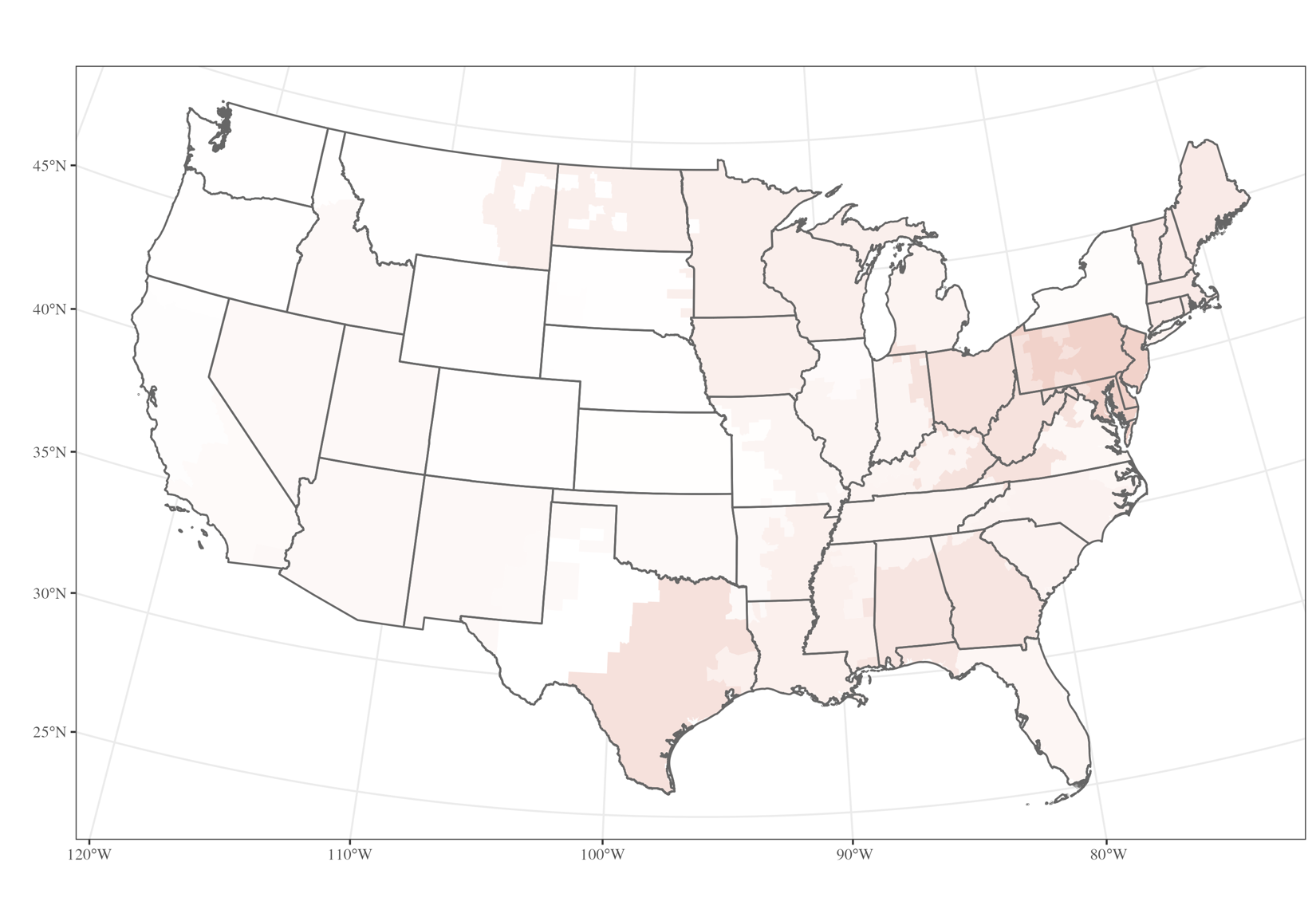} & 
            \includegraphics[width=\linewidth]{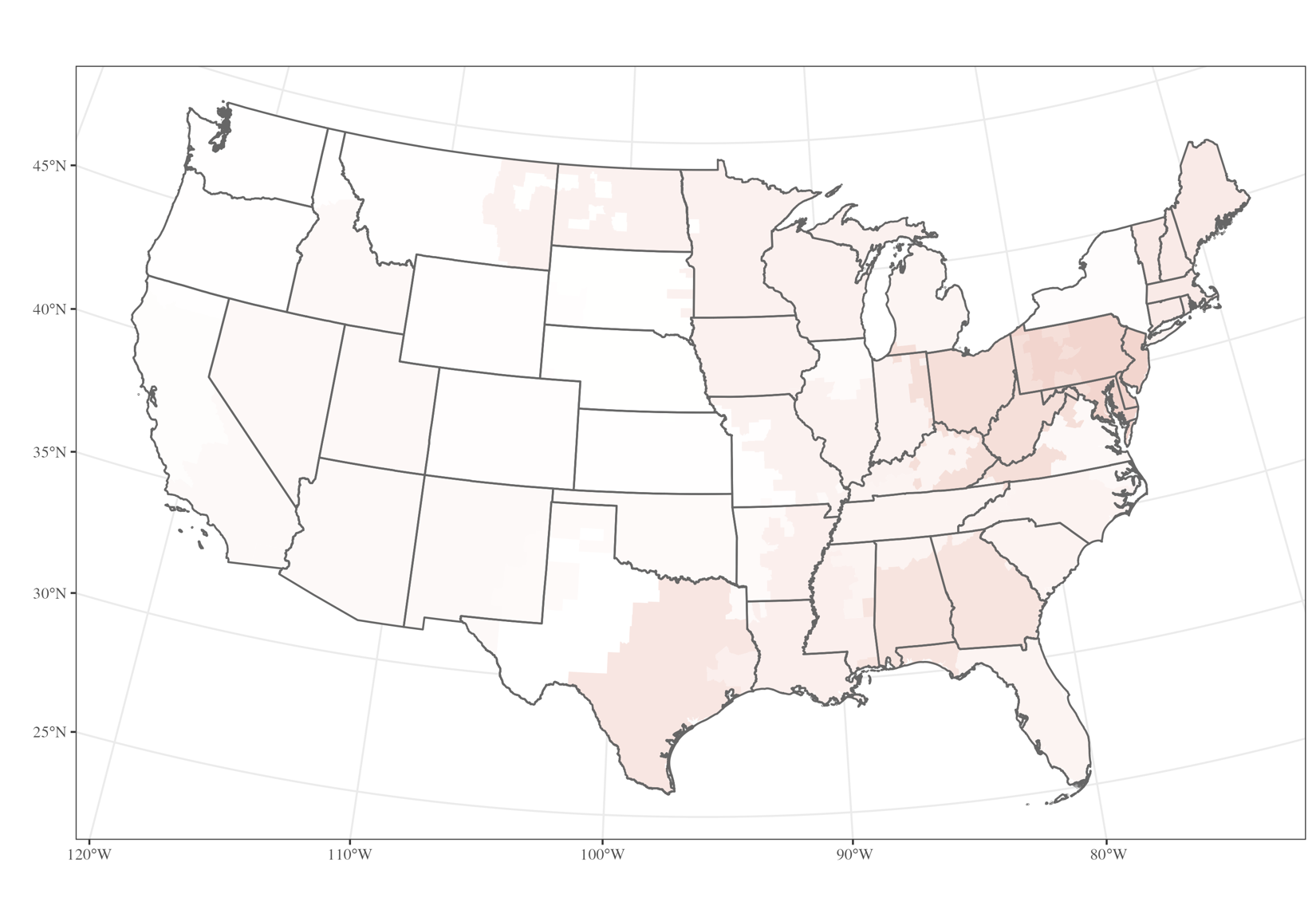} \\
        
            & {\footnotesize Emiss: 1.6e8 ,MSE: 4.4e5 tons CO$_2$} &
            {\footnotesize Emiss: 1.4e8 ,MSE: 2.0e5 tons CO$_2$} &
            {\footnotesize Emiss: 1.2e8 ,MSE: 3.9e5 tons CO$_2$} &
            {\footnotesize Emiss: 1.4e8 ,MSE: 1.5e5 tons CO$_2$} &
            {\footnotesize Emiss: 1.4e8 ,MSE: 0 tons CO$_2$} \\

            & & & & & \includegraphics[width=\linewidth]{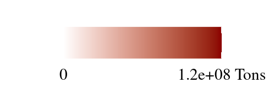}
    
        \end{tabular}
        \captionof{figure}{Systemwide CO$_2$ emissions by spatial and temporal resolution. \Coc case. Results show decreasing total emissions as spatial and temporal resolution increases. Mean squared error (MSE) of location of emissions correlates inversely with model resolution. $MSE_{emiss} = \sqrt{\sum{(emiss - emiss_{HRB})^2}} \div 26$\label{map_emiss}. All emissions shown here must be abated at \$200 per ton.}
    
    \end{table}
    
\end{landscape}

\subsection{Optimization Phase Comparison} \label{results:case}
Operational inaccuracies occur because low resolution \opone~systems cannot predict their resources' optimal operations when high resolution operational constraints are enforced. Here we compare predicted \opone~phase operations with final operations from the \optwo~phase to explore why coarse models proposed the resource mixes that they did.

\begin{figure}[H]
    \begin{center}
    \subfloat[Spatial / operational resolutions \label{phase_comparison_spat}]{\includegraphics[width=\linewidth]{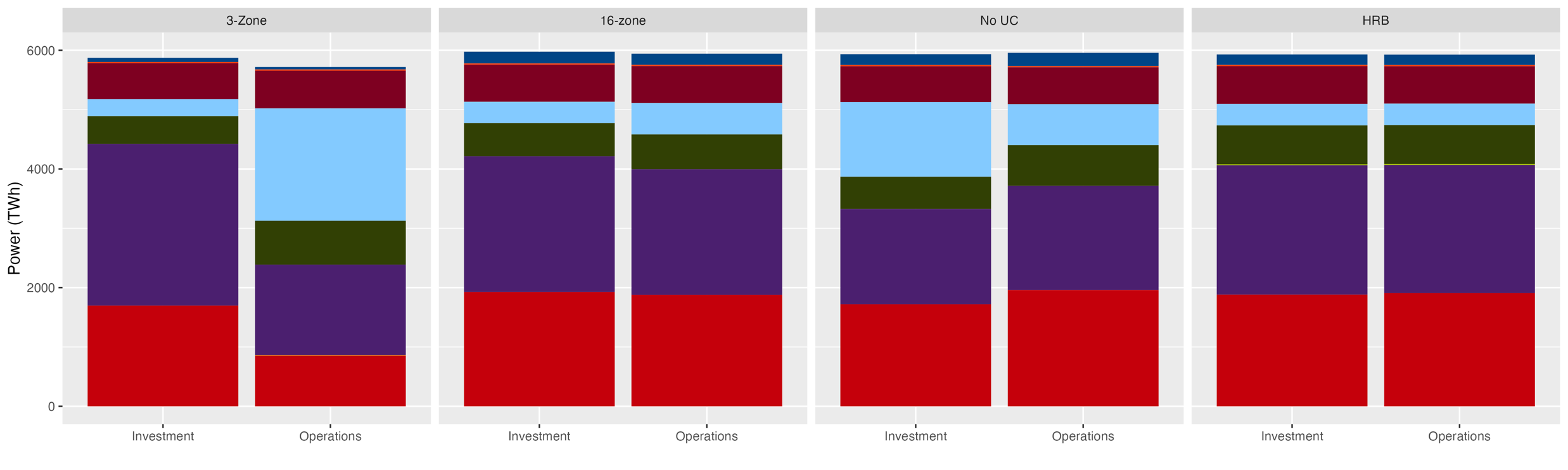}} \\
    \subfloat[Temporal resolutions\label{phase_comparison_temp}]{\includegraphics[width=\linewidth]{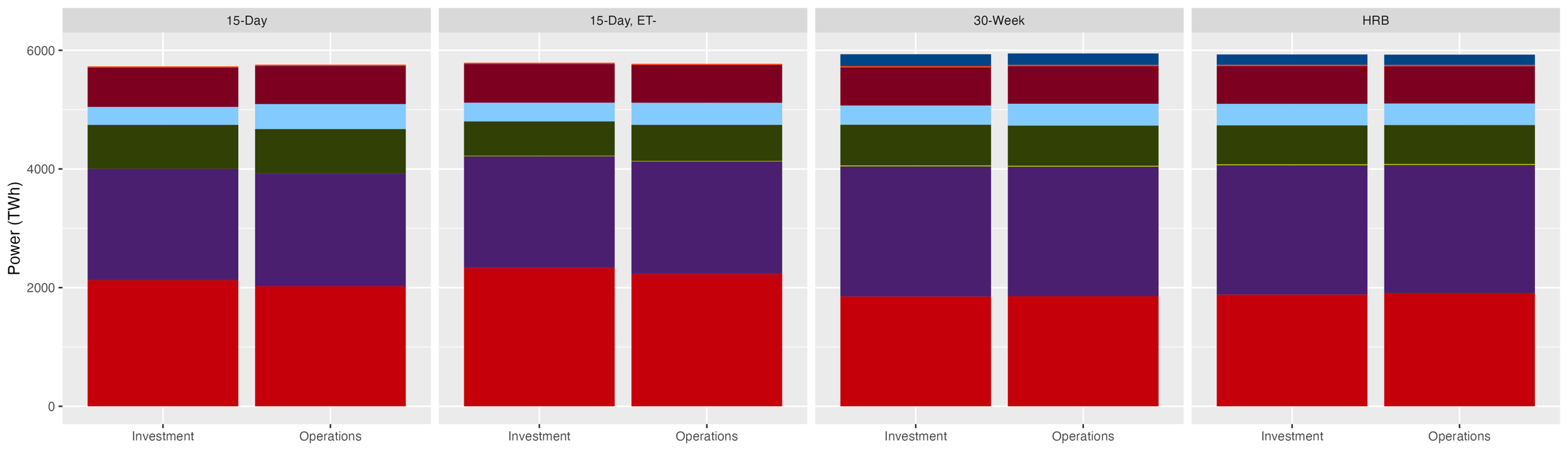}} \\
    \subfloat{\includegraphics[width=\linewidth]{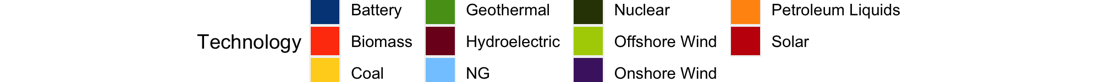}}
    \caption{Generation by technology. Figure compares operations from the \opone~and \optwo~phases. For low resolution cases, first phase optimizations are misinformed about how much they can mobilize resources. Results here are for the \coc case. Results for the \refc case show similar trends and are included as Fig~\ref{phase_comparison_si} in the SI. HRB: High resolution baseline. UC: Unit commitment.\label{phase_comparison}}
    \end{center}
\end{figure}  

Spatially coarse models overestimate VRE mobilization (Fig~\ref{phase_comparison_spat}.) These systems poorly captured weather and demand patterns, leading to altered VRE availability at high resolution. Furthermore, they ignored intraregional transmission across large areas, failing to predict bottlenecks. As a result of poor siting with concentrated areas of VRE, low resolution systems provide recommendations with low VRE consistency. Due to failure to predict bottlenecks, these systems overestimate deliverability. Impacts decrease by 16-zones. 

Omission of UC leads \opone~models to overestimate NG generation (Fig~\ref{phase_comparison_spat}). In GenX, thermal resources with minimum run constraints can power down fully with UC. Without UC, they are constrained to run at their minimum output value. This is the cause of the VRE curtailment seen in the first phase non-UC optimization: NG under minimum run constraints is unable to shut down and power up intermittently to accommodate times of high VRE generation. We confirm that NG with carbon capture (with minimum run constraints) replaces VRE in times of curtailment. We also confirm that plants with nonzero generation in the non-UC case always have capacity factor greater than the minimum output, and that this does not hold in the UC optimization.

Temporally coarse models have marginally higher VRE operations in their investment phase (Fig~\ref{phase_comparison_temp}.) This is because they fail to predict weather patterns, leading to suboptimal siting within regions. Consistent with the rest of our results, temporally coarse models show smaller error than spatially coarse ones with smaller discrepancies between optimization phases. Impacts are largely masked by 30 weeks.

\begin{figure}[h]
    \begin{center}
    \subfloat[\refca case, zonal.\label{cost_comp_ref_z}]{\includegraphics[width=0.5\linewidth]{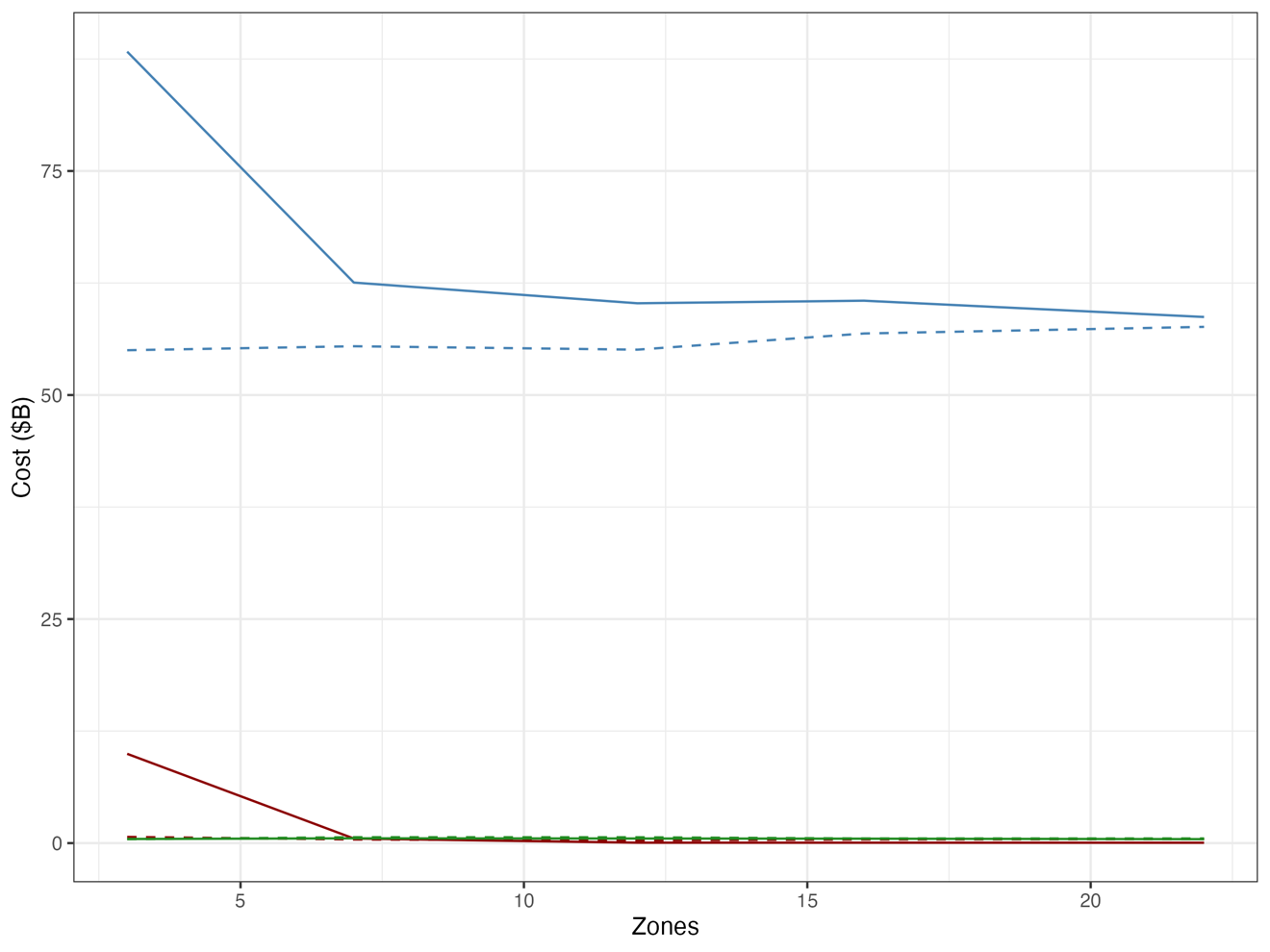}} 
    \subfloat[\coca case, zonal.\label{cost_comp_co2_z}]{\includegraphics[width=0.5\linewidth]{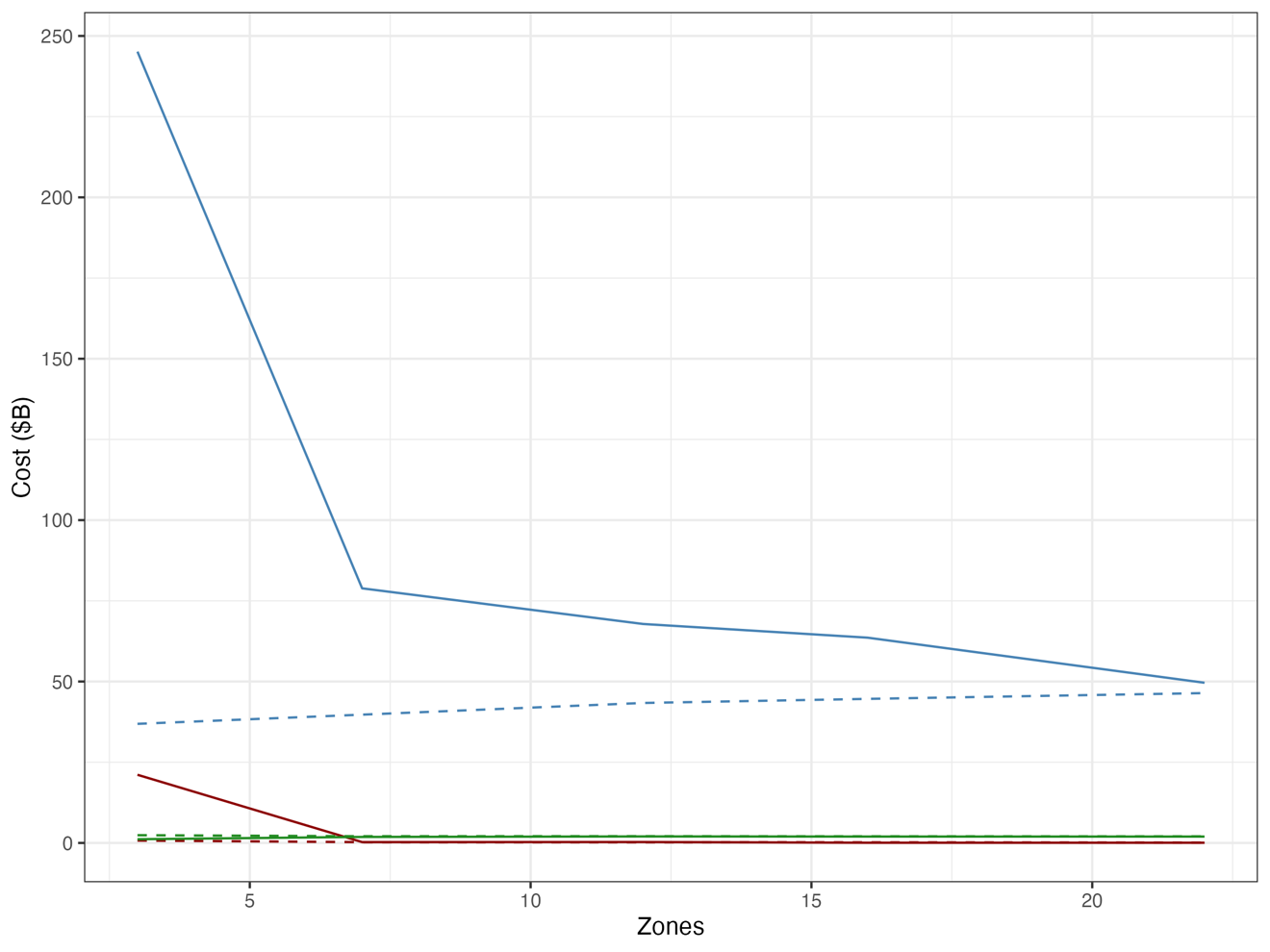}} \\
    \subfloat[\refca case, temporal.\label{cost_comp_ref_t}]{\includegraphics[width=0.5\linewidth]{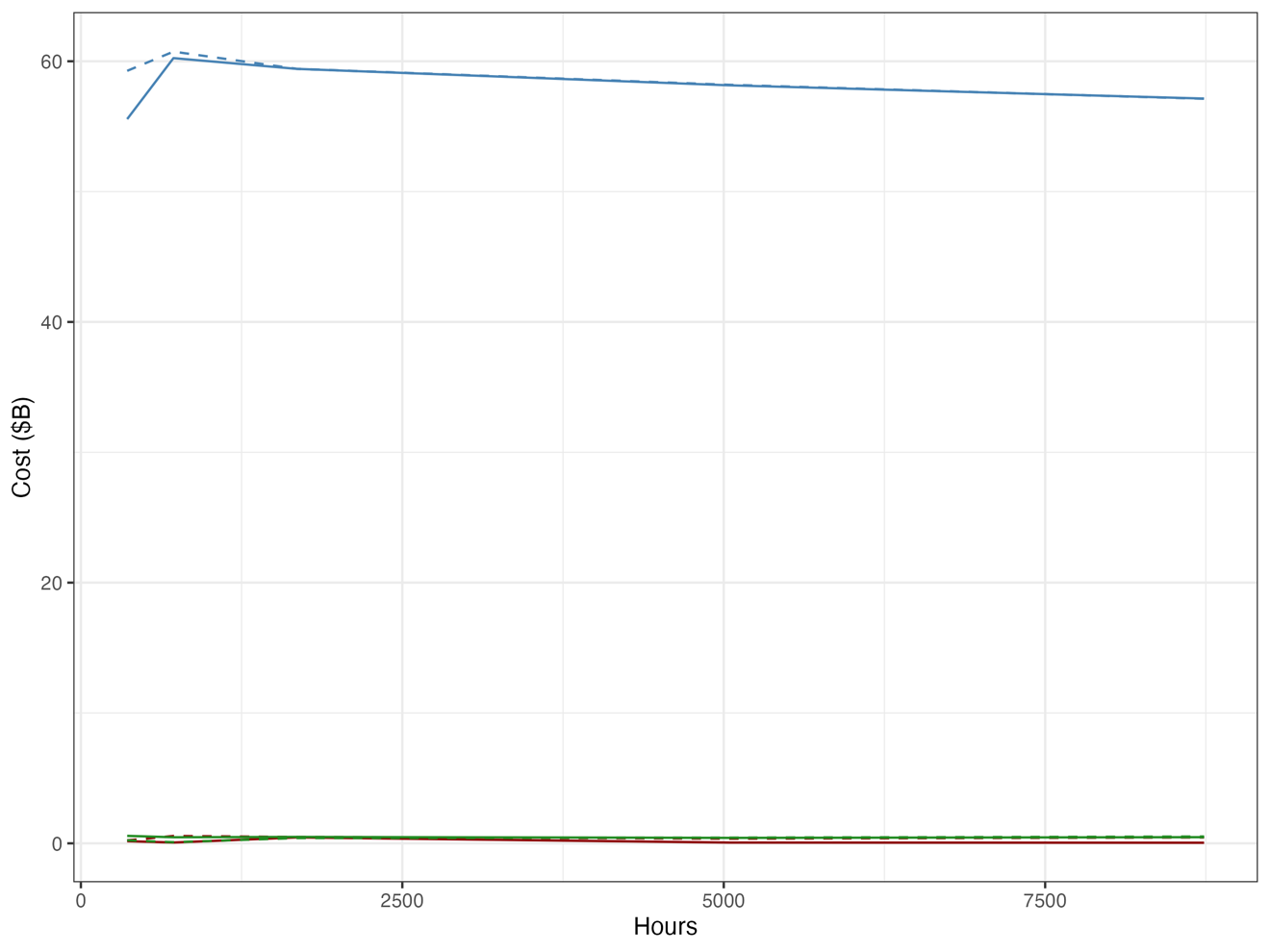}} 
    \subfloat[\coca case, temporal.\label{cost_comp_co2_t}]{\includegraphics[width=0.5\linewidth]{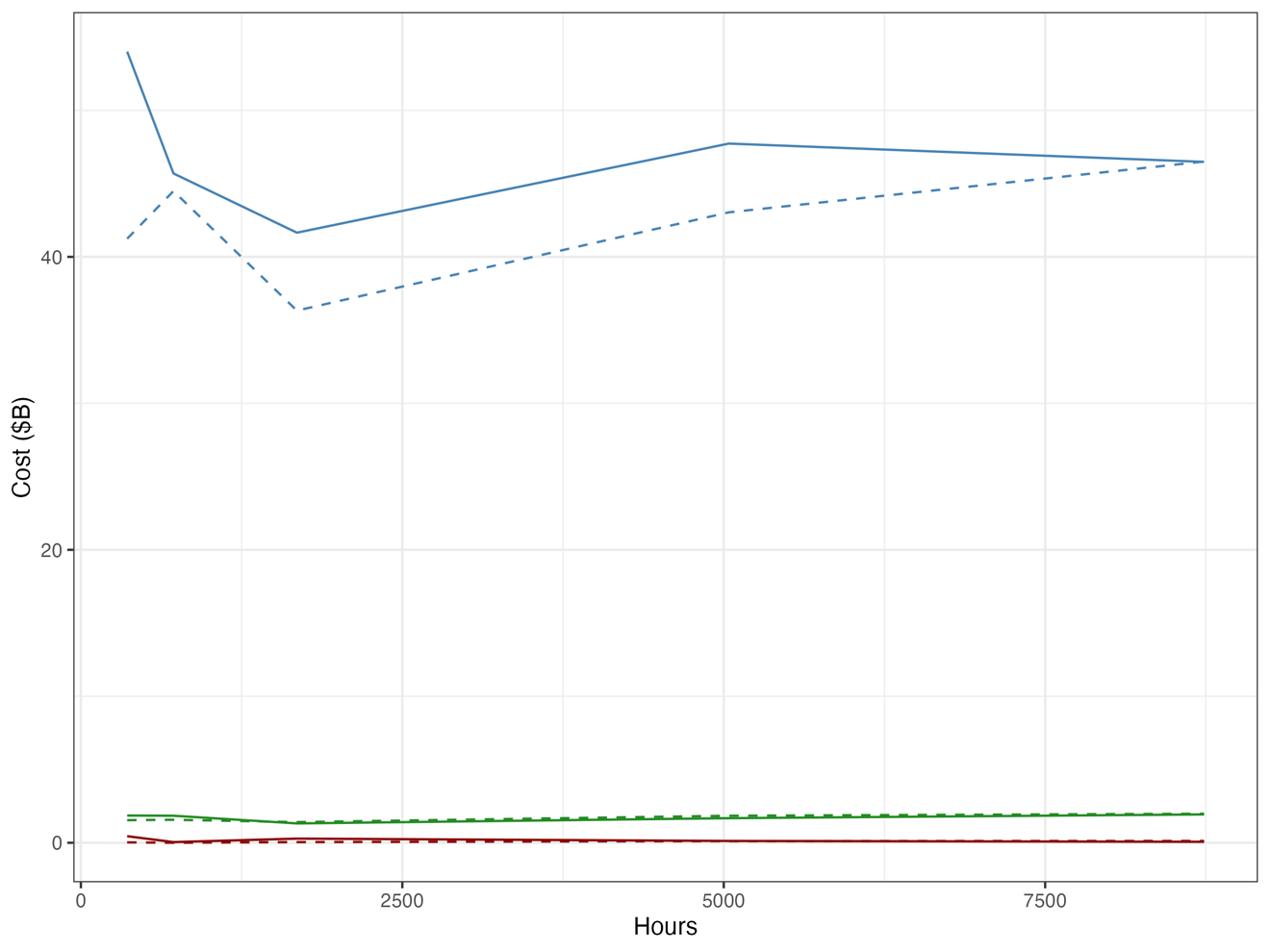}} \\
    \subfloat{\includegraphics[width=\linewidth]{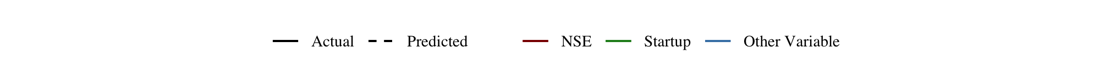}}
    \end{center}
    \caption{Comparison of operational costs between the abstracted models of \opone's prediction of varaible costs, and actual costs as output from \optwo~phase. Data show that operational models consistently underestimated cost for operations when compared to actual system costs. Trends are stronger for zonal abstractions.\label{cost_comp}}
\end{figure}

Low resolution models consistently underestimate variable costs, SI Fig~\ref{cost_comp}. High resolution in the \optwo~phase of optimization is critical for predicting cost trends; coarse models should not be trusted for cost measurements due to their misrepresentation of operational constraints.

\section{Discussion}
\label{discussion}

Mathematical modeling is key in creating energy portfolios that improve human welfare while mitigating emissions. While coarse ESMs are known to be less accurate \cite{brinkerink2024role,frysztacki2021strong,frysztacki2023inverse,frew2016temporal,jacobson2023benders}, it is difficult to assert an upper-bound on degree of error from low granularity\cite{frew2016temporal,jacobson2023benders}. We leverage modern mathematical techniques to create new, high resolution systems to better quantify errors resulting from low granularity.

Higher resolution increases accuracy and improves outcomes for prices, CO$_2$ emissions, reliability, and siting for renewables. Improvement for reliability (Fig~\ref{map_nse_co2}) tapers asymptotically with resolution, as our systems incur high penalties from lost load and prioritize reducing it. For other metrics, like siting accuracy or emissions, improvements are nonasymptotic due to features of the system: 26-zones is still coarse relative to a real-world transmission network with thousands of nodes, such that even our higher granularity systems saw benefits from adding regions. Because improvements persist at all levels we examine (SI Table~\ref{model_table_b},) we conclude that modelers should always increase resolution where possible, confirming the utility of recent algorithmic advances.

Spatial resolution is more impactful than temporal resolution at the scales tested here. In spatially coarse models, systems are unable to recapture weather patterns for optimal siting and also are missing information on the structure of the transmission system. As a result, binding network constraints at higher levels of resolution may preclude the operations envisioned by the coarser model. Temporally coarse models with high spatial resolution have trouble recapitulating weather patterns but do not misrepresent the feasibility of delivering power. In other words, so long as extreme periods are appropriately captured in the temporal aggregation methods, a temporally coarse model is likely to be suboptimal, but operations are likely to be feasible. In contrast, a spatially coarse model may omit key binding network constraints that may render planned systems both suboptimal and far more costly or even infeasible to operate than envisioned by the planning stage. That said, even our ``high resolution baseline'' model includes only one representative weather year, leading to relatively simple demands on the temporal downscaling methodologies. A study with more temporal data may see more impact of temporal resolution, as more information would be needed to recapture the full gamut of VRE availability and load curves across multiple futures. We therefore do not conclude temporal granularity to be strictly less consequential than spatial outside the scope of this study.

We show models to be impacted by their lowest level of granularity between spatial and temporal (Figs~\ref{map_wind}~-~\ref{map_emiss}, Fig \ref{isoquant}, SI Table~\ref{model_table_b}.) A model that has lost accuracy due to low resolution in one dimension cannot recover it using another; computational resources put into high granularity may be wasted if any aspect of the model is left unduly coarse. 

Due to computational limitations, virtually all studies are forced to decrease resolution. We include some sample studies and their resolutions employed in Table~\ref{model_table_a}, but note that downscaling is ubiquitous outside of the ones shown here. We encourage the interested reader to closely examine the methodology sections of their favorite studies for more examples. While papers provide valuable insights on decarbonization regardless of resolution employed, planners should be conscious of the structural uncertainty involved when relying on investment recommendations from any individual study. Multi-model analyses to mitigate uncertainties should be used where possible.

\begin{table}[H]
    \centering
    \footnotesize
    \renewcommand{\arraystretch}{1.75}
    \begin{tabular}{L{0.4\linewidth}L{0.1\linewidth}L{0.1\linewidth}L{0.1\linewidth}L{0.2\linewidth}}
        \toprule
        Study & Year & Model & Number of Regions & Temporal Resolution \\
        \midrule
        \textit{Actions for Reducing US Emissions at Least 50\% by 2030}\cite{bistline2022actions} & 2022 & EDF-NEMS & 25 & 9 time slices \\ 
        '' \ \ '' & & GCAM-USA-AP & 51 & Annual. Four non-sequential representative time blocks (electricity module) \\
        '' \ \ '' & & LBNL & 134 & 17 (investments) full annual (operations) \\
        '' \ \ '' & & PATHWAYS & 16 & 960 hours \\
        '' \ \ '' & & REGEN & 16 & Full annual hourly \\
        '' \ \ '' & & USREP-ReEDS & 12 & 17 time slices \\
        \textit{Annual Decarbonization Perspective: Carbon-Neutral Pathways for the United States}\cite{adp2023} & 2023 & RIO & 27 & 40 days \\
        \textit{Carbon Pricing and Energy Efficiency: Pathways to Deep Decarbonization of the US Electric Sector}\cite{brown2019carbon} & 2019 & NEMS & 22$^\ddagger$ & 9 slices$^\ddagger$  \\
        \textit{Evaluating Impacts of the Inflation Reduction Act and Bipartisan Infrastructure Law on the US Power System}\cite{steinberg2023evaluating} & 2023 & ReEDS & 134 & 17 slices of 4 - 40 hours\\
        \textit{Impact of Carbon Dioxide Removal Technologies on Deep Decarbonization of the Electric Power Sector}\cite{bistline2021impact} & 2021 & US-REGEN & 16 & Full hourly annual \\
        \textit{Net Zero America: Potential Pathways, Infrastructure, and Impacts, Final Report Summary}\cite{nzap} & 2021 & RIO & 16 & 41 days, hourly resolution \\
        \textit{Quantifying the Challenge of Reaching a 100\% Renewable Energy Power System for the United States}\cite{cole2021quantifying} & 2021 & ReEDS & 134 & 17 slices of 4 - 40 hours\\
        \textit{Robust Decarbonization of the US Power Sector: Policy Options}\cite{stock2021robust} & 2021 & ReEDS & 134 & 17 slices of 4 - 40 hours\\
        \textit{What is Different about Different Net-zero Carbon Electricity Systems?}\cite{baik2021different} & 2021 & RESOLVE & 3\textsuperscript{*} & 37 days \\
        '' \ \ '' & & urbs & 12\textsuperscript{*} & Full annual hourly\\
        '' \ \ '' & & GenX & 9\textsuperscript{$\dagger$} & 16 weeks \\
        \bottomrule
    \end{tabular}
    \caption{Some example studies using ESMs with the parameter settings that they were run under. Most models represent CONUS, though two (*) represent California only, and one represents WECC ($\dagger$). These systems may require less spatial resolution for accuracy, accordingly. $\ddagger$ Metrics are shown for the NEMS Electricity Market Module in particular.}
    \label{model_table_a}
\end{table}

Decisionmakers should be informed on their models' structure and associated implications vis-\'a-vis uncertainty: how many spatial regions are in their system relative to their real-world transmission topology? How many weather years were sampled, with how many hours per year? How were timesteps selected? Are physical systems represented faithfully with high operational resolution? Users should determine a tolerance for error and valuable accuracy metrics based on model use. A planner interested in systemwide investment costs, for example, does not need fine-grained details on operations or site locations. This user may be satisfied with our high level aggregate results (Section~\ref{results:aggregate}) to explore granularity needed for their work. To a local planner who needs confidence that spending in their county is faithfully modeled in a nationwide study, our maps (Figs~\ref{map_solar}~-~\ref{map_emiss}) will prove more informative in determining minimum appropriate modeling resolution.

It would be reductive to assume that models can be summarized by their spatial and temporal resolution, and inclusion of unit commitment. No two models are alike: Some do not incorporate consecutive hours\cite{brown2020regional}, some run in a few soft-linked sectoral modules\cite{brown2020regional,nems_docs} while others are single-sector\cite{genX_github,brown2020regional}, some are highly user-configurable\cite{temoa_docs,brown2020regional,switch_github,genX_github}. Due to endless examples of different design decisions made by separate models, we cannot extrapolate numerical bounds on error from this study to others. Still, the trend that models are impacted by their lowest dimension of granularity should give pause to anyone relying on studies that have ``put all of their eggs in one basket'' in  allocation of computational resources to a single dimension of resolution (spatial, temporal, operational.) Models that allocate granularity in a balanced manner are likely to reduce overall error.

Our new model formulation\cite{jacobson2023benders,genX_github,pecci2024regularized} allows higher resolution simulations than would have been possible previously. Despite mathematical advances, it is impossible to increase resolution \textit{ad infinitum}. The value of this writeup is in its explorations of multiple dimensions of resolution and the relationships between them to explore levels of granularity sufficient to support policy and other decisionmaking. Future studies should leverage mathematical advances on the horizon to more deeply probe the space of granularity with even higher resolution once larger systems are tractable.

\section{Conclusion}
This paper used new, high resolution modeling for a more in-depth exploration of ESM granularity than was previously feasible. Locational accuracy was more vulnerable to uncertainty in resolution than aggregate installed capacity or cost. Spatial resolution was more impactful than temporal at the scales tested here. For many metrics, there was no asymptotic behavior to accuracy improvement, implying that more resolution is always preferable: methods for accelerating model performance are confirmed to be critical tools in ensuring accurate recommendations. Because models' accuracy is limited by their lowest-resolution dimension, users are remiss to undervalue any individual dimension (spatial, temporal, operational) of their model. In the face of unavoidable computational limitations, users should carefully balance their allocation of model granularity.

\section{Acknowledgements}
The simulations and writing comprising this writeup were designed and completed by Anna Jacobson. Jesse Jenkins and Denise Mauzerall provided guidance for the experimental setup and format of the writeup, as well as edits to drafts once written.

The experiments run here would not have been possible without the modification of GenX coded by Filippo Pecci according to the publication jointly written by him and Anna Jacobson.

Additional thanks are due to the ZERO Lab and to Anna Jacobson's thesis committee (professors Simon Levin, Stephen Pacala, Jonathan Levine, Jesse Jenkins, and Denise Mauzerall) who provided feedback throughout the process of this work.

\subsection*{Funding}
This work was supported by the High Meadows Environmental Institute (HMEI) through its fellowship in Science, Technology, and Environmental Policy (HMEI-STEP) and by the Princeton Carbon Mitigation Initiative through a gift from BP.

The authors have no other pertinent conflicts of interest to share relating to the content of this report.

\bibliographystyle{plain}
\bibliography{bibliography} 

\newpage

\section{Supplemental Information}
\label{supplemental}

\begin{table}[H]
    \centering
    \footnotesize
    \renewcommand{\arraystretch}{1.45}
    \begin{tabular}{l|l|llllll}
        \toprule
        Abbrev & Region & 3z & 7z & 12z & 16z & 22z & 26z \\
        \midrule
        BASN & WECC-Basin & & & & & \checkmark & \checkmark \\
        CA & California & & & \checkmark & \checkmark & & \\
        CANO & California-North & & & & & \checkmark & \checkmark \\
        CASO & California-South & & & & & \checkmark & \checkmark \\
        EASC & Eastern Central & & & \checkmark & & & \\
        EIC & Eastern Interconnect & \checkmark & & & & & \\
        EICW & Eastern Interconnect-West & & \checkmark & & & & \\
        FRCC & Florida Reliability Coordinating Council & & & & & & \checkmark \\
        ISNE & Independent System Operator New England & & & \checkmark & \checkmark & \checkmark & \checkmark \\
        MISC & Midcontinent Independent System Operator-Central & & & & \checkmark & \checkmark & \checkmark\\
        MISE & Midcontinent Independent System Operator-East & & & & & \checkmark & \checkmark\\
        MISW & Midcontinent Independent System Operator-West & & & \checkmark & \checkmark & \checkmark  & \checkmark\\
        MISS & Midcontinent Independent System Operator-South & & & \checkmark & \checkmark & \checkmark & \checkmark\\ 
        NE & Northeast & & \checkmark  & & & & \\
        NWPP & Northwest Power Pool & & & \checkmark & \checkmark & \checkmark & \checkmark \\
        NY & New York & & & \checkmark & & & \\
        NYUP & New York Upstate & & & & \checkmark & \checkmark & \checkmark\\
        NYCW & New York City-West & & & & \checkmark & \checkmark & \checkmark\\ 
        PJMC & PJM-Commonwealth Edison & & & & & \checkmark & \checkmark \\
        PJMD & PJM-Dominion & & & & & \checkmark & \checkmark \\
        PJME & PJM-East & & & & & \checkmark & \checkmark \\
        PJMW & PJM-West & & & & & \checkmark & \checkmark \\
        PJM & Pennsylvania-New Jersey-Maryland Interconnection & & \checkmark & \checkmark & \checkmark & & \\
        RMRG & WECC-Rockies & & & & & \checkmark & \checkmark \\
        SOU & South & & \checkmark & \checkmark & \checkmark & \checkmark & \\
        SPP & Southwest Power Pool & & \checkmark & \checkmark & \checkmark &\checkmark  & \\
        SPPN & Southwest Power Pool-North & & & & & & \checkmark \\
        SPPC & Southwest Power Pool-Central & & & & & & \checkmark \\
        SPPS & Southwest Power Pool-South & & & & & & \checkmark \\
        SRCA & SERC Reliability Corporation-East & & & & \checkmark & \checkmark & \checkmark \\
        SRCE & SERC Reliability Corporation-Central & & & & \checkmark & \checkmark & \checkmark \\
        SRSE & SERC Reliability Corporation-Southeast & & & & & & \checkmark \\
        SRSG & Southwest Reserve Sharing Group & & & & \checkmark & \checkmark & \checkmark \\
        TRE & Texas Reliability Entity & \checkmark & \checkmark & \checkmark & \checkmark & \checkmark & \checkmark$^\ast$ \\
        TREW & Texas Reliability Entity-West & & & & & & \checkmark \\
        WECC & Western Electricity Coordinating Council & \checkmark & \checkmark & & & & \\
        WECCC & Western Electricity Coordinating Council-Central & & & & \checkmark & & \\
        WECCE & Western Electricity Coordinating Council-East & & & \checkmark & & & \\
        \bottomrule
    \end{tabular}
    \caption{Glossary of region names. $\ast$ TRE represents a smaller spatial extent in the 26-zone case than in the 3 - 22-zone cases (Fig~\ref{geography}.)\label{region_glossary}}
\end{table}

\subsection{Supplemental Methodology\label{supplemental:methodology}}
Systems are broken into discrete zones, each representing region in CONUS with existing generation, storage, and transmission capacity; VRE sites and availability; and demand. Regions are based on input for the Integrated Planning Model (IPM), which subdivides CONUS based on structures established for system operation within the United States.

System aggregations (Fig~\ref{geography}) were plugged PowerGenome. PowerGenome uses the Annual Technology Baseline (ATB) from the National Renewable Energy Lab (NREL), along with data from the Energy Information Agency (EIA) and the Public Utility Data Liberation Project (PUDL)\cite{schivley2021powergenome}. Output from PowerGenome was plugged into the GenX\cite{genX_github} for \opone~and subsequently \optwo~phase models.

Within each zone for the systems (Fig~\ref{geography}) tested, investment sites are divided into separate resource clusters based on LCOE and capacity factor (CF). Clustering sites into resource clusters involves:

\begin{enumerate}
    \item Determining and including spurline costs per-site given the predicted cost of connecting to local urban areas.
    \item \label{step_cpa} Selecting candidate sites for the resource (e.g., solar photovoltaic cells) within a given region using a set of filters, see Table~\ref{filter_table}.
    \item \label{step_cluster} Grouping sites based on LCOE and CF.
\end{enumerate}

Because zones are included as monoliths within our model scheme, (demand and VRE within a zone has no geographic location,) it is not possible to include intraregional transmission explicitly. Spurline costs attempt to overcome this omission.

Urban areas with population greater than 1 million, as well as the largest one per region, are assumed to have infinite demand such that every site will has somewhere to send its generation. Increasing the level of aggregation decreases the number of infinite sinks across the system, thereby altering where sites can connect. This is a prime source of differences across different spatial aggregations.
   
For our work, sites with LCOE over 200 - 300 \$/MW (technology-dependent) are excluded. Remaining sites are put into bins by LCOE (weighted by capacity) and CF. For offshore wind turbines, only preferential sites given sociopolitical constraints are considered (Table~\ref{filter_table}).

\begin{table}[h!]
    \centering
    \footnotesize
    \renewcommand{\arraystretch}{1.45}
    \begin{tabular}{lllll}
        \toprule
        Technology & LCOE Filter & LCOE Bins (\#) & CF Bins (\#)  & Total Bins (\#) \\
        \cmidrule(lr){1-1} \cmidrule(lr){2-2} \cmidrule(lr){3-5}
        Utility PV & $\leq$ 200 \$/MW & 5 & 3 & 15 \\
        Onshore Wind & $\leq$ 200 \$/MW & 5 & 3 & 15 \\
        Offshore Wind (Floating) & $\leq$ 300 \$/MW & 3 & 2 & 6 \\
        Offshore Wind (Fixed) & $\leq$ 300 \$/MW & 3 & 2 & 6 \\
        \bottomrule
    \end{tabular}
    \caption{Clustering and filter parameters used to create the resource clusters plugged into sections \ref{genx_inv} - \ref{genx_op}.}
    \label{filter_table}
\end{table}

\subsection{Supplemental Results}

\begin{table}[p]
    \centering
    \setlength\tabcolsep{0pt}
    \begin{tabular}{cM{0.3\linewidth}M{0.3\linewidth}M{0.3\linewidth}}
        & \large 3-zone & \large 16-zone & \large 26-zone \\

        \rotatebox[origin=c]{90}{Onshore Wind, \refca} &
        \includegraphics[width=\linewidth]{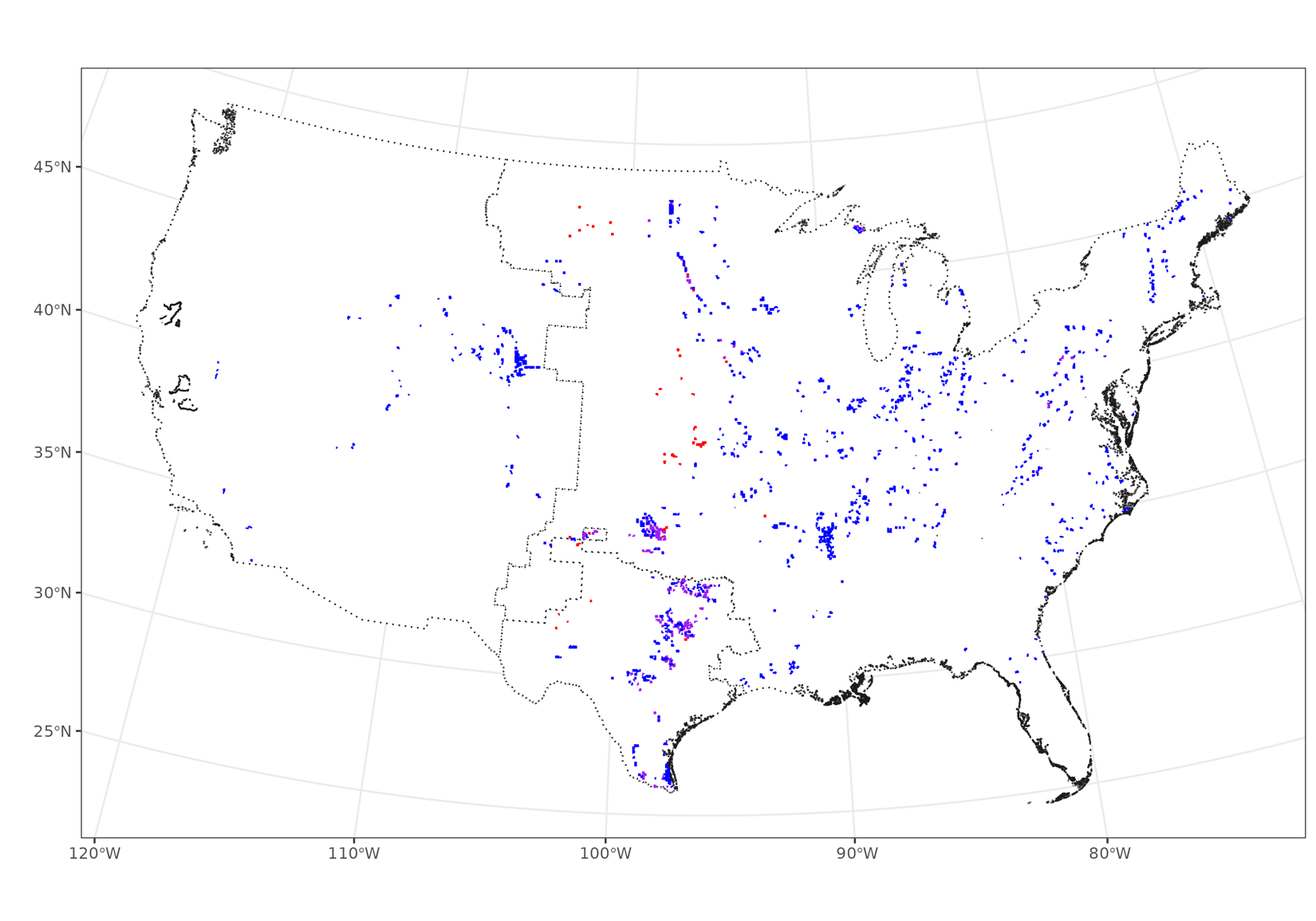} & 
        \includegraphics[width=\linewidth]{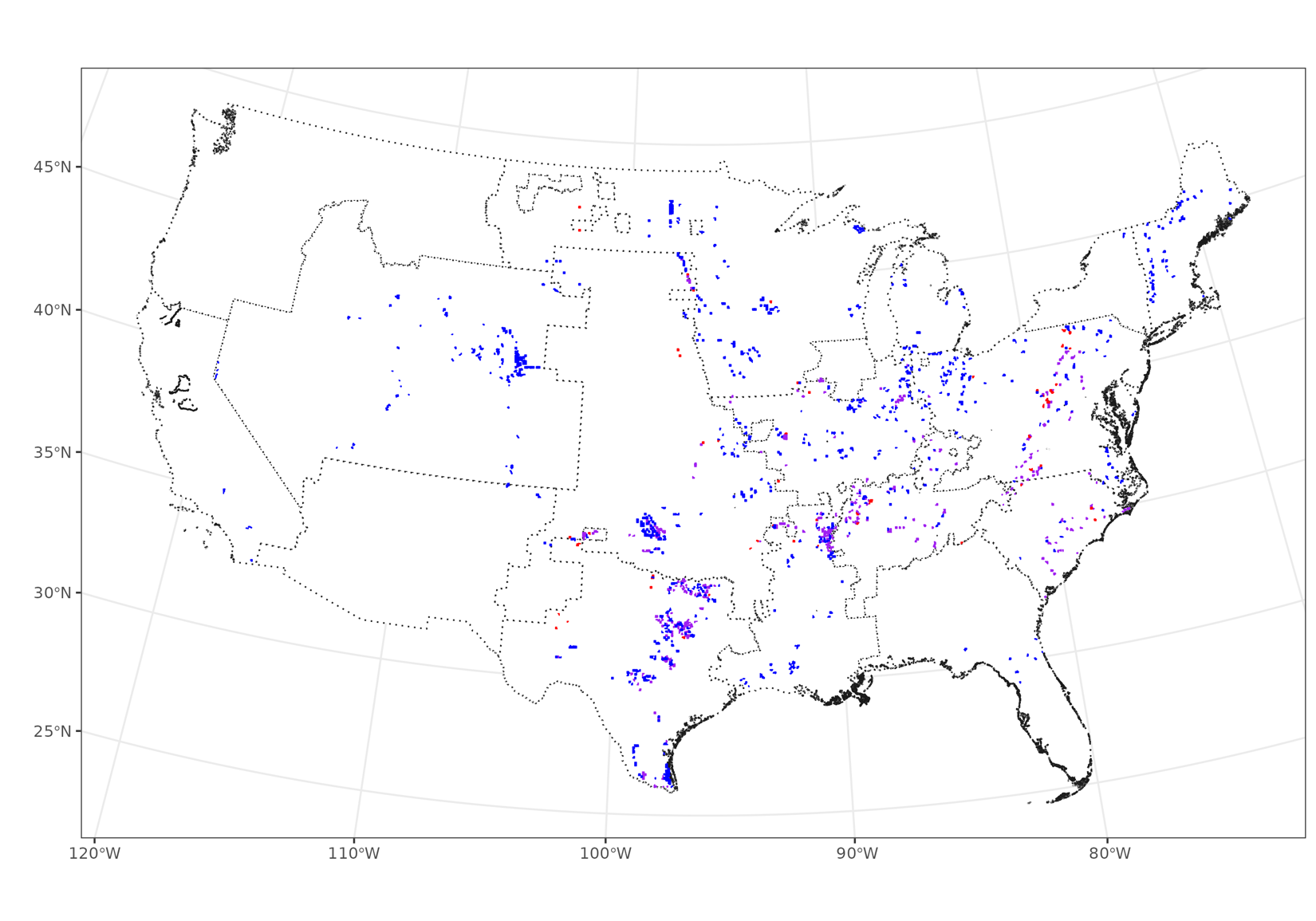} &
        \includegraphics[width=\linewidth]{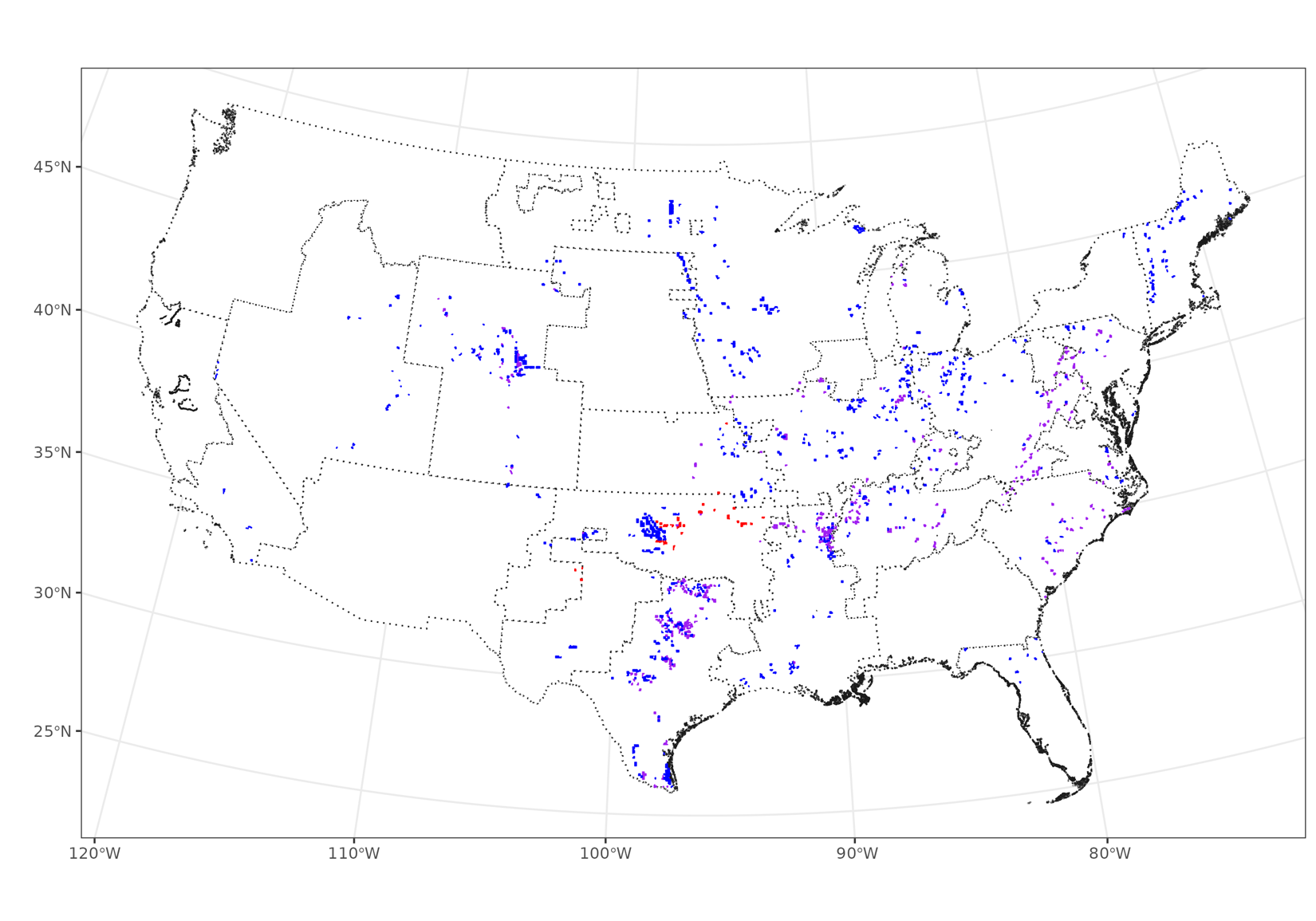} \\ 

        & {\footnotesize 16.5\% SCO} &
        {\footnotesize 28.4\% SCO} &
        {\footnotesize 31\% SCO}\\

        \rotatebox[origin=c]{90}{Solar, \refca} &
        \includegraphics[width=\linewidth]{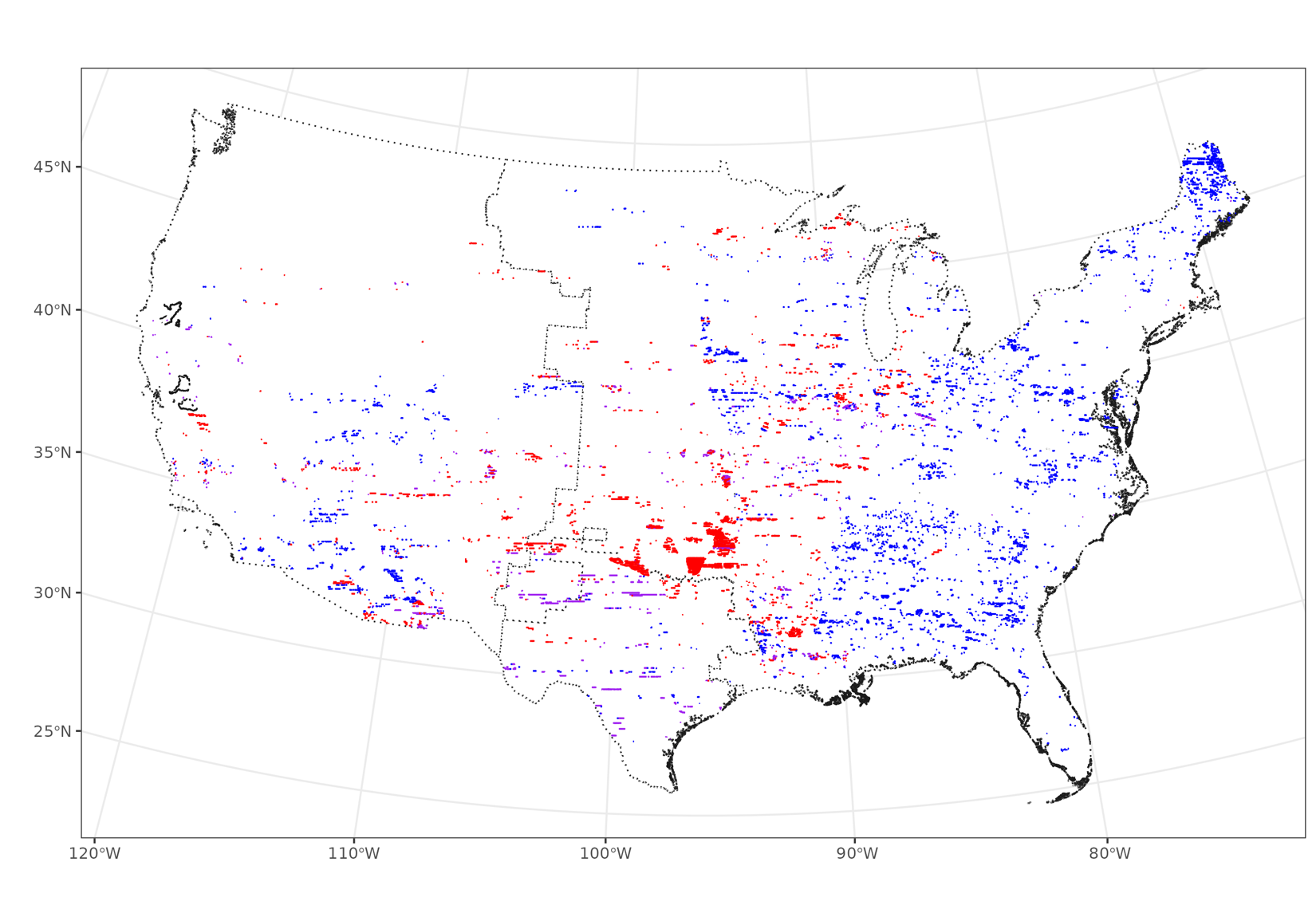} & 
        \includegraphics[width=\linewidth]{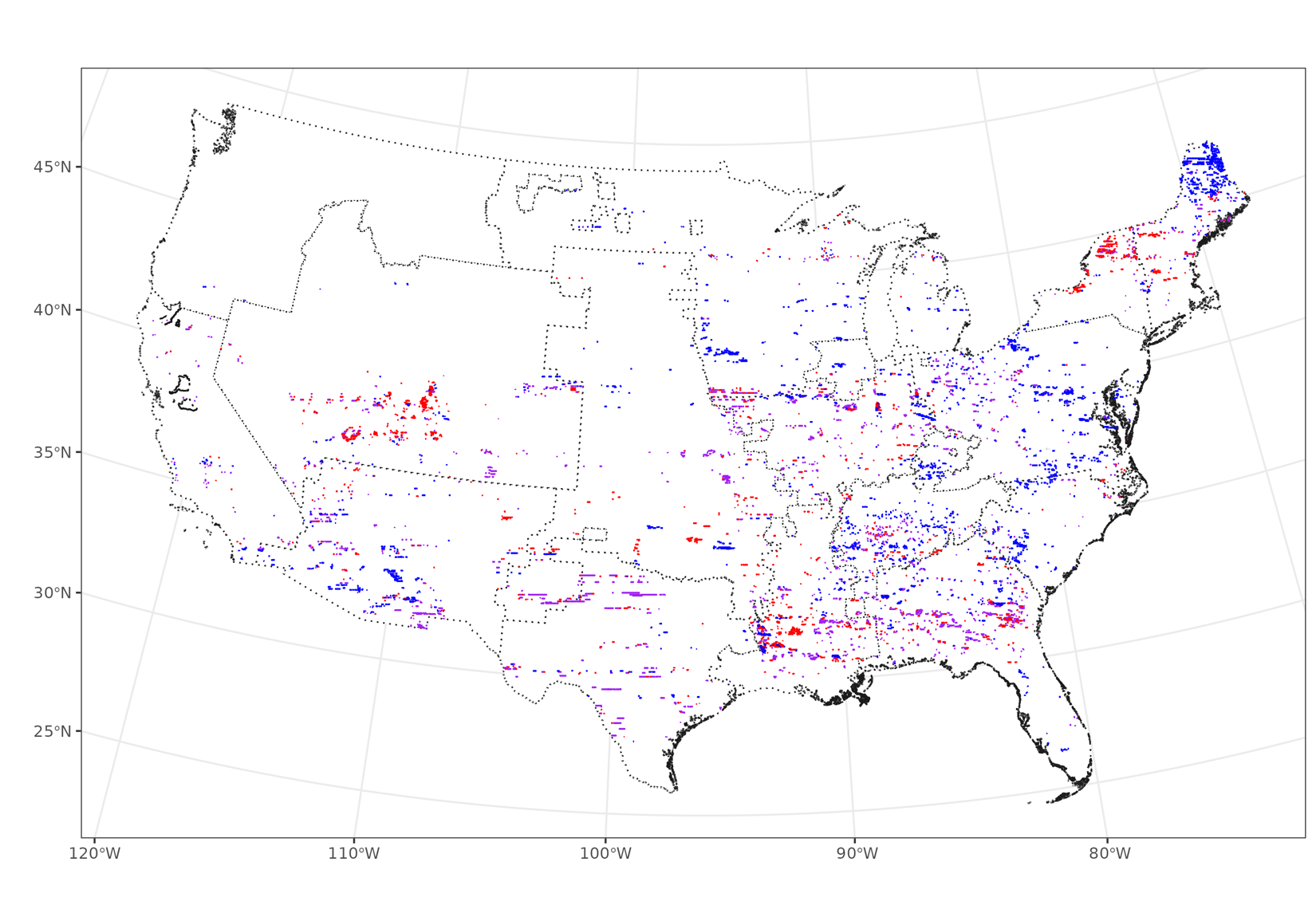} &
        \includegraphics[width=\linewidth]{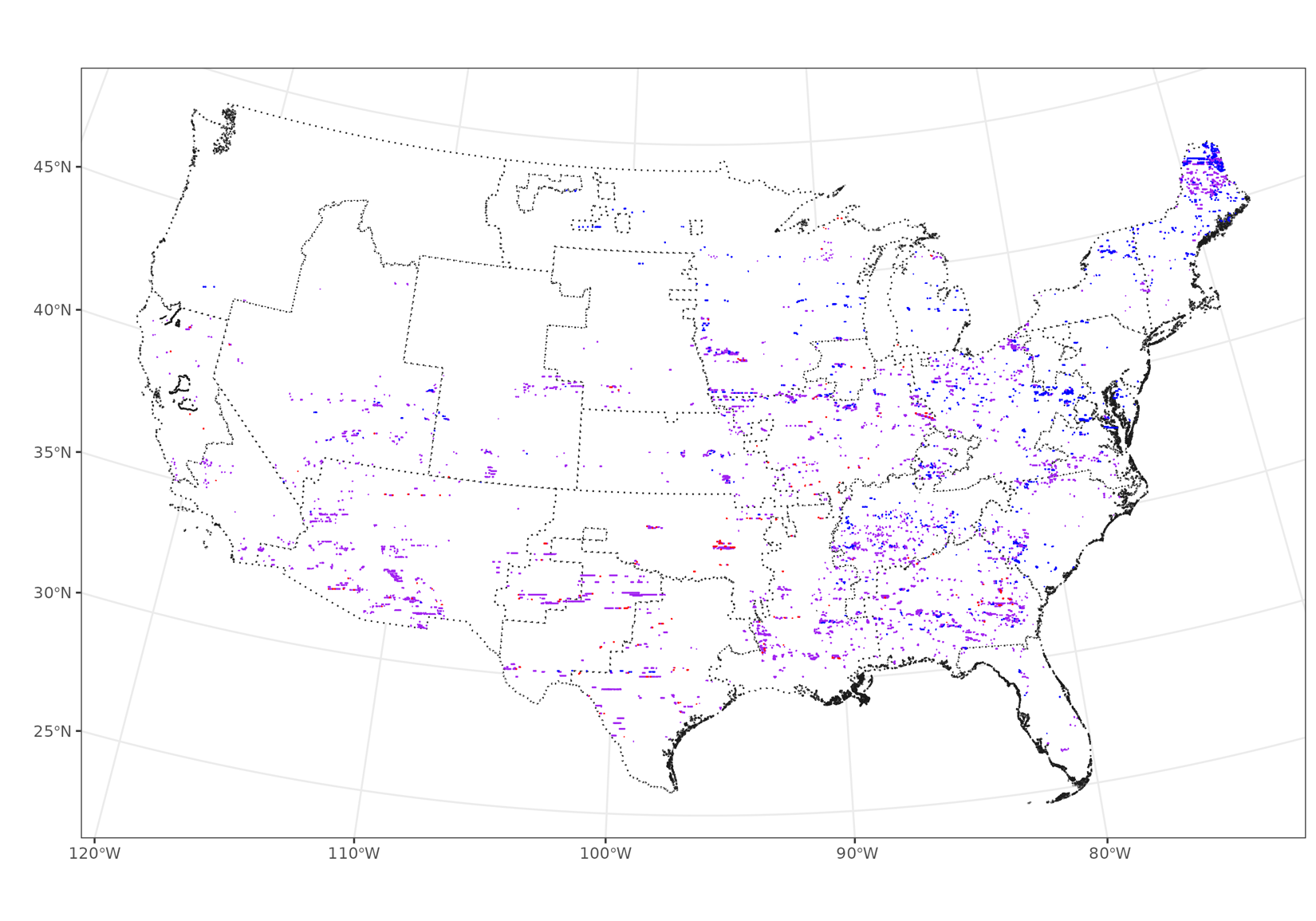} \\ 

        & {\footnotesize 14\% SCO} &
        {\footnotesize 36.3\% SCO} &
        {\footnotesize 75.1\% SCO}\\

        \rotatebox[origin=c]{90}{Onshore Wind, \coca} &
        \includegraphics[width=\linewidth]{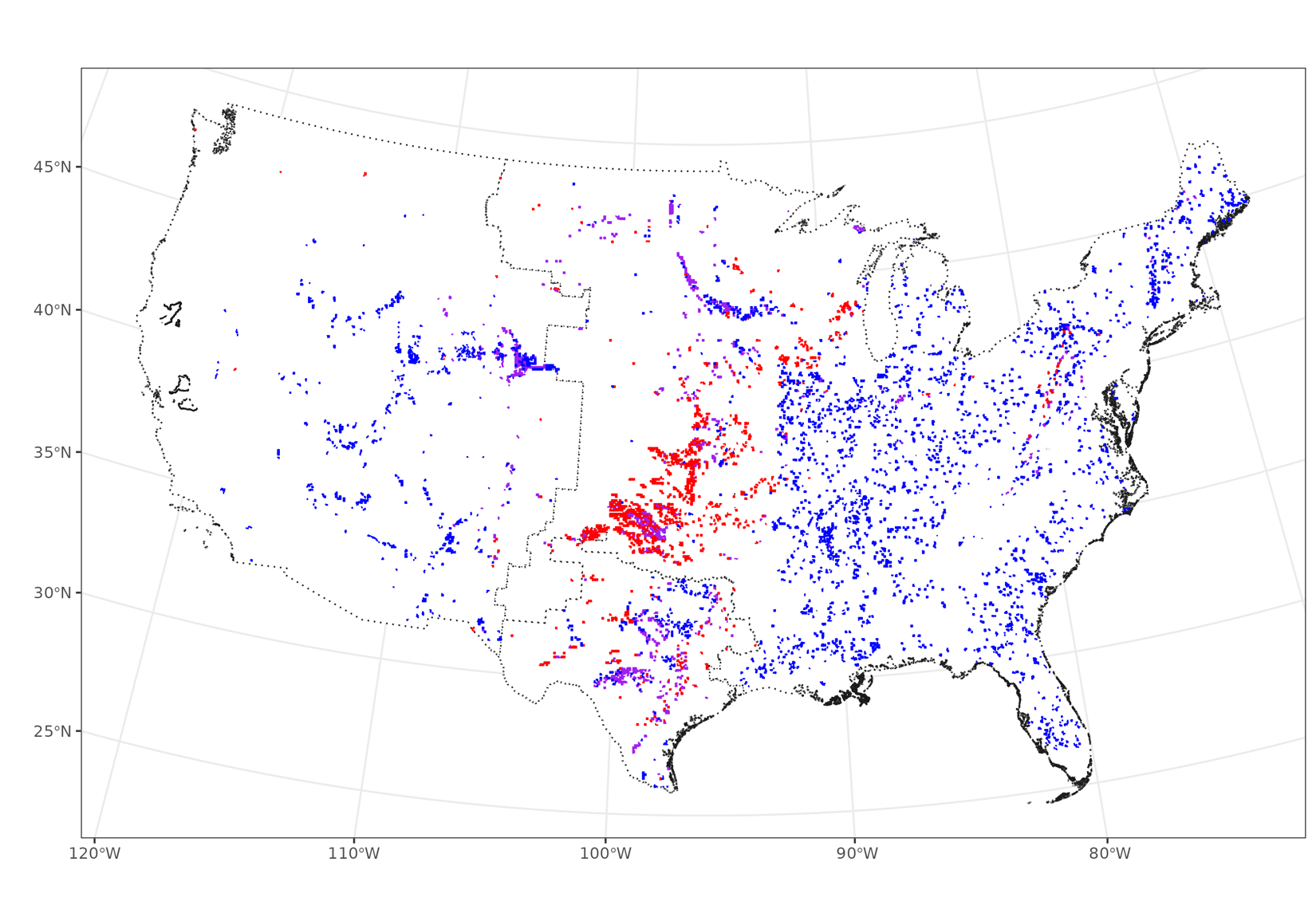} & 
        \includegraphics[width=\linewidth]{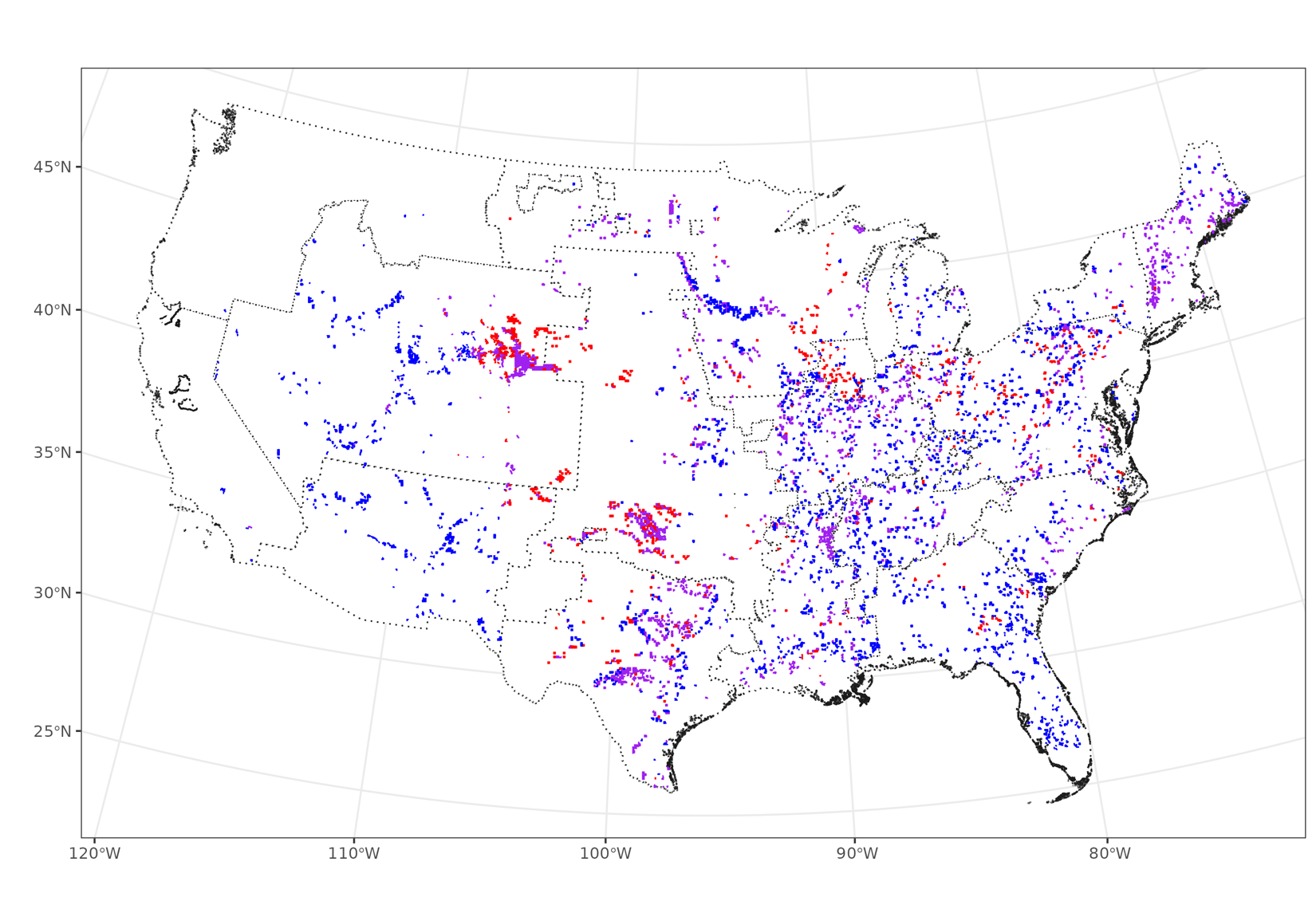} &
        \includegraphics[width=\linewidth]{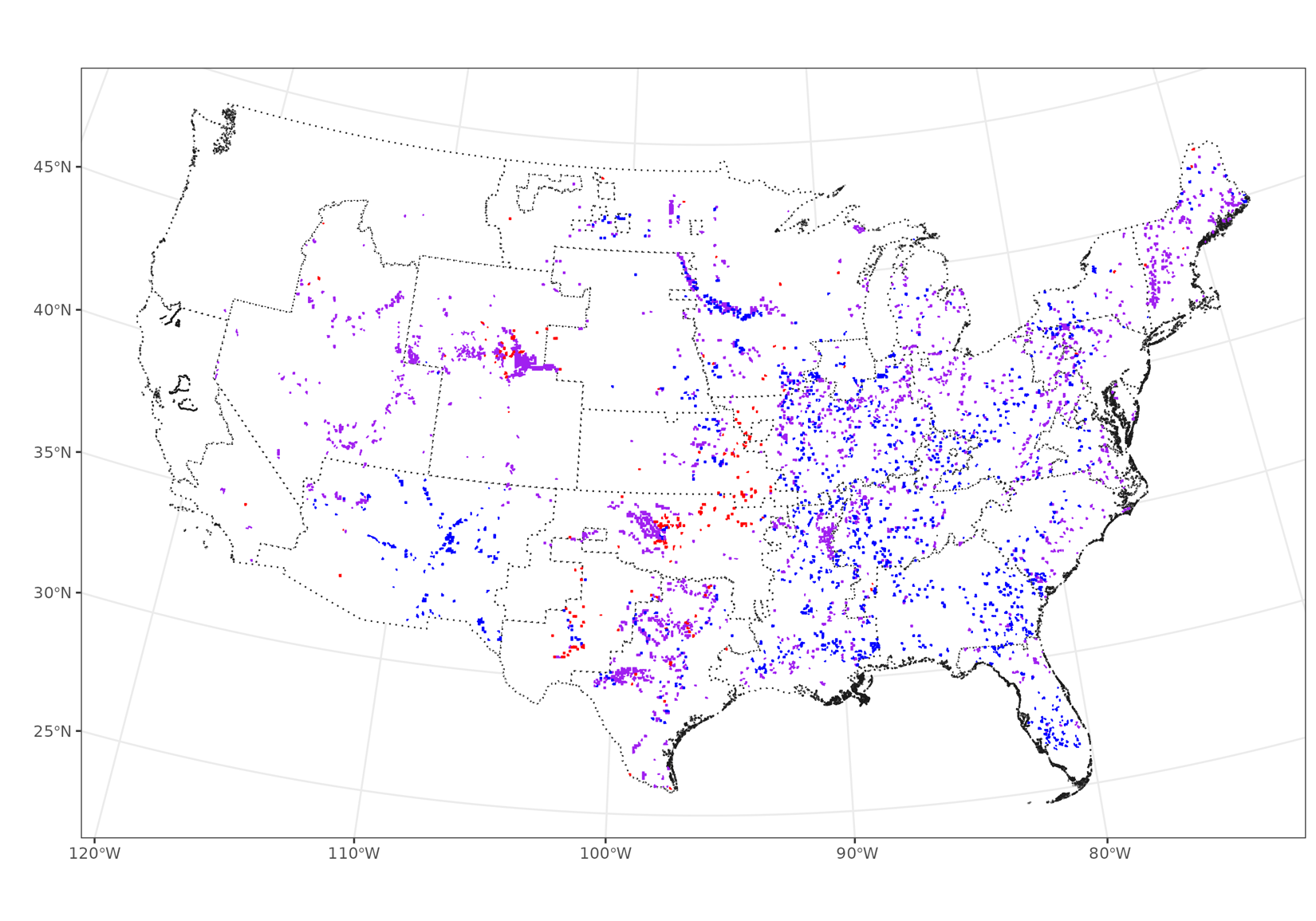} \\ 

        & {\footnotesize 19.5\% SCO} &
        {\footnotesize 38.3\% SCO} &
        {\footnotesize 60.5\% SCO}\\

        \rotatebox[origin=c]{90}{Solar, \coca} &
        \includegraphics[width=\linewidth]{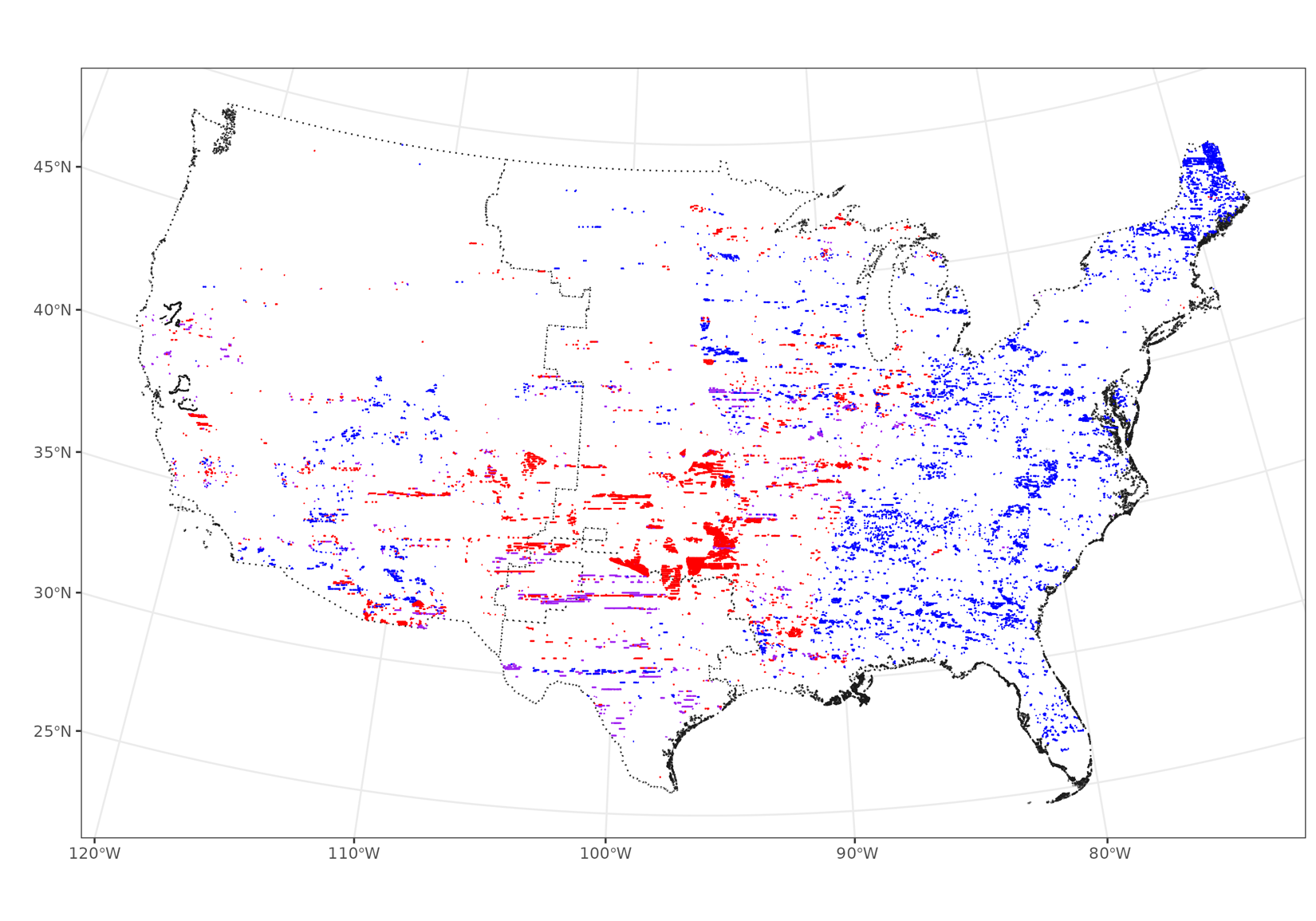} & 
        \includegraphics[width=\linewidth]{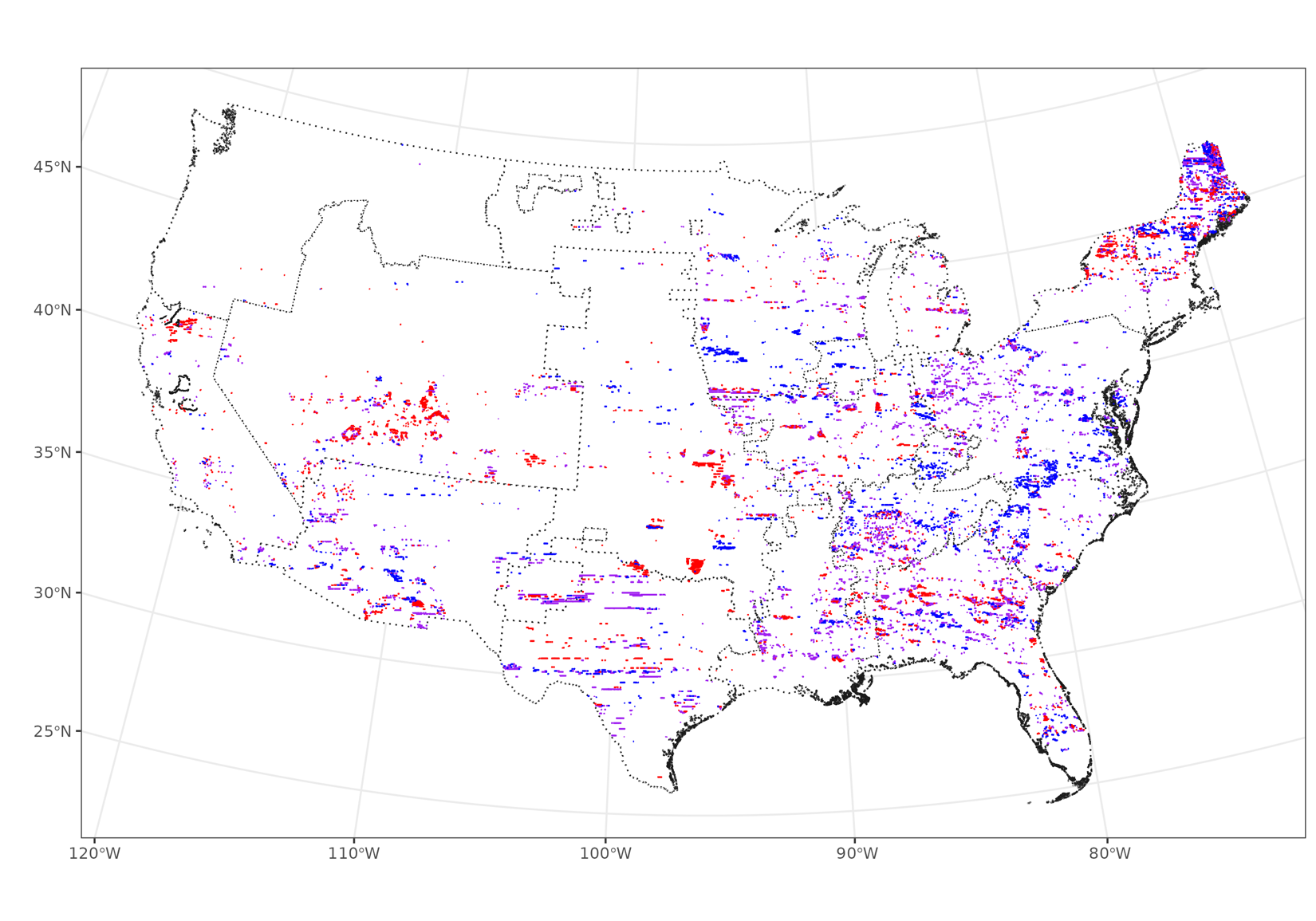} &
        \includegraphics[width=\linewidth]{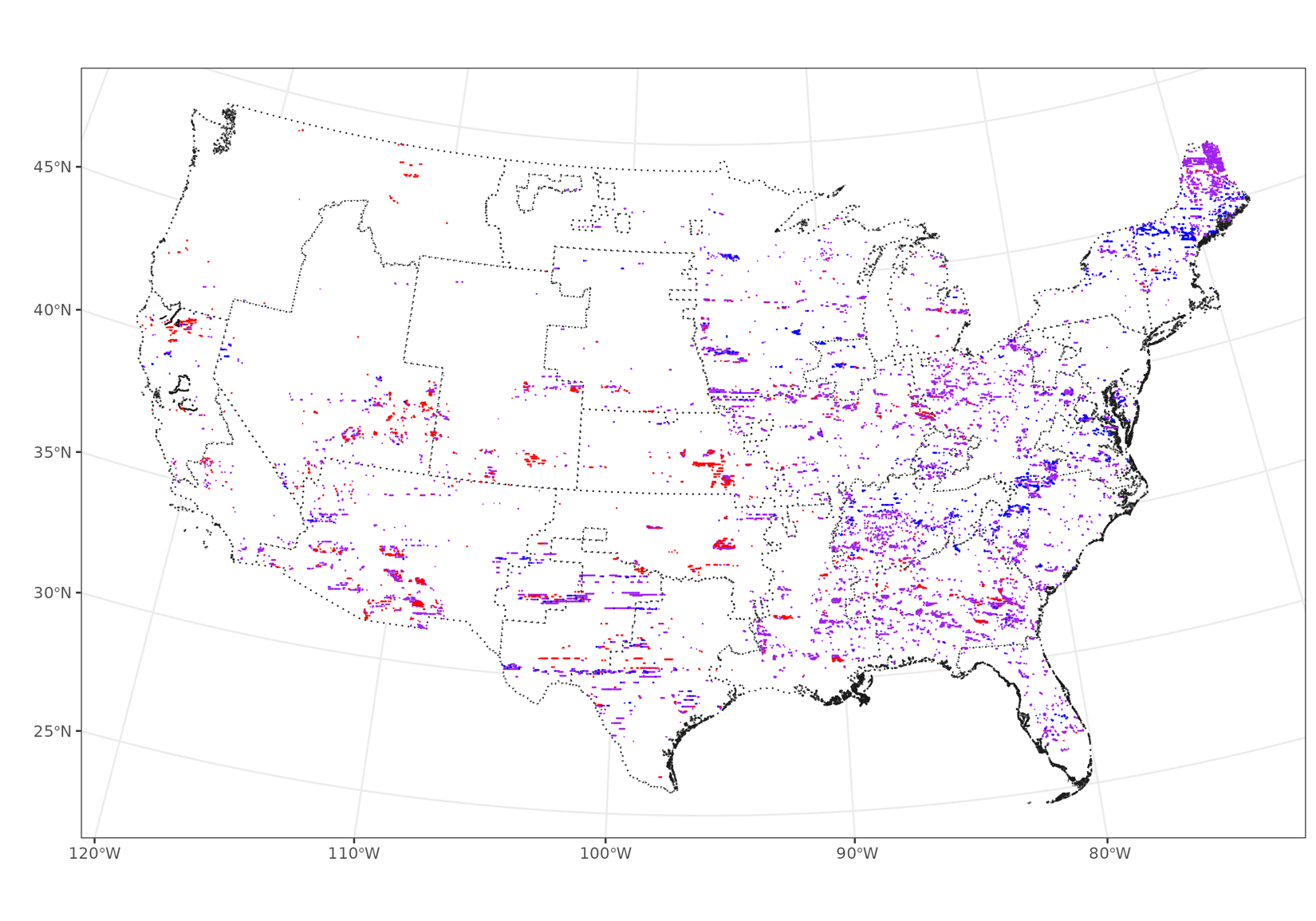} \\ 

        & {\footnotesize 14.3\% SCO} &
        {\footnotesize 45.5\% SCO} &
        {\footnotesize 72.8\% SCO}\\

    \end{tabular}
    \captionof{figure}{Difference plot for VRE capacity for the cases without unit commitment (UC) constraints for both the \refc (\refca\unskip) and \coc (\coca\unskip) cases. Sites selected by coarse cases are highlighted in red, sites selected by the highest resolution baseline (HRB) case are highlighted in blue. Overlapping sites are highlighted in purple. Let \textit{site capacity overlap} (SCO) be the percentage of capacity that is invested in by both the HRB and coarse-resolution case. $SCO = 100\% \cdot \frac{sites_{HRB} \cap sites}{sites_{HRB} \cup sites}$ Cases with increased spatial resolution and operational resolution are better able to recapitulate siting results.}
    \label{map_uc}
    
\end{table}

\begin{figure}[h]
    \begin{center}
    \subfloat[Zonal, high temporal resolution\label{iso_sol_z}]{\includegraphics[width=0.25\linewidth]{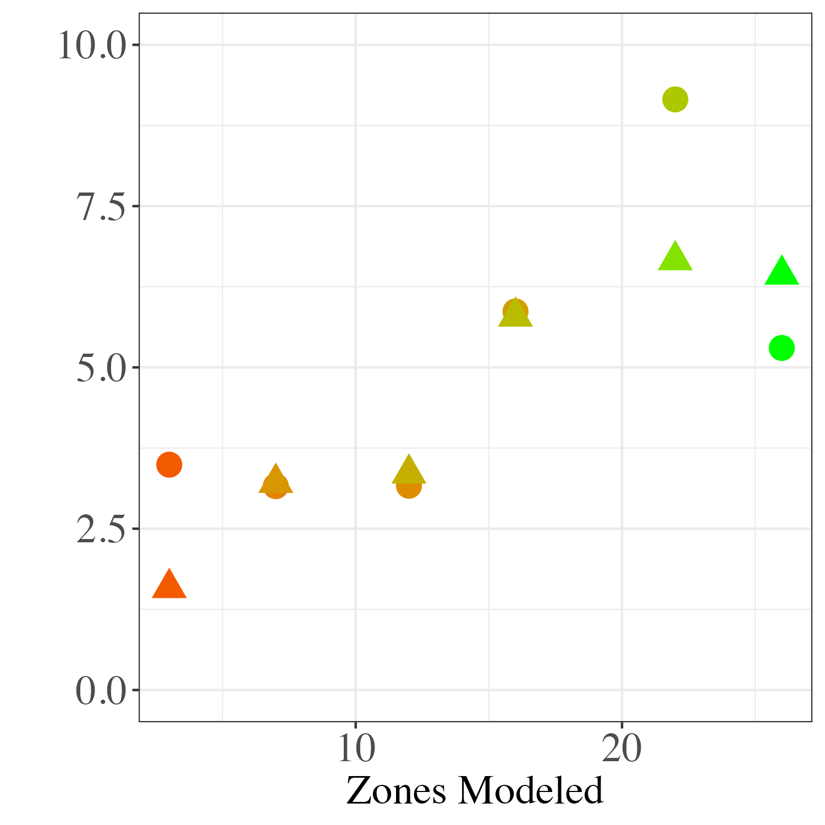}}
    \subfloat[Temporal, high zonal resolution\label{iso_sol_t}]{\includegraphics[width=0.25\linewidth]{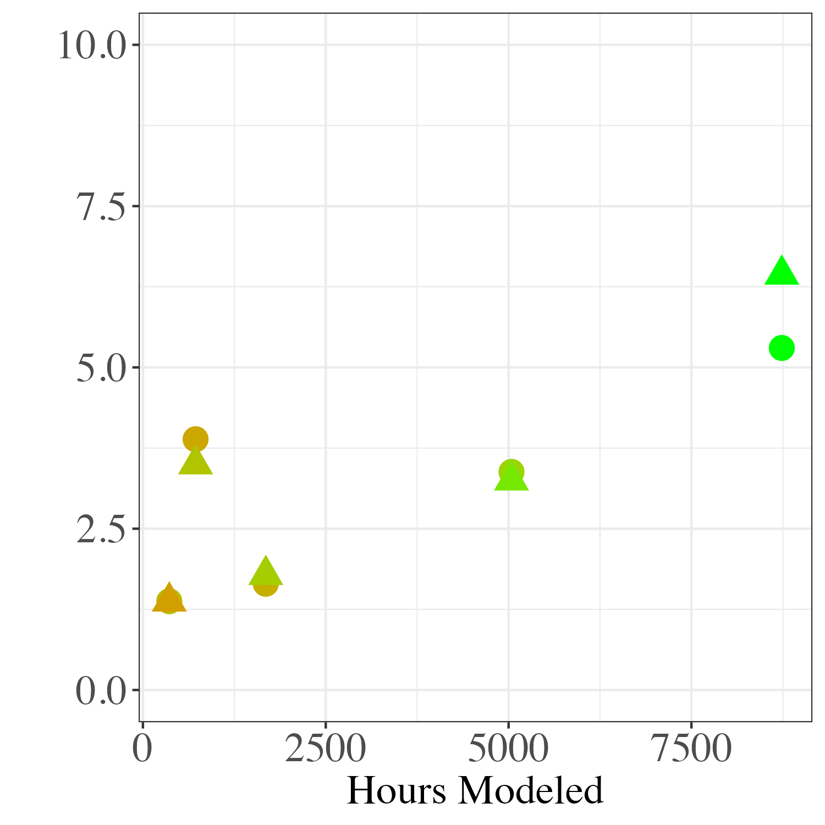}}
    \subfloat[Zonal, low temporal resolution\label{iso_sol_zl}]{\includegraphics[width=0.25\linewidth]{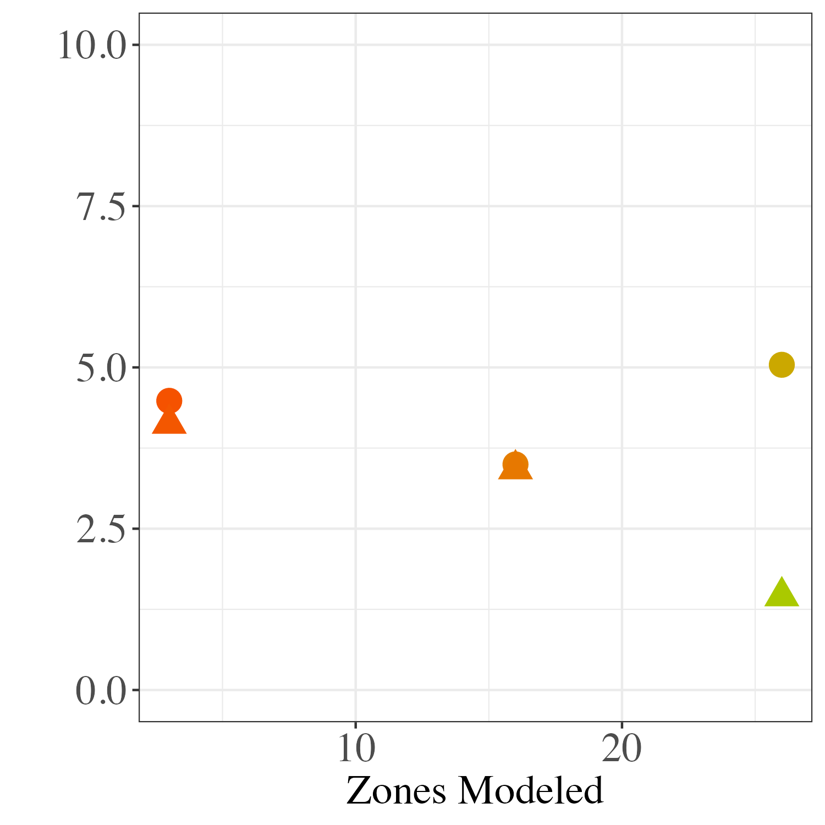}}
    \subfloat[Temporal, low zonal resolution\label{iso_sol_tl}]{\includegraphics[width=0.25\linewidth]{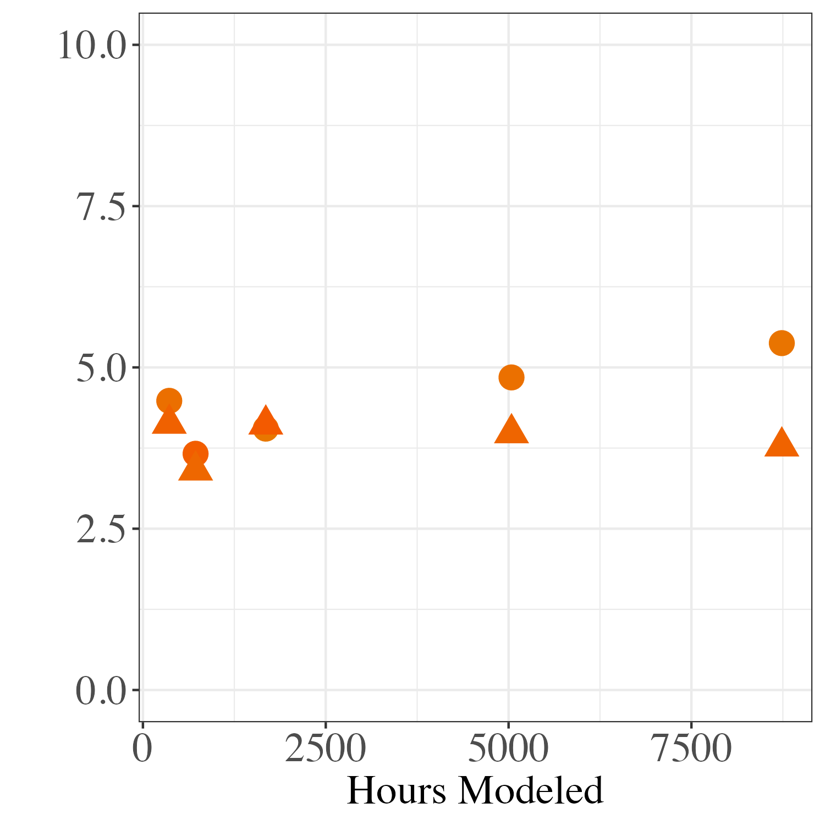}} \\
    \subfloat{\includegraphics[width=\linewidth]{Aggregate/Isoquant/isoquant_legend.png}}
    \end{center}
    \caption{Resolution vs runtime in hours. Points are colored reflecting the percentage of site capacity overlap for VRE technologies as a metric for accuracy, (see figures~\ref{map_solar} and~\ref{map_wind}. $SCO = 100\% \cdot \frac{sites_{HRB} \cap sites}{sites_{HRB} \cup sites}$) Figures include data for both the \coc (\coca\unskip) and \refc (\refca\unskip) cases. Figures \ref{iso_sol_z}~and~\ref{iso_sol_t} have highest resolution (26 zones or 8736 hours) for the dimension not included in the x-axis. Figures \ref{iso_sol_zl}~and~\ref{iso_sol_tl} have lowest resolution (3 zones or 15 days) for the dimension not included in the x-axis. Increasing resolution improves accuracy, though this trend is severely masked in cases where the non-considered dimension of resolution is too low. An equivalent figure for onshore wind is included in the main text as Fig~\ref{isoquant}.}
    \label{isoquant_sol}
\end{figure}

\begin{figure}[H]
    \begin{center}
    \subfloat[Runtime by temporal aggregation\label{rt_t}]{\includegraphics[width=0.5\linewidth]{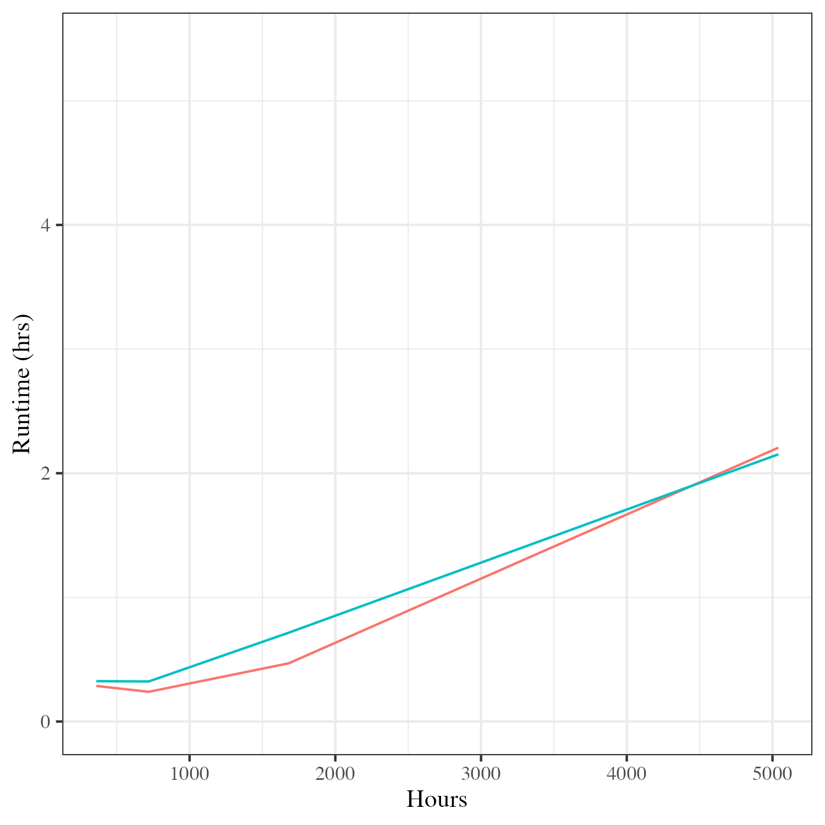}}
    \subfloat[Runtime by zonal aggregation\label{rt_z}]{\includegraphics[width=0.5\linewidth]{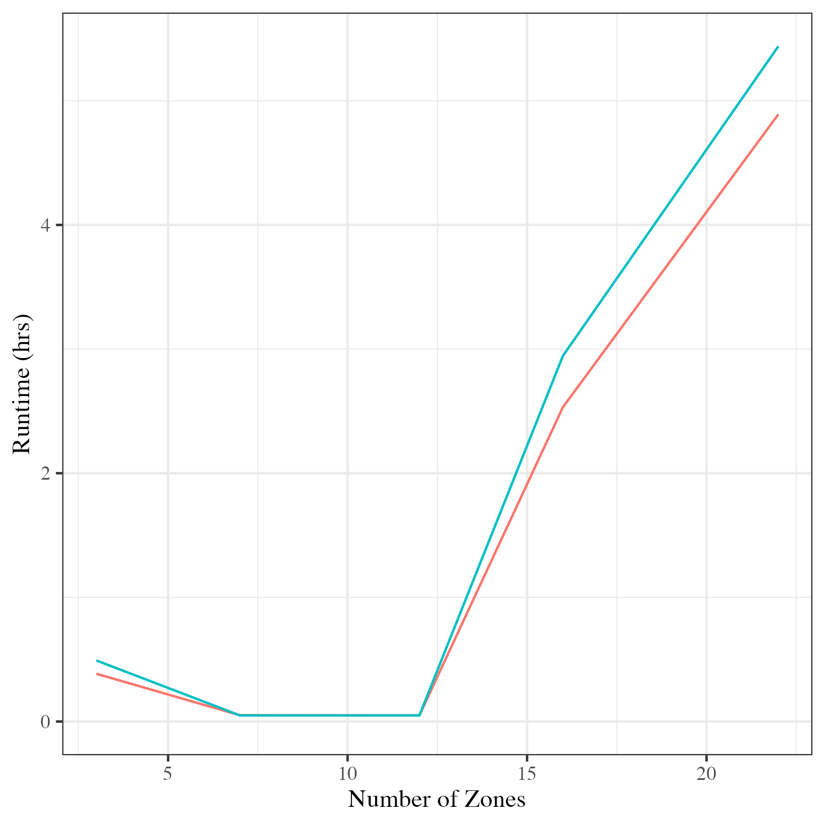}} \\
    \subfloat{\includegraphics[width=0.3\linewidth]{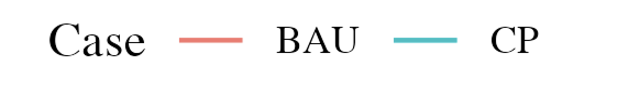}}
    \end{center}
    \caption{Runtime by zonal (\ref{rt_z}) and temporal aggregation (\ref{rt_t}) for the \opone~optimizations of varying resolution. Figure shows increasing runtime alongside resolution with much stronger impact of zonal resolution, in accordance with expectations laid out in~\cite{jacobson2023benders}.}
    \label{rt}
\end{figure}

\begin{landscape}
    \newpage
    \begin{figure}
        \begin{center}
            \subfloat[Average energy price by hour, ISONE. Zonal aggregations.\label{price_isone_zone}]{\includegraphics[width=0.5\linewidth]{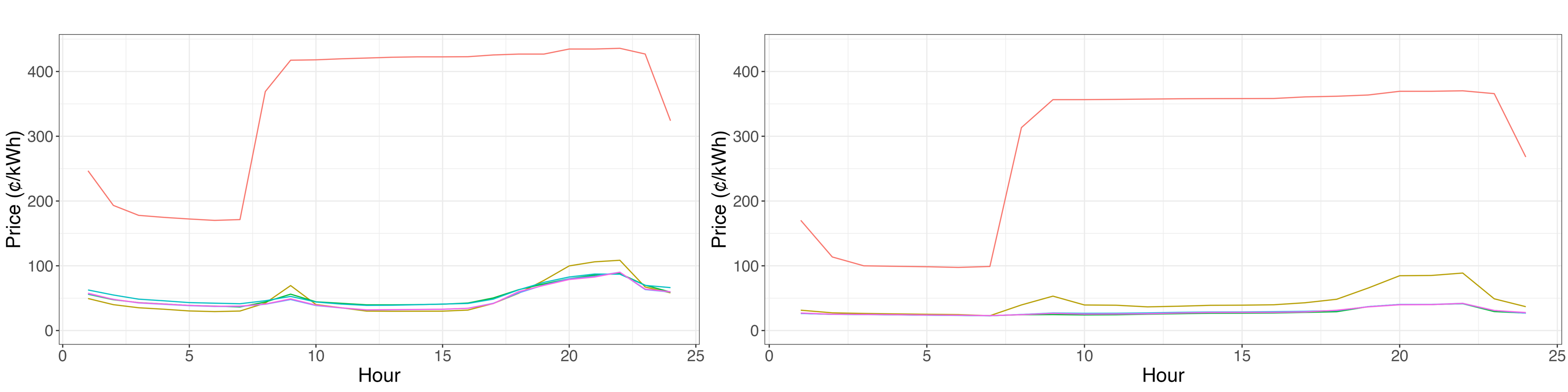}}
            \subfloat[Average energy price by hour, ISONE. Temporal / operational aggregations.\label{price_isone_time}]{\includegraphics[width=0.5\linewidth]{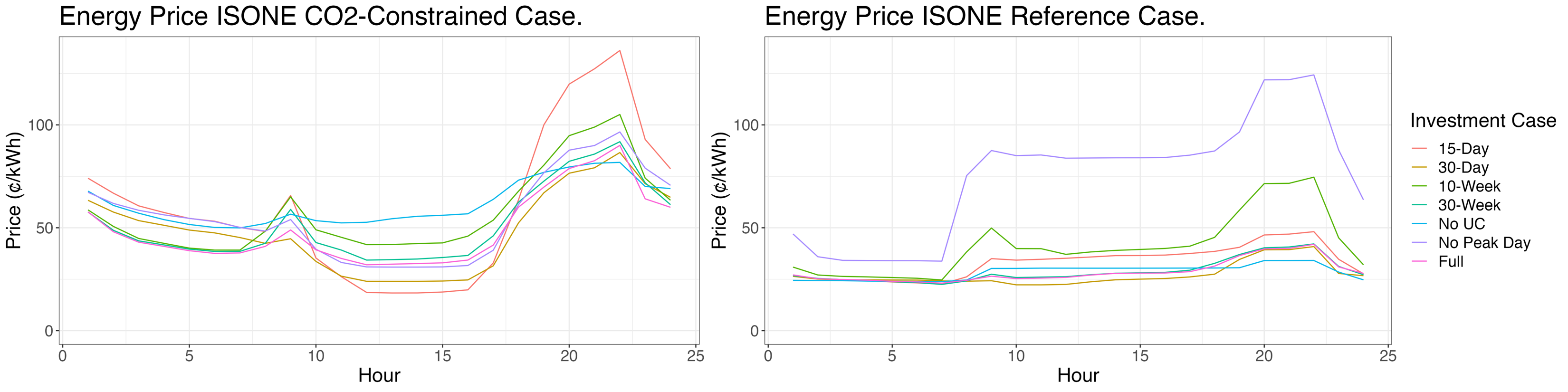}} \\
            \subfloat[Average energy price by hour, NWPP. Zonal abstractions.\label{price_nwpp_zone}]{\includegraphics[width=0.5\linewidth]{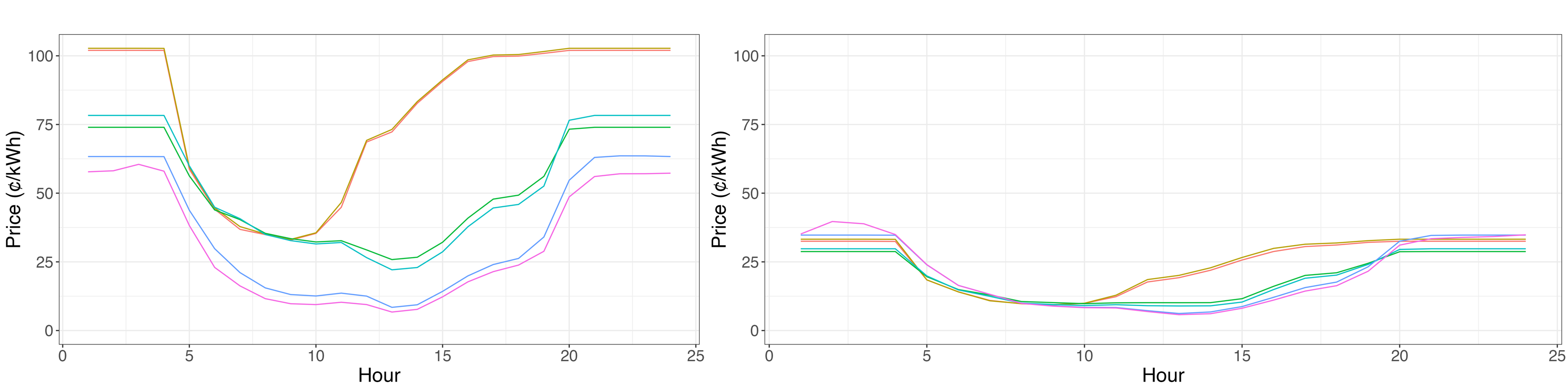}}
            \subfloat[Average energy price by hour, NWPP. Temporal / operational abstractions.\label{price_nwpp_time}]{\includegraphics[width=0.5\linewidth]{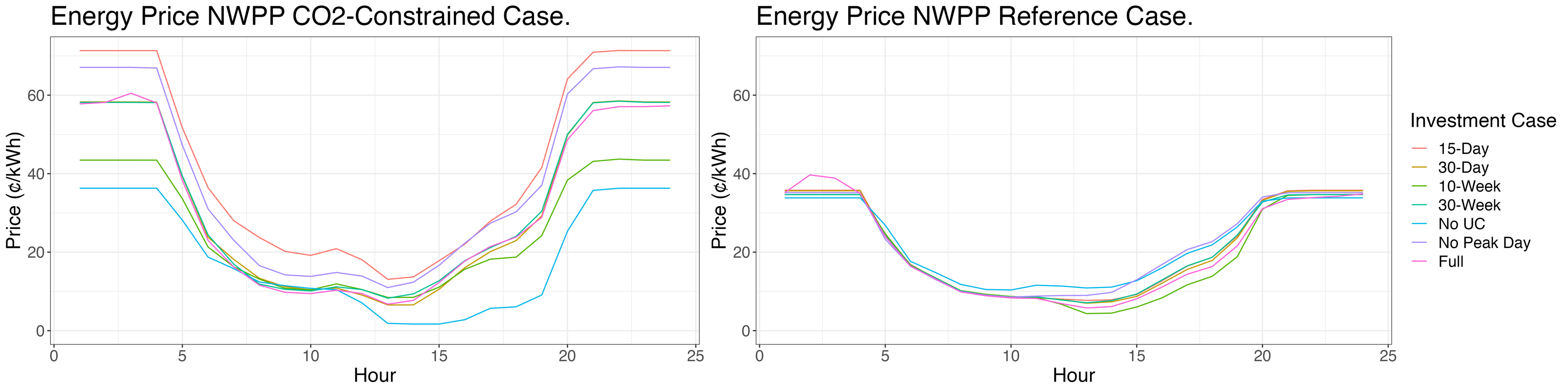}} \\
            \subfloat[Average energy price by hour, FRCC. Zonal abstractions.\label{price_frcc_zone}]{\includegraphics[width=0.5\linewidth]{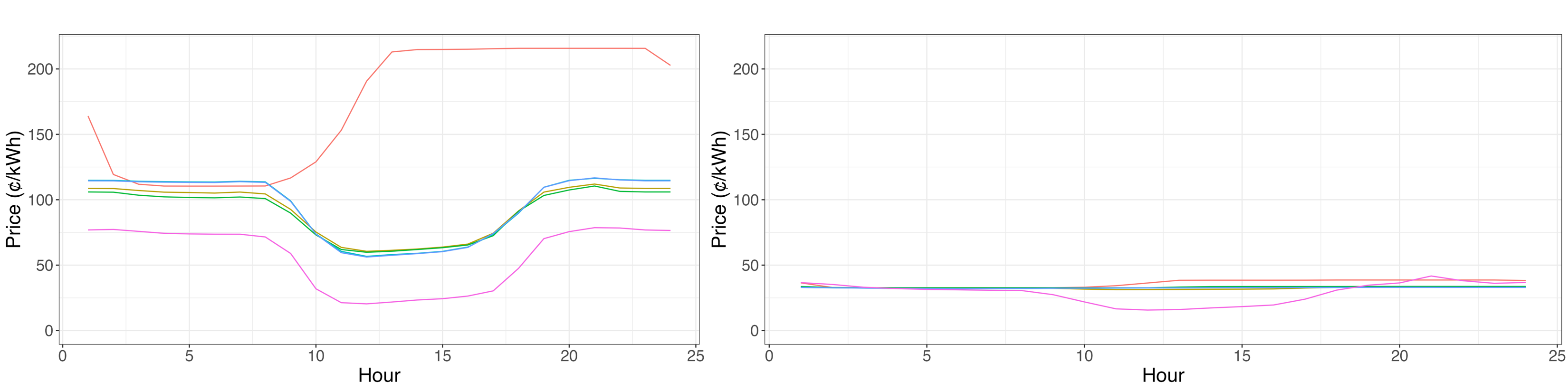}}
            \subfloat[Average energy price by hour, FRCC. Temporal / operational abstractions.\label{price_frcc_time}]{\includegraphics[width=0.5\linewidth]{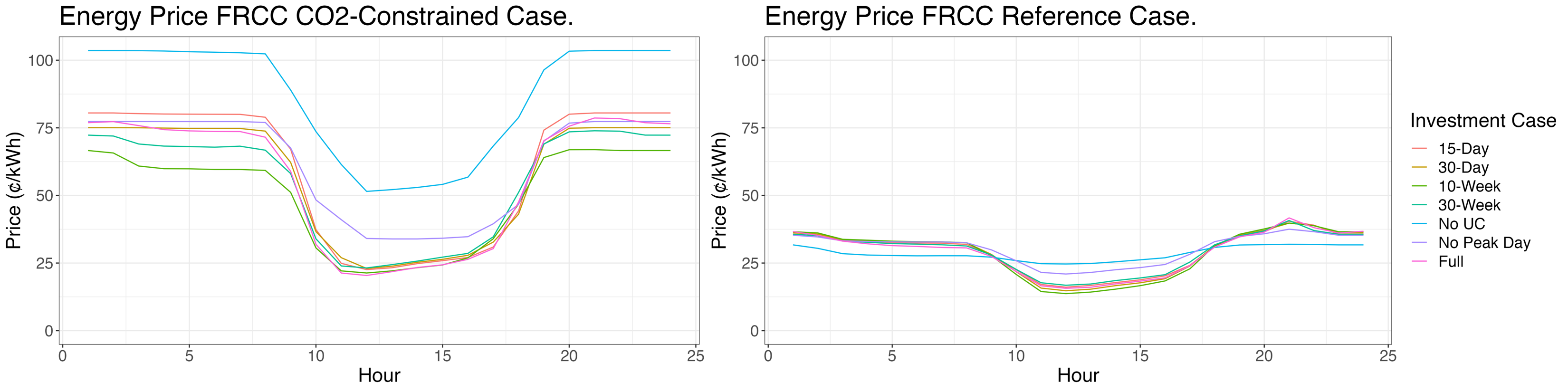}} \\
            \subfloat{\includegraphics[width=0.35\linewidth]{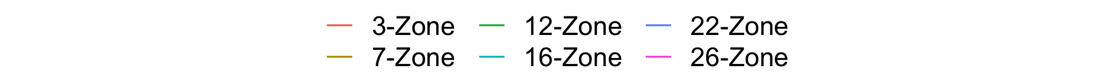}}\\
            \caption{Average energy price by hour across the entire timeseries (time of day) for four selected regions (see figure~\ref{geography}). Data shown for Northeast (\ref{price_isone_zone},\ref{price_isone_time}) northwest (\ref{price_nwpp_zone},\ref{price_nwpp_time}) and Florida (\ref{price_frcc_zone},\ref{price_frcc_time}.) Figure shows that higher granularity models tend to have cheaper energy for consumers. Impact is region-specific and abstraction-specific in strength.\label{price}}
        \end{center}
    \end{figure}  
\end{landscape}

\begin{figure}[H]
    \begin{center}
    \subfloat[Spatial / operational resolutions \label{phase_comparison_spat_si}]{\includegraphics[width=\linewidth]{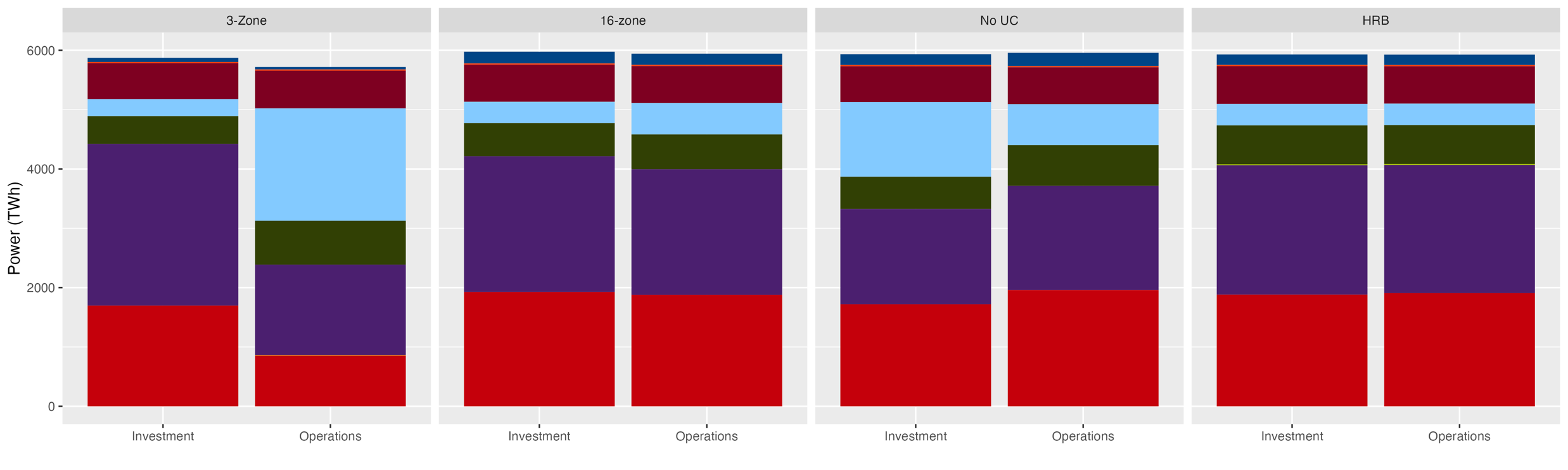}} \\
    \subfloat[Temporal resolutions\label{phase_comparison_temp_si}]{\includegraphics[width=\linewidth]{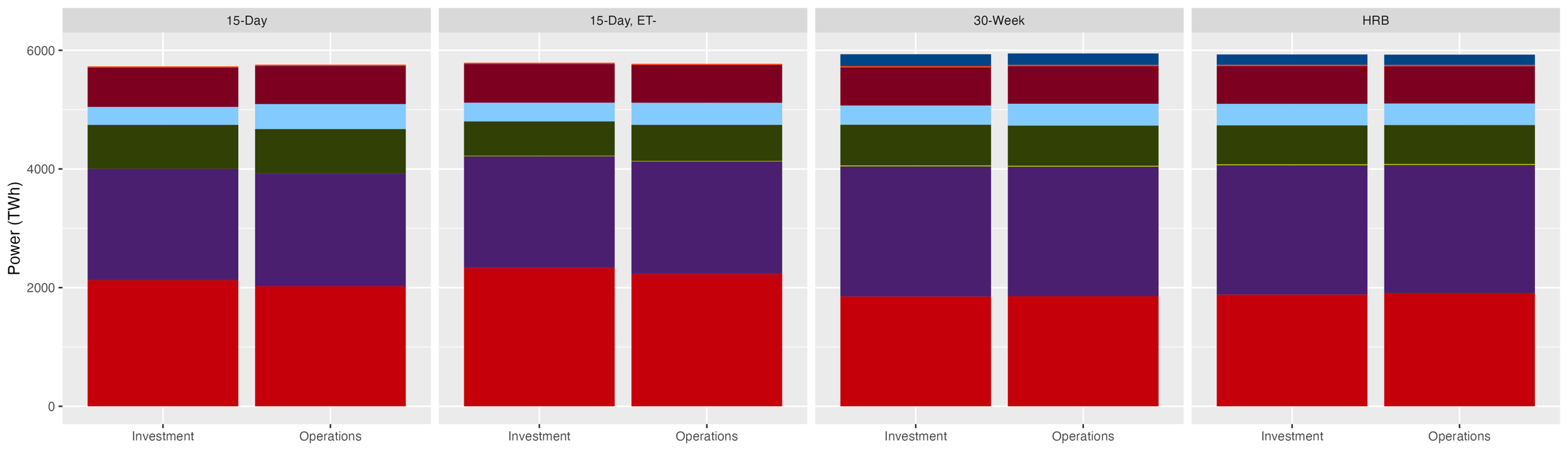}} \\
    \subfloat{\includegraphics[width=0.75\linewidth]{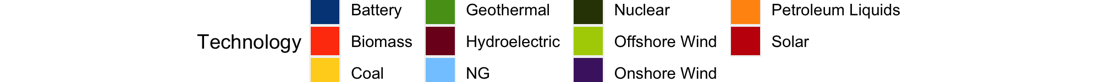}}
    \caption{Generation by technology. Figure compares operations from the \opone~phase from operations with those from the \optwo~phase. For low resolution cases, first phase optimizations (missing information about systems and their physical constraints) are misinformed about how much they can mobilize resources when operating them. Results here are for the \refc case. Results for the \coc case are in the main text. HRB: High resolution baseline. UC: Unit commitment.\label{phase_comparison_si}}
    \end{center}
\end{figure} 

\begin{table}[h]
    \centering
    \setlength\tabcolsep{0pt}
    \begin{tabular}{cM{0.195\linewidth}M{0.195\linewidth}M{0.195\linewidth}M{0.195\linewidth}M{0.195\linewidth}}
        \toprule
        & 15-day & 30-day & 15-week & 30-week & 52-week \\
        
        \midrule
        \multirow{4}{*}{\rotatebox[origin=c]{90}{3-Zone}} &
        {\footnotesize Solar SCO: 13.6\%} &
        {\footnotesize Solar SCO: 13.9\%} &
        {\footnotesize Solar SCO: 7.1\%} &
        {\footnotesize Solar SCO: 13.2\%} &
        {\footnotesize Solar SCO: 14.5\%} \\
        
        &
        {\footnotesize Onshore SCO: 17.1\%} &
        {\footnotesize Onshore SCO: 19.1\%} &
        {\footnotesize Onshore SCO: 14.6\%} &
        {\footnotesize Onshore SCO: 18.5\%} &
        {\footnotesize Onshore SCO: 17.8\%} \\

        &
        {\footnotesize Profit MSE: \$5.7 B}&
        {\footnotesize Profit MSE: \$7.2 B}&
        {\footnotesize Profit MSE: \$7.8 B}&
        {\footnotesize Profit MSE: \$5.8 B}&
        {\footnotesize Profit MSE: \$7.1 B}\\

        &
        {\footnotesize CO$_2$ MSE: 7.3e6 T}&
        {\footnotesize CO$_2$ MSE: 8.2e6 T}&
        {\footnotesize CO$_2$ MSE: 7.7e6 T}&
        {\footnotesize CO$_2$ MSE: 7.4e6 T}&
        {\footnotesize CO$_2$ MSE: 8.0e6 T}\\

        \midrule
        
        \multirow{4}{*}{\rotatebox[origin=c]{90}{16-Zone}} &
        {\footnotesize Solar SCO: 25.8\%} &
        {\footnotesize Solar SCO: 35.9\%} &
        {\footnotesize Solar SCO: 28.2\%} &
        {\footnotesize Solar SCO: 39.3\%} &
        {\footnotesize Solar SCO: 59.1\%} \\
        
        &
        {\footnotesize Onshore SCO: 37.1\%} &
        {\footnotesize Onshore SCO: 37\%} &
        {\footnotesize Onshore SCO: 42.9\%} &
        {\footnotesize Onshore SCO: 46.2\%} &
        {\footnotesize Onshore SCO: 48.2\%} \\

        &
        {\footnotesize Profit MSE: \$1 B}&
        {\footnotesize Profit MSE: \$1.2 B}&
        {\footnotesize Profit MSE: \$1.1 B}&
        {\footnotesize Profit MSE: \$1.1 B}&
        {\footnotesize Profit MSE: \$1 B}\\

        &
        {\footnotesize CO$_2$ MSE: 1.9e6 T}&
        {\footnotesize CO$_2$ MSE: 2.8e6 T}&
        {\footnotesize CO$_2$ MSE: 2.4e6 T}&
        {\footnotesize CO$_2$ MSE: 1.8e6 T}&
        {\footnotesize CO$_2$ MSE: 1.6e6 T}\\

        \midrule

        \multirow{4}{*}{\rotatebox[origin=c]{90}{26-Zone}} &
        {\footnotesize Solar SCO: 67.1\%} &
        {\footnotesize Solar SCO: 61.8\%} &
        {\footnotesize Solar SCO: 57.1\%} &
        {\footnotesize Solar SCO: 64.2\%} &
        {\footnotesize Solar SCO: 100\%} \\
        
        &
        {\footnotesize Onshore SCO: 43.8\%} &
        {\footnotesize Onshore SCO: 63.7\%} &
        {\footnotesize Onshore SCO: 70\%} &
        {\footnotesize Onshore SCO: 85.7\%} &
        {\footnotesize Onshore SCO: 100\%} \\

        &
        {\footnotesize Profit MSE: \$0.3 B}&
        {\footnotesize Profit MSE: \$0.2 B}&
        {\footnotesize Profit MSE: \$0.4 B}&
        {\footnotesize Profit MSE: \$0.2 B}&
        {\footnotesize Profit MSE: \$0 B}\\

        &
        {\footnotesize CO$_2$ MSE: 4.4e5 T}&
        {\footnotesize CO$_2$ MSE: 2.0e5 T}&
        {\footnotesize CO$_2$ MSE: 3.9e5 T}&
        {\footnotesize CO$_2$ MSE: 1.5e5 T}&
        {\footnotesize CO$_2$ MSE: 0.0e0 T}\\
        \bottomrule

    \end{tabular}
    \caption{A summary of some of the error metrics included in section~\ref{results}. When combined with model information in table~\ref{model_table_a}, this table illustrates the magnitude of some of the errors potentially being used in policy relevant modeling. Data includes site capacity overlap (SCO) for VRE ($SCO = 100\% \cdot \frac{cap_{full} \cap cap_{abstracted}}{cap_{full} \cup cap_{abstracted}}$) and per-region mean squared error (MSE) CO$_2$ emissions and generator profit.\label{model_table_b}}
\end{table}

\end{document}